\documentclass[12pt]{article}
\usepackage[top=2cm,bottom=3cm,left=2cm,right=2cm]{geometry}
\usepackage{amssymb} 
\usepackage{amsthm}
\usepackage{hyperref}
\usepackage{wrapfig}
\usepackage{mathtools}
\usepackage[mathscr]{euscript}
\usepackage{relsize}

\numberwithin{equation}{section}
\numberwithin{figure}{section}

\newtheorem{exercise}{Exercise}[section]
\newtheorem{conjecture}{Conjecture}
\newtheorem{claim}{Claim}[section]

\def\eq#1{(\ref{eq:#1})}
\def\lineup{\!\!\!\!\!\!\!\! &&}

\newcommand{\Tr}{\mathop{\rm Tr}\nolimits}

\newcommand{\dash}{\operatorname{\!-\!}}
\def\d{\partial}
\def\eps{\epsilon}

\def\zbar{\overline{z}}
\def\H{\mathcal{H}}
\def\BCFT{\mathrm{BCFT}}
\def\tv{\mathrm{tv}}
\def\sigmabar{\overline{\sigma}}
\def\Sigmabar{\overline{\Sigma}}
\def\bigstar{\mathlarger{\mathlarger{\star}}}

\begin{document}
\begin{titlepage}
\rightline\today

\begin{center}
\vskip 3.5cm

{\large \bf{Four Lectures on Analytic Solutions\\ in Open String Field Theory}}

\vskip 1.0cm

{\large {Theodore Erler\footnote{tchovi@gmail.com}}}

\vskip 1.0cm

{\it CEICO, FZU - Institute of Physics of the Czech Academy of Sciences}\\
{\it Na Slovance 2, 182 21 Prague 8, Czech Republic}\\

\vskip 2.0cm

{\bf Abstract} 

\end{center}

The following notes derive from review lectures on the subject of analytic solutions in open string field theory, given at the School for String Field Theory and String Phenomenology at the Harish-Chandra Research Institute in February 2018. 

\end{titlepage}

\tableofcontents

\section{Preface}

These are notes for four lectures on the topic of analytic solutions in Witten's open bosonic string field theory \cite{Witten} (open SFT). The subject originates in efforts to understand tachyonic instabilities of certain D-brane systems \cite{Sen_BAB,Sen_universality,SenZwiebach}. Here it was realized that open SFT gives the most complete, if not necessarily most accessible, formalism for addressing such questions. It was also suggested that open SFT (perhaps in a supersymmetric version) could provide a nonperturbative and background independent definition of string theory. While numerical approaches will always be indispensable, progress in this direction requires some conceptual and analytic understanding of how the theory encodes nonperturbative physics. At the classical level, this requires understanding how D-brane vacua appear as solutions of the field equations. The major breakthrough was Schnabl's analytic solution describing the endpoint of tachyon condensation~\cite{Schnabl}, which demonstrated the utility of the subalgebra of wedge states with insertions as a controlled setting for exploring the space of nonperturbative vacua. These lectures give a relatively complete account of this technology and the results that have been obtained. We limit the discussion to bosonic strings, though many ideas are applicable to open superstrings. 

The lectures were prepared for students and assume little knowledge beyond the first few chapters of Polchinski~\cite{Polchinski}.  The first lecture is a serviceable introduction to open string field theory in general, though on certain points is optimized for later discussion of analytic solutions. We take a ``modern" approach to the subject from the point of view of path integrals, correlation functions, and noncommutative algebra in the spirit of Okawa \cite{Okawa}. For a dedicated introduction to the physics of tachyon condensation see  \cite{SenRev,TaylorZwiebachRev}.  The thesis of Kudrna \cite{Kudrna} describes the state of the art in the numerical approach to string field theory solutions. Another useful reference on analytic solutions is the review of Okawa \cite{OkawaRev}.  

\subsubsection*{Conventions}

We assume $\alpha'=1$, mostly plus metric, and set the open string field coupling constant to unity. Commutators are always graded with respect to Grassmann parity, and we use the left handed star product convention.

\section{Introduction}

\subsection{First look}

Open bosonic SFT is the field theory of fluctuations of a D-brane in bosonic string theory. The fluctuations of a D-brane are characterized by the open strings which attach to that D-brane.

Consider for example a D$p$-brane in bosonic string theory. An open string attached to this D$p$-brane can mimic an infinite variety of particle states, depending on how the string vibrates. The lowest modes of vibration give a spin $0$ tachyon, a $p+1$-dimensional spin 1 photon, and $25-p$ massless spin $0$ particles, where $25+1=26$ is the dimension of spacetime in bosonic string theory. The higher vibrations give an infinite tower of massive particle states of higher spin. We can easily infer what kind of fields should enter the SFT Lagrangian to create this spectrum of particle states:
\begin{eqnarray}
\phantom{\bigg)}\text{spin 0 tachyon}\ \ \ \ \ \ \ \ \lineup \longrightarrow\ \ \ \ \ \ \ \ \ \ \ \ \ \ \ \ \ \ \ \ \ \ \ \ \text{tachyonic scalar field, } T(x);\nonumber\\
\phantom{\bigg)}\text{spin 1 photon}\ \ \ \ \ \ \ \ \ \lineup\longrightarrow\ \ \ \ \ \ \ \ \, p+1\text{-dimensional gauge field, }A_\mu(x),\ \ \mu=0,1,...,p;\nonumber\\
\phantom{\bigg)}{25-p \text{ massless}\atop\text{spin 0 particles}}\ \ \ \ \ \ \ \ \lineup \longrightarrow\ \ \ \ \ \ \ \ 25-p\text{ massless scalar fields, } \phi_a(x),\ \ a=1,...,25-p;\nonumber\\
\phantom{\bigg)}\vdots\ \ \ \ \ \ \ \ \ \ \ \ \ \ \ \ \ \ \lineup\ \ \ \ \ \ \ \ \ \ \ \ \ \ \ \ \ \ \ \ \ \ \ \ \ \ \ \ \ \ \ \ \ \ \ \ \ \ \ \ \ \ \ \ \ \ \ \  \vdots\nonumber\\
\phantom{\bigg)}{\text{massive\ particles}\atop\text{of higher spin}}\ \ \ \ \ \ \ \ \lineup \longrightarrow\ \ \ \ \ \ \ \ \ \ \ \ \ \ \ \ \ \ \ \ \ \ \ \ \  \text{higher rank tensor fields}.\nonumber
\end{eqnarray}
Note that the coordinate $x\in\mathbb{R}^{1,p}$ refers to a point on the worldvolume of the D$p$-brane. Since open strings are attached to the D-brane, the fields do not depend on spacetime coordinates away from the D-brane worldvolume. With this we can at least begin to write an action for the fluctuations of the D-brane. We have
\begin{eqnarray}
S\lineup = -\int d^{p+1}x\left(\frac{1}{2}\d^\mu T\d_\mu T-\frac{1}{2}T^2+\frac{1}{4}F^{\mu\nu}F_{\mu\nu}+\frac{1}{2}\d^\mu\phi_a\d_\mu\phi_a+{\text{massive}\atop\text{fields}}\right)+\text{interactions}.\ \ \ \ \ \  
\end{eqnarray}
With knowledge of higher spin Lagrangians it is in principle possible to write down kinetic terms for the massive fields. The form of the interactions, however, is almost impossible to guess. The formulation of interactions depends heavily on the conformal field theory description of the string worldsheet, which we discuss later. In any case, the interactions must be constructed in such a way that the Feynman diagrams of the SFT action compute open string S-matrix elements on the D-brane. The kinetic term defines a propagator; the interactions define a cubic vertex, a quartic vertex, and so on as is necessary to get the right scattering amplitudes. So, for example, the color ordered $4$-string amplitude will be expressed as a sum of an $s$-channel, $t$-channel, and quartic vertex contributions:
\begin{wrapfigure}{l}{1\linewidth}
\centering
\vspace{-.5cm}
\resizebox{5in}{1.2in}{\includegraphics{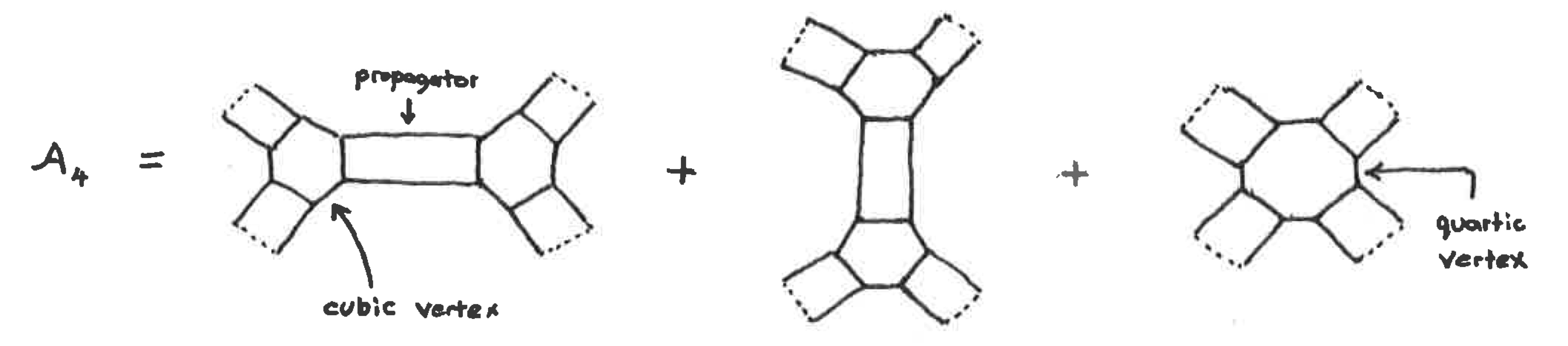}}
\vspace{-.5cm}
\end{wrapfigure}\\ \\ \\ \\ \\ \\ \\ 
This may seem a little uncomfortable. One of the nicest things about string scattering is that each amplitude is represented by a single diagram; the interaction is a global property of the diagram, not a process inside vertices in a part of the diagram. There is nothing inconsistent about this, however. The three diagrams represent integration over different portions of the moduli space of disks with four boundary punctures; the single string diagram we are accustomed to visualizing represents integration over the entire moduli space. It is possible to slice the moduli spaces of Riemann surfaces into Feynman graph components in many ways. Different decompositions correspond to different SFT actions, but since the actions produce the same scattering amplitudes, they should be related by field redefinition. The field redefinition ambiguity is not something special to string theory, but is present in all Lagrangian field theories. The reason you do not hear about it more often is that, for the field theories we are accustomed to dealing with, there is a canonical or ``best possible" formulation of the Lagrangian---or at least a finite number of useful alternatives. From this point of view, one can articulate the discomfort with the string field theory concept as an impression that there is no preferred way to decompose string diagrams into Feynman diagrams, so there should not be a ``natural" formulation of the Lagrangian.  Surprisingly, this impression is incorrect. For open bosonic strings, there is clearly a best possible Lagrangian, and this defines Witten's open bosonic string field theory. There is an analogous Lagrangian for closed bosonic strings based on Riemann surfaces endowed with minimal area metrics \cite{Zwiebach}, but in this case it is not quite as clear that the advantages of the Lagrangian are decisive. For superstrings much less is understood. There is a very nice formulation of open superstring interactions in the NS sector due to Berkovits \cite{Berkovits}. Recently this has been extended to include the Ramond sector by Kunitomo and Okawa \cite{complete}.

\subsection{Classical solutions}

In these lectures we discuss classical solutions in open bosonic SFT. Interest of this subject in large part originates in the problem of background independence in string theory. In string theory, we always start with the action for a relativistic string moving in some background---a spacetime, perhaps with D-branes, fluxes, orbifolds or orientifolds. Quantizing the string action gives a perturbative description of string scattering in that background. {\it A priori}, each background represents a different version of string theory. But the widespread assumption is that all of these string theories are equivalent, in the sense that they represent perturbative expansions of the same fundamental theory around different vacua. One way to see that this is likely is that the spectrum of the string always includes particle states which represent linearized deformations in the choice of background. The most famous example of course is the graviton, which represents linearized deformations in the shape of spacetime. In the context of perturbative string scattering, however, the background cannot change. Understanding the relation between different backgrounds requires something more powerful than the usual formulation of string theory. This is one of the motivations for string field theory. In SFT, string backgrounds are related as classical solutions of an underlying field equation. This is exactly analogous to how, in general relativity, physical spacetimes are represented as solutions of Einstein's equations.

Presently we are concerned with open bosonic strings. So the question is whether different D-brane configurations in bosonic string theory, for a fixed spacetime (or closed string) background, can be described as solutions to the field equations of open bosonic SFT. Describing shifts in the closed string background either requires closed SFT or a much better understanding of quantum effects in open SFT. At present both approaches seem difficult. We will have enough work understanding the open string sector.

Let us return to open bosonic SFT of a D$p$-brane. The configuration where all fluctuation fields of the D$p$-brane vanish represents the D$p$-brane itself. This is called the {\it perturbative vacuum}. If we turn on the gauge field $A_\mu$, we obtain a new background corresponding to a D$p$-brane with nontrivial Maxwell field. If we turn on the massless scalars $\phi_a$, we obtain a new background where the D$p$-brane has been displaced from its initial position. If $x^a,\ a=1,...,25-p$ represent coordinates transverse to the D$p$-brane, and the D-brane is initially located at 
\begin{equation}x^a = 0,\end{equation}
after giving a constant expectation value to $\phi_a$, the new D$p$-brane is located at
\begin{equation}x^a = \frac{1}{\sqrt{T_p}}\phi_a,\end{equation}
assuming that $\phi_a$ is small enough that we can ignore nonlinear corrections to the field 
\begin{wrapfigure}{l}{.23\linewidth}
\centering
\resizebox{1.6in}{1.3in}{\includegraphics{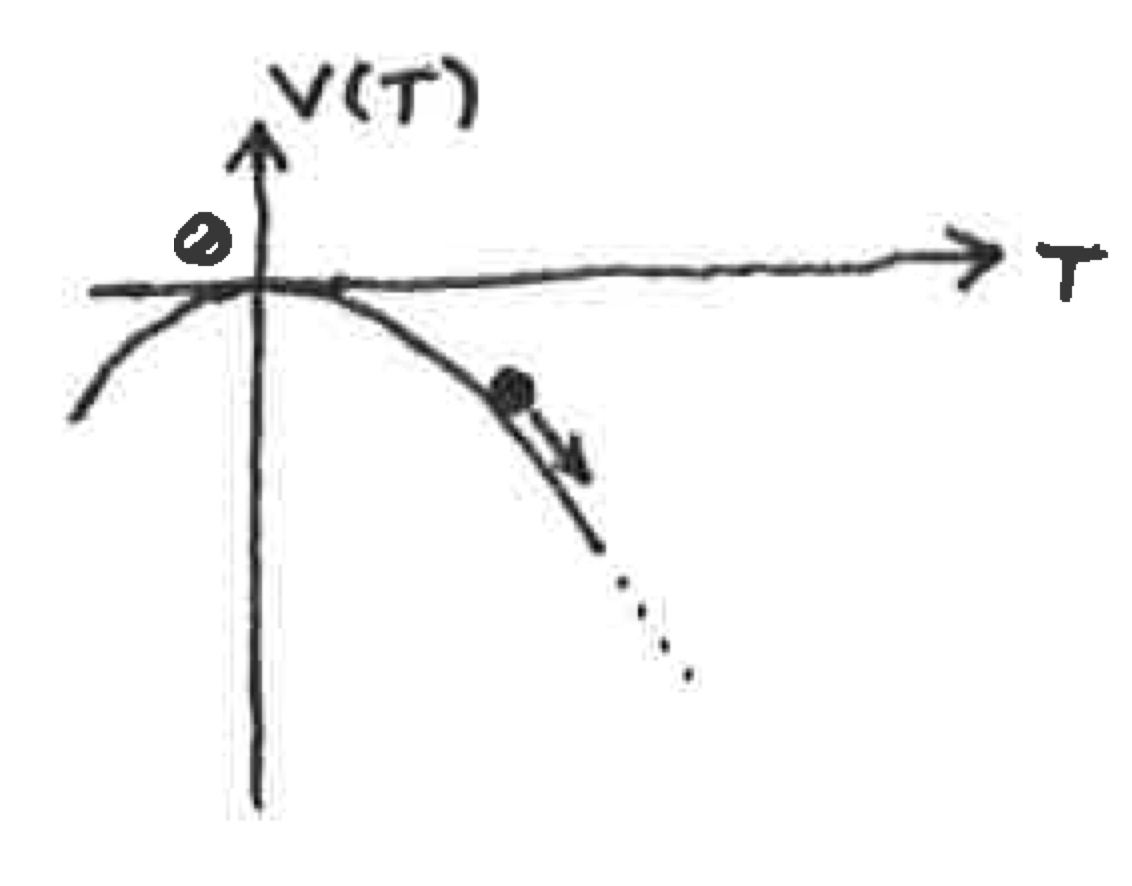}}
\end{wrapfigure}
equations. The number $T_p$ is the tension of the D$p$-brane; with our conventions, it is given by 
\begin{equation}T_p = \frac{1}{2\pi^2}.\end{equation}
See \cite{TaylorZwiebachRev} for a derivation. Finally, we can give an expectation value to the tachyon. Since the 
tachyon field is pulled by an ``upside down" harmonic oscillator potential $V(T) = -\frac{1}{2}T^2+...$, it cannot remain constant. Instead, it will roll down the potential with exponentially increasing expectation value. From this we see that the initial
configuration, where all fluctuations of the D$p$-brane vanish, is unstable. In other words, the D$p$-brane is itself unstable. The fate of this instability is unclear since the tachyon expectation value becomes large and nonlinear terms in the equations of motion dominate; the perturbative description of the D$p$-brane breaks down. This is the  problem  of {\it tachyon condensation}.

A physical understanding of tachyon condensation in open bosonic SFT emerged from work of Sen and others in the early 2000s \cite{Sen_BAB,Sen_universality,SenZwiebach}. The upshot is as follows:\ \ \ \ \ \ \ \ \ \ \ \ \ \ \ \ \ \ \ \ \ \ \ \ \ \ \ \ \ \ \ \ \ \ \ \ \ \ \ \ 
\begin{wrapfigure}{l}{.25\linewidth}
\centering
\resizebox{1.9in}{1.4in}{\includegraphics{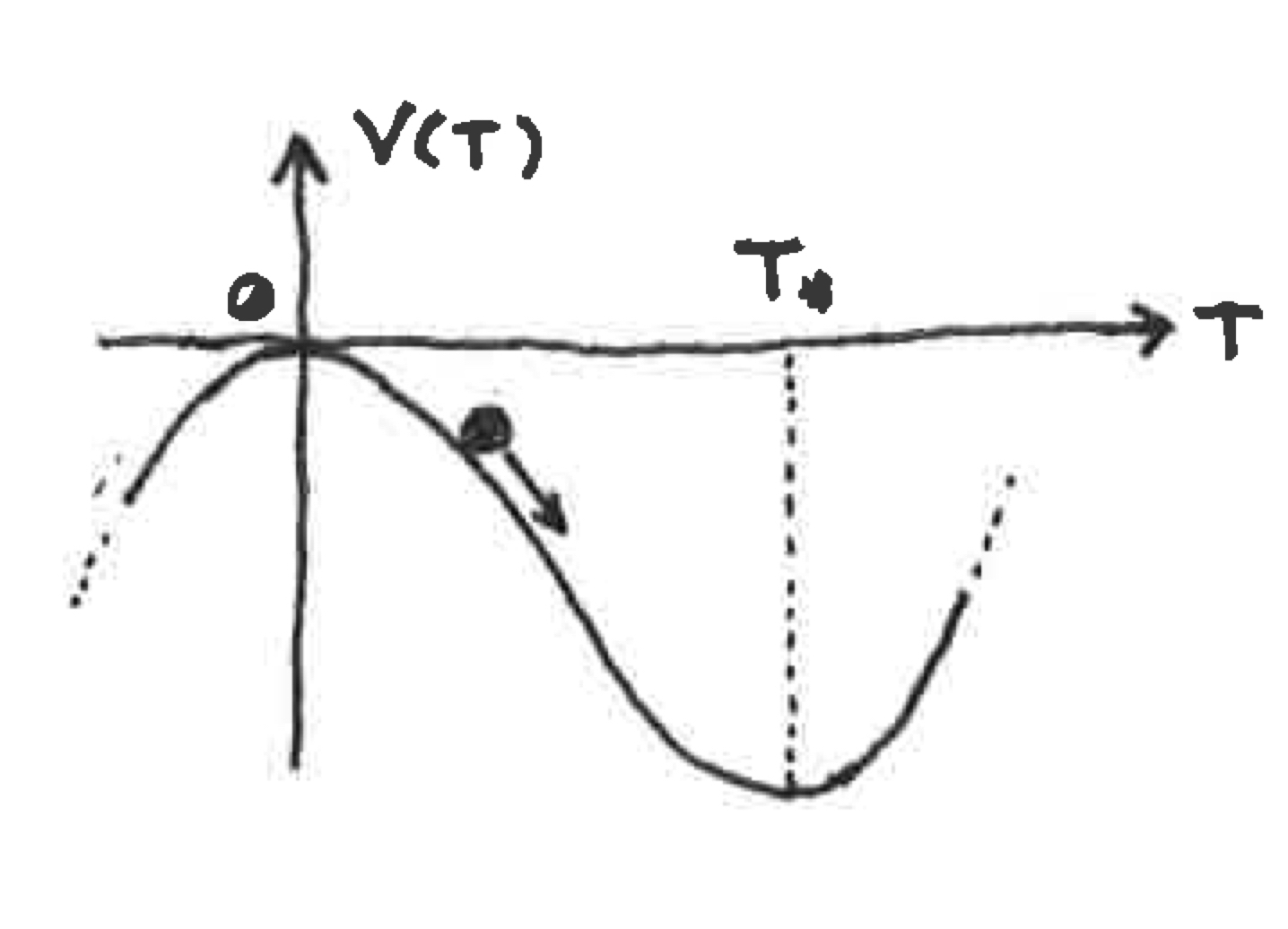}}
\end{wrapfigure}
\begin{itemize}
\item Given the action of open bosonic SFT, one can define a tachyon effective potential $V(T)$ (an energy density) by integrating out all of the massive fields using the equations of motion. The claim is that this potential has a local minimum at $T=T_*$ representing the endpoint of tachyon condensation. The local minimum represents a highly nontrivial solution to the equations of motion of open bosonic SFT, and is called the {\it tachyon vacuum}.
\end{itemize}

\begin{itemize}
\item The tachyon vacuum represents a configuration where the D$p$-brane has disappeared, and we are left with empty space without D-branes or open strings. This has two important consequences:
\begin{description}
\item{(a)} The shift in the potential between the perturbative vacuum and the tachyon vacuum is given by the D$p$-brane tension
\begin{equation}V(0)-V(T_*)=T_p =\frac{1}{2\pi^2}.\end{equation}
In other words, the missing energy density at the tachyon vacuum is precisely accounted for by the fact that the D$p$-brane has disappeared.
\item{(b)} There are no physical excitations around the tachyon vacuum. This reflects the fact that there are no D-branes at the tachyon vacuum, and therefore no open strings. 
\end{description}
\end{itemize}
Points (a) and (b) are specific predictions that can be confirmed by detailed calculations in open bosonic SFT. Similar predictions also exist for unstable D-branes in superstring field theory. Traditionally, these are known as {\it Sen's conjectures}. For the bosonic string, conjecture (a) was effectively proven in 2005 when M. Schnabl found an exact solution for the tachyon vacuum \cite{Schnabl}. A proof of (b) soon followed \cite{EllwoodSchnabl}.

Before Schnabl's result in 2005, the main approach to solving the open SFT equations of motion was {\it level truncation}. The idea is to approximate the action by dropping all fields with mass$^2$ above a fixed integer, and solve the resulting equations of motion numerically. This strategy is still actively pursued \cite{Kudrna}. The success of the level truncation technique forms the empirical basis for the belief that it is meaningful to look for nonperturbative vacua in string field theory. With level truncation it is also possible to construct backgrounds whose worldsheet description is not exactly known. Excepting perhaps perturbative solutions, this is not yet possible with analytic techniques. 

Besides the tachyon vacuum, other classical solutions which have been widely studied include:
\begin{itemize}
\item {\it Marginal deformations.} These solutions correspond to turning on finite expectation values for the massless fields on the D-brane. Such solutions can describe, for example, translations of a D$p$-brane over a finite distance. From the worldsheet perspective, these solutions represent deformation of the worldsheet conformal field theory by an exactly marginal boundary operator. Such solutions have been constructed approximately in level truncation \cite{SenZwiebach_marginal} and analytically soon after Schnabl's result for the tachyon vacuum \cite{KORZ,Schnabl_marg,FKP,KO}.
\item {\it Lump solutions.} Given a scalar field with a potential containing local maxima and minima, it is possible to construct solitonic solutions in the form of ``kinks" or ``lumps." The same is true for the tachyon in open bosonic SFT. In this case, lumps of the tachyon field are believed to describe lower dimensional D-branes in the field theory of a higher dimensional D-brane. From a worldsheet point of view, such solutions represent the infrared fixed point of a renormalization group flow given by perturbing the worldsheet conformal field theory by a relevant boundary operator. Lump solutions were constructed in level truncation in the early 2000s \cite{MoellerSenZwiebach}, but an analytic solution appeared only recently \cite{KOSsing}.
\end{itemize}
While lumps and marginal deformations cover a large class of interesting solutions, there are many open string backgrounds that cannot be described in this way. For example, starting from the fluctuations of a D$p$-brane, can we describe the formation of a D$(p+1)$-brane? If the transverse dimensions are large, the D$(p+1)$-brane will have higher energy than the D$p$. Therefore, such a configuration must be generated by giving expectation values to the massive modes of the string. It is hard to tell, however, which among the infinite number of massive fields should be the most important for this purpose. From the worldsheet point of view, we would need to perturb the worldsheet theory by some combination of irrelevant boundary operators, and try to run the renormalization group flow ``backwards" to reconstruct the ultraviolet fixed point. Needless to say, this seems difficult. The construction of higher energy vacua has become one of the major outstanding problems in the subject since the formulation of Sen's conjectures.

These questions point to a broad generalization of Sen's conjectures: 
\begin{conjecture} {\bf Background Independence.} Open string field theory on a reference D-brane supports classical solutions representing all D-brane systems which share the same closed string background. Moreover, any two D-brane systems which share the same closed string background define string field theories which are related by field redefinition. 
\end{conjecture}
\begin{wrapfigure}{l}{.35\linewidth}
\centering
\resizebox{2.4in}{1.7in}{\includegraphics{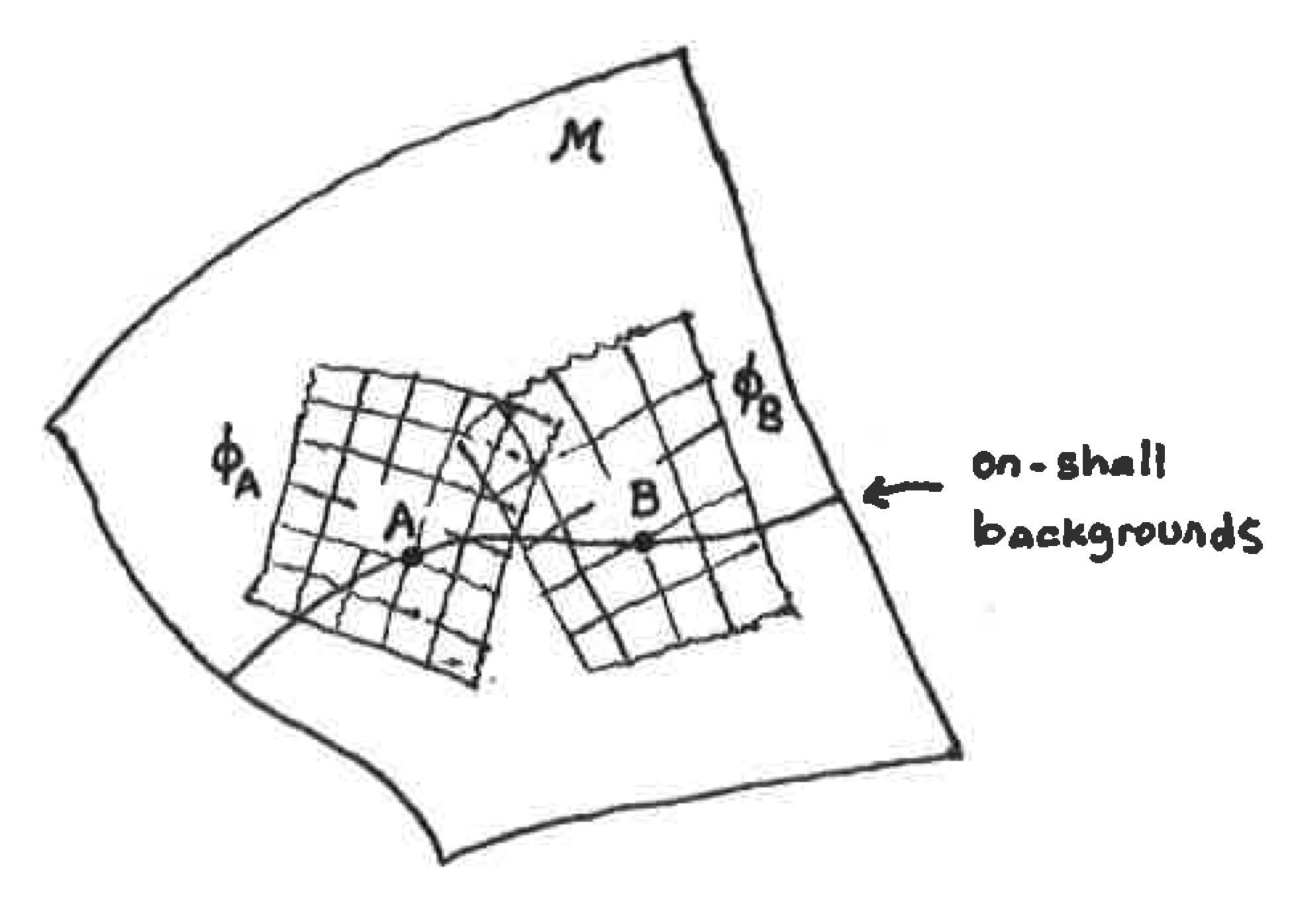}}
\end{wrapfigure}
\noindent Imagine a manifold $\mathcal{M}$ representing the space of on-shell and off-shell configurations of 
classical open string theory. Embedded in $\mathcal{M}$ is a subspace representing the on-shell configurations, i.e. the set of D-brane vacua. Each D-brane vacuum comes with a natural set of fluctuation fields which define a coordinate system on $\mathcal{M}$ in the vicinity of that vacuum. The above conjecture posits that each local coordinate system defined in this way extends to cover all of $\mathcal{M}$. Furthermore, let $\phi^{(A)},\phi^{(B)}$ represent fluctuation fields around backgrounds labeled $A$ and $B$, respectively. Then there should be a coordinate transformation, 
\begin{equation}\phi^{(A)} = f^{(AB)}[\phi^{(B)}],\end{equation}
which relates the actions of the backgrounds $A$ and $B$:
\begin{equation}S^{(A)}[\phi^{(A)}]=S^{(A)}\Big[f^{(AB)}[\phi^{(B)}]\Big] = S^{(B)}[\phi^{(B)}]+\text{constant}.\end{equation}
Logically it is possible that the fluctuation fields of a given D-brane can only rearrange themselves into other configurations that are sufficiently ``close by."  That is, the local coordinate system of each background may only extend a limited distance from the origin, and to cover the whole configuration space we would have to work in patches. Numerical results in level truncation appear consistent with this scenario. Solutions representing higher energy vacua, very large marginal deformations of the perturbative vacuum, or multiple coincident D-branes were for many years not found. Recent improvements in computational power and efficiency have lead to discovery of solutions with at least somewhat higher energy \cite{Kudrna,Rapcak}, and there are hints that the complete moduli space of marginal deformations can be reproduced \cite{MaccaferriMatjej}. But without a major breakthrough it seems unlikely that the set of vacua seen in level truncation can be expanded much further. The situation in the analytic approach seems much better. Recent results indicate that, in a fixed closed string background, all D-brane vacua of bosonic string theory can be represented as classical solutions in the SFT of a reference D-brane.  However, the relation between field variables of different D-brane systems may be more subtle than a strict isomorphism~\cite{KOSsingII}. 

We mention one more conjecture. Following studies of open string tachyon condensation in the early 2000s, it has become customary to assume that open SFT solutions represent D-branes. A D-brane is defined by a choice of boundary conditions for the worldsheet  fields  in the 2-dimensional conformal field  theory of an open string worldsheet. With the boundary conditions specified, the open string worldsheet theory becomes what is called  a {\it boundary conformal field theory} (BCFT).  So the conjecture can be  stated as  follows::
\begin{conjecture} {\bf SFT/BCFT correspondence:} Every classical solution in open string field theory represents a boundary conformal field theory of an open string attached to a D-brane. 
\end{conjecture}
\noindent The circumstantial evidence in favor of this conjecture is very strong, but the conceptual understanding is poor. In \cite{boundary1},  and later  in \cite{boundary2} by a different method, it was shown that every classical solution in open SFT can be associated  to a  closed   string state which, in  known cases, is  equal to  the so-called {\it boundary state} of the BCFT represented by the solution. The boundary state is a fundamental object in BCFT, and in string theory represents how a D-brane is seen when probed with an off-shell closed  string. However,  it has not been proven that the closed string states of \cite{boundary1,boundary2} are necessarily boundary states. In fact it is not fully clear what defines an acceptable boundary state, but there are a number of consistency  conditions, such from modular invariance, which  in principle may be  shown to follow from the open SFT equations of  motion. This remains to be seen.

\section{Lecture 1: Open string field theory}

\subsection{Conformal field theory} 

We begin by reviewing some facts about two dimensional conformal field theory. The following is not really intended as an introduction to the subject. The purpose is to touch on certain points which in a general treatment may not be useful to emphasize, but in the context of open SFT are important.  See for example \cite{Polchinski,Ginsparg,Francesco} for dedicated introduction. Also \cite{Kudrna} has a good discussion of more advanced aspects that are important in open SFT.

A background of string theory is characterized by a worldsheet conformal field theory---for open strings, specifically a boundary conformal field theory (BCFT). A BCFT is a conformal field theory on a 2-dimensional manifold $\Sigma$ which is topologically a disk. The boundary of $\Sigma$ maps to the worldlines swept out by the endpoints of an open string attached to a D-brane; the interior of $\Sigma$ maps to the worldsheet swept out by the interior 
of the open string in spacetime. Since all disks are conformally equivalent, without loss of generality we can formulate BCFT on the upper half 
\begin{wrapfigure}{l}{.28\linewidth}
\centering
\resizebox{2in}{1.4in}{\includegraphics{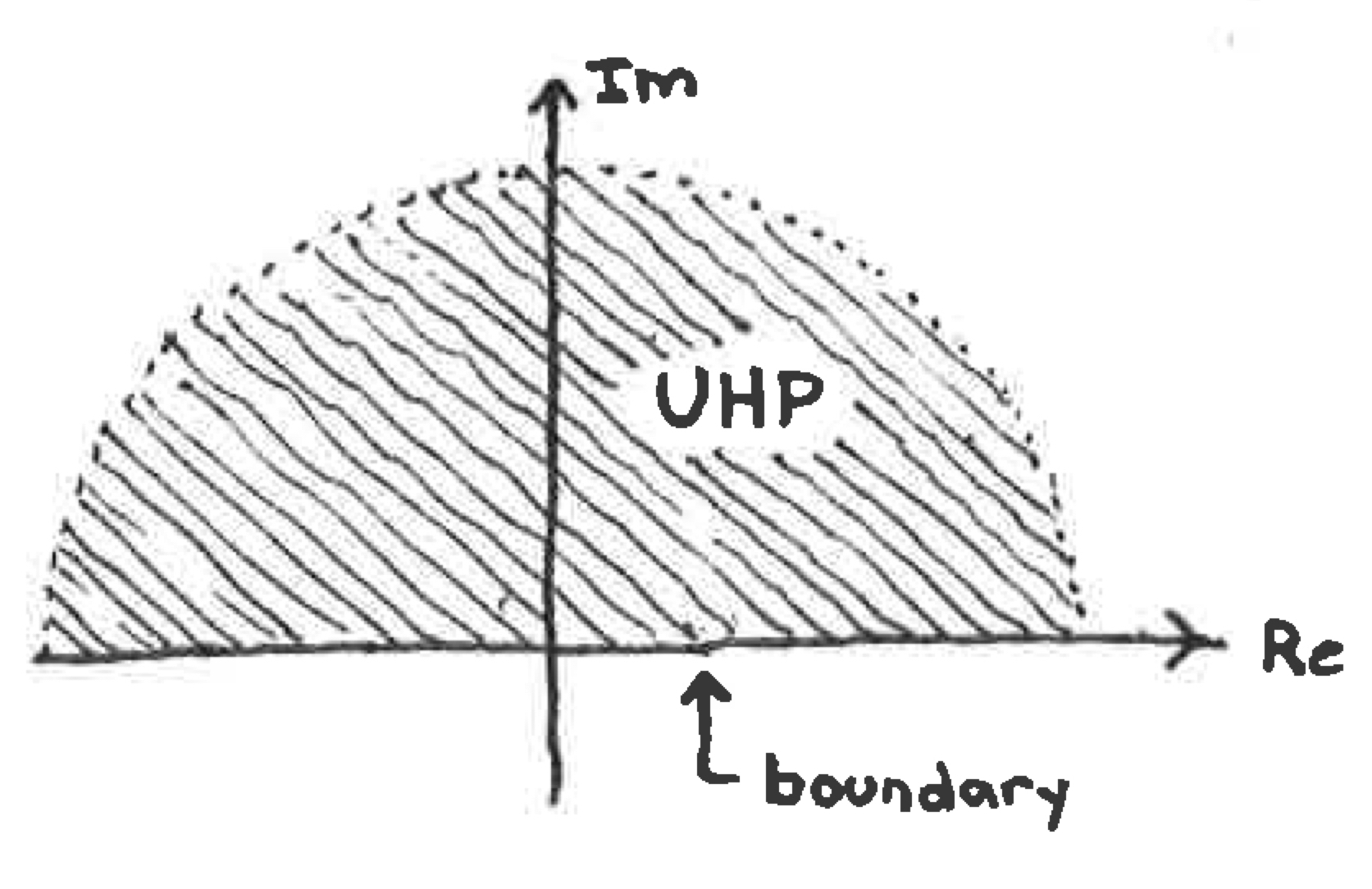}}
\end{wrapfigure}
plane (UHP):
\begin{equation}\text{UHP}:\ \ \ z\in\mathbb{C}\cup\{\infty\},\ \ \ \mathrm{Im}(z)\geq 0.\end{equation}
The real axis is the boundary, and including the point at $\infty$ is topologically a circle. A BCFT comes with two kinds of local operators: {\it bulk} operators $\mathcal{O}(z,\zbar)$ which can be inserted in the interior of the UHP, and {\it boundary} operators $\mathcal{O}(x)$ which can be inserted on the real axis. Generally, the two kinds of operators are different. Correlation functions of bulk operators $\mathcal{O}(z,\zbar)$ diverge as $(z,\zbar)$ approaches the real axis, and correlation functions of boundary operators $\mathcal{O}(x)$ do not have a natural analytic continuation for $x$ not real. An important conceptual point is that the set of local operators of a quantum field theory represents the space of possible local deformations of the theory. In our case, given a bulk operator we can deform the worldsheet action by adding a term $\int_\text{UHP}d^2 z\,\mathcal{O}(z,\zbar)$; given a boundary operator we can deform the worldsheet action with a boundary coupling $\int_{-\infty}^\infty dx\,\mathcal{O}(x)$. Generally, such deformations will not preserve conformal invariance and therefore will not define a string background. You can think of such deformations as creating a hypothetical background which does not satisfy the equations of motion---an ``off-shell" configuration of string theory. To leading order, conformal invariance requires that $\mathcal{O}(z,\zbar)$ is a bulk primary field of weight $(1,1)$, and $\mathcal{O}(x)$ is a boundary primary of weight $1$. In this case, the operators generate what is known as a ``marginal deformation" of the BCFT. From the SFT perspective, such deformations correspond to giving expectation values for the massless fields on the background. Note that bulk operators deform the background as seen by the interior of the string. These are deformations of the closed string background. Boundary operators deform  the background as seen from the endpoints of the open string. These are deformations of the D-brane system in a fixed closed string background. At least classically, open  SFT describes the later deformations, but not the former.

A point in the UHP can be described by two real numbers $(x,y)$ with $y\geq 0$. These may be represented as a holomorphic combination $z = x+i y$ or as an antiholomorphic combination $\bar{z} = x-i y$. Note that $\zbar$ describes the same point $(x,y)$ as $z$ does. In particular $\zbar$ should be interpreted as a point in the upper half plane.  Often we are interested in correlation functions of purely holomorphic or antiholomorphic operators on the UHP. Consider a holomorphic operator $\phi(z)$, satisfying $\overline{\d}\phi(z)=0$. Since a correlation function
\begin{equation}\langle \phi(z)\,...\,\rangle_\text{UHP}\end{equation}
is holomorphic in $z$, generally it can be analytically continued to the lower half plane with $\mathrm{Im}(z)<0$. Now we also have a corresponding correlation function with the anti-holomorphic operator $\overline{\phi}(\zbar)$ satisfying $\d\overline{\phi}(\zbar)=0$, which can also be analytically continued to the lower half plane. Typically, on the real axis there is a relation between $\phi$ and $\overline{\phi}$, called a {\it gluing condition}. In the simplest case, $\phi(x)=\overline{\phi}(x)$, which holds for example for the energy-momentum tensor. In this case we can conclude that 
\begin{equation}\langle \overline{\phi}(\zbar)\,...\,\rangle_\text{UHP} = \left.\langle\phi(z)\,...\,\rangle_\text{UHP}\right|_{z\to z^*}.\end{equation}
The left hand side is a correlation function of an anti-holomorphic operator on the UHP, and on the right is a correlation function of a holomorphic operator which has been analytically continued from the UHP to the lower half plane and evaluated at the point $z^*$,  which is the complex conjugate of $z$. In this way, we represent the UHP with a holomorphic copy of the entire plane; we cut our work in half by discussing holomorphic operators on the entire plane instead of holomorphic and anti-holomorphic operators on the UHP. This is called the {\it doubling trick}. However, when working with correlation functions of operators which are neither holomorphic nor anti-holomorphic, it may be more convenient to stick to the UHP visualization.

For some purposes it is useful to describe BCFT in a state/operator formalism. The relation to correlation functions is given by
\begin{equation}\langle 0|\mathcal{O}_1(z_1,\zbar_1)\,...\,\mathcal{O}_n(z_n,\zbar_n)|0\rangle =  \langle\mathcal{O}_1(z_1,\zbar_1)\,...\,\mathcal{O}_n(z_n,\zbar_n) \rangle_\text{UHP},\ \ \ \ \ \infty>|z_1|>...>|z_n|>0.\label{eq:op_corr}
\end{equation}
On the right hand side is a BCFT correlation function on the UHP. On the left hand side, $|0\rangle$ is a special state of the BCFT called the $SL(2,\mathbb{R})$ {\it vacuum}, and $\mathcal{O}_1(z_1,\zbar_1)...\mathcal{O}_n(z_n,\zbar_n)$ are interpreted as operators, in the sense of the canonical formalism, acting on the state space $\H$ of the BCFT. The operators on the left hand side are ordered from left to right in sequence of decreasing distance to the origin (radial ordering). The $SL(2,\mathbb{R})$ vacuum is called this way since it is invariant under the $SL(2,\mathbb{R})$ subalgebra of the Virasoro algebra:
\begin{equation}[L_1,L_0]=L_1,\ \ \ [L_1,L_{-1}]=2L_0,\ \ \ [L_{-1},L_0]=-L_{-1},\end{equation}
where $L_n$ are the Virasoro operators, appearing in the mode expansion of the energy-momentum tensor:
\begin{equation}
T(z) = \sum_{n\in\mathbb{Z}}\frac{L_n}{z^{n+2}},\ \ \ \ \ L_n = \oint_0\frac{dz}{2\pi i}z^{n+1}T(z).
\end{equation}
The part of the contour in the lower half plane represents $\overline{T}(\zbar)$ via the doubling trick. Using \eq{op_corr} it is easy to show that 
\begin{equation}L_{-1}|0\rangle=0,\ \ \ \ L_0|0\rangle=0,\ \ \ \ L_1|0\rangle = 0,\end{equation}
implying invariance under $SL(2,\mathbb{R})$. More generally, let $\phi(z)$ be a holomorphic primary operator of weight $h$ with mode expansion
\begin{equation}
\phi(z) = \sum_{n\in\mathbb{Z}}\frac{\phi_n}{z^{n+h}},\ \ \ \ \ \phi_n = \oint_0\frac{dz}{2\pi i}z^{n+h-1}\phi(z),
\end{equation}
where the index $n$ labeling the modes is chosen so that
\begin{equation}[L_0,\phi_n]=-n\phi_n.\end{equation}
One can show that 
\begin{equation}\phi_n|0\rangle = 0,\ \ \ n>-h.\end{equation}
Usually we think of a state as representing the configuration of a quantum system at $t=0$; in radial quantization, this corresponds to $|z|=1$. It is therefore natural to interpret the vacuum expectation value as an inner product between an ``out" state and an ``in" state:
\begin{equation}
\underbrace{\langle 0|\mathcal{O}_1(z_1,\zbar_1)\,...\,\mathcal{O}_i(z_i,\zbar_i)}_{\langle\text{out}|}\,\underbrace{\mathcal{O}_{i+1}(z_{i+1},\zbar_{i+1})\,...\,\mathcal{O}_n(z_n,\zbar_n)|0\rangle}_{|\text{in}\rangle},\ \ \ |z_i|>1,\ \ |z_{i+1}|<1,
\label{eq:inout}\end{equation}
where the ``out" state contains operators with $|z|>1$ and the ``in" state contains operators with $|z|<1$. The vector space of ``in" states defines the state space $\H$ of the BCFT.

Given a state $|A\rangle\in\H$ we can define a corresponding boundary operator $V_A(0)$, called the {\it vertex operator}, so that 
\begin{equation}|A\rangle = V_A(0)|0\rangle.\end{equation}
\begin{wrapfigure}{l}{.2\linewidth}
\centering
\resizebox{1.5in}{1in}{\includegraphics{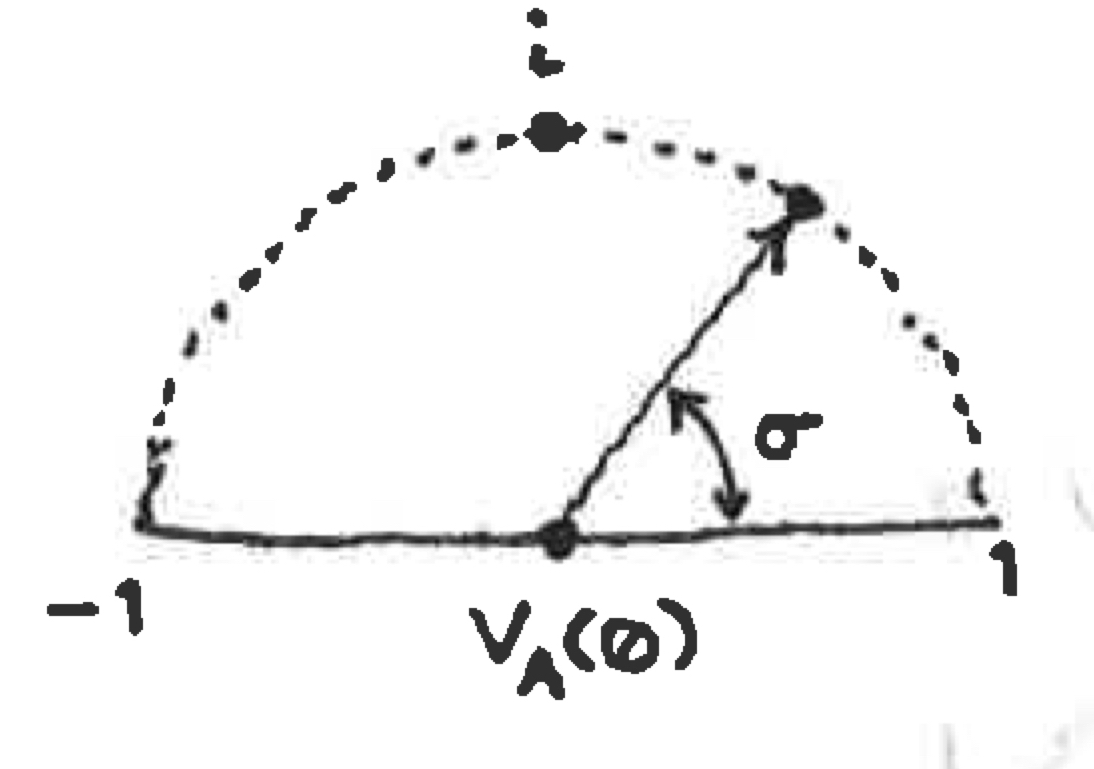}}
\end{wrapfigure}
We can visualize this state as a portion of the UHP consisting of the unit half-disk $|z|<1$ with the vertex operator $V_A(0)$ inserted at the origin. The unit half-circle at the boundary of the half-disk can be parameterized by an angle $\sigma\in[0,\pi]$. The $SL(2,\mathbb{R})$ vacuum is a half-disk without a vertex operator; equivalently, it is the half-disk with an insertion of the identity operator. Given a dual state $\langle A|\in\H^\bigstar$, we can define a corresponding boundary vertex operator at infinity so that
\begin{equation}\langle A| = \langle 0|V_A(\infty).\end{equation}
We can visualize this as a portion of the UHP with the unit half-disk removed and $V_A(\infty)$ inserted at infinity. The unit half-circle at the boundary of this region can be parameterized by an angle $\sigma\in[0,\pi]$. To compute the overlap $\langle A| B\rangle$, we glue the surface of $\langle A|$ to the surface of $| B\rangle$ so that the angle $\sigma$ along the half-circle of $| B\rangle$ is identified with the angle $\sigma$ along the half-circle of $\langle A|$. This effectively glues the unit half-disk to its complement so as to form the entire UHP. The overlap is then given by $\langle A| B\rangle = \langle V_A(\infty)V_ B(0)\rangle_\text{UHP}$. 

\begin{wrapfigure}{l}{.35\linewidth}
\centering
\resizebox{2.5in}{1in}{\includegraphics{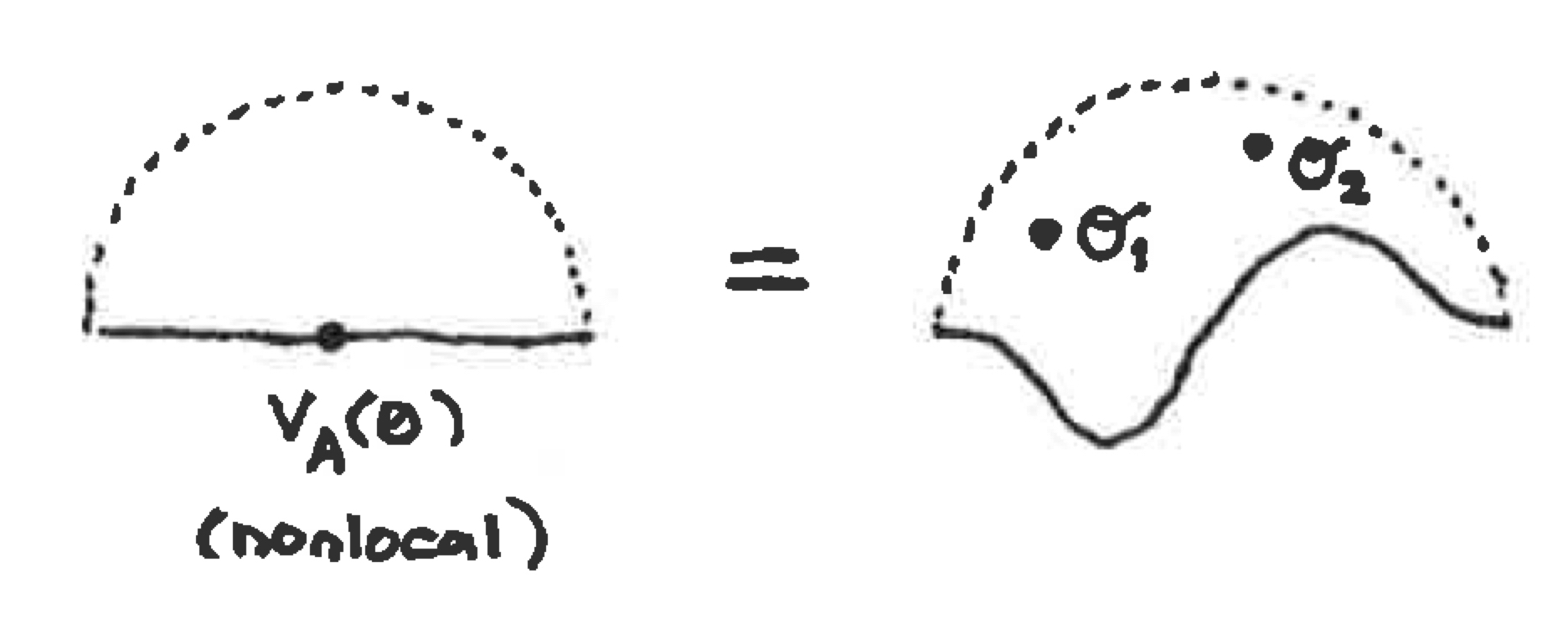}}
\end{wrapfigure}
Conventionally, a vertex operator $V_A(0)$ is imagined as a local operator inserted at the origin of the half-disk. This is the right picture for a so-called {\it Fock space state}---a state built as a linear combination of states whose {\it level} is bounded from above. {\it Level} refers to the $L_0$ eigenvalue, possibly with a shift so that the relevant ground state(s) of the system have level $0$. In string field theory we will need to think about states whose level is not bounded from above, and vertex operators will often be nonlocal. For example, this occurs for states carrying operators displaced from the origin of the half-disk, as  in \eq{inout}. Vertex operators may also contain an infinite number of insertions of the energy-momentum tensor which have the cumulative effect of changing the shape of the half-disk. The assumption, however, is that we can construct a basis for $\H$ using states of definite level whose vertex operators are local at the origin. This is often called a {\it Fock space basis}, and the expression of a state in this basis is called the {\it Fock space expansion}. A more or less equivalent term (though with slightly different emphasis) is {\it level expansion}. Nonlocal vertex operators can appear if we allow infinite sums of states in a  Fock space basis. To give an example which illustrates the essential point, consider a delta function $\delta(x)$, which has support at $x=0$. All derivatives of the delta function also have support only at $x=0$, but the infinite sum,
\begin{equation}\sum_{n=0}^\infty \frac{a^n}{n!}\delta^{(n)}(x)=\delta(x+a),\end{equation}
has support at $x=-a$. Though a vertex operator may be nonlocal, it cannot be arbitrarily nonlocal. It must still be localized within the unit half-disk. What this means concretely is that, in the context of a correlator of bulk local operators and $V_A(0)$ on the UHP, the OPE of any pair of bulk local operators will converge inside a circle which extends (at minimum) to the nearest other local operator, to the real axis, or to the unit half-disk, whichever is closest. 

\begin{wrapfigure}{l}{.35\linewidth}
\centering
\resizebox{2.5in}{1.9in}{\includegraphics{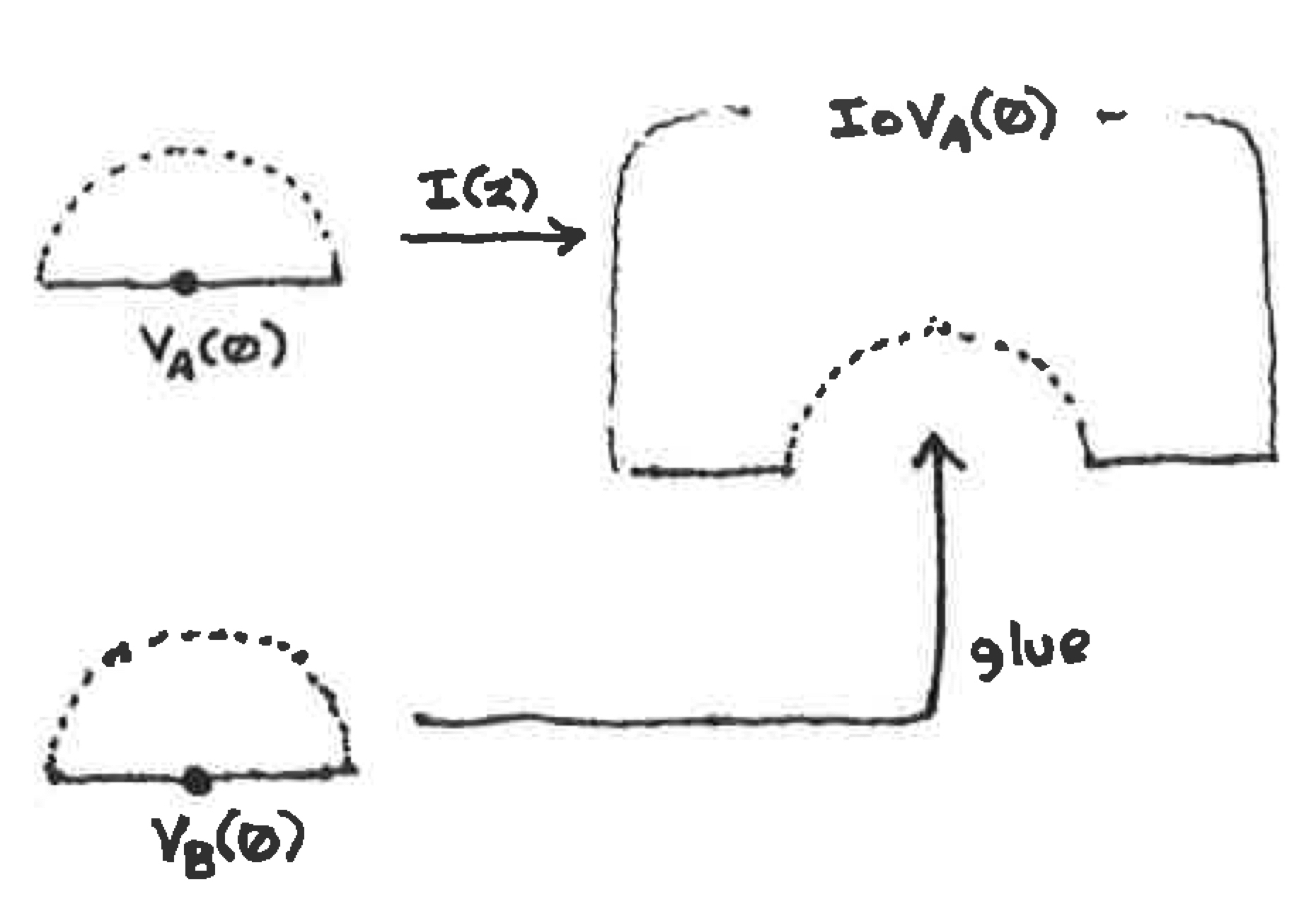}}
\end{wrapfigure}
The interior and exterior of the unit half-disk can be interchanged by a conformal transformation
\begin{equation}I(z)=-\frac{1}{z}.\end{equation}
This allows us to define an isomorphism between states in $\H$ and dual states in $\H^\bigstar$ through a bilinear form, called the {\it BPZ inner
product} (after Belavin, Polyakov, and Zamolodchikov):
\begin{equation}\langle A, B\rangle = \Big\langle \big(I\circ V_A(0)\Big) V_ B(0)\Big\rangle_\text{UHP}, \ \ \ \ \ |A\rangle,| B\rangle\in\H.\end{equation}
The right hand side is a correlation function on the upper half plane of two vertex operators, the first vertex operator having been transformed with $I(z)$. This effectively maps the vertex operator at the origin, which defines a state, into a vertex operator at infinity, which defines a dual state. We use the notation $f\circ...$ to denote conformal transformation of an operator by a map $f(z)$. For a (spinless) boundary primary $\phi(x)$ of weight $h$ and a bulk primary $\phi(z,\zbar)$ of weight $(h,\overline{h})$ the conformal transformation law is
\begin{eqnarray}
f\circ \phi(x) \lineup = \left|\frac{d f(x)}{d x}\right|^h \phi\big(f(x)\big),\\ f\circ\phi(z,\zbar)\lineup =\left(\frac{d f(z)}{dz}\right)^h\left(\overline{\frac{d f(z)}{dz}}\right)^{\overline{h}}\phi\big(f(z),\overline{f(z)}\big).
\end{eqnarray}
Note that because 
\begin{equation}I(e^{i\sigma})= e^{i(\pi-\sigma)},\end{equation}
the BPZ inner product glues the point at an angle $\sigma$ on the unit half-circle of $A$ to the point at an angle $\pi-\sigma$ on the unit half circle of $ B$. The BPZ inner product is graded symmetric: 
\begin{eqnarray}
\langle A, B\rangle \lineup = \Big\langle I\circ\Big(\big(I\circ V_A(0)\big)V_ B(0)\Big)\Big\rangle_\text{UHP}\nonumber\\
\lineup = \big\langle  V_A(0) I\circ V_ B(0)\big\rangle_\text{UHP}\nonumber\\
\lineup = (-1)^{|A|| B|}\big\langle  \big(I\circ V_ B(0)\big)V_A(0)\big\rangle_\text{UHP}\nonumber\\
\lineup =  (-1)^{|A|| B|}\langle B,A\rangle.
\end{eqnarray}
In the first step we used $SL(2,\mathbb{R})$ invariance of UHP correlation functions to transform operator insertions with $I(z)$, and in the second step we used that $I(z)$ composed with itself gives the identity. The sign appears if the vertex operators are anticommuting. The BPZ inner product is also nondegenerate. That is, if $\langle A, B\rangle = 0$ for all $|A\rangle\in \H$, we can conclude $| B\rangle=0$. This follows from the fact that an operator which has vanishing 2-point function with itself and every other operator can be taken to vanish. Note that the BPZ ``inner product" is actually a bilinear form. In the usual meaning, an inner product is a sesquilinear form, in the sense that it takes the complex conjugate of the first argument. The BPZ inner product is different from the usual inner product which defines the Hilbert space of a quantum system, and is a structure specific to the state space of a conformal field theory.

The above allows us to define the notion of {\it BPZ conjugation}. Given an operator $\mathcal{O}$ acting on $\H$, the BPZ conjugate operator $\mathcal{O}^\bigstar$ is defined by
\begin{equation}\langle A,\mathcal{O} B\rangle = (-1)^{|\mathcal{O}||A|}\langle \mathcal{O}^\bigstar A, B\rangle,\end{equation}
where the sign may appear if $\mathcal{O}$ anticommutes with the vertex operator $V_A(0)$. It is  natural  to extend BPZ  conjugation to act on states and dual states. The BPZ conjugate  of a  state $|A\rangle$ is a dual  state written as  $|A\rangle^\bigstar  = \langle A^\bigstar|$. It is defined so  that  
\begin{equation}\langle A^\bigstar|B\rangle  =  \langle A,B\rangle.\end{equation}
The BPZ  conjugate of a dual state $\langle A|$ is  a  state  written as  $\langle  A|^\bigstar =  |A^\bigstar\rangle$. It is defined  so that
\begin{equation}\langle A|B\rangle  =\langle A^\bigstar,B\rangle.\end{equation}
We further extend  BPZ conjugation to  act on scalars by defining $a^\bigstar = a$ for $a\in \mathbb{C}$.  With these definitions  we have the formal properties
\begin{eqnarray}
X^{\bigstar\bigstar} \lineup =  X,\\
(X+Y)^\bigstar  \lineup  = X^\bigstar+Y^\bigstar,\\
(XY)^\bigstar \lineup = (-1)^{|X||Y|}Y^\bigstar X^\bigstar,
\end{eqnarray}
where  $X$ and $Y$  are scalars, operators, vectors or  dual vectors (in the  last equation grammatically composed).  Sometimes operators have definite parity under BPZ conjugation:
\begin{equation}\mathcal{O}^\bigstar = \pm \mathcal{O}.\end{equation}
If we have the plus sign, the operator is called {\it BPZ even}, and the minus sign, {\it BPZ odd}.   The BPZ conjugate of the modes of an $SL(2,\mathbb{R})$ primary field are given by
\begin{equation}\phi_n^\bigstar = (-1)^{n+h}\phi_{-n}.\label{eq:phibpzconj}\end{equation}
The zero mode $\phi_0$ is BPZ even (odd) if the conformal weight is even (odd). For example, $L_0$ is BPZ even.  

Physical string backgrounds should correspond to ``real" solutions of the underlying equations of string theory. How this condition is precisely articulated in string field theory will be explained later, but it starts with an assumption that ``physical" BCFTs are preserved by complex conjugation of correlation functions. This means is that every local operator $\mathcal{O}(z,\overline{z})$ has a ``conjugate" local operator $\mathcal{O}^*(z,\overline{z})$ so that the following relation holds
\begin{equation}
\bigg[\Big\langle \mathcal{O}_1(z_1,\overline{z}_1)\, ...\, \mathcal{O}_n(z_n,\overline{z}_n)\Big\rangle_\text{UHP}\bigg]^* = \Big\langle \mathcal{O}_n^*(-z_n^*,-\overline{z}_n^*)\, ...\, \mathcal{O}_1^*(-z_1,-\overline{z}_1^*)\Big\rangle_\text{UHP}.
\end{equation}
In words, complex conjugation of  a  correlation function of local  operators  produces another correlation function of local  operators (in the  same theory) whose positions  are reflected through the imaginary axis. For  anticommuting  operators, note that the ordering on the right hand side  is reversed. We  further  assume  that  the energy-momentum tensor  is its own conjugate:  
\begin{equation}
T^*(z)=T(z)\label{eq:Treal}
\end{equation}
This structure allows us to define the notion of {\it reality conjugation}. The reality  conjugate of an operator $\mathcal{O}$ will be  written as $\mathcal{O}^\ddag$. For local operators it can be computed as  
\begin{equation}\mathcal{O}(z,\overline{z})^\ddag = \mathcal{O}^*(-z^*,-\overline{z}^*).\end{equation}
The  reality  conjugate  of a state $|A\rangle$ is another state written  as $|A\rangle^\ddag = |A^\ddag\rangle$. It is defined so that
\begin{equation}|A^\ddag\rangle  = V_A^*(0)|0\rangle\ \ \mathrm{if}\ \ |A\rangle = V_A(0)|0\rangle.\end{equation}
The  reality  conjugate  of a dual state $\langle A|$ is another dual state written  as $\langle A|^\ddag = \langle A^\ddag|$. It is defined so that
\begin{equation}\langle A^\ddag| = \langle 0| V_A^*(\infty)\ \ \mathrm{if}\ \ \langle A|  = \langle 0|V_A(\infty).\end{equation}
We extend reality conjugation to act on scalars by defining  $a^\ddag = a^*$ for  $a\in \mathbb{C}$. We have formal properties
\begin{eqnarray}
X^{\ddag\ddag} \lineup =  X,\\
(X+Y)^\ddag  \lineup  = X^\ddag+Y^\ddag,\\
(XY)^\ddag \lineup = (-1)^{|X||Y|}X^\ddag Y^\ddag,
\end{eqnarray}
where  $X$ and $Y$  are scalars, operators, vectors or  dual vectors (in the  last equation grammatically composed). Note that reality conjugation maps states into states, and therefore endows the state space of the BCFT with a real structure.

\begin{exercise} Show that reality conjugation and BPZ  conjugation commute, and moreover $\langle A,B\rangle^*  = \langle B^\ddag,A^\ddag\rangle $.\end{exercise}
\noindent Reality conjugation together with the BPZ conjugation implies  the  more  familiar  notion of {\it Hermitian conjugation}. The Hermitian conjugate of an operator $\mathcal{O}$ will be denoted $\mathcal{O}^\dag$; the Hermitian conjugate of a state $|A\rangle$ is a dual state denoted $|A\rangle^\dag = \langle A^\dag|$; the Hermitian conjugate of a dual state $\langle A|$ is a state denoted $\langle A|^\dag = |A^\dag\rangle$; the Hermitian conjugate of a scalar is $a^\dag  = a^*$ for $a\in \mathbb{C}$.  Hermitian conjugation is defined so that the following relation holds: 
\begin{equation}X^\ddag   = X^{\dag\bigstar}.\end{equation} 
We have formal properties
\begin{eqnarray}
X^{\dag\dag} \lineup =  X,\\
(X+Y)^\dag  \lineup  = X^\dag+Y^\dag,\\
(XY)^\dag \lineup = Y^\dag X^\dag,
\end{eqnarray}
where  $X$ and $Y$  are scalars, operators, vectors or  dual vectors (in the  last equation grammatically composed). With this we can define a sesquilinear form on the BCFT state space,
\begin{equation}\langle A^\dag|B\rangle^* = \langle B^\dag|A\rangle,\end{equation}
which, in comparison to the BPZ inner product, is more analogous the inner product of quantum mechanics. This does not imply that the BCFT state space is a Hilbert space, since the sesquilinear form is not necessarily positive definite. In fact, it is not positive definite for the matter/ghost CFTs which appear in covariant string field theory. 
\begin{exercise}Using \eq{Treal}, show that Hermitian conjugation negates mode number of the Virasoro generators, i.e. $L_n^\dag = L_{-n}$.\end{exercise}

\subsection{Worldsheet theory of open bosonic string}
\label{subsec:openBCFT}

We  now describe how the worldsheet theory of an open bosonic string is realized as a BCFT. We will mostly just state the results without derivation. For more, see any modern string theory textbook, for example \cite{Polchinski}.

The worldsheet theory of an open bosonic string is a tensor product of ``matter" and ``ghost" BCFTs:
\begin{equation}\text{BCFT}= \text{BCFT}^\text{m}\otimes\text{BCFT}^\text{gh}.\end{equation}
We will use superscripts $\text{m}$ and $\text{gh}$ to denote objects defined in the matter and ghost BCFTs respectively. The ghost factor is a $bc$ system with central charge $-26$. It is characterized by anticommuting, holomorphic worldsheet fields $b(z),c(z)$, with antiholomorphic counterparts we can account for with the doubling trick, satisfying
\begin{eqnarray}
b(z)\lineup = \text{primary of dimension } (2,0),\\
c(z)\lineup = \text{primary of dimension }(-1,0),\\
b(x)\lineup = \overline{b}(x),\ \ \ \ c(x)=\overline{c}(x),\ \ \ \ x\in\mathbb{R},\\
b(z)c(w)\lineup = \frac{1}{z-w}+\text{regular}\ .\label{eq:bcOPE}
\end{eqnarray}
The ghost factor of the BCFT is the same for all backgrounds of the open bosonic string. The information about the background is contained in the matter BCFT. The main condition on the matter BCFT is that it has central charge $+26$, so the total matter/ghost BCFT has central charge $+26-26=0$. Additionally, the matter BCFT should be chosen so that the Hermitian inner product of on-shell states is positive definite, as we will explain shortly. Our main example of an open string background will be a D$p$-brane in flat space, by which we mean no orbifolds, orientifolds, compactification, or background fields other than the flat metric. In this case the matter BCFT consists of $p+1$ free bosons $X^\mu(z,\zbar),\ \mu=0,...,p$ subject to Neumann boundary conditions, and $25-p$ free bosons $X^a(z,\zbar),\ a=1,...,25-p$ subject to Dirichlet boundary conditions:
\begin{eqnarray}
\d X^\mu(z)\lineup = \text{primary of dimension }(1,0),\\
\d X^a(z)\lineup = \text{primary of dimension }(1,0),\\
\d X^\mu(x)\lineup =\overline{\d}X^\mu(x),\ \ \ \ \ \ x\in\mathbb{R}\ \ \ \ \text{(Neumann b.c.)},\\
\d X^a(x)\lineup =-\overline{\d}X^a(x),\ \ \ \ x\in\mathbb{R}\ \ \ \ \text{(Dirichlet b.c.)},\\
\d X^\mu(z)\d X^\nu(w) \lineup = -\frac{1}{2}\frac{\eta^{\mu\nu}}{(z-w)^2}+\text{regular}\ ,\label{eq:dXmuOPE}\\
\d X^a(z)\d X^b(w) \lineup = -\frac{1}{2}\frac{\delta^{ab}}{(z-w)^2}+\text{regular}\ .\label{eq:dXaOPE}
\end{eqnarray}
We also have antiholomorphic operators $\overline{\d} X^\mu(\zbar)$ and $\overline{\d}X^a(\zbar)$ which we can account for with the doubling trick. Note that in the Dirichlet case the gluing condition for $\d X$ comes with a sign. 

It is useful to list mode expansions of the $bc$ ghosts and free scalars:
\begin{eqnarray}
b(z)\lineup = \sum_{n\in\mathbb{Z}}\frac{b_n}{z^{n+2}},\ \ \ \ \ \ \ \ \ \ \ \ \, b_n|0\rangle = 0\ \ \text{for}\  n\geq -1,\\
c(z)\lineup = \sum_{n\in \mathbb{Z}}\frac{c_n}{z^{n-1}},\ \ \ \ \ \ \ \ \ \ \ \ \, c_n|0\rangle = 0\ \ \text{for}\ n\geq 2,\\
\d X^\mu(z)\lineup = -\frac{i}{\sqrt{2}}\sum_{n\in \mathbb{Z}}\frac{\alpha_n^\mu}{z^{n+1}},\ \ \ \ \ \alpha_n^\mu|0\rangle = 0\ \ \text{for}\ n\geq 0,\\
\d X^a(z)\lineup = -\frac{i}{\sqrt{2}}\sum_{n\in \mathbb{Z}}\frac{\alpha_n^a}{z^{n+1}},\ \ \ \ \ \alpha_n^a|0\rangle = 0\ \ \text{for}\ n\geq 0.
\end{eqnarray}
The normalization in front of the mode expansion for $\d X$ is chosen so that the matter oscillators obey the commutation relations
\begin{equation}[\alpha_m^\mu,\alpha_{-n}^\nu]=m\delta_{mn}\eta^{\mu\nu},\ \ \ \ [\alpha_m^a,\alpha_{-n}^b]=m\delta_{mn}\delta^{ab},\end{equation}
as follows from the OPEs \eq{dXmuOPE} and \eq{dXaOPE}. Moreover,
\begin{equation}[b_n,c_{-n}] =\delta_{mn},\end{equation}
as follows from the OPE \eq{bcOPE}. We use the bracket $[\cdot,\cdot]$ to denote a graded commutator, that is, it  is a commutator between commuting objects and an anticommutator between anticommuting objects. The modes have Hermitian and BPZ conjugation properties
\begin{eqnarray}
b_n^\dag \lineup = b_{-n},\ \ \ \ b_n^\bigstar = (-1)^n b_{-n},\\
c_n^\dag \lineup = c_{-n},\ \ \ \ c_n^\bigstar = (-1)^{n+1} c_{-n},\\
(\alpha^\mu_n)^\dag \lineup = \alpha^\mu_{-n},\ \ \ \ (\alpha^\mu_n)^\bigstar = (-1)^{n+1} \alpha^\mu_{-n},\\
(\alpha^a_n)^\dag \lineup = \alpha^a_{-n},\ \ \ \ (\alpha^a_n)^\bigstar = (-1)^{n+1} \alpha^a_{-n}.
\end{eqnarray}
The operators $\d X^\mu$ and $\d X^a$ are given by holomorphic derivatives of the string embedding coordinates $X^\mu$ and $X^a$, which have mode expansions
\begin{eqnarray}
X^\mu(z,\zbar) \lineup = x^\mu - p^\mu \ln|z|^2+\frac{i}{\sqrt{2}}\sum_{n\in\mathbb{Z}-\{0\}}\frac{\alpha_n^\mu}{n}\left(\frac{1}{z^n}+\frac{1}{\zbar^n}\right),\label{eq:Xmumode}\\
X^a(z,\zbar) \lineup = x^a +\frac{i}{\sqrt{2}}\sum_{n\in\mathbb{Z}-\{0\}}\frac{\alpha_n^a}{n}\left(\frac{1}{z^n}-\frac{1}{\zbar^n}\right),
\end{eqnarray}
We have three new objects here, $p_\mu,x^\mu$ and $x^a$. The operator $p_\mu$ is the momentum of the open string along the Neumann directions of the D-brane, and $x^\mu$ is its conjugate  position:
\begin{equation}[x^\mu,p_\nu] = i\delta^\mu_\nu.\end{equation}
The position zero mode $x^\mu$ is Hermitian. The momentum is related to the zeroth oscillator through
\begin{equation}\alpha_0^\mu = \sqrt{2}p^\mu.\end{equation}
Since $\alpha_0^\mu|0\rangle = 0$, the $SL(2,\mathbb{R})$ vacuum carries zero momentum. To get a state of nonzero momentum $k_\mu$, we can ``translate" the $SL(2,\mathbb{R})$ vacuum in momentum space using the position operator:
\begin{equation}|k_\mu\rangle = e^{ik\cdot x}|0\rangle = e^{i k\cdot X(0,0)}|0\rangle,\label{eq:kmu}\end{equation}
where $e^{ik\cdot X(0,0)}$ is a boundary normal ordered plane wave vertex operator. It is a boundary primary operator of weight $k^2$.
In the Dirichlet directions, $x^a$ is the position of the D$p$ brane and is a number, not an operator. The zeroth oscillator vanishes in the Dirichlet directions
\begin{equation}\alpha_0^a=0.\end{equation}
since the open string does not carry a conserved momentum orthogonal to the D-brane.

The matter/ghost form of the open string BCFT provides additional structure and properties not present for a generic BCFT:
\begin{description}
\item{(1)} Let $T^\text{m}$ and $T^\text{gh}$ be the energy momentum tensors of the matter and ghost factors of the BCFT. Since the central charge of the total BCFT vanishes, the total energy-momentum tensor,
\begin{equation}T(z) = T^\text{m}(z)+T^\text{gh}(z),\end{equation}
is a primary of dimension $(2,0)$. From here on we will always use $T(z)$ to denote the total energy-momentum tensor. Correlation functions are identically conformally invariant
\begin{equation}\langle ...\rangle_\Sigma = \langle f\circ(...)\rangle_{f\circ \Sigma},\end{equation}
where $\Sigma$ is a 2-dimensional Riemann surface with the topology of a disk (not necessarily the UHP), and $f\circ\Sigma$ is the surface obtained after applying the conformal transformation $f$. For a generic BCFT, this equality will hold only up to a (typically singular) proportionality factor generated by the Weyl anomaly \cite{Polchinski}. 
\item{(2)} The set of operators in the theory has a $\mathbb{Z}_2$ grading according to whether they are commuting or anticommuting. Commuting operators are said to be {\it Grassmann even}, anticommuting operators {\it Grassmann odd}, and the $\mathbb{Z}_2$ grading is called {\it Grassmann parity}. The Grassmann parity of an operator $\mathcal{O}$ is denoted $|\mathcal{O}|$.  In addition, the set of operators carries a $\mathbb{Z}$ grading called {\it ghost number}, which counts the number of $c$ minus the number of $b$ factors contained in the operator. Thus
\begin{eqnarray}
\d X^\mu(z) \lineup = \text{Grassmann even, ghost number }0,\\
b(z)\lineup = \text{Grassmann odd, ghost number } -1,\\
c(z)\lineup = \text{Grassmann odd, ghost number }1.
\end{eqnarray}
The space of states $\H$ is also graded by Grassmann parity and ghost number, according to the Grassmann parity and ghost number of the corresponding vertex operators. The Grassmann parity of a state $|A\rangle$ is denoted $|A|$.  For ordinary backgrounds, $b$ and $c$ are the only anticommuting operators of the worldsheet theory, which leads to an identification between Grassmann parity and ghost number
\begin{equation}\text{Grassmann parity} = \text{ghost number mod }\mathbb{Z}_2.\end{equation}
In defining the SFT path integral it is necessary to consider states multiplied by formal anticommuting parameters, and in this context Grassmann parity and ghost number may not be related. In this lecture we are concerned only with the classical theory, and this identification holds. 
\item{(3)} The theory comes with a dimension $(1,0)$ holomorphic primary field called the {\it BRST current}:
\begin{equation}j_B(z)= cT^\text{m}(z)+:bc\d c(z):+\frac{3}{2}\d^2 c(z).\end{equation}
This is Grassmann odd and ghost number 1. There is also an antiholomorphic counterpart $\overline{j}_B(\zbar)$ of dimension $(0,1)$. On the real axis, we have the gluing condition
\begin{equation}j_B(x) = \overline{j}_B(x),\ \ \ \ x\in\mathbb{R},\end{equation}
which allows us to describe the antiholomorphic current with the doubling trick. The zero mode of the BRST current defines the {\it BRST operator}
\begin{equation}Q:\mathcal{H}\to\mathcal{H},\ \ \ \ Q= \oint_0 \frac{dz}{2\pi i}j_B(z).\end{equation}
The BRST operator is nilpotent 
\begin{equation}Q^2=0,\end{equation}
and is Hermitian and BPZ odd:
\begin{equation}Q^\dag = Q,\ \ \ \ Q^\bigstar = -Q.\end{equation}
We also introduce the notion of the {\it BRST variation} of a local operator. For a boundary operator $\mathcal{O}(x)$, the BRST variation is defined
\begin{equation}Q\cdot\mathcal{O}(x) = \oint_x\frac{dz}{2\pi i}j_B(z)\mathcal{O}(x).\end{equation}
where the contour surrounds the point $x$ where the operator is located. The BRST variation of a bulk operator $\mathcal{O}(z,\zbar)$ is defined
\begin{eqnarray}
Q\cdot\mathcal{O}(z,\zbar) =\oint_z\frac{dz'}{2\pi i}j_B(z')\mathcal{O}(z,\zbar) - \oint_{\zbar}\frac{d\zbar'}{2\pi i}\overline{j}_B(\zbar')\mathcal{O}(z,\zbar).
\end{eqnarray}
where the contours surround the point $z$ and $\zbar$ where the operator is located. Since the BRST current is a weight 1 primary operator, the BRST variation is preserved by conformal transformation 
\begin{equation}f\circ\Big(Q\cdot\mathcal{O}\Big) = Q\cdot\Big(f\circ\mathcal{O}\Big).\end{equation}
Computing the BRST variation in general can be complicated, but for most of our computations it will be enough  to know the relations
\begin{eqnarray}
Q\cdot b(z) \lineup = T(z),\\
Q\cdot T(z)\lineup =0,\\
Q\cdot c(z) \lineup = c\d c(z),\\
Q\cdot\mathcal{O}^\text{m}(z)\lineup = c\d \mathcal{O}^\text{m}(z)+h\d c\mathcal{O}^\text{m}(z),
\end{eqnarray}
where $\mathcal{O}^\text{m}(z)$  is a holomorphic matter primary operator of weight $h$.
\item{(4)} A {\it physical state} is a BRST invariant state of the BCFT at ghost number 1:
\begin{equation}\text{physical state:}\ \ \ \ Q|\Psi\rangle = 0,\ \ \ \ |\Psi\rangle = \text{ghost number }1.\end{equation}
A BRST invariant state is equivalently said to be {\it BRST closed}. Physical states are defined to be equivalent if they differ by the BRST variation of a state at ghost number 0:
\begin{equation}
\text{physical equivalence:} \ \ \ \ |\Psi'\rangle = |\Psi\rangle + Q|\Lambda\rangle,\ \ \ \ |\Lambda\rangle = \text{ghost number }0.
\end{equation}
A state which can be written as the BRST variation of something else is trivially BRST closed due to $Q^2=0$. Such a state is called {\it BRST exact}. The space of inequivalent physical states is then defined by the space of BRST closed states modulo the addition of BRST exact states at ghost number 1. This defines the {\it cohomology} of $Q$ at ghost number 1, denoted $H^1(Q)$. Note that the distinction between ``physical" and ``unphysical" states does not originate from the BCFT itself; while the matter/ghost form of the BCFT implies the existence of the BRST operator and an associated cohomology, the interpretation of this cohomology originates elsewhere. Ultimately, it comes from the fact that the BCFT description of the worldsheet theory arises after gauge fixing the reparameterization and Weyl symmetries of the Polyakov action. The statement that physical states of the worldsheet theory should be gauge invariant translates, after gauge fixing, to the statement that physical states of the BCFT should be BRST invariant. The BRST operator has cohomology at other ghost numbers, but they do not represent physical  states of the open string. A basis for the cohomology at each ghost number is given by vertex operators 
\begin{eqnarray}
\text{ghost number }0:\lineup\ \ \ 1,\\
\text{ghost number }1:\lineup\ \ \ c V^\text{m}(0),\label{eq:coh1}\\
\text{ghost number }2:\lineup\ \ \ c\d c V^\text{m}(0),\label{eq:coh2}\\
\text{ghost number }3:\lineup \ \ \ c\d c\d^2 c(0),
\end{eqnarray}
where $V^\text{m}(0)$ is a boundary primary in the matter factor of the BCFT of dimension 1. In total, all of these vertex operators are primaries of weight 0. The cohomology at ghost number $g$ is isomorphic to the cohomology at ghost number $3-g$, in interesting analogy to Poincar\'{e} duality of the de Rham cohomology of differential forms in three dimensions. The cohomology at ghost numbers greater than three or less than zero is trivial, in the sense that all BRST closed states are BRST exact. We can define an inner product on the ghost number 1 cohomology 
\begin{equation}\langle \Psi_1,\Psi_2\rangle_{H^1(Q)} = \mathcal{N}\langle\Psi_1^\dag|c_0|\Psi_2\rangle,\label{eq:prod_coh}\end{equation}
where on the right hand side $|\Psi_1\rangle,|\Psi_2\rangle$ are representatives of the cohomology in $\mathcal{H}$ annihilated by $b_0$. The constant $\mathcal{N}$ is a normalization. For a D$p$-brane in flat space, the normalization needs to be vanishing to cancel a $\delta(0)$ arising from a timelike momentum delta function, which is singular since the momentum of physical states is constrained to the mass shell. The inner product on the cohomology must be positive definite so that physical string states can be interpreted quantum mechanically. 
\item{(5)} The correlation functions of the BCFT are nonvanishing only if the ghost number of all operator insertions adds up to 3. Using Wick's theorem, all correlation functions in the ghost sector can be reduced to a correlator with three $c$-ghost insertions:
\begin{equation}\langle c(z_1)c(z_2)c(z_3)\rangle_\text{UHP}^\text{gh} = (z_1-z_2)(z_1-z_3)(z_2-z_3).\label{eq:3cs}\end{equation}
Correlation functions with $\overline{c}(\zbar)$ are given by the doubling trick.
\end{description}

\subsection{The string field}
\label{subsec:Psi}

We now pass from the first quantized worldsheet theory to the classical field theory of fluctuations of a D-brane. The first step is to specify the nature of the fluctuation fields. It is convenient to consider the set of fluctuation fields together as a single object, called the {\it string field}. We make the following claim:

\begin{quote}
{\it A string field is an element of the state space $\H$ of the worldsheet BCFT of an open bosonic string attached to a given D-brane. }
\end{quote}
                                                  
\noindent At first this statement might be confusing. An element of $\H$ is a quantum state of the string, but now we are claiming that it also represents a classical fluctuation of a D-brane. There are a couple of ways to understand this. The first is that it follows a general rule about the correspondence between first quantized theories and classical field theories: Namely, the wavefunction of a first quantized theory can be interpreted as a spacetime field of an equivalent classical field theory. This connection is sometimes under-emphasized since most quantum field theories do not have a simple first quantized description. The main complication is formulating worldline actions for particles with spin. But we illustrate the concept without considering spin. A nonrelativistic free particle has a worldline action
\begin{equation}S  = \frac{m}{2}\int dt \left(\frac{dx}{dt}\right)^2.\end{equation}
Quantization proceeds by passing to the Hamiltonian formalism and interpreting position and  momentum as operators  satisfying  canonical commutation relations. The quantum particle is then described by a quantum state $|\psi\rangle$ which evolves in time according to the Schr\"{o}dinger equation:
\begin{equation}i\frac{\d}{\d t}|\psi(t)\rangle = \frac{p^2}{2m}|\psi(t)\rangle.\end{equation}
The wavefunction is given by expressing $|\psi\rangle$ in the position basis:
\begin{equation}\psi(x,t) = \langle x|\psi(t)\rangle,\end{equation}
where the Schr\"{o}dinger equation reads
\begin{equation}i\frac{\d}{\d t}\psi(x,t)=-\frac{1}{2m}\frac{\d^2}{\d x^2}\psi(x,t).\end{equation}
Now if we forget where this equation came from, there is nothing contradictory about interpreting $\psi(x,t)$ as a classical complex scalar field subject to a nonrelativistic wave equation. In fact, the wave equation can be derived by variation of a field theory action
\begin{equation}
S = \int dx \,dt\left[i\psi^*(x,t)\frac{\d}{\d t}\psi(x,t) -\frac{1}{2m}\frac{\d}{\d x}\psi^*(x,t)\frac{\d}{\d x}\psi(x,t)\right].
\end{equation}
From this point of view, $\psi(x,t)$ is just a complex scalar field, and there is no reason to interpret it as a probability amplitude. However, this classical field theory is equivalent to the first quantized theory in the following sense: If we start from the action for $\psi(x,t)$ and follow the usual recipe for canonical quantization of a classical field theory, we find a Fock space of multiparticle states given by acting creation operators on the vacuum. The Hamiltonian of the resulting QFT implies that the wavefunction for the single particle state inside this Fock space will evolve according to the Schr\"{o}dinger equation for a free, nonrelativistic particle. So we are back to where we started, only we have a formalism describing many and variable number of indistinguishable nonrelativistic particles.  Presently we  are proceeding in an analogous way in string theory. We  start  with a classical worldsheet  action (the Polyakov action), quantize it to obtain a BCFT, and then we interpret the resulting quantum states as classical fields. 

There is a second, perhaps more physical justification for the definition of the string field. From the state-operator mapping of BCFT, we know that every state $|A\rangle\in\H$ has a corresponding boundary vertex operator $V_A(0)$. As mentioned before, the set of boundary operators corresponds to the set of possible boundary deformations of the BCFT. This, in turn, corresponds to the space of deformations (or fluctuations) of the D-brane system defining the open bosonic string BCFT.

It is important to distinguish between a generic string field and the particular kind of string field which enters the action and equations of motion---the {\it dynamical string field}. In a similar way, in gauge theories we have Lie algebra valued differential forms---including the 2-form field strength---but the dynamical variable of the theory is a 1-form---the gauge potential. The dynamical field in open bosonic SFT is the same kind of state in $\H$ where we impose the physical state condition, namely it is Grassmann odd and ghost number $1$. Just as the Schr\"{o}dinger equation of the nonrelativistic particle can be interpreted as a wave equation for a complex scalar field, the physical state condition is interpreted as a linearized field equation:
\begin{equation}Q\Psi = 0,\ \ \ \ \Psi=\text{Grassmann odd, ghost number }1.\label{eq:linEOM}\end{equation}
The equivalence of physical states is interpreted as a linearized gauge invariance:
\begin{equation}\Psi' = \Psi+Q\Lambda,\ \ \ \ \Lambda = \text{Grassmann even, ghost number }0.\end{equation}
Henceforth we will mostly drop the ket around $\Psi$. This is to emphasize that $\Psi$ is a classical field. We will not try to interpret it as a quantum state.

In fact, unlike a quantum state, the dynamical string field should in a sense be ``real." One way to see that this is true is that the most basic excitation of a D-brane---a photon---is obtained by quantization of the Maxwell potential, which is a real field. Fortunately, we have seen the state space of a BCFT naturally comes with a real structure, which suggests the following reality condition on the dynamical string field: 
\begin{equation}\Psi^\ddag = \Psi.\end{equation}
One can show that 
\begin{equation}(Q A)^\ddag = (-1)^{|A|+1}Q (A^\ddag),\end{equation}
since the BRST operator is Hermitian and BPZ odd. To preserve the reality condition, the gauge parameter $\Lambda$ must therefore satisfy 
\begin{equation}\Lambda^\ddag = -\Lambda,\end{equation}
and is in a sense ``imaginary."

To see that \eq{linEOM} makes sense as a linear field equation, it is helpful to give a more concrete description of the string field as an expansion in eigenstates of $L_0$. Let us do this for the D$p$-brane. The dynamical string field can be represented as a sum of states created by acting mode oscillators on the momentum eigenstate $|k_\mu\rangle$ in \eq{kmu}. Arranging these states in sequence of increasing $L_0$ eigenvalue for a given momentum, and recalling that the dynamical string field carries ghost number $1$, we find the expansion 
\begin{equation}
\Psi = \int\frac{d^{p+1}k}{(2\pi)^{p+1}}\Big[\underbrace{T(k)c_1\phantom{\Big(}\!\!}_{L_0=k^2-1}+\underbrace{A_\mu(k)\alpha_{-1}^\mu c_1 +\phi_a(k)\alpha_{-1}^a c_1 +\frac{i}{\sqrt{2}}\beta(k)c_0}_{L_0=k^2}+\underbrace{\phantom{\Big(}\ \ .\ .\ .\ \ \ }_{L_0=k^2+n,\ n\geq 1}\Big]|k_\mu\rangle.
\end{equation}
The coefficient functions $T(k)$ etc. are an infinite list of ordinary spacetime fields---the fluctuation fields of the D$p$-brane---expressed in momentum space. As you can probably anticipate, $T(x)$ is the tachyon on the D$p$-brane, $A_\mu(x)$ is the Maxwell gauge potential, and $\phi_a(x)$ are the massless scalars representing transverse displacement of the D$p$-brane. We will see the role of $\beta(x)$ in a moment. The reality condition implies that the coefficient fields are real. Plugging this into $Q\Psi=0$ implies a set of linearized field equations:
\begin{eqnarray}
(\Box+1)T \lineup = 0,\\
\Box A_\mu -\d_\mu\beta \lineup = 0,\\
\Box\phi_a \lineup = 0,\\
\beta -\d^\mu A_\mu \lineup = 0,\\
\vdots\ \ \ \lineup \ \ \ \ \ .\nonumber
\end{eqnarray}
We can similarly expand the gauge parameter:
\begin{equation}\Lambda = \int\frac{d^{p+1}k}{(2\pi)^{p+1}}\Big[\underbrace{\, i\lambda(k)\,}_{L_0=k^2}+\underbrace{\ \ .\ .\ .\ \ }_{L_0=k^2+n, n\geq 1}\Big]|k_\mu\rangle .\end{equation}

\pagebreak

\noindent The $i$ in front of $\lambda(k)$ ensures that the gauge parameter is imaginary if $\lambda(x)$ is real. The linearized gauge transformation of $\Psi$ translates to
\begin{eqnarray}
T \lineup = \text{invariant},\\
A_\mu'\lineup = A_\mu+\d_\mu\lambda,\\
\phi_a\lineup =\text{invariant},\\
\beta' \lineup = \beta+\Box\lambda,\\
\lineup \vdots\nonumber\ \ \ \ \ .
\end{eqnarray}
\begin{exercise} Derive these equations by computing $Q\Psi = 0$ and $\Psi'=\Psi+Q\Lambda$.\end{exercise}
\noindent The gauge potential has the expected Maxwell gauge invariance. The field $\beta$ does not carry any physical degrees of freedom, since its equation of motion is zeroth order in derivatives. It simply fixes $\beta$ to be $\d^\mu A_\mu$. Substituting into the field equation for $A_\mu$ implies
\begin{equation}
\Box A_\mu -\d_\mu(\d^\nu A_\nu) = \d^\nu(\d_\nu A_\mu-\d_\mu A_\nu) = \d^\nu F_{\nu\mu} = 0,
\end{equation}
which is Maxwell's equation. At higher mass level, the number of fields like $\beta(x)$ which carry no degrees of freedom proliferates. One way to understand this is that $\Lambda$ contains an infinite tower of gauge parameters, ascending in order of increasing $L_0$ eigenvalue for a given momentum. But naively one does not expect massive higher spin fields to need gauge symmetry. The only ``true" gauge invariance of the string spectrum is that of the photon. This implies that the dynamical string field must contain additional variables to soak up superfluous gauge symmetries at higher mass level. It should be possible to formulate a string field theory where all auxiliary fields like $\beta(x)$ have been integrated out, and the only remaining gauge invariance is that of Maxwell theory. However, it seems complicated to do this. A deeper issue is that, while the Maxwell gauge symmetry is sufficient to describe perturbative physics of the D$p$-brane, one would like to think that the string field theory has a classical solution describing, for example, a pair of D$p$-branes with a non-Abelian $U(2)$ gauge symmetry. It is hard to see how that gauge invariance could be captured by expanding a theory with Maxwell gauge symmetry around a nonzero vacuum solution. Therefore, the gratuitous redundancy of $\Psi$ may be an important hint as to its capacity to describe physics which is far removed from that of the reference D-brane.

Since the spectrum of $L_0$ is degenerate, there is a lot of freedom in the choice of Fock space basis. The basis discussed above consists of $\d X$ and $b,c$ ghost oscillators acting on the vacuum at some momentum. This is often called the {\it oscillator basis}. It is very simple but has a number of drawbacks. First, its definition is specific to free boson BCFTs. If we are interested in open bosonic SFT on more abstract backgrounds, we might not have a $\d X$ to furnish oscillators. Second, the basis is often inefficient when looking for classical solutions. A solution may generate expectation values for all component fields in the oscillator basis, but the physics of the new background may imply hidden linear relations between the fields. In a more efficient Fock space basis, one might see that certain fields could be set to zero even before solving the equations of motion. Let us describe a basis which solves the first problem and may partly address the second. Consider a boundary primary operator $\phi$ and the associated state $\phi(0)|0\rangle$. By acting negative mode Virasoro operators we can generate other states called {\it descendants}. The subspace of states generated by $\phi(0)|0\rangle$ together with all of its descendants is often called the {\it Verma module} of $\phi$. If the matter BCFT is unitary, we can form a Fock space basis for the matter part of the string field by collecting the Verma modules corresponding to the complete set of primary operators $\phi_i(0)$ of the matter BCFT. A generic string field $A$ can then be expanded 
\begin{eqnarray}
A = \sum_i \ \ \sum_{l_1>...>l_L\geq 2}\ \ \sum_{m_1>...m_M\geq -1}\ \ \lineup\sum_{n_1\geq...\geq n_N\geq 1} \ A^{i;\, l_1,..,l_L;\,m_1,...,m_M;\,n_1,...n_N}\nonumber\\
\lineup \times \Big(b_{-l_1}..b_{-l_L}\Big)\Big(c_{-m_1}...c_{-m_M}\Big)\Big(L^\text{m}_{-n_1}...L^\text{m}_{-n_N}\phi_i(0)|0\rangle\Big),\phantom{\bigg)}\ \ \ \ \ \ \ 
\label{eq:Verma_basis}\end{eqnarray}
where $L_{-n}^\mathrm{m}$ are Virasoros of the matter part of the BCFT. The numbers $A^{i;\, l_1,..,l_L;\,m_1,...,m_M;\,n_1,...n_N}$ are the component fields in this basis. The basis elements are eigenstates of $L_0$ with eigenvalue
\begin{equation}l_1+...+l_L + n_1+...n_N+m_1+...+m_M+h_i,\end{equation}
where $h_i$ is the dimension of the matter primary operator $\phi_i(0)$. Unlike the oscillator basis, the Verma module basis only assumes ingredients which are available for any matter BCFT. A complication, however, is that it may be overcomplete. There can be nontrivial linear combinations of states in the Verma module, called {\it null states}, which turn out to vanish identically. Whether this occurs depends on the central charge and the dimension of the primary generating the Verma module, as described by the Kac determinant. The simplest example of a null state is $L_{-1}|0\rangle=0$, which appears in the Verma module of the identity operator. There are other variations of the Verma module basis one can consider. For example, one can replace the matter Virasoros with total Virasoros, or if the matter BCFT breaks into a tensor product of components, one can construct a basis from the Verma modules of each factor. Verma modules do not work as well in the ghost sector due to the non-unitary nature of the $bc$ system. For example the state $b_{-2}c_1|0\rangle$ is neither a primary nor a descendant of a primary.  However, more sophisticated choices of basis in the ghost sector have been considered. For a more complete account of these matters, see~\cite{Kudrna}.

An important subspace of string fields is generated by acting ghost oscillators and matter Virasoros on the $SL(2,\mathbb{R})$ vacuum. This basis is called a {\it universal basis}, and the subspace it generates is called the {\it universal sector}. States in the universal sector are called {\it universal}. Since the $SL(2,\mathbb{R})$ vacuum, energy momentum tensor and ghosts exist for all open string BCFTs, the universal sector is characterized independently of the D-brane system which defines the string field theory. It was shown in \cite{Sen_universality} that the universal sector contains  solutions for the tachyon vacuum, so that all D-brane systems in bosonic string theory support tachyon vacuum solutions which take the same form when expressed in a universal basis. The universal basis is also much more efficient in describing tachyon condensation than the oscillator basis. An explicit comparison of the number of fields in the universal basis and the oscillator basis at a given level can be found in table~1~of~\cite{MoellerTaylor}.

The above summarizes most of what we need to know about the Fock space expansion. We now turn to a different representation of the string field: the {\it Schr\"{o}dinger representation}. In quantum mechanics, the wavefunction is derived by contracting $|\psi(t)\rangle$ with a position eigenstate. We can do a similar thing for the string field, contracting with an eigenstate of the curve of the string. The result is a {\it Schr{\"o}dinger functional} of a curve in spacetime. This representation can be defined for any free boson $\BCFT$. Even the dependence on ghosts can be described through a ``curve" of formal Grassmann odd variables. But for simplicity we consider the dependence of the string field on a single spacelike free boson $X(z,\overline{z})$ subject to Neumann boundary conditions. We wish to obtain a functional of the eigenvalues of $X(z,\overline{z})$ at $|z|=1$ (corresponding to $t=0$ in radial quantization). From \eq{Xmumode} we obtain a mode expansion
\begin{equation}\widehat{x}(\sigma) = \widehat{X}(e^{i\sigma},e^{-i\sigma}) = \widehat{x}+2\sum_{n=1}^\infty \widehat{x}_n \cos(n\sigma),\ \ \ \ \widehat{x}_n = \frac{i}{\sqrt{2}}\frac{\widehat{\alpha}_n-\widehat{\alpha}_{-n}}{n}, \ \ n\geq1.\label{eq:xsigma}\end{equation}
We temporarily place a hat over operators to distinguish them from their eigenvalues. $\widehat{x}(\sigma)$ is an operator representing the curve of an open string, and $\widehat{x}_n$s are position mode operators. The cosine expansion reflects the fact that the boundary conditions are Neumann. Now we can consider a basis of dual eigenvectors of the position mode operators:
\begin{equation} \langle x(\sigma)|\widehat{x}_n =x_n\langle x(\sigma)|,\end{equation}
where the eigenvalue $x_n$ is related to the curve $x(\sigma)$ through the mode expansion \eq{xsigma} with hats deleted.
\begin{exercise}Show that
\begin{equation}
\langle x(\sigma)| = \int \frac{d k}{2\pi}e^{i k x_0}\langle k|\prod_{n=1}^\infty \frac{1}{\sqrt[4]{\pi}}\exp\left[-\frac{1}{2}x_n^2 +i\sqrt{2}x_n\widehat{\alpha}_n +\frac{1}{2n}\widehat{\alpha}_n^2\right],
\end{equation}
where $\langle k|$ is the dual momentum eigenstate.
\end{exercise}
\noindent The overlap
\begin{equation}A[x(\sigma)]=\langle x(\sigma)|A\rangle\end{equation}
can be interpreted as a scalar field which depends on a curve $x(\sigma)$ in spacetime. This is natural. Just as an ordinary field depends on a point $x$ in spacetime, representing a possible location of a point particle, the string field depends on a curve in spacetime, representing a possible configuration of a string. As we have seen, the string field can also be understood in a Fock space basis as an infinite tower of ordinary fields. This description makes manifest that a theory of a free string is indistinguishable from a theory with an infinite tower of free particles of a particular kind. At the interacting level, however, string theory is very different from particle theory. For this reason, interactions are rather opaque when formulated in terms of the infinite tower of ordinary fields. In Witten's open bosonic string field theory, they appear as an array of cubic nonlocal couplings with obscure relative coefficients. In the Schr{\" o}dinger representation, however, interactions are easy to understand.

The Hermitian and BPZ conjugation properties of the mode operators imply 
\begin{equation}\widehat{x}(\sigma)^\dag = \widehat{x}(\sigma),\ \ \ \ \widehat{x}(\sigma)^\bigstar = \widehat{x}(\pi-\sigma),\end{equation}
from which we learn that the eigenstates have conjugation properties
\begin{equation}\langle x(\sigma)|^\dag = |x(\sigma)\rangle,\ \ \ \ \langle x(\sigma)|^\bigstar = |x(\pi-\sigma)\rangle,\end{equation}
where $|x(\sigma)\rangle$ is the eigenstate of $\widehat{x}_n$ with eigenvalue $x_n$. With this we can derive the reality conjugate Schr\"{o}dinger functional of $A$:
\begin{eqnarray}
A^\ddag[x(\sigma)] \lineup = \langle x(\sigma)|A^\ddag\rangle\nonumber\\
\lineup = \langle x(\sigma)|A^{\bigstar\dag}\rangle\nonumber\\
\lineup = \langle A^\bigstar| x (\sigma)\rangle^*\nonumber\\
\lineup =  \langle x(\pi-\sigma)|A\rangle^*\nonumber\\
\lineup = A[x(\pi-\sigma)]^*.\label{eq:Sch_conj}
\end{eqnarray} 
Thus reality conjugation takes the complex conjugate of the functional and reverses the parameterization of the string. We can describe the BPZ inner product in the Schr\"{o}dinger representation by inserting a resolution of the identity,
\begin{equation}1=\int [dx(\sigma)]|x(\sigma)\rangle\langle x(\sigma)|,\end{equation}
with the appropriately defined functional integral measure. We find
\begin{eqnarray}
\langle A, B\rangle\lineup = \langle A^\bigstar| B\rangle\nonumber\\
\lineup = \int [dx(\sigma)]\langle A^\bigstar| x(\sigma)\rangle\langle x(\sigma)| B\rangle\nonumber\\
\lineup = \int [dx(\sigma)]\langle x(\pi-\sigma)|A\rangle\langle x(\sigma)| B\rangle\nonumber\\
\lineup = \int [dx(\sigma)]A[x(\pi-\sigma)] B[x(\sigma)].\label{eq:BPZ_Sch}
\end{eqnarray}
The reversal of the parameterization of the string in the first functional is connected to the fact that the BPZ inner product glues a point at an angle $\pi-\sigma$ on the unit half circle of the first state to the point at an angle $\sigma$ on the unit half circle of the second state.

It will be helpful to further articulate the connection between the Schr\"{o}dinger representation and the visualization of a state as the unit half-disk carrying a vertex operator. The BPZ inner product can be computed as a correlation function of vertex operators in the UHP:
\begin{equation}\langle A, B\rangle = \big\langle( I\circ V_A(0))V_ B(0)\big\rangle_\text{UHP}.\end{equation}
This, in turn, can be computed by a path integral over the worldsheet fields in the UHP. Again considering the spacelike free boson subject to Neumann boundary conditions, we obtain
\begin{equation}\langle A, B\rangle = \int[d X(z,\zbar)]_{(z,\zbar)\in\text{UHP}}\Big( I\circ V_A(0)\Big) V_ B(0)e^{-S}.\end{equation}
In the integrand, the vertex operators and the worldsheet action are understood as functionals of $X(z,\zbar)$. We now factorize the integration into three components. First, we integrate $X(e^{i\sigma},e^{-i\sigma})=x(\sigma)$ on the unit half-circle $|z|=1$. Second we integrate $X(z,\zbar)$ outside the unit half circle $|z|>1$, with the boundary condition that $X(z,\zbar)$ must be equal to $x(\sigma)$ at $|z|=1$. Third we integrate $X(z,\zbar)$ inside the unit half circle $|z|<1$, again with the boundary condition that $X(z,\zbar)$ must be equal to $x(\sigma)$ at $|z|=1$. Therefore we have
\begin{eqnarray}\langle A, B\rangle \lineup = \int [dx(\sigma)] \int_{X(e^{i\sigma},e^{-i\sigma})=x(\sigma)}[dX(z,\zbar)]_{|z|>1}\nonumber\\ \lineup\ \ \ \ \ \ \ \ \ \ \ \ \ \ \ \ \ \ \ \times \int_{X(e^{i\sigma},e^{-i\sigma})=x(\sigma)}[dX(z,\zbar)]_{|z|<1} \Big( I\circ V_A(0)\Big) V_ B(0)e^{-S}.\end{eqnarray}
Recall that vertex operators need not be local, but must be localized within the unit half-disk. This means that $V_B(0)$ will only depend on the worldsheet fields inside the unit half-circle, and $I\circ V_A(0)$ will only depend on the worldsheet fields outside.  Furthermore, locality of the worldsheet theory implies that the exponential of the action factorizes into pieces which depend respectively on worldsheet fields inside and outside the unit half-circle. Therefore the integrand can be factorized:
\begin{eqnarray}
\langle A, B\rangle \lineup = \int [dx(\sigma)]\left(\int_{X(e^{i\sigma},e^{-i\sigma})=x(\sigma)}[dX(z,\zbar)]_{|z|>1} I\circ V_A(0)e^{-S}\right)\nonumber\\
\lineup\ \ \ \ \ \ \ \ \ \ \ \ \ \ \ \ \ \ \ \ \ \ \ \ \ \ \ \ \ \ \times\left(\int_{X(e^{i\sigma},e^{-i\sigma})=x(\sigma)}[dX(z,\zbar)]_{|z|<1} V_ B(0)e^{-S}\right).
\end{eqnarray}
The path integral over the exterior of the half circle can be rewritten as a path integral over the interior after making a conformal transformation  $I(z)=-1/z$. Accounting for the boundary conditions then gives 
\begin{eqnarray}
\langle A, B\rangle \lineup = \int [dx(\sigma)]\left(\int_{X(e^{i\sigma},e^{-i\sigma})=x(\pi-\sigma)}[dX(z,\zbar)]_{|z|<1} V_A(0)e^{-S}\right)\nonumber\\
\lineup\ \ \ \ \ \ \ \ \ \ \ \ \ \ \ \ \ \ \ \ \ \ \ \ \ \ \ \ \ \ \times\left(\int_{X(e^{i\sigma},e^{-i\sigma})=x(\sigma)}[dX(z,\zbar)]_{|z|<1} V_ B(0)e^{-S}\right).
\end{eqnarray}
Comparing to \eq{BPZ_Sch} it is clear that this is the BPZ inner product of the Schr{\"o}dinger functionals 
\begin{eqnarray}
A[x(\sigma)]\lineup =\int_{X(e^{i\sigma},e^{-i\sigma})=x(\sigma)}[dX(z,\zbar)]_{|z|<1} V_A(0)e^{-S},\nonumber\\
B[x(\sigma)]\lineup =\int_{X(e^{i\sigma},e^{-i\sigma})=x(\sigma)}[dX(z,\zbar)]_{|z|<1} V_B(0)e^{-S}.
\end{eqnarray}
Therefore the Schr{\"o}dinger functional can be derived as a path integral on the unit half-disk with the appropriate vertex operator at the origin and fixed boundary conditions on the worldsheet fields at $|z|=1$, representing the configuration of a string.

To make this more concrete, let us use the path integral to evaluate the Schr{\"o}dinger functional of the $SL(2,\mathbb{R})$ vacuum. A similar calculation (for the closed string) is discussed in chapter 2 of Polchinski \cite{Polchinski}. We write the $SL(2,\mathbb{R})$ vacuum with the Greek letter omega,
\begin{equation}\Omega[x(\sigma)] = \langle x(\sigma)|0\rangle,\end{equation}
in order to avoid the confusing notation $0[x(\sigma)]$. The vertex operator in this case is the identity, so we only need to evaluate
\begin{equation}
\Omega[x(\sigma)] =\int_{X(e^{i\sigma},e^{-i\sigma})=x(\sigma)}[dX(z,\zbar)]_{|z|<1}\, e^{-S}.
\end{equation}
The key idea is to make a change of variables in the path integral 
\begin{equation}X(z,\zbar) = x(z,\zbar)+Y(z,\zbar),\end{equation}
where $x(z,\zbar)$ is a solution to the Laplace equation on the unit half-disk
\begin{equation}\d\overline{\d} x(z,\zbar)=0,\end{equation}
subject to the boundary conditions
\begin{equation}x(z,\zbar)|_{z=e^{i\sigma}}=x(\sigma),\ \ \ \ \d x(z,\zbar)|_{\mathrm{Im}(z)=0}=\overline{\d}x(z,\zbar)|_{\mathrm{Im}(z)=0}.
\end{equation}
The first says that $x(z,\zbar)$ matches the argument of the wavefunctional on the unit half-circle, and the second says that $x(z,\zbar)$ satisfies Neumann boundary conditions on the real axis. In particular, this implies that $Y(z,\zbar)$ must vanish on the unit half-circle. We now change the integration variable from  $X(z,\zbar)$ to  $Y(z,\zbar)$; since they differ  through a shift by a fixed function, the measure is unchanged and we have 
\begin{equation}
\Omega[x(\sigma)] =\int_{Y(e^{i\sigma},e^{-i\sigma})=0}[dY(z,\zbar)]_{|z|<1}\, \exp\Big(-S\big[x(z,\zbar) +Y(z,\zbar)\big]\Big) .
\end{equation}
Since $x(z,\zbar)$ satisfies the classical equations of motion with consistent boundary conditions we can show that
\begin{equation}
S\big[x(z,\zbar) +Y(z,\zbar)\big] = S[x(z,\zbar)]+S[Y(z,\zbar)],
\end{equation}
so
\begin{eqnarray}
\Omega[x(\sigma)] \lineup =\int_{Y(e^{i\sigma},e^{-i\sigma})=0}[dY(z,\zbar)]_{|z|<1}\, e^{-S[x(z,\zbar)]-S[Y(z,\zbar)]} \nonumber\\
\phantom{\Bigg)}\lineup = \mathcal{N} e^{-S[x(z,\zbar)] },
\end{eqnarray}
where $\mathcal{N}$ is a constant determined by evaluating the path integral over $Y(z,\zbar)$, and is completely independent of $x(\sigma)$. Next we must compute the worldsheet action evaluated on the classical solution $x(z,\zbar)$. In our conventions ($\alpha'=1$) the Polyakov action is
\begin{equation}S = \frac{1}{2\pi}\int d^2 z\, \d X (z,\zbar)\overline{\d}X(z,\zbar),\end{equation}
where $d^2z = 2dx dy$ if $z=x+iy$. The solution of Laplace's equation with the assumed boundary conditions can be expressed in terms of the position  modes of $x(\sigma)$
\begin{equation}x(z,\zbar) = x_0 + \sum_{n=1}^\infty x_n(z^n+\zbar^n).\end{equation}
Plugging in gives 
\begin{equation}\Omega[x(\sigma)] = \mathcal{N}\exp\left[-\frac{1}{2}\sum_{n=1}^\infty n x_n^2\right].\label{eq:SL2R_Sch}\end{equation}
\begin{exercise} Do this calculation. \end{exercise}
\noindent This is a Gaussian in the space of string position modes. The position zero mode does not appear, consistent with the fact that the $SL(2,\mathbb{R})$ vacuum has vanishing momentum. The Gaussian has a maximum when the entire curve $x(\sigma)$ shrinks to a point. 

One way to check this calculation is to note that the $SL(2,\mathbb{R})$ vacuum can be viewed as a tensor product of an infinite number of harmonic oscillator vacua, one for each mode oscillator $\alpha_n$. The string mode oscillators and the position modes, however, are not canonically normalized. To relate to the conventional harmonic oscillator we should identify
\begin{equation}a^\dag \sim \frac{\alpha_{-n}}{\sqrt{n}},\ \ \ \ a\sim \frac{\alpha_n}{\sqrt{n}},\ \ \ \ x\sim \sqrt{n}x_n,\ \ \ \  (n\geq 1).\end{equation}
The ground state wavefunction for a harmonic oscillator is $e^{-\frac{1}{2}x^2}$. Substituting $\sqrt{n}x_n$ for $x$ and taking the product over all $n$ gives \eq{SL2R_Sch}.

\begin{exercise}
\label{ex:vac_functional}
Write the $SL(2,\mathbb{R})$ vacuum functional explicitly in terms of the curve $x(\sigma)$ by finding an integral kernel $\omega(\sigma_1,\sigma_2)$ such that
\begin{equation}\Omega[x(\sigma)] = \mathcal{N}\exp\left[-\frac{1}{2}\int_0^\pi d\sigma_1\int_0^\pi d\sigma_2\, x(\sigma_1)x(\sigma_2)\omega(\sigma_1,\sigma_2)\right].\end{equation}
This representation is somewhat tricky since the integral kernel is a nontrivial distribution. To confirm that the distribution has been correctly defined, check your result by substituting the mode expansion of $x(\sigma)$ to recover \eq{SL2R_Sch}.
\end{exercise}

To get the complete $SL(2,\mathbb{R})$ vacuum functional on a D$p$-brane, we must include additional factors for the Dirichlet coordinates and the ghosts. We will not discuss this further because the Schr{\"o}dinger representation is awkward for explicit calculations. Its main utility is to provide conceptual justification for the procedure of cutting and gluing surfaces to form string vertices and Feynman diagrams, a fundamental idea in the development of string field theory.

\subsection{Witten's open bosonic SFT}

Now we are ready to define Witten's open bosonic SFT. The task is to find a nonlinear extension of the linearized equations of motion $Q\Psi=0$ and define an appropriate action principle. First we note an analogy between string fields and gauge fields formulated in the language of differential forms:
\begin{eqnarray}
{\text{rank of a} \atop \text{form}}\lineup\ \ \  \longleftrightarrow \ \ \ \text{ghost number},\nonumber\\
\text{exterior derivative }d\lineup\ \ \ \longleftrightarrow\ \ \ \text{BRST operator }Q,\phantom{\bigg)}\nonumber\\
\text{gauge potential }A\lineup \ \ \ \longleftrightarrow\ \ \ {\text{dynamical string} \atop \text{field }\Psi}.
\end{eqnarray}
This analogy suggests a nonlinear gauge invariance of the string field 
\begin{equation}\Psi'  = \Psi + Q\Lambda + [\Psi,\Lambda],\end{equation}
where $\Lambda$ is an infinitesimal gauge parameter. The proposed gauge symmetry requires defining a product between the string fields $\Psi$ and $\Lambda$. This is Witten's open string star product. Sometimes this is written $A*B$, but usually we will simply write $AB$. Let us assume for the moment that a suitable product has been defined and proceed. A gauge covariant nonlinear extension of the equations of motion is given by
\begin{equation}Q\Psi + \Psi^2= 0.\end{equation}
These resemble the equations of motion of Chern-Simons theory. This leads to an action
\begin{equation}S = \Tr\left(-\frac{1}{2}\Psi Q\Psi-\frac{1}{3}\Psi^3\right),\end{equation}
for an appropriately defined trace operation. The consistency of this action relies on the following ``axioms:"
\begin{eqnarray}
\lineup (1)\ Grading:\ \ \ \ \ \, \mathrm{gh}(QA) = \mathrm{gh}(A)+1,\phantom{\Big)}\nonumber\\
\lineup \ \ \ \ \ \ \ \ \ \ \ \ \ \ \ \ \ \ \ \ \ \ \ \ \mathrm{gh}(AB) = \mathrm{gh}(A)+\mathrm{gh}(B),\nonumber\\
\lineup \ \ \ \ \ \ \ \ \ \ \ \ \ \ \ \ \ \ \ \ \ \ \ \ \Tr(A)=0\text{ if gh}(A)\neq3,\phantom{\Big)}\nonumber\\
\lineup\ \ \ \ \ \text{Here ``gh" refers to ghost number. Mod $\mathbb{Z}_2$, the first two }\nonumber\\
\lineup\ \ \ \ \ \text{properties also hold for Grassmann parity.} \nonumber\\
\lineup(2)\ Nilpotency:\ \ \ \ Q^2=0.\phantom{\Bigg)}\nonumber\\
\lineup (3)\ Derivation\ property:\ \ \ \ Q(AB) = (QA)B+(-1)^{|A|}AQB.\nonumber\\
\lineup (4)\ Associativity:\ \ \ \ A(BC) = (AB)C.\phantom{\Bigg)}\nonumber\\
\lineup (5)\ Integration\ by\ parts:\ \ \ \ \Tr(QA) = 0.\nonumber\\
\lineup (6)\ Cyclicity:\ \ \ \ \Tr(AB) = (-1)^{|A||B|}\Tr(BA).\phantom{\Bigg)}\nonumber\\
\lineup (7)\ Nondegeneracy:\ \ \ \ \text{If }\Tr(AB)\text{\ vanishes for all }A\text{, then }B=0.\nonumber
\end{eqnarray}
These properties imply that the state space $\mathcal{H}$ of the BCFT has been endowed with the structure of a cyclic, graded differential associative algebra. The same structure applies to matrix valued forms on a 3-manifold, which allows the definition of Chern-Simons theory.

Let us give a rough picture of how the product and trace should be defined. The product is associative, and all associative products are, in some way or another, matrix products. The string field in the Schr{\"o}dinger representation is a functional of a curve,
\begin{equation}\Psi[x(\sigma)],\end{equation}
\begin{wrapfigure}{l}{.23\linewidth}
\centering
\vspace{-.5cm}
\resizebox{1.5in}{1.2in}{\includegraphics{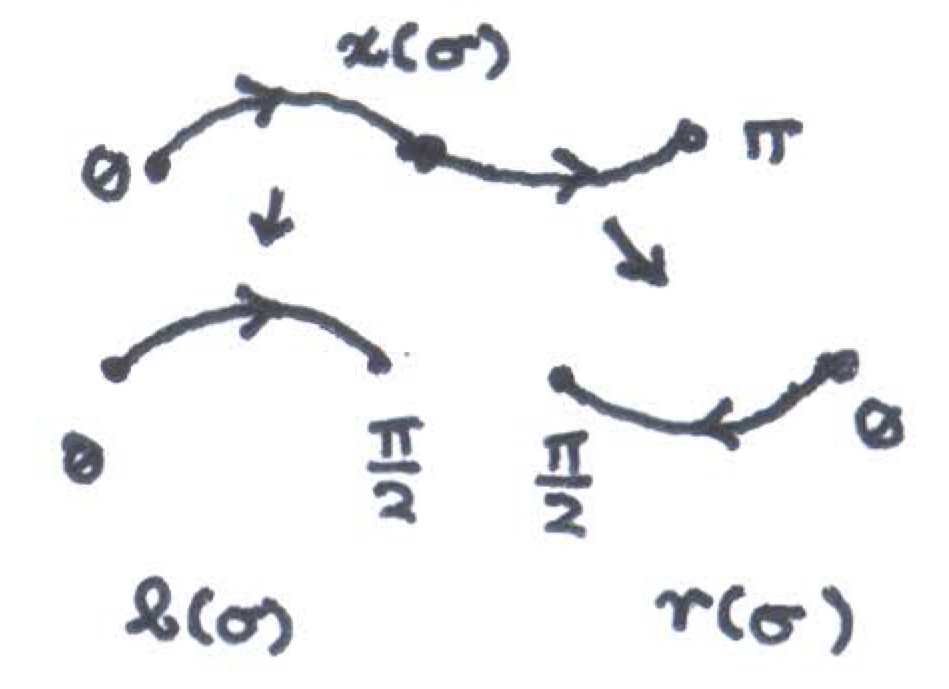}}
\vspace{-1cm}
\end{wrapfigure}
\noindent and it is natural to interpret the curve as representing matrix indices, in some sense. However, a matrix should have two indices, and there is only one curve $x(\sigma)$. We can deal with this by regarding the full curve as a pair of half-curves
\begin{eqnarray}
l(\sigma)\lineup = x(\sigma),\ \ \ \ \ \ \ \ \ \, \sigma\in[0,\pi/2],\\
r(\sigma)\lineup =x(\pi-\sigma),\ \ \ \ \sigma\in[0,\pi/2].
\end{eqnarray}
$l(\sigma)$ is the ``left half" of the string, and $r(\sigma)$ is the ``right half." The left and right halves join at a common point,
\begin{equation}l\left(\frac{\pi}{2}\right)=r\left(\frac{\pi}{2}\right)=x\left(\frac{\pi}{2}\right),\end{equation}
called the ``midpoint." If we view the string field as a functional of the left and right halves of~the

\begin{wrapfigure}{l}{.18\linewidth}
\centering
\resizebox{1.3in}{1.3in}{\includegraphics{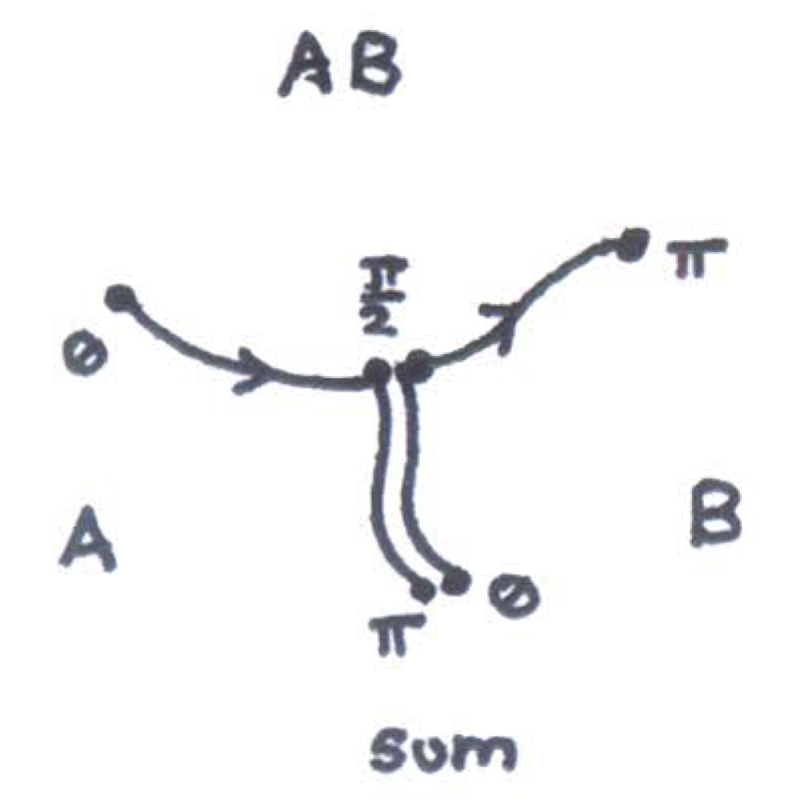}}
\end{wrapfigure}
\noindent string,
\begin{equation}
\Psi[l(\sigma),r(\sigma)],
\end{equation}
we have a matrix. This is often referred to as the {\it half string}, or {\it split string} representation of the Schrodinger functional \cite{half,split}. The associative product of string fields may then be defined
\begin{equation}
AB[l(\sigma),r(\sigma)]= \int [dw(\sigma)]A[l(\sigma),w(\sigma)]B[w(\sigma),r(\sigma)].
\end{equation}
In words, one equates the right half curve in $A$ with the left half curve in $B$ and sums over the common half curve to derive $AB$. This is a functional integral version of a matrix product. In a similar way, we can define the trace operation
\begin{equation}\Tr[A] = \int [dw(\sigma)] A[w(\sigma),w(\sigma)].\end{equation}
\noindent So far this is schematic. To make it concrete, we would need to account for ghosts and learn how to precisely evaluate the functional integrals. We will not try to develop the theory in this direction. Instead, our strategy will be to rephrase things so that all basic operations of the theory are given by routine evaluation of BCFT correlation functions. It should also be mentioned that the split string representation is not a completely correct description of the string field, since the Hilbert space of curves does not decompose into a tensor product of Hilbert spaces of half curves~\cite{fresh}. Essentially, the left and right halves cannot be regarded as independent ``matrix indices" since they must tie together smoothly at the midpoint. We will not need to deal with this problem since we do not develop the half string representation in detail.

\begin{wrapfigure}{l}{.18\linewidth}
\centering
\resizebox{1.2in}{1.15in}{\includegraphics{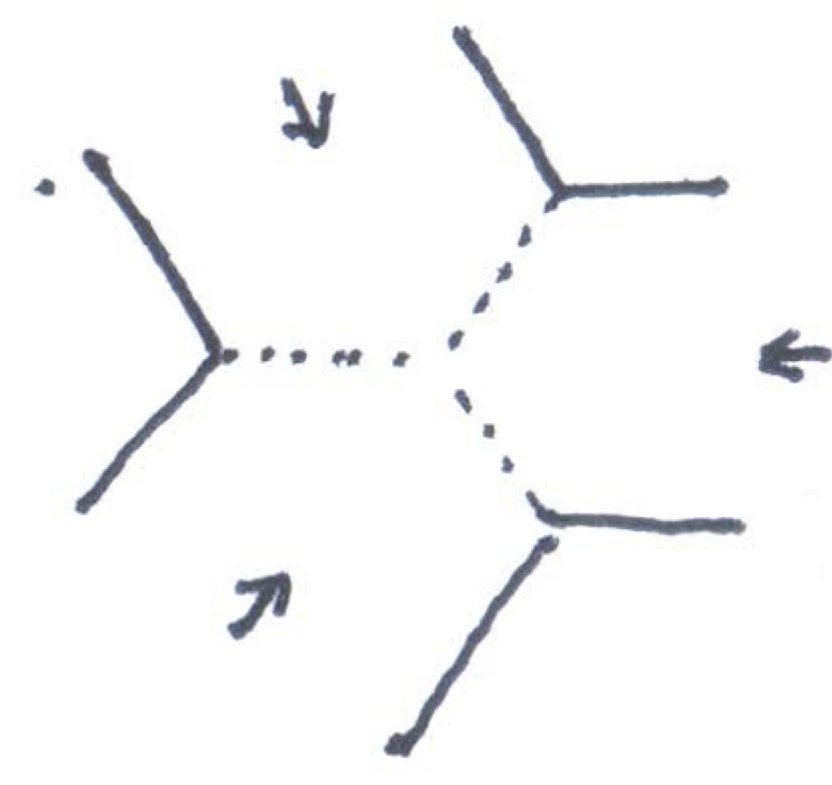}}
\end{wrapfigure}
The product and trace together define a cubic vertex $\Tr[\Psi^3]$. In Feynman diagrams, the cubic vertex can be visualized as a process where three incoming strings collide and join along their halves. Since the action is cubic, gluing propagators together with this vertex generates all Feynman diagrams needed to compute tree level open string amplitudes. It is not obvious that these diagrams lead to open string amplitudes in the conventional form as integrals of the appropriate measure over the moduli spaces of disks with boundary punctures. We will  not prove that this is the case, but it turns out to be true. The result is not surprising since Witten's theory gives consistent interactions to open strings, by virtue of the fact that the framework is gauge invariant. It is hard to imagine that open strings can consistently interact in any  other way than  in the conventional  string theory. 

What is more surprising is that loop diagrams in Witten's open string field theory continue to implement the correct integration over the moduli spaces of higher genus Riemann surfaces~\cite{Zwiebach_Witten}. In higher genus amplitudes, closed string states must  appear in handles, which suggests we 
\begin{wrapfigure}{l}{.23\linewidth}
\centering
\resizebox{1.5in}{.9in}{\includegraphics{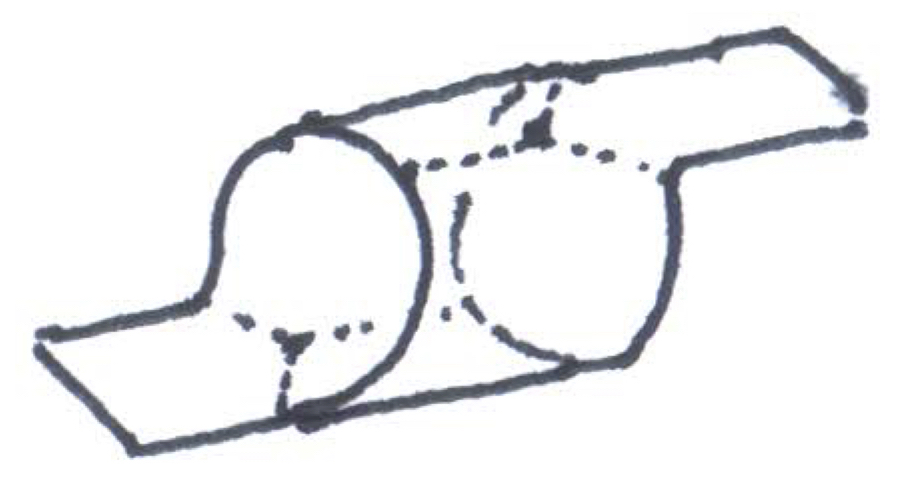}}
\vspace{.2cm}
\end{wrapfigure}
\noindent will need closed string propagators, closed string fields, and an array of new closed string and open/closed string vertices. Indeed, these  things are needed to compute higher genus open string amplitudes  in lightcone  gauge string field  theory. But in Witten's string field theory, it appears that closed string states are already accounted for as ``bound states" of open strings. 
This can be seen, for example, in the non-planar 1-loop 2-point function. The corner of the moduli space where the open string propagators in the loop shrink to zero length (the ultraviolet from the open string perspective) is equivalent to the corner of moduli space where a tube of worldsheet becomes  infinitely long, and the closed string states inside the tube must be on-shell. However, there are difficulties with the quantum theory which have not been resolved. The absence of closed string fields makes it unclear how closed strings appear as asymptotic~states. New instabilities appear at the quantum level due
to the closed string tachyon that cannot be handled in perturbation theory. There is no known systematic procedure for canceling infrared divergences related to the closed string tachyon. There is hope these problems would be better addressed in the context of open superstring field theory. The status of present understanding of quantum effects can mostly be found in \cite{Giddings,Thorn,Taylor}.

 \begin{wrapfigure}{l}{.5\linewidth}
\centering
\resizebox{2.8in}{3in}{\includegraphics{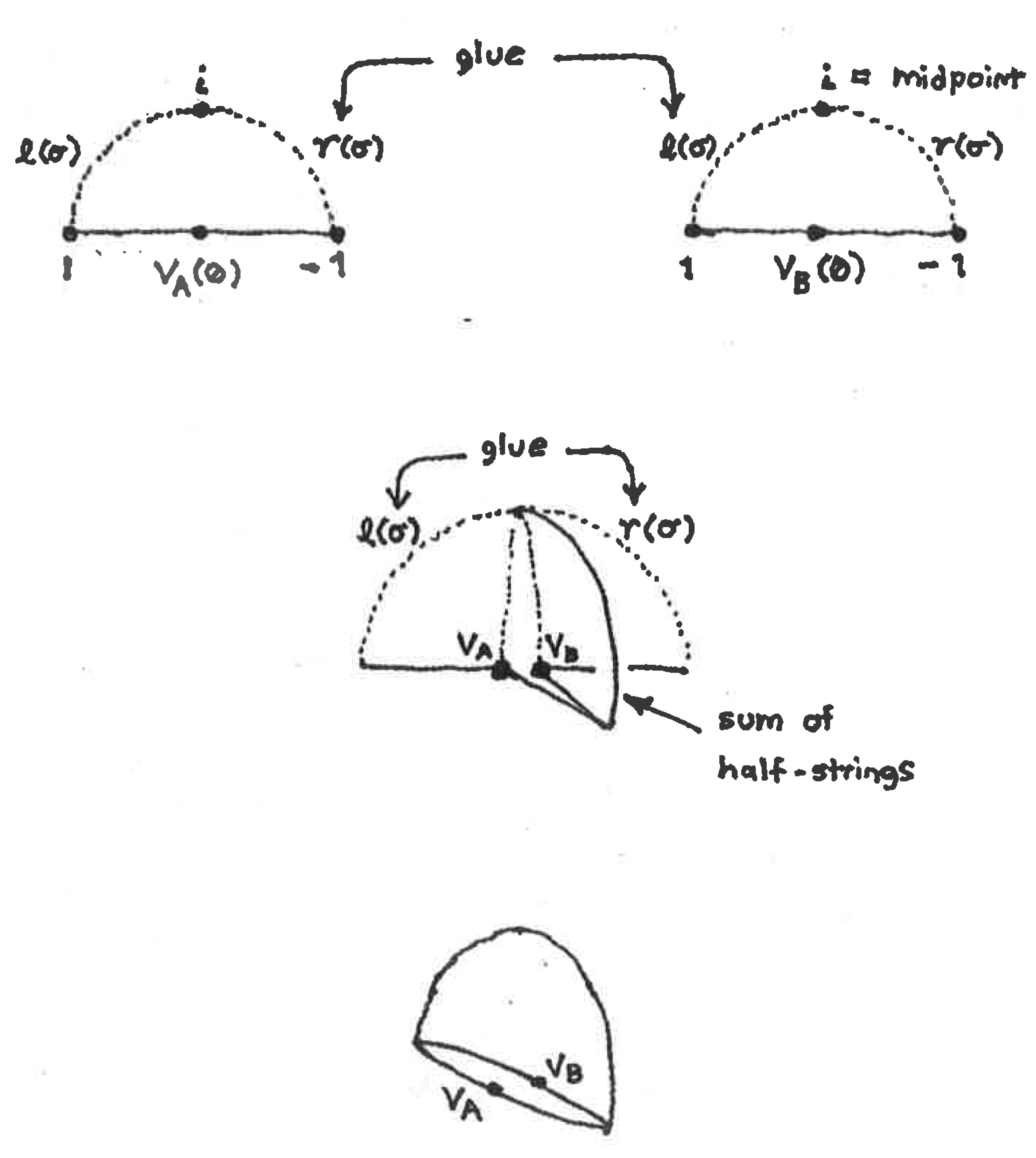}}
\resizebox{3.5in}{3.2in}{\includegraphics{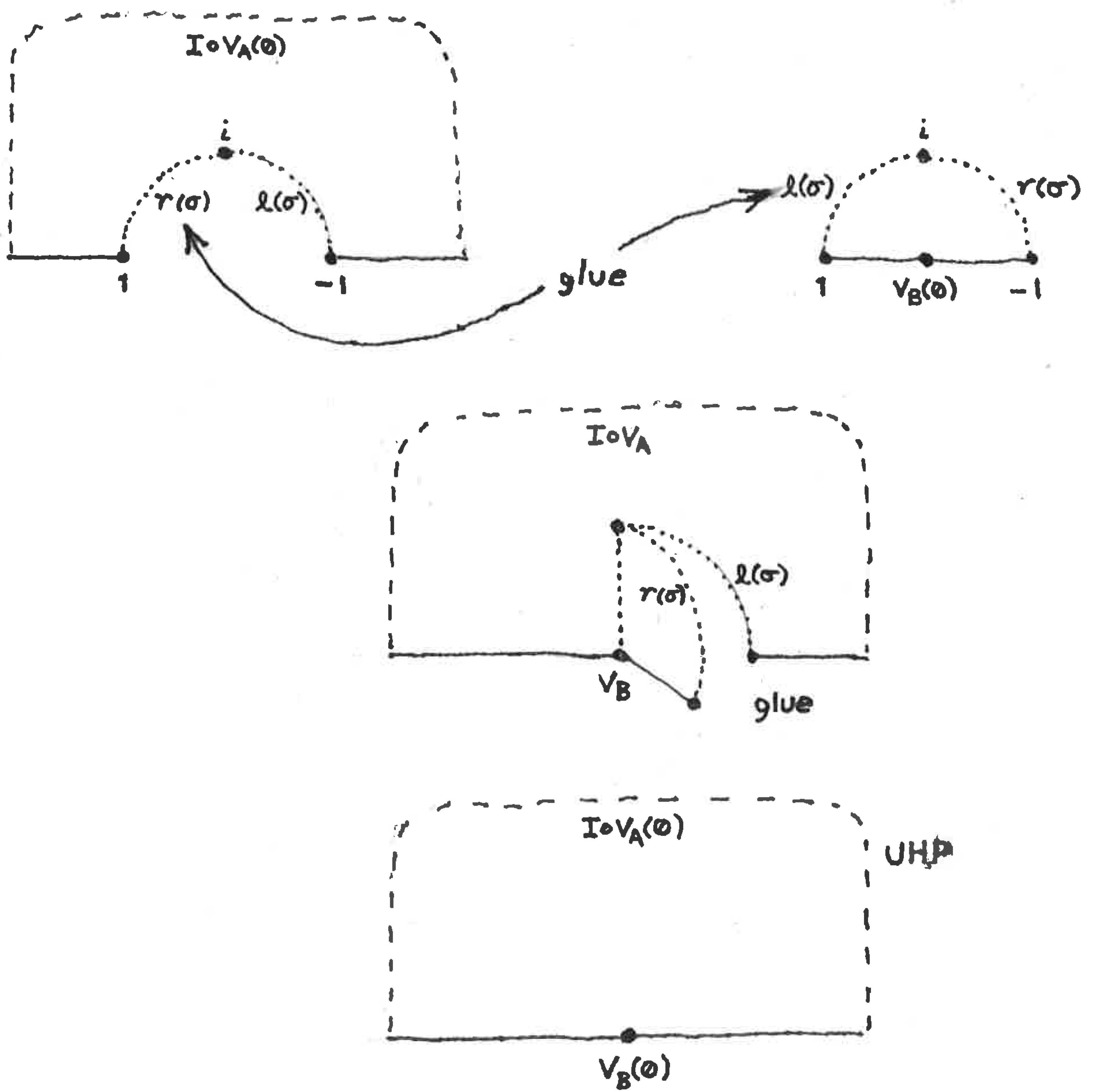}}
\vspace{.5cm}
\end{wrapfigure}
\noindent \ \ \ Let us return to the question of giving a more precise definition of the product and trace. First consider the 2-string vertex 
\begin{equation}\Tr[AB].\end{equation}
For this we need to evaluate the product of $A$ and $B$ followed by the trace. The states $A$ and $B$ can be represented as half-disks with corresponding vertex operator insertions. To compute the product $AB$, we should form a new surface by gluing the right quarter-circle
on the half-disk of $A$ to the left quarter-circle on the half-disk of $B$. On the remaining~quarter-circles we associate boundary~conditions~corresponding  to the left and right halves of the string. The  worldsheet path integral on this new surface, with vertex operator insertions and these boundary conditions, is supposed to give the Schr{\"o}dinger functional of the product $AB$. Note that the path integral includes integration over the worldsheet variables on the quarter-circle where the half-disks of $A$ and $B$ have been joined. This is the sum over half-string matrix indices described above. To further evaluate the trace, we should glue the left quarter-circle on the half-disk of $A$ to the right quarter-circle on the half-disk of~$B$. The result is a BCFT correlation function of two vertex operators on a ``pita" shaped surface, as shown above left. To express this as a correlation function  on the upper half plane, it is helpful to represent the state $A$ in the BPZ dual coordinate system consisting of the upper half plane with the unit half-disk removed, and the vertex operator inserted at infinity with 
the conformal transformation $I(z)=-1/z$. As illustrated to the above, gluing quarter circles as prescribed by the product and trace effectively joins the half-disk of $B$ to the complementary half-disk of $A$ to form a correlation function on the upper half plane
\begin{equation}\Tr[AB] = \langle I\circ V_A(0) V_B(0)\rangle_\mathrm{UHP}.\end{equation}
This is the same as the BPZ inner product. We therefore find
\begin{equation}\Tr[AB] = \langle A,B\rangle.\end{equation}
Note that cyclicity and nondegeneracy of the trace is implied by symmetry and nondegeneracy of the BPZ inner product.

Let us mention a visual problem which leads to an important issue of conventions. One might notice that the {\it left} half of the string $\sigma\in[0,\pi/2]$ maps to the {\it right} portion of the unit circle $\mathrm{Re}[e^{i\sigma}]>0$. Meanwhile, the {\it right} half of the string $\sigma\in[\pi/2,\pi]$ maps to the {\it left} portion of the unit circle $\mathrm{Re}[e^{i\sigma}]<0$. 
\begin{center}\resizebox{1.3in}{1.5in}{\includegraphics{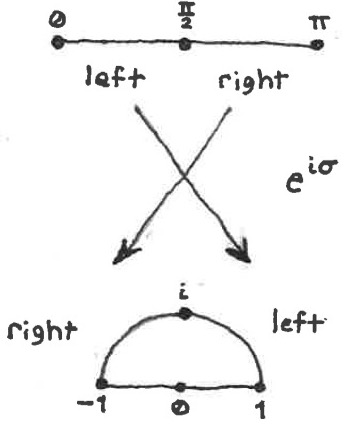}}\end{center}
Thus what we are calling left and right is backwards from the point of view of the 
complex plane. This can create confusion when gluing surfaces to~form the star product. Gluing the right  half~of~the  string of $A$ to the left half of the string of $B$ corresponds to gluing the half-disks together in the opposite order. 
\begin{center}\resizebox{4in}{1.2in}{\includegraphics{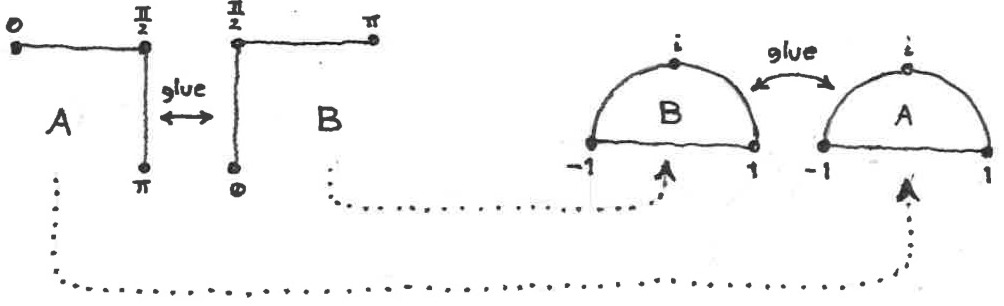}}\end{center}
For this reason, Okawa \cite{Okawa} introduced a different definition of the star product where the left half of the string of the first state is identified with the right half of the string of the second state. This may be called the {\it right handed convention} for the open string star product. This convention finds widespread use in the literature. However, we define the star product using the {\it left handed convention}. The left handed convention appears more commonly in older literature, such as the papers of Witten~\cite{Witten} and Schnabl \cite{Schnabl}. When using the left handed convention the backwards gluing of surfaces can turn into an annoyance. We deal with this by adopting an unconventional visualization of the complex plane. We simply draw the positive real axis so that it increases towards the left; the positive imaginary axis still increases upwards. In this visualization, the left half of the string sits on the left portion of the unit half circle. The star product of surfaces also appears in the natural order:
\begin{center}\resizebox{4in}{1.3in}{\includegraphics{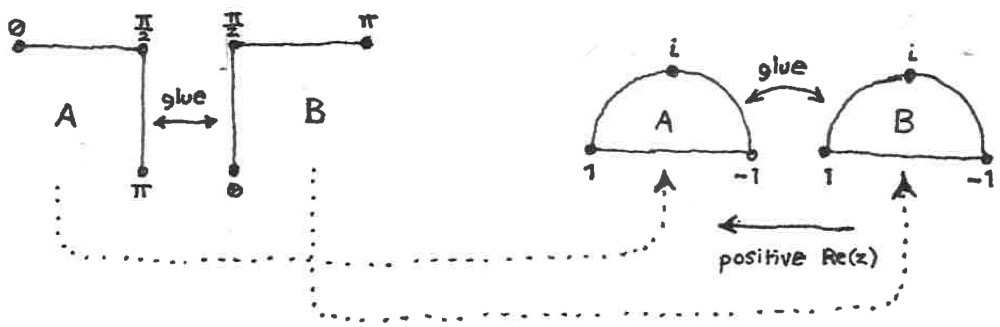}}\end{center}
Note that in this picture the standard orientation of contour integrals is {\it clockwise}. This motivates the term ``left handed." A distinguishing feature of the left handed convention is that the sign of the tachyon field at the tachyon vacuum is positive. In the right handed convention, it is negative. 

\begin{wrapfigure}{l}{.16\linewidth}
\resizebox{1.1in}{.9in}{\includegraphics{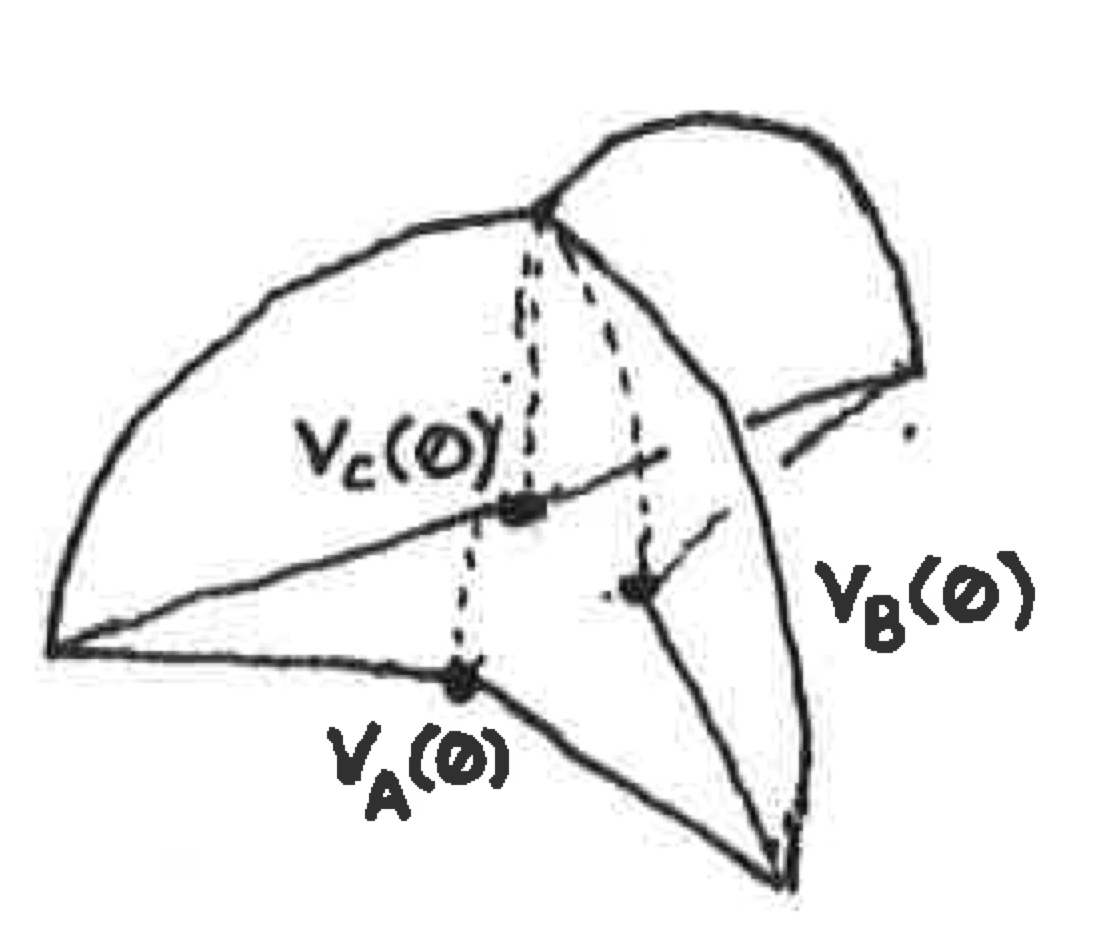}}
\end{wrapfigure}
Our next task is to define the cubic vertex $\Tr(ABC)$. Gluing the half-disks of $A,B$ and $C$ together along quarter-circles, as prescribed by the product and trace, gives a correlation function of three vertex operators on the surface shown left. We transform this to the UHP following a specific visualization which is important in the study of analytic solutions. Each unit half-disk can be expressed in terms of a local coordinate $\xi$ satisfying $\mathrm{Im}(\xi)\geq 0$ and $|\xi|\leq 1$. We perform a conformal transformation to a new  coordinate $z$

\begin{wrapfigure}{l}{.4\linewidth}
\centering
\resizebox{2.7in}{1.5in}{\includegraphics{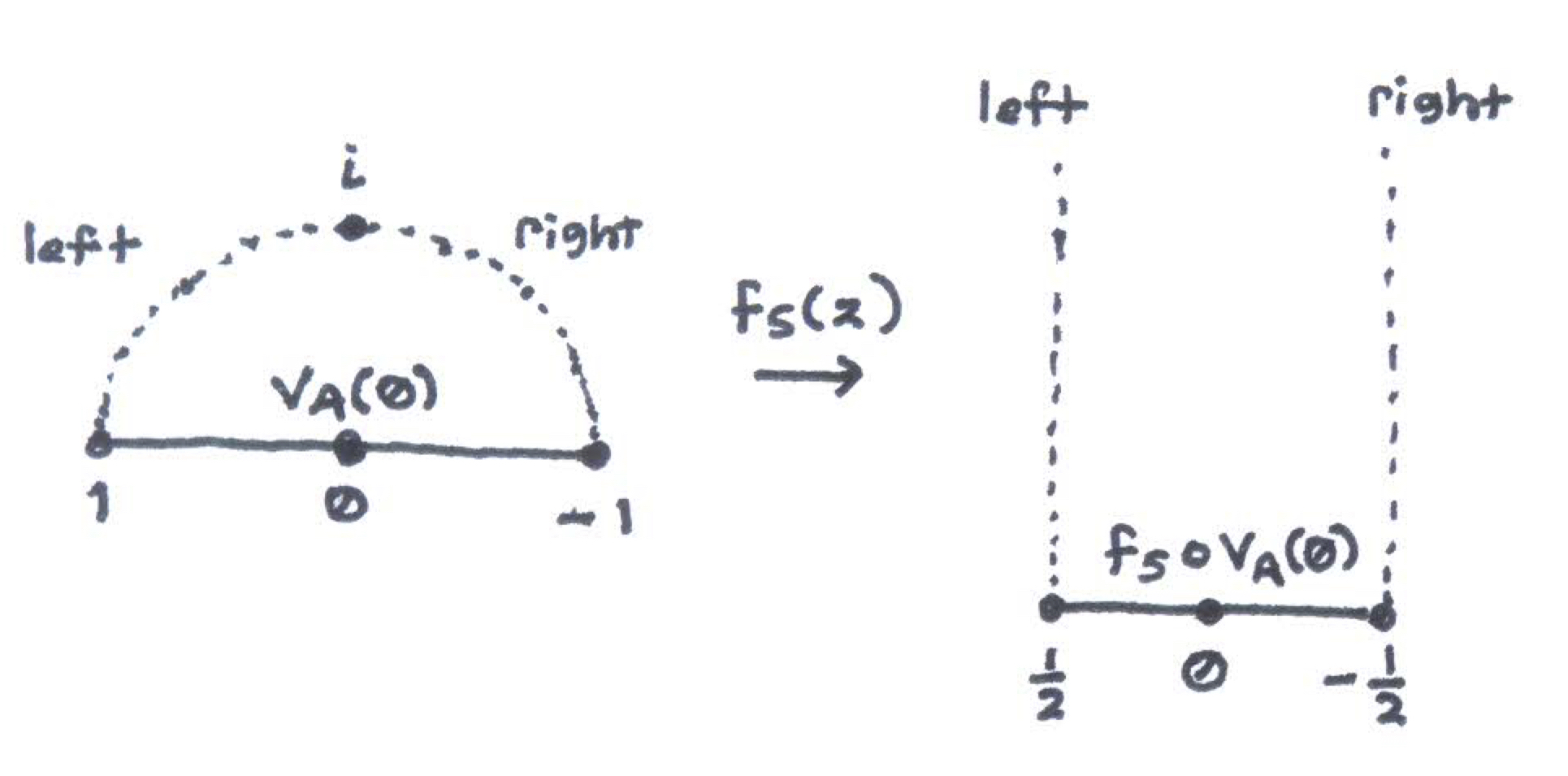}}
\end{wrapfigure}
\noindent 
\begin{equation}z=f_\mathcal{S}(z)=\frac{2}{\pi}\tan^{-1} \xi.\end{equation}
The image of the half-disk in this coordinate system is given by a semi-infinite vertical strip  $\mathrm{Im}(z)\geq 0$ and $\frac{1}{2}\geq \mathrm{Re}(z)\geq -\frac{1}{2}$. This is called the {\it sliver coordinate frame}; the conformal transformation $f_\mathcal{S}(z)$ is called the {\it sliver coordinate map}. The interval $[1,-1]$ on the real axis of the half-disk is mapped to the interval \ $[1/2,-1/2]$ \ on the real axis of the vertical strip; the left  half of the string $e^{i\sigma},\ \sigma\in[0,\pi/2]$ is mapped to the positive facing vertical edge of the strip $\frac{1}{2}+iy,\ y\geq 0$ and the right half of the string $e^{i(\pi-\sigma)},\ \sigma\in[0,\pi/2]$ is mapped to the negative facing vertical edge of the strip $-\frac{1}{2}+iy,\ y\geq 0$. The worldsheet path integral on the half-disk and on the semi-infinite vertical strip define the same Schr{\"o}dinger functional provided that the boundary conditions on the half-circle correspond to those on the vertical lines. Specifically, if the string coordinate takes a certain value at an angle $\sigma$ on the left portion of the unit half circle, it should take the same value at a point $y$ above the real axis on the positive facing boundary of the strip, where $y$ and $\sigma$ are related by
\begin{equation}\frac{1}{2}+iy = \frac{2}{\pi}\tan^{-1}e^{i\sigma},\end{equation}
which leads to
\begin{equation}\mathrm{gd}(\pi y) = \sigma,\end{equation}
where $\mathrm{gd}$ is the Gudermannian function
\begin{equation}\mathrm{gd}x  = 2\tan^{-1}\left(\tanh\frac{x}{2}\right).\end{equation}
As $\sigma$ ranges from $0$ to $\pi/2$, $y$ ranges from $0$ to infinity. The Gudermannian function is known for its relation to the Mercator projection; the relation between the angle with respect to the equator and vertical displacement on the map is the same as that between the angle $\sigma$ on the unit half circle and the coordinate $y$ on the edge of the strip. The midpoint $\sigma=\pi/2$ plays the role of the ``north pole," and is mapped to $+i\infty$ in the sliver coordinate frame. When using the doubling trick, the unit half-disk is replaced with a holomorphic copy of the unit disk $|\xi|\leq 1$; correspondingly, the semi-infinite strip of the sliver coordinate frame is represented as a holomorphic copy of the full infinite strip $-\frac{1}{2}\leq \mathrm{Re}(z)\leq\frac{1}{2}$.

The sliver coordinate frame makes it natural to visualize the star product in terms of gluing strips. To find the product $AB$, we glue the right edge of the strip of $A$ to the left edge of
 \begin{wrapfigure}{l}{.35\linewidth}
\centering
\resizebox{2.4in}{1.5in}{\includegraphics{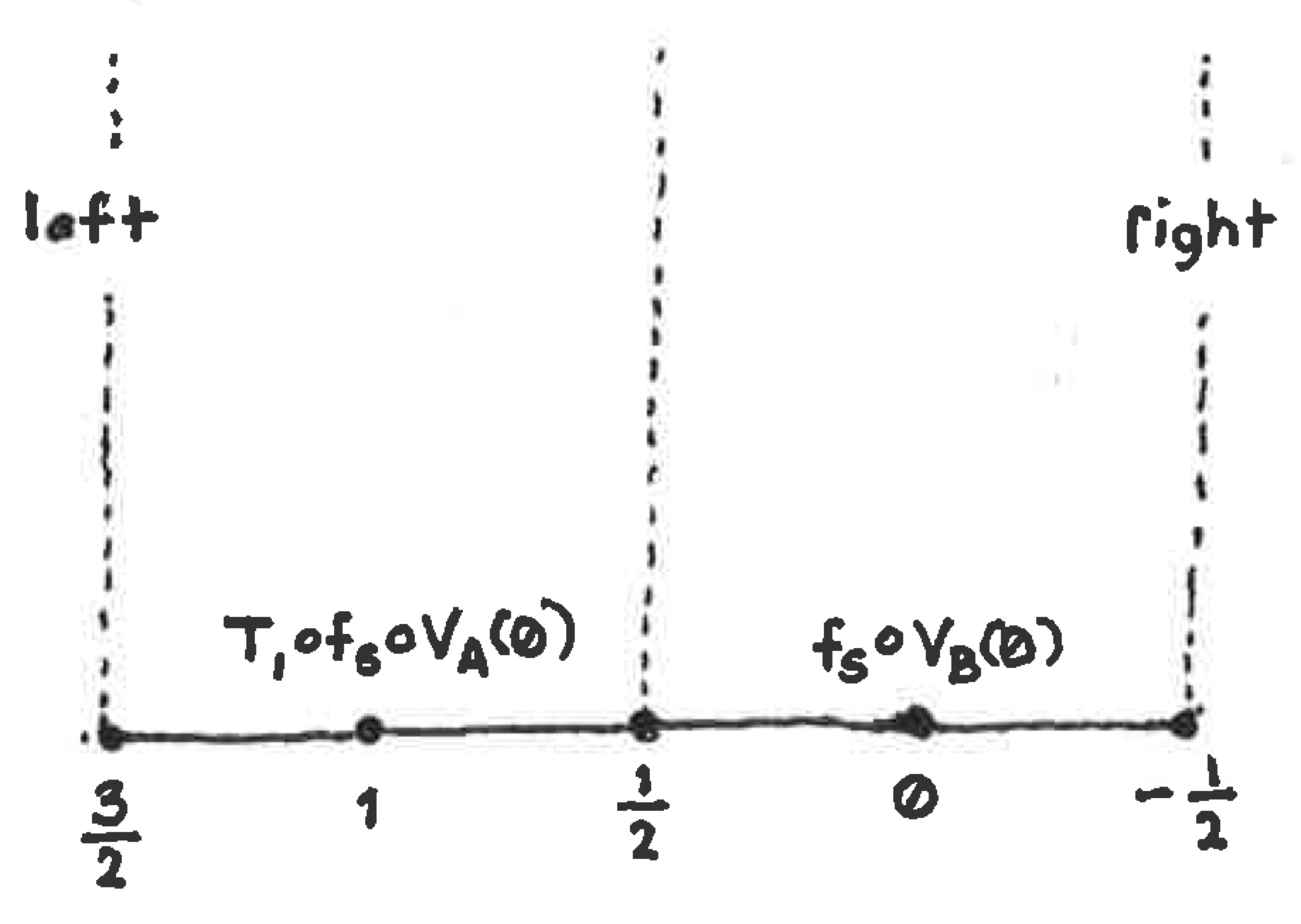}}
\end{wrapfigure}
the strip of $B$. This creates a semi-infinite strip of width $2$ carrying operator insertions
\begin{equation}\big (T_1\circ f_\mathcal{S}\circ V_A(0)\big)\big( f_\mathcal{S}\circ V_B(0)\big),\end{equation}
where $T_a$ is the translation map
\begin{equation}T_a(z) = z+a,\label{eq:transTa}\end{equation} 
and we fix the origin on the double strip to coincide with the location of $V_B$. Imposing the appropriate  boundary conditions on the left and right edges of the double strip and performing the  worldsheet path integral in the interior defines the Schr{\"o}dinger functional of the product $AB$. It is worth pointing out that the vertex operator of the state $AB$ is nonlocal. It effectively inserts a whole new piece of surface between $1$ and $0$ and places vertex operators at the edges of this region. Therefore the star product does not multiply inside the subspace of Fock space states.

Now consider the 3-string vertex $\Tr(ABC)$. To compute this, we place the strips of $A,B$ and $C$ side by side to form a strip of width $3$ with insertions
\begin{equation}\big(T_2\circ f_\mathcal{S}\circ V_A(0)\big)\big(T_1\circ f_\mathcal{S}\circ V_B(0)\big)\big(f_\mathcal{S}\circ V_C(0)\big).\end{equation}

\begin{wrapfigure}{l}{.5\linewidth}
\centering
\resizebox{3.5in}{1.3in}{\includegraphics{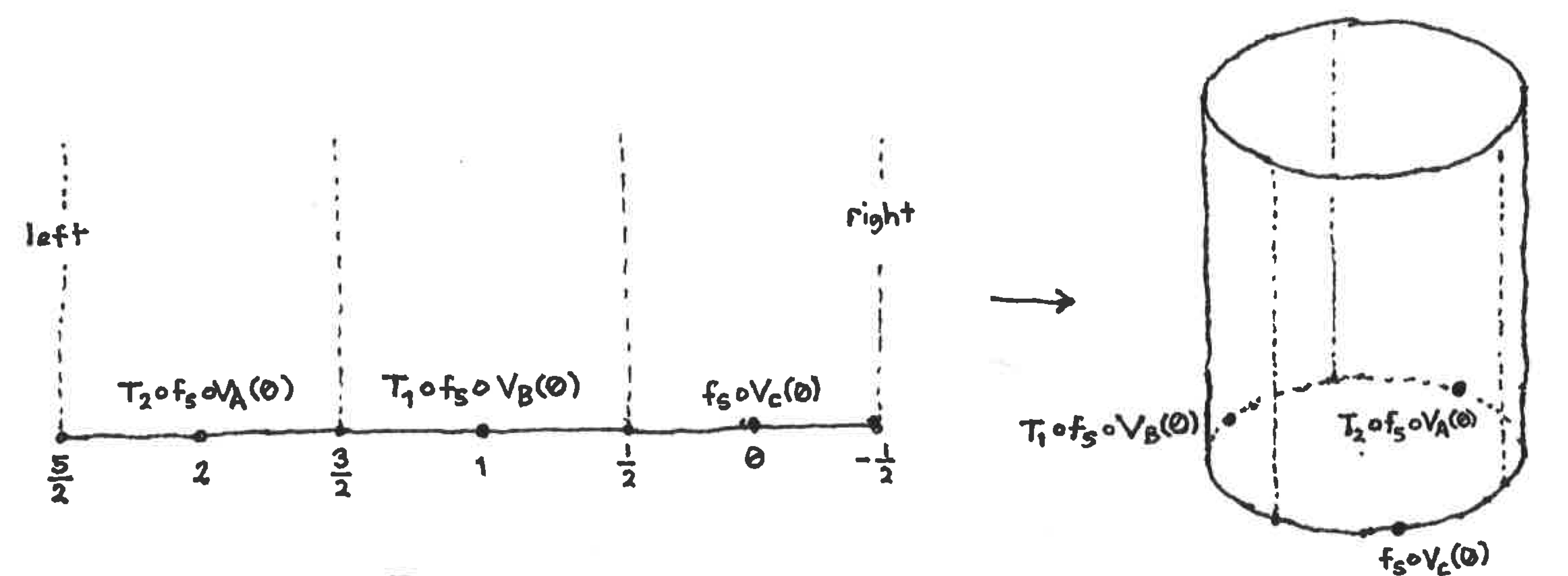}}
\end{wrapfigure}
\noindent The trace then glues the left and right edges of this strip to form a correlation function on a cylinder of circumference $3$. A cylinder of circumference $L$ can be mapped into the upper half plane using the conformal transformation
\begin{equation}C_L^{-1}(z) = \frac{L}{\pi}\tan\frac{\pi z}{L}.\label{eq:CL}\end{equation}
Correlation functions on a cylinder of circumference $L$ will be denoted $\langle ...\rangle_{C_L}$, and can be defined in terms of correlation functions on the UHP:
\begin{equation}\langle...\rangle_{C_L} = \langle C_L^{-1}\circ(...)\rangle_\mathrm{UHP}.\end{equation}
This gives an explicit definition of the cubic vertex in terms of the correlation function
\begin{equation}\Tr(ABC) = \Big\langle\big(T_2\circ f_\mathcal{S}\circ V_A(0)\big)\big(T_1\circ f_\mathcal{S}\circ V_B(0)\big)\big(f_\mathcal{S}\circ V_C(0)\big)\Big\rangle_{C_3}.\end{equation}
In a similar way, the 2-string vertex (a.k.a. the BPZ inner product) can be written as a correlation function on a cylinder of circumference $2$,
\begin{equation}\Tr(AB) = \langle A,B\rangle = \Big\langle \big(T_1\circ f_\mathcal{S}\circ V_A(0)\big)\big(f_\mathcal{S}\circ V_B(0)\big)\Big\rangle_{C_2},\end{equation}
and the 1-string vertex is a correlation function on a cylinder of circumference 1,
\begin{equation}\Tr(A) = \Big\langle f_\mathcal{S}\circ V_A(0)\Big\rangle_{C_1}.\end{equation}
This generalizes in the obvious way to the trace of a product of any number of string fields.

Finally, we would like to give a definition of the star product which does not assume the Schr{\"o}dinger representation. This is important both as a matter of practical computation, and also because the star product should be defined for arbitrary BCFTs where the Schr{\"o}dinger representation might not be available. Suppose that we want to express the string field in a basis of states $|\phi_i\rangle$, where $i$ ranges over an index set which may include discrete and continuous labels. Typically we are interested in a Fock space basis. We introduce a dual basis $|\phi^i\rangle$ with the property that 
\begin{equation}\langle\phi^i,\phi_j\rangle = \delta^i_j.\end{equation}
We operate on the star product $AB$ with a resolution of the identity,
\begin{equation}1=\sum_i |\phi_i\rangle\langle (\phi^i)^\bigstar|,\end{equation}
to obtain 
\begin{eqnarray}
AB \lineup = \sum_{i}|\phi_i\rangle\langle(\phi^i)^\bigstar| AB\rangle\nonumber\\
\lineup =\sum_{i}|\phi_i\rangle \langle \phi^i, AB\rangle\nonumber\\
\lineup = \sum_{i}|\phi_i\rangle\Tr(\phi^i AB)\nonumber\\
\lineup =  \sum_{i}|\phi_i\rangle\Big\langle\big(T_2\circ f_\mathcal{S}\circ \phi^i(0)\big)\big(T_1\circ f_\mathcal{S}\circ V_A(0)\big)\big(f_\mathcal{S}\circ V_B(0)\big)\Big\rangle_{C_3}.
\label{eq:starC3}
\end{eqnarray}
In this way the star product is characterized in terms of 3-point functions on a cylinder of circumference 3. We have  thus completed the task of defining all basic operations of the theory in terms of BCFT correlation functions. Still, the right hand side of \eq{starC3} has a lot of 3-point functions. Depending on the setting there may be special techniques to ease their evaluation.  In the oscillator basis, the  star product may be evaluated using the squeezed state oscillator vertex \cite{Gross1,Gross2}. In a Verma module basis, the star  product  may be  evaluated with  the  help  of  (Virasoro) conservation laws \cite{RZ}. However, \eq{starC3} will not be a very useful definition of the star product for our purposes. The reason is that it is really designed for computing the star product of two Fock space states. Since Fock space states can be used to form a basis, this is in principle enough, but in our development this way of thinking is not sufficiently flexible. To illustrate this  point, suppose we wish to compute the star product of a Fock space state $A$ with another state which itself is given by the star product of two Fock space states $BC$. Using \eq{starC3} the result would be given as
\begin{eqnarray}
ABC\lineup=\sum_i\sum_j |\phi_i\rangle \Big\langle\big(T_2\circ f_\mathcal{S}\circ \phi^i(0)\big)\big(T_1\circ f_\mathcal{S}\circ V_A(0)\big)\big(f_\mathcal{S}\circ \phi_j(0)\big)\Big\rangle_{C_3}\nonumber\\
\lineup\ \ \ \ \ \ \ \ \ \ \ \ \ \ \ \ \ \ \ \ \ \ \ \ \ \ \ \times\Big\langle\big(T_2\circ f_\mathcal{S}\circ \phi^j(0)\big)\big(T_1\circ f_\mathcal{S}\circ V_B(0)\big)\big(f_\mathcal{S}\circ V_C(0)\big)\Big\rangle_{C_3}.
\label{eq:starstarC3C3}
\end{eqnarray}
Here we are tasked with computing a sum over intermediate basis states $|\phi_j\rangle$, and associativity of the product is not at all manifest. On the other hand, by gluing strips side-by-side it is clear that the answer should be given by a correlation function on a cylinder of circumference 4:
\begin{equation}
ABC = \sum_i|\phi_i\rangle\Big\langle\big(T_3\circ f_\mathcal{S}\circ \phi^i(0)\big)\big(T_2\circ f_\mathcal{S}\circ V_A(0)\big)\big(T_1\circ f_\mathcal{S}\circ V_B(0)\big)\big(f_\mathcal{S}\circ V_C(0)\big)\Big\rangle_{C_4}.
\label{eq:starstarC4}
\end{equation}
Explicitly demonstrating the equivalence of \eq{starstarC3C3} and \eq{starstarC4} is not trivial. The  result follows from  the so-called  {\it generalized gluing and resmoothing theorem}, formulated in \cite{Leclair1,Leclair2} and further generalized to arbitrary CFTs in \cite{SchwarzSen}. We will take these things for granted and deduce star products using the intuitive operation of gluing strips, as suggested by the Schr{\"o}dinger representation.

There is an additional set of axioms in Witten's SFT relating to reality of the action. This assumes the existence of a conjugation in the space of string fields satisfying the following identities
\begin{eqnarray}
(QA)^\ddag\lineup = (-1)^{|A|+1}QA^\ddag,\\
(AB)^\ddag \lineup = B^\ddag A^\ddag,\\
\Tr(A)^*\lineup = \Tr(A^\ddag).
\end{eqnarray}
As the notation suggests, this structure can be identified with {\it reality conjugation} of the BCFT state space.
\begin{exercise}Show that reality conjugation satisfies these relations.\end{exercise}
\noindent From this it follows that the action is a real number if the dynamical string field satisfies the reality condition
\begin{equation}\Psi^\ddag=\Psi.\end{equation}
Moreover, nonlinear infinitesimal gauge transformations preserve the reality condition if the gauge parameter satisfies $\Lambda^\ddag = -\Lambda$. The reality condition at the nonlinear level is therefore the same as what we have already discussed for the linearized equations of motion. Incidentally, from \eq{Sch_conj} we can deduce the reality conjugate of the Schr{\"o}dinger functional in the split string representation:
\begin{equation}A^\ddag[l(\sigma),r(\sigma)]=  A[r(\sigma),l(\sigma)]^*.\end{equation}
This is analogous to the conjugate transpose of a matrix. 

In principle, one final structure is needed to complete the definition of Witten's open bosonic string field theory---the {\it open string star algebra}. This should be a space of string fields which is closed under star multiplication and  where the Chern-Simons axioms are certain to hold. Ideally, it should be a topological algebra which  allows discussion of continuity and convergence. Unfortunately, the definition of the open string star algebra is not known, though it is common to refer to it as an informal notion. This  question has greater importance than one might expect. A major aspect of the study of analytic solutions is understanding the space of string fields in controlled settings. The fact that it is possible to perform unambiguous and physically meaningful computations in Witten's SFT is evidence that some ``open string star algebra" is holding everything together.

Let us briefly describe physical observables. They can be categorized as~follows:
\begin{description}
\item{(1)} {\it The space of solutions modulo gauge transformation.} This is a rather abstract notion, but a special instance deserves more discussion here. This is the space of inequivalent  linearized fluctuations around a classical solution. Consider a classical solution $\Psi_*$ and expand the string field 
\begin{equation}\Psi = \Psi_*+\varphi,\end{equation}
where $\varphi$ is a fluctuation of $\Psi_*$. The action can be expanded
\begin{equation}S[\Psi_*+\varphi] = \underbrace{S[\Psi_*]}_{\mathrm{constant}}+S_{\Psi_*}[\varphi],\end{equation}
where $S_{\Psi_*}[\varphi]$ takes the form
\begin{equation}S_{\Psi_*}[\varphi] = \Tr\left(\frac{1}{2}\varphi Q_{\Psi_*}\varphi+\frac{1}{3}\varphi^3\right),\end{equation}
and
\begin{equation}Q_{\Psi_*}=Q+[\Psi_*,\cdot].\end{equation}
One can show that $Q_{\Psi_*}$ is nilpotent due to the equations of motion of $\Psi_*$, and satisfies the same axioms as $Q$. From this it follows that the linearized equations of motion for the fluctuation field are
\begin{equation}Q_{\Psi_*}\varphi=0.\end{equation}
Solutions must be identified modulo linearized gauge transformations
\begin{equation}\varphi'= \varphi +Q_{\Psi_*}\Lambda.\end{equation}
The space of physical linearized fluctuations of $\Psi_*$ is then given by the cohomology of $Q_{\Psi_*}$ at ghost number $1$. 
\item{(2)} {\it Scattering amplitudes}. Computing scattering amplitudes generally requires fixing a gauge to determine the propagator. The most conventional choice is {\it Siegel gauge}
\begin{equation}b_0\Psi = 0,\end{equation}
where the propagator (around the perturbative vacuum) takes the form
\begin{equation}\frac{b_0}{L_0} = b_0 \int_0^\infty dt\, e^{-t L_0}.\end{equation}
The Siegel gauge propagator is often visualized in the conformal frame $w=\ln \xi$, where the coordinate $\xi$ on the unit half-disk is mapped to a coordinate $w$ on a semi-infinite horizontal strip of height~$\pi$ 
\begin{equation}\mathrm{Re}(w)\leq 0, \ \ \ 0\leq \mathrm{Im}(w) \leq \pi .\end{equation}
The image of the unit half circle is the line segment between $0$ and $i\pi$ in this coordinate system. 
The operator $e^{-t L_0}$ glues a horizontal strip to this line segment of height $\pi$ and length $t$. The propagator further integrates over the length $t$ of the strip, and inserts a vertical contour integral of the $b$ ghost representing $b_0$. We can then derive scattering amplitudes through Feynman diagrams by gluing the propagator strips together through cubic vertices. The length $t$ amounts to a coordinate on part of the moduli space of the Riemann surface represented by a diagram where the propagator appears, and $b_0$ provides the measure.  Following the discovery of Schnabl's solution, there has also been some effort to understand scattering amplitudes in {\it Schnabl gauge} \cite{Fuji,RZVeneziano,KSZ,KZloop}, defined by the condition
\begin{equation}\mathcal{B}_0\Psi = 0,\ \ \ \ \ \ \mathcal{B}_0 = f_\mathcal{S}\circ b_0.\end{equation}
This leads to much simpler expressions for off-shell amplitudes and more explicit parameterizations of the moduli space, but there are subtleties with defining the propagator which have not been fully understood. It is also possible to consider scattering amplitudes around a nontrivial classical solution $\Psi_*$, where the Siegel gauge propagator takes the form
\begin{equation}\frac{b_0}{[Q_{\Psi_*},b_0]} = b_0 \int_0^\infty dt\, e^{-t [Q_{\Psi_*},b_0]}.\end{equation}
Typically it is hard to say much concrete about perturbation theory around classical solutions due to the complicated form of the propagator. 

\vspace{-.15cm}

\ \ \ \ A special amplitude is the {\it closed string tadpole}, representing emission or absorption of a single on-shell closed string off a D-brane. This amplitude does not contain moduli, and therefore does not involve propagators. The procedure for extracting it is indirect, but the surprisingly simple result was given by Ellwood \cite{tadpole}. It is determined by  what is now often called the {\it Ellwood invariant}
\begin{equation}\Tr_\mathcal{V}(\Psi),\end{equation}
where $\Tr_\mathcal{V}$ denotes the trace accompanied by an insertion of a BRST invariant closed string vertex operator $\mathcal{V}(z,\zbar)$ of weight $(0,0)$ at the midpoint. Concretely, if $V_\Psi(0)$ is the vertex operator for the state $\Psi$ on the unit half-disk, the Ellwood invariant can be computed as a correlator on the cylinder of circumference 1:
\begin{equation}\Tr_\mathcal{V}(\Psi) = \Big\langle \mathcal{V}(i\infty,\overline{i\infty})f_\mathcal{S}\circ V_\Psi(0)\Big\rangle_{C_1}.\end{equation}
The closed string vertex operator is inserted on the ``top" of the cylinder at $i\infty$. Inserting the closed string vertex operator at $i\infty$ does not break the rotational symmetry of the cylinder, and in this way we can see that $\Tr_\mathcal{V}(\cdot)$ is cyclic; moreover, since $\mathcal{V}$ is BRST invariant, $\Tr_\mathcal{V}(\cdot)$ is vanishing on BRST exact states. From this it follows that the Ellwood invariant is unchanged by infinitesimal gauge transformation of $\Psi$; it is a gauge invariant observable. Suppose we have a classical solution $\Psi_*$  describing $\BCFT_*$. The Ellwood invariant evaluated on $\Psi_*$  is believed to be equal  to the shift in the closed string tadpole amplitude between the background $\BCFT_*$ and perturbative vacuum:
\begin{equation}\Tr_\mathcal{V}(\Psi_*) = \mathcal{A}_*(\mathcal{V})-\mathcal{A}_0(\mathcal{V}).\end{equation} 
The closed string tadpole is given by a correlation function in the matter component of the BCFT on the unit disk:
\begin{equation}
\mathcal{A}(\mathcal{V}) = \frac{1}{2\pi i}\Big\langle V^\text{m}(0,0)\Big\rangle^\text{m}_\mathrm{disk},
\end{equation}
where the matter vertex operator $V^\text{m}$ is a primary of weight $(1,1)$ related to $\mathcal{V}$ through
\begin{equation}\mathcal{V} = c\overline{c}V^\text{m}.\end{equation}
The Ellwood invariant has been generalized in a couple of ways to give information about the boundary state of the BCFT represented by the classical solution \cite{boundary1,boundary2}. This in principle provides a direct map from a classical solution into BCFT data, which in the approach of \cite{boundary2} is useful even for approximate solutions derived in level truncation. The discovery of the Ellwood invariant and its applications is one of the most important developments in the study of analytic solutions.
\item{(3)} {\it The classical action}. The action evaluated on a solution is gauge invariant, but typically its value is divergent due to the infinite volume of the D-brane. However, for time independent solutions the action is equal to minus the energy of the solution times the volume of the time coordinate
\begin{equation}S = -E\cdot\mathrm{vol}_{X^0}.\end{equation}
\item{(4)} {\it Other observables} include boundary condition changing projectors derived from singular  gauge transformations, described in subsection \ref{subsec:singularGT}. In superstring field theory there we expect to find gauge invariants representing charges of topological solitons, though presently it is unclear what form they take. 
\end{description}
The most important classical solution in open bosonic SFT is the tachyon vacuum, $\Psi_\tv$. Sen's conjectures make the following prediction about the above gauge invariants:
\begin{description}
\item{(1)} Since the tachyon vacuum describes a configuration without D-branes or open strings, the cohomology of $Q_{\Psi_\tv}$ should be empty; all linearized fluctuations of the tachyon vacuum are pure gauge.
\item{(2)} Since there are no D-branes around the tachyon vacuum, the closed string tadpole should vanish. Therefore, the Ellwood invariant evaluates to minus the tadpole amplitude around the perturbative vacuum:
\begin{equation}\Tr_\mathcal{V}(\Psi_\tv) = -\mathcal{A}_0(\mathcal{V}).\end{equation}
\item{(3)} The action divided by the volume should give the tension of the reference D-brane:
\begin{equation}\frac{S[\Psi_\tv]}{\mathrm{vol}} = \frac{1}{2\pi^2}.\end{equation}
\end{description}

\begin{exercise}
The zero momentum sector of the string field can describe translationally invariant vacua of SFT. As an approximation to the full string field in this sector, consider the zero momentum tachyon state 
\begin{equation}T c_1|0\rangle.\end{equation}
Substitute this into the action to determine the resulting approximation to the tachyon effective potential. Show that the potential has a stationary point with energy density 
\begin{equation}E = -\frac{2^{11}}{3^{10}}.\end{equation}
This is the first approximation to the tachyon vacuum in the level truncation scheme. Compare the energy density to the value predicted by Sen's conjecture.
\end{exercise}

\section{Lecture 2: Algebraic setup}

Having described the basics of Witten's open bosonic string field theory, we now turn to the main topic of these lectures, which is solving the equations of motion
\begin{equation}Q\Psi +\Psi^2 = 0.\end{equation}
In the analytic approach, the idea is to look for solutions within a tractable but sufficiently rich subalgebra of states. For the solutions considered here, the subalgebras are always defined by adding certain operator insertions to the subalgebra of {\it wedge states}, which we describe now. 

\subsection{Wedge states} 
\label{subsec:wedge}

The simplest state in the BCFT is the $SL(2,\mathbb{R})$ vacuum
\begin{equation}\Omega = |0\rangle.\end{equation}
In the sliver frame, $\Omega$ is represented by a semi-infinite strip of width $1$ carrying no operator insertions (or, equivalently, an insertion of the identity operator). We can multiply $\Omega$ with itself to give the state $\Omega^2$. This corresponds to gluing two semi-infinite strips of width 1 side-by-side to 
\begin{wrapfigure}{l}{.5\linewidth}
\centering
\resizebox{3.5in}{1.3in}{\includegraphics{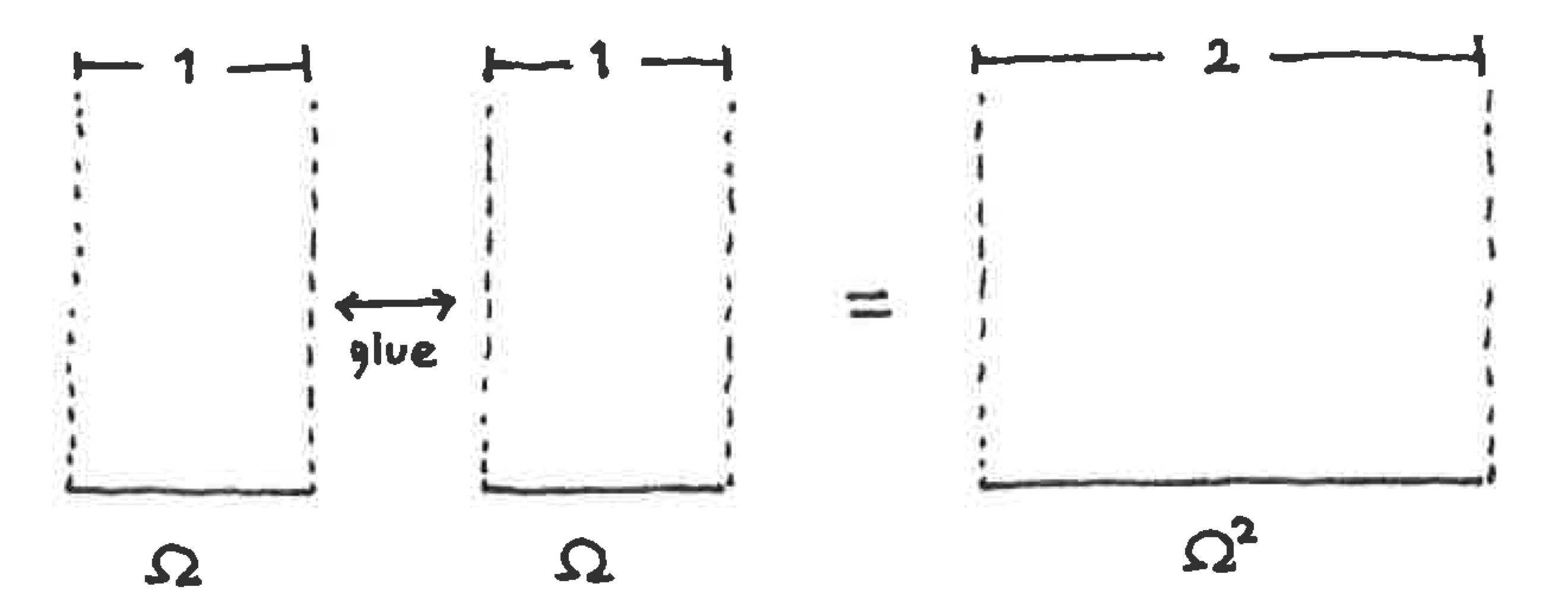}}
\vspace{-1cm}
\end{wrapfigure}
form a semi-infinite strip of width $2$. Now it might appear that a strip of width 2 is not really different from a strip of width 1; they can be related by conformal transformation, specifically a scaling by a factor of $\frac{1}{2}$. The point, however, is that in this conformal transformation we must account for the boundary conditions of the path integral on the left and right vertical edges of the strip, which represent the left and right halves of the string in the Schr{\"o}dinger functional. For the free boson subject to Neumann boundary conditions, we have seen how to represent the $SL(2,\mathbb{R})$ vacuum as a functional of $x(\sigma)$, which in the split string representation can be written as a functional of the left and right halves of the string:
\begin{equation}\Omega[x(\sigma)] = \Omega[l(y),r(y)].\end{equation}
For convenience, we parameterize the left and right halves of the string in terms of the height $y$ on a vertical edge of the strip
\begin{equation}
{l(y) = x(\sigma),\phantom{\Big)}\ \ \ \ \ \atop  r(y) = x(\pi-\sigma), \phantom{\Big)}} \ \ \ \ \mathrm{for}\ \ \sigma\in[0,\pi/2]\ \ \mathrm{and}\ \ \mathrm{gd}(\pi y)=\sigma.
\end{equation}
Now $l(y)$ gives the boundary condition for the path integral at a point $y$ above the real axis on the left vertical edge of the strip of width $1$, while $r(y)$ gives the boundary condition at the corresponding point on the right vertical edge. When we compute $\Omega^2$, the boundary conditions on the left and right edges are the same, but the strip over which we compute the path integral has doubled in width. If we scale by a factor of $\frac{1}{2}$, we are back to a strip of width $1$, but the boundary conditions on the left and right edges have also been scaled. Now the boundary condition at a point $y$ above the left vertical edge of the strip of width $1$ should be $l(2y)$, and similarly on the right edge. This implies that the Schr{\"o}dinger functional of $\Omega^2$ should be related to that of the $SL(2,\mathbb{R})$ vacuum through
\begin{equation}\Omega^2[l(y),r(y)] = \Omega[l(2y),r(2y)].\label{eq:Om2Om}\end{equation}
One might note that $(l(y),r(y))$ and $(l(2y),r(2y))$ actually represent the same curves in spacetime. But they are different as parameterized curves, and the $SL(2,\mathbb{R})$ vacuum functional is not invariant under reparameterizations of $x(\sigma)$. Therefore, $\Omega^2$ is really a different state from $\Omega$.

Continuing, we may construct $\Omega^3$ by gluing three strips of unit width side-by-side; the result is a strip of width $3$. Similarly $\Omega^4$ is a strip of width $4$ and so on for any positive integer $n$. It is clear that there is nothing particularly special about positive integer powers of the $SL(2,\mathbb{R})$ vacuum. We may generalize to any positive real power, defining $\Omega^\alpha$ as a semi-infinite strip of width $\alpha$ containing no operator insertions: \ \ \ \ \ \ \ \ \ \ \ \ \ \ \ \ \ \ \ \ \ \ \ \ \ \ \  \ \ \ \ \ \ \ \ \ \ \ \ \ \ \ \ \ \ \ \ \ \ \ \ \ \ \ \ \ \ \ \ \ \ \ \ \ \ \ \ \ \ \ \ \ \ \ \ \ \ \ \ \ \ \ \ \ \ \ \ \ \ \ \ 
\begin{wrapfigure}{l}{1\linewidth}
\centering
\vspace{-.7cm}
\resizebox{2.2in}{.9in}{\includegraphics{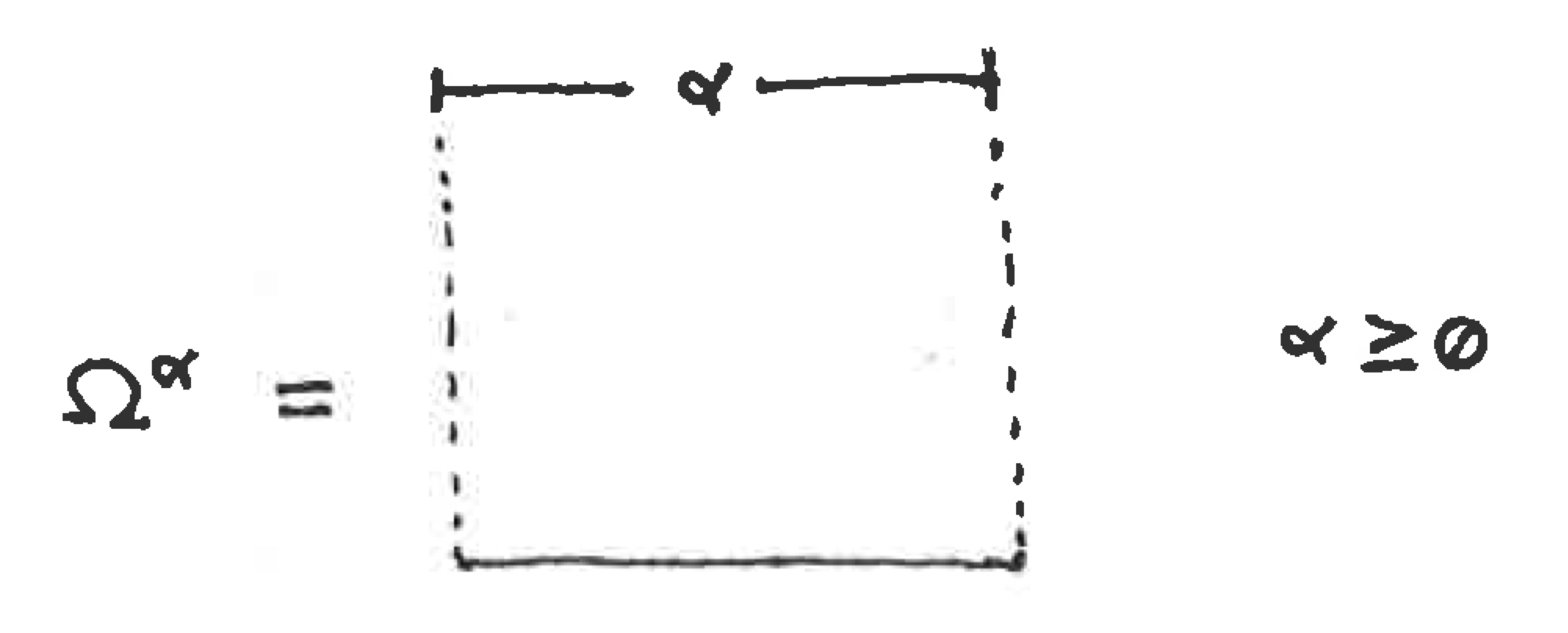}}
\vspace{-.5cm}
\end{wrapfigure}\\ \\ \\ \\ \\
$\Omega^\alpha$ is called a {\it wedge state}, and $\alpha$ is called the {\it wedge parameter} \cite{RZ}. Sometimes $\alpha$ is referred to as the {\it width} of the wedge state. It is immediately clear from gluing strips that multiplication of~wedge

\begin{wrapfigure}{l}{.48\linewidth}
\centering
\resizebox{3.3in}{.8in}{\includegraphics{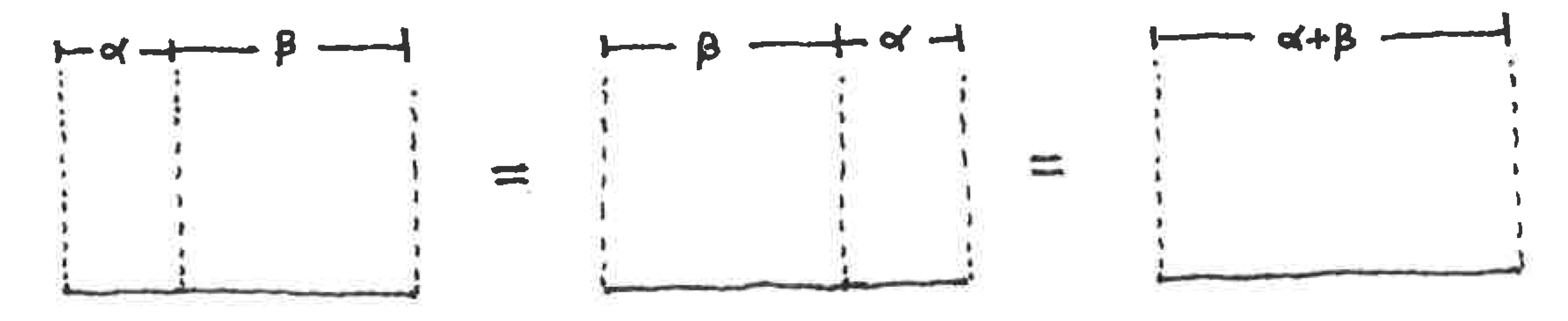}}
\end{wrapfigure} 
\noindent  states is Abelian
\begin{equation}\Omega^\alpha\Omega^\beta = \Omega^\beta\Omega^\alpha = \Omega^{\alpha+\beta}.\end{equation}
Geometrically, the restriction $\alpha\geq 0$ seems natural, but this deserves some comment. From the above discussion of $\Omega^2$ it is clear that all wedge states are related to the $SL(2,\mathbb{R})$ vacuum by a scale transformation of $y$, which amounts to a reparameterization of $\sigma$. This implies that $\Omega^\alpha$ is a Gaussian functional of $x(\sigma)$ for $\alpha\geq 0$. We can analytically continue this functional to complex $\alpha$. If $\mathrm{Re}(\alpha)<0$, it turns out that the functional is an inverted Gaussian $e^{x^2}$, and is therefore not normalizable. If $\mathrm{Re}(\alpha)>0$ but complex, the functional is a Gaussian with complex width, and appears to be normalizable. However, it is not clear how to think about such states in terms of a strip of worldsheet. In most concrete settings it  is enough to assume that the wedge parameter is real and positive. 

The picture of wedge states as semi-infinite strips is closely related to the Schr{\"o}dinger representation. However, we can readily translate to other descriptions. First let us describe them as correlation functions derived by computing the overlap with a test state. In the sliver coordinate frame, a test state can be represented by a semi-infinite strip of unit width carrying a conformally transformed vertex operator. Gluing this strip to the left/right edges of the strip of a wedge state forms a correlation function on the cylinder of circumference $\alpha+1$, where $\alpha$ is the wedge parameter:
\begin{equation}\langle \phi,\Omega^\alpha\rangle = \langle f_\mathcal{S}\circ \phi(0)\rangle_{C_{\alpha+1}}.\label{eq:ws}\end{equation} 
This description adheres to the standard presentation of what is called a {\it surface state}. For present purposes, a surface state $|S\rangle$ is characterized by a surface $\mathscr{S}$ with boundary, a global complex coordinate $z$ on $\mathscr{S}$, and a conformal transformation
\begin{equation}z=f(\xi),\ \ \ \ |\xi|\leq 1, \mathrm{Im}(\xi)>0,\end{equation}
which maps the unit half disk one-to-one into the coordinate $z$ in such a way that the real axis on the half disk maps to the boundary of $\mathscr{S}$. The image of $\xi=0$ on $\mathscr{S}$ is called the {\it puncture}, $\xi$ is referred to as the {\it local coordinate} around the puncture, $f(\xi)$ is called the {\it local coordinate map}, and the image of the unit half-disk in $\mathscr{S}$ is called the {\it local coordinate patch}. The surface state is then defined by a correlation function
\begin{equation}\langle\phi, S\rangle = \langle f\circ\phi(0)\rangle_\mathscr{S},\end{equation}
where operators inside the correlation function are expressed in the coordinate $z$ on $\mathscr{S}$. For wedge states as characterized in \eq{ws}, we take the surface to be the cylinder $C_{\alpha+1}$, the global coordinate $z$ to be the upper half plane with the identification $z\sim z+(\alpha+1)$, and the local coordinate map is $f_\mathcal{S}(\xi)$. The puncture sits at $z=0$, and the local coordinate patch is the semi-
infinite strip on the cylinder between $\mathrm{Re}(z)=-1/2$ and $\mathrm{Re}(z)=1/2$. The cylinder is often the 
most convenient representation of wedge states, but a few other descriptions are useful to know. We may also describe wedge states in the upper half plane 
\begin{equation}\langle\phi,\Omega^\alpha\rangle = \langle f_\text{UHP}^{(\alpha)}\circ\phi(0)\rangle_\text{UHP},\end{equation}
where 
\begin{equation}f_\text{UHP}^{(\alpha)}(\xi) = C_{1+\alpha}^{-1}\circ f_\mathcal{S}(\xi),\end{equation}
\begin{wrapfigure}{l}{.4\linewidth}
\centering
\resizebox{2.9in}{1.6in}{\includegraphics{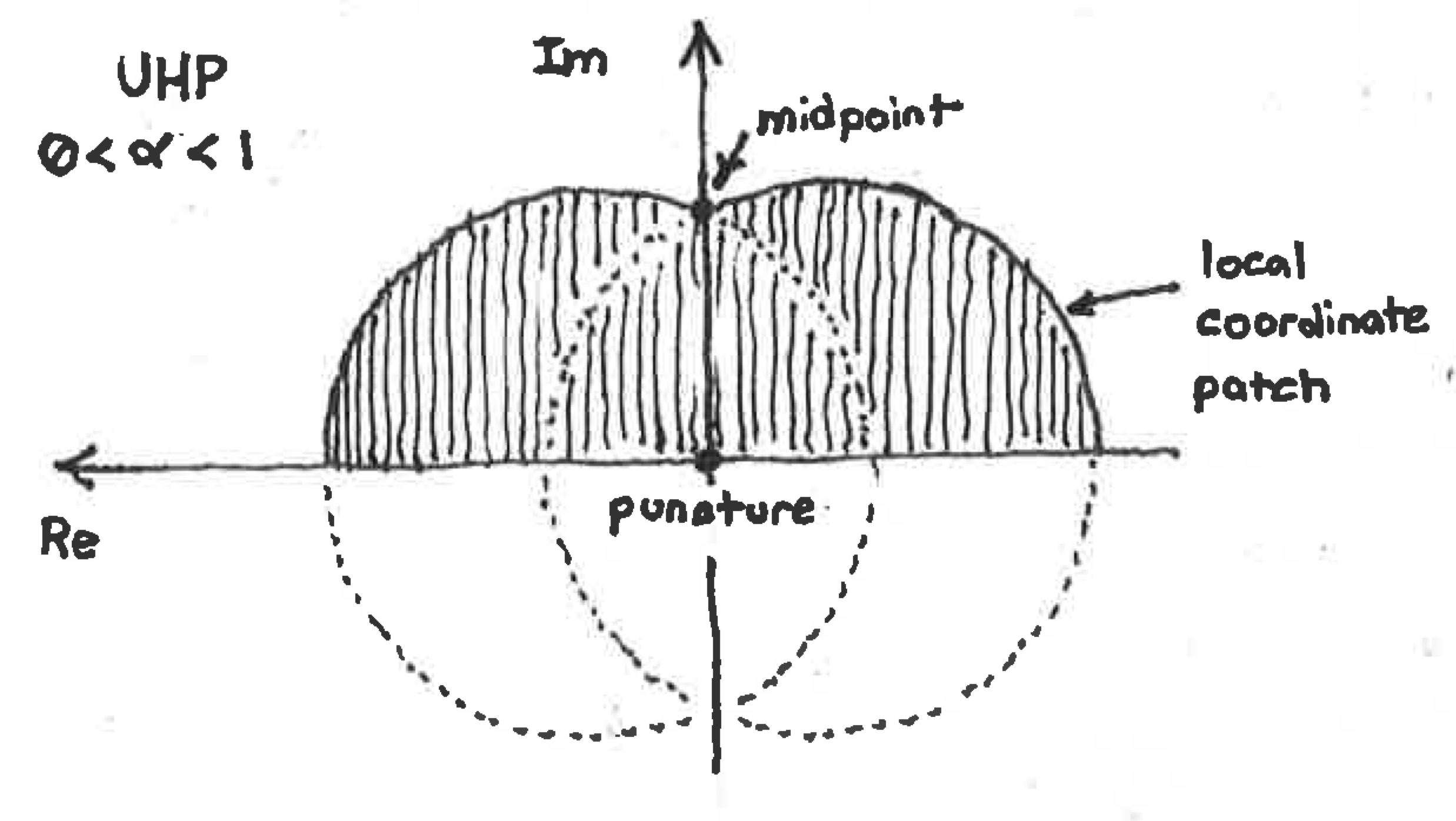}}\\
\resizebox{2.5in}{1.6in}{\includegraphics{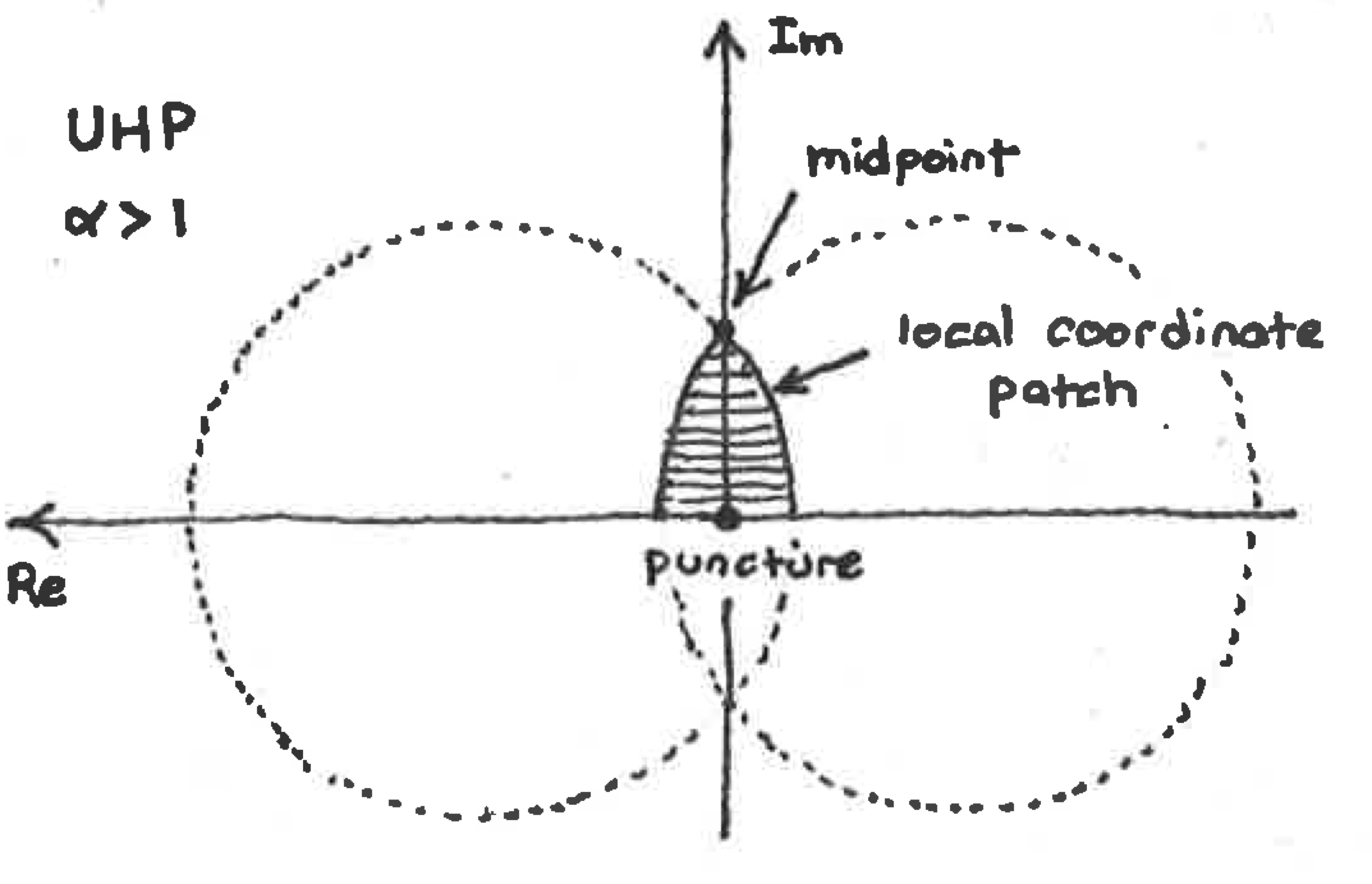}}
\vspace{-1cm}
\end{wrapfigure}
and $C_L^{-1}$ is given in \eq{CL}.
In this representation the surface is simple, but the local coordinate patch is more complicated. It is a region defined by  a pair of circles with centers at 
\begin{equation}c =\pm \frac{\alpha+1}{2\pi}\left(\tan \frac{\pi}{2(\alpha+1)}-\cot\frac{\pi}{2(\alpha+1)}\right),\end{equation}
with a common radius
\begin{equation}R = \frac{\alpha+1}{2\pi}\left(\tan \frac{\pi}{2(\alpha+1)}+\cot\frac{\pi}{2(\alpha+1)}\right).\end{equation}
 If $0<\alpha<1$ the local coordinate patch is the union of the interior of these circles, and if $\alpha>1$ it is the intersection of the interior of these circles. The image of the midpoint is located at the intersection of the circles on the imaginary axis:
\begin{equation}\ \ \ \ \ \ \ \ \ \ \ \ \ \ \ \ \ \ \ \ \ \ \ \ \ \ \ \ \ \ \ \ \ \ \ \ \ \ \ \ f_\text{UHP}^{(\alpha)}(i) = \frac{i(1+\alpha)}{\pi}.\end{equation}

\begin{wrapfigure}{l}{.3\linewidth}
\vspace{0cm}
\centering
\resizebox{1.6in}{1.4in}{\includegraphics{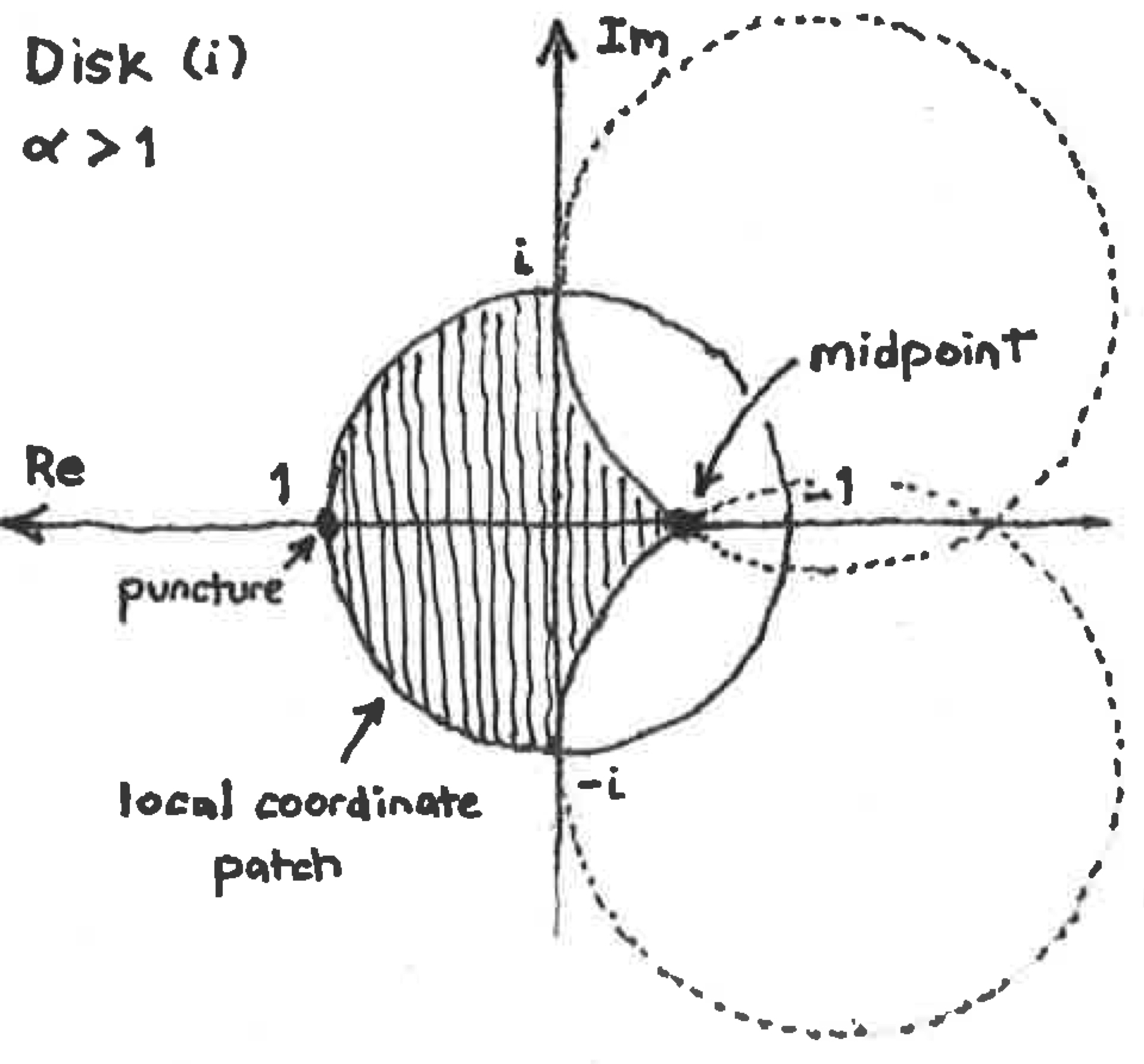}}\\
\resizebox{1.6in}{1.4in}{\includegraphics{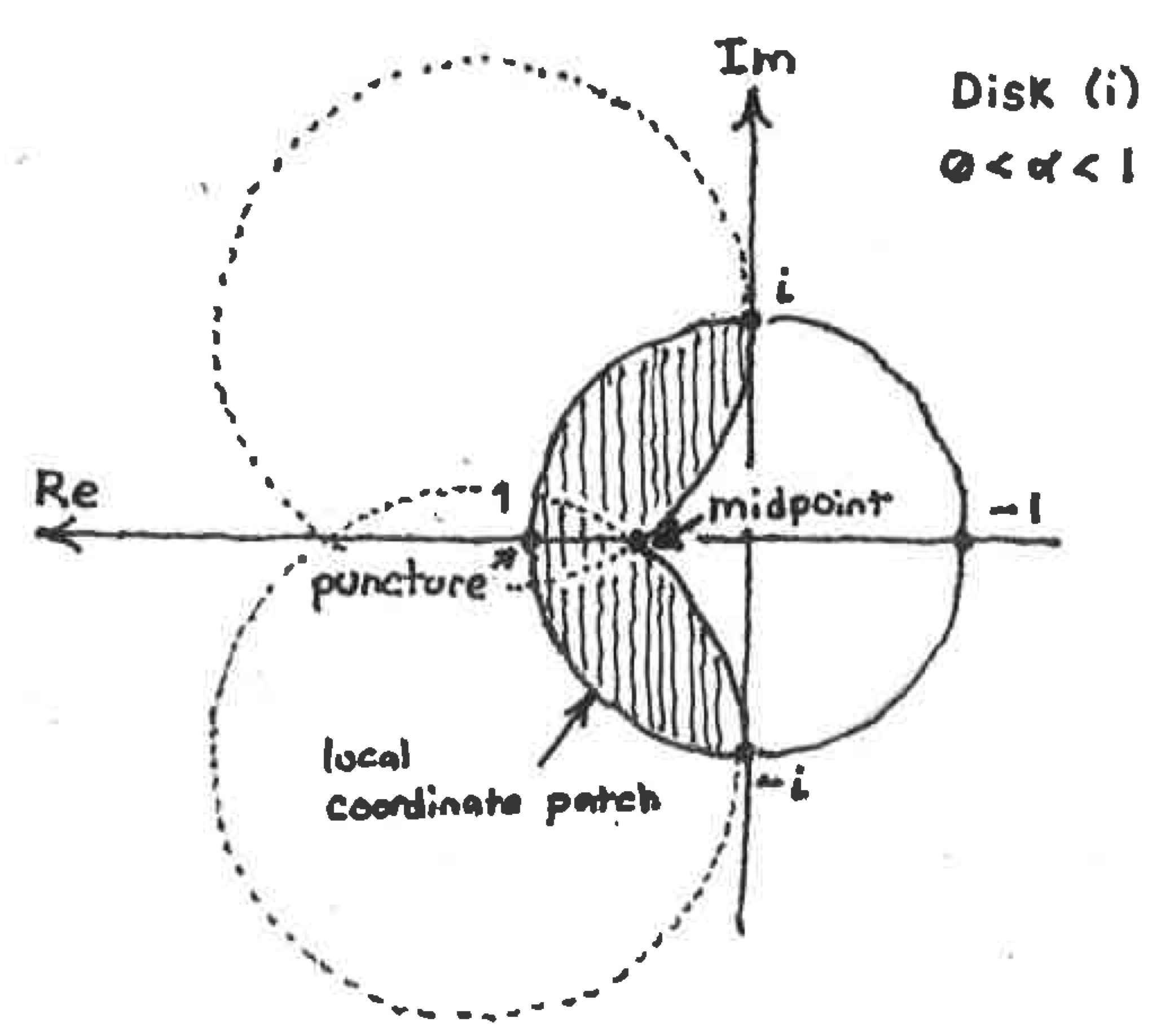}}\\
\resizebox{1.6in}{1.4in}{\includegraphics{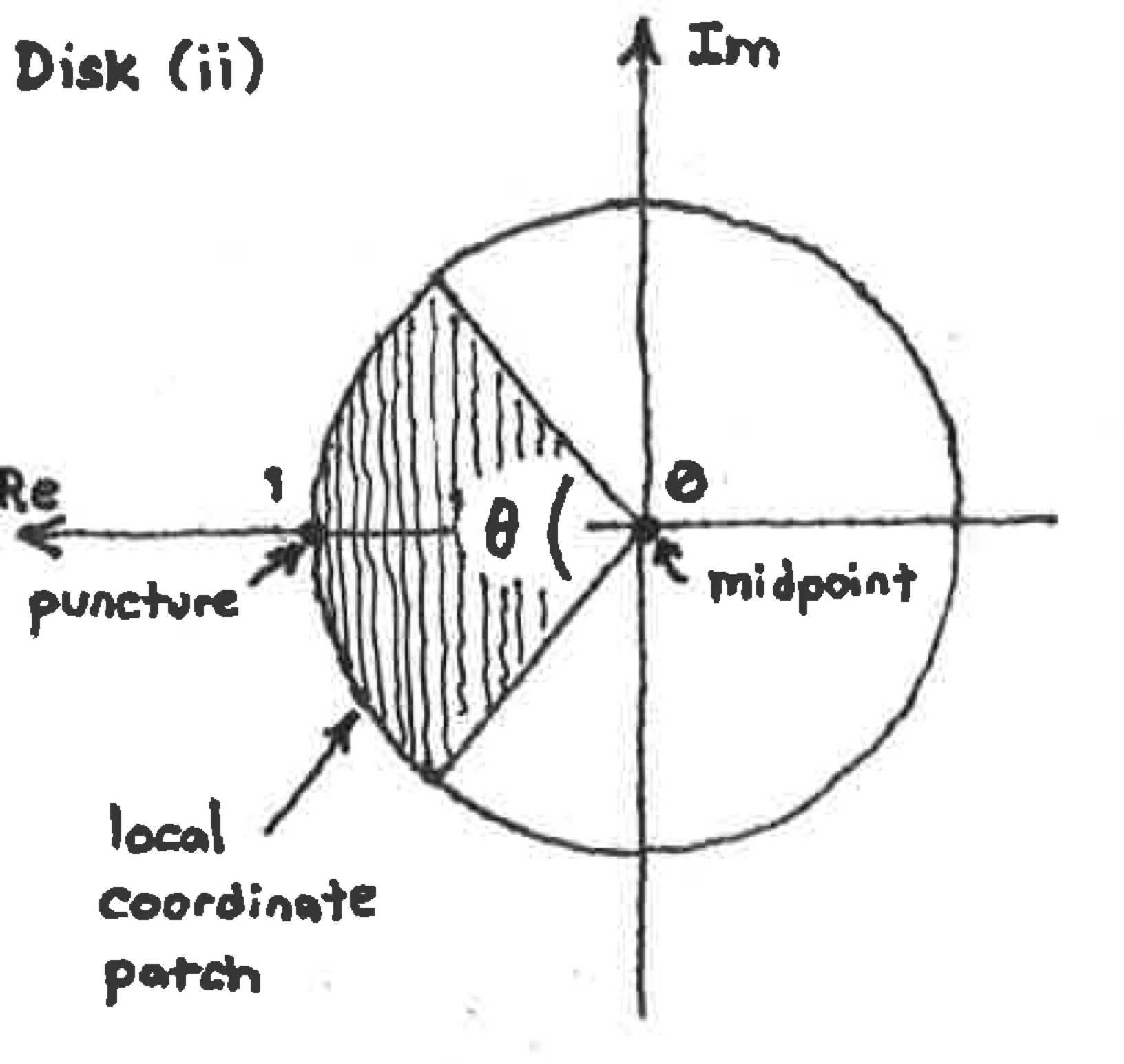}}
\vspace{-2cm}
\end{wrapfigure}
\noindent We may also represent wedge states as correlation functions on the unit disk. The unit disk is obtained by a conformal transformation of the upper half plane
\begin{equation}
D_2(u) = \frac{1+i u}{1-i u}.
\end{equation}
By applying a scale transformation to the upper half plane before mapping to the unit disk, we find 
two notable presentations:
\begin{equation}\langle \phi,\Omega^\alpha\rangle = \Big\langle f_\text{disk\,(i)}^{(\alpha)}\circ\phi(0) \Big\rangle_{D_2}, \  \  \  \langle \phi,\Omega^\alpha\rangle = \Big\langle f_\text{disk\,(ii)}^{(\alpha)}\circ\phi(0) \Big\rangle_{D_2},
\end{equation}
where 
\begin{eqnarray}f_\text{disk\,(i)}^{(\alpha)}(\xi) \lineup =  \frac{i \tan\left(\frac{\pi}{2(\alpha+1)}\right)-\tan\left(\frac{2}{\alpha+1}\tan^{1}\xi\right)}{i \tan\left(\frac{\pi}{2(\alpha+1)}\right)+\tan\left(\frac{2}{\alpha+1}\tan^{1}\xi\right)},\label{eq:(i)}\\
f_\text{disk\,(ii)}^{(\alpha)}(\xi) \lineup = \left(\frac{1+i\xi}{1-i\xi}\right)^{\frac{2}{\alpha+1}}.\label{eq:(ii)}
\end{eqnarray}
In the first case, the local coordinate patch on the unit disk is bounded by two circles (related by M{\"o}bius transformation to the circles in the upper half plane) placed diametrically above and 
below the real axis. The boundary of the local coordinate patch intersects the boundary of the unit disk on the imaginary axis at $\pm i$. In the second case the local coordinate patch is simpler. It is bounded by two radially directed rays extending from the origin  diametrically above and below the real axis containing an angle 
\begin{equation}\theta = \frac{2\pi}{\alpha+1}.\end{equation}
The origin on the unit disk is the image of the midpoint. Therefore, the local coordinate patch looks like a ``wedge" inside the unit disk. This is the origin of the term ``wedge state"~\cite{RZ}.

The most explicit description of wedge states is in the Fock space expansion. The best Fock space basis for present purposes (that is, the basis which produces the fewest nonvanishing coefficients at a fixed level) is given by total Virasoro descendants of the vacuum. One finds an expression of the form
\begin{equation}
|\Omega^\alpha\rangle = \sum_{n_N\geq...\geq n_2\geq n_1\geq 2}P_{n_N,...,n_2,n_1}\left(\frac{1}{\alpha+1}\right)L_{-n_N}...L_{-n_2}L_{-n_1}|0\rangle,
\end{equation}
where $P_{n_N,...,n_2,n_1}(x)$ are polynomials of order $n_1+n_2+...+n_N$. Up to level four, the nonvanishing polynomials are
\begin{eqnarray}
P\lineup =1,\ \ \ \ \ \ \ \ \ \ \ \ \ \ \ \ \ \ \ \ \ \ \ \ \ \ \ \ \ \ \ \ \ \ \ \ \ \ \ \ P_2(x) = -\frac{1}{3}+\frac{4}{3}x^2,\nonumber\\
P_{2,2}(x)\lineup = \frac{1}{18}-\frac{4}{9}x^2 + \frac{16}{18}x^4 = \frac{1}{2!}P_2(x)^2,\ \ \ \ P_4(x) = \frac{1}{30}-\frac{16}{30}x^4,
\end{eqnarray}
so that
\begin{eqnarray}
|\Omega^\alpha\rangle\lineup = \underbrace{|0\rangle }_{\text{level }0}+\underbrace{ P_2\!\left(\frac{1}{1+\alpha}\right)L_{-2}|0\rangle}_{\text{level }2}+\underbrace{\frac{1}{2!}\left(\! P_2\!\left(\frac{1}{1+\alpha}\right)\!\right)^2(L_{-2})^2|0\rangle + P_4\!\left(\frac{1}{1+\alpha}\right)L_{-4}|0\rangle}_{\text{level } 4}\nonumber\\
\lineup\ \ \ \ \ \ \ +\ \text{higher levels.}
\end{eqnarray}
The derivation of these polynomials requires expressing the local coordinate map for wedge states in the upper half plane as an infinitely nested composition of finite conformal transformations generated by the individual Virasoro generators $L_{-n}$. For more details, see appendix A of \cite{Schnabl}. No closed form expression for the polynomials is known, but their computation can be easily automated on a computer. Note that wedge states are {\it universal}: Their definition only requires the $SL(2,\mathbb{R})$ vacuum and Virasoros, which exist for any open string background. In fact, all surface states are universal.

The wedge parameter has two special limits. The limit $\alpha\to 0$ defines the {\it identity string field}
\begin{equation}\Omega^0 = 1.\end{equation}
Sometimes the identity string field is written as $I$ or $|I\rangle$, but usually we will simply denote it as~$1$. 
The identity string field is characterized by a strip of vanishing width, and formally acts as the identity of the open string star product:
\begin{equation}I*A = A*I = A.\end{equation}
\begin{wrapfigure}{l}{.5\linewidth}
\centering
\resizebox{3.4in}{1.2in}{\includegraphics{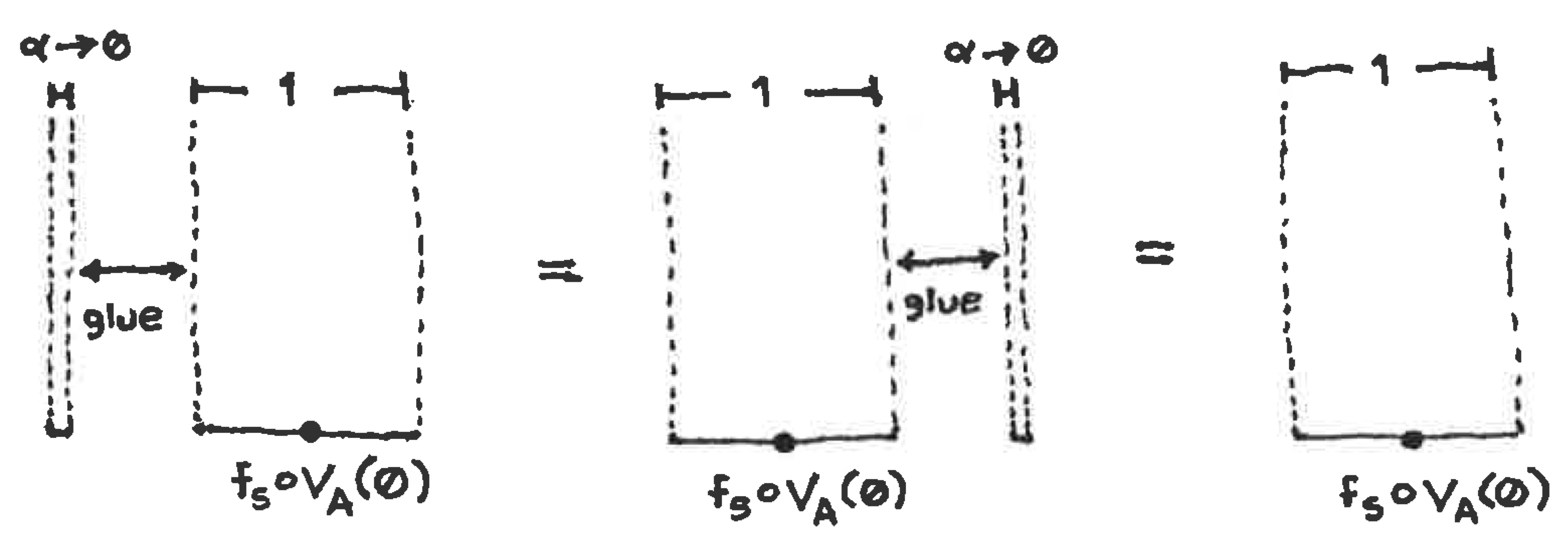}}
\vspace{-.5cm}
\end{wrapfigure}
This can be seen by viewing a generic state $A$ as a strip of unit width with vertex operator insertion. 
Multiplying by $1$ amounts to attaching a strip of vanishing width to either side, which leaves the strip unchanged. Another way to understand this is that the path integral on a strip of vanishing width is zero unless the boundary conditions on the left and right vertical edges match. Therefore in the split string representation the identity string field  is a delta functional between the left and right halves of the string:
\begin{equation}I[l(\sigma),r(\sigma)] = \delta[l(\sigma)-r(\sigma)].\label{eq:ISch}\end{equation}
This is the functional equivalent of the Kronecker delta defining the identity matrix. Via the BPZ inner product, the existence of the identity string field implies the existence of the trace operation in Witten's open bosonic SFT through 
\begin{equation}\Tr(A) = \langle I,A\rangle.\end{equation}
In the past there has been some question as to whether the identity  string field ``exists," in the sense that the open string star algebra really possesses an identity element. As a delta functional, the identity string field is more singular than wedge states of strictly positive width. But the doubts revolve more specifically around the question of whether the identity string field really multiplies like the identity. One problem is that the ghost oscillator $c_0$ is a derivation of the star product, but does not annihilate the identity string field \cite{RZ}. In our discussion it will be convenient (though perhaps not strictly necessary) to assume that the identity string field exists. Presently, there is no strong reason to assume that the open string star algebra must be closed under the action of~$c_0$. For more discussion see \cite{RZ,Sch_wedge}. 

The opposite limit $\alpha\to\infty$ defines the {\it sliver state} $\Omega^\infty$ \cite{RSZ_classical}. This corresponds to a strip of infinite width. The meaning of this is clearer if we characterize the sliver state through its overlap with a test state. For finite wedge parameter this is given by a correlation function on a cylinder of circumference $\alpha+1$. As $\alpha$ tends to infinity the cylinder unfolds and we obtain a correlation function on the upper half plane:
\begin{equation}\langle\phi,\Omega^\infty\rangle = \langle f_\mathcal{S}\circ\phi(0)\rangle_\mathrm{UHP}.\end{equation}
Since the width of a strip cannot increase past infinity, the sliver state is invariant under 
multiplication with other wedge states 
\begin{equation}\Omega^\alpha\Omega^\infty = \Omega^\infty\Omega^\alpha = \Omega^\infty.\end{equation}
and formally invariant under multiplication with itself:
\begin{equation}\Omega^\infty*\Omega^\infty =\Omega^\infty.\end{equation}
In this sense the sliver state behaves like a projection operator. To gain further insight into this interpretation, it is useful to visualize 
the sliver state on the unit disk in the representation  \eq{(i)}. In the sliver limit, the midpoint 
on the local coordinate patch touches the boundary of the unit disk. If 
\begin{wrapfigure}{l}{.35\linewidth}
\centering
\resizebox{2.2in}{2in}{\includegraphics{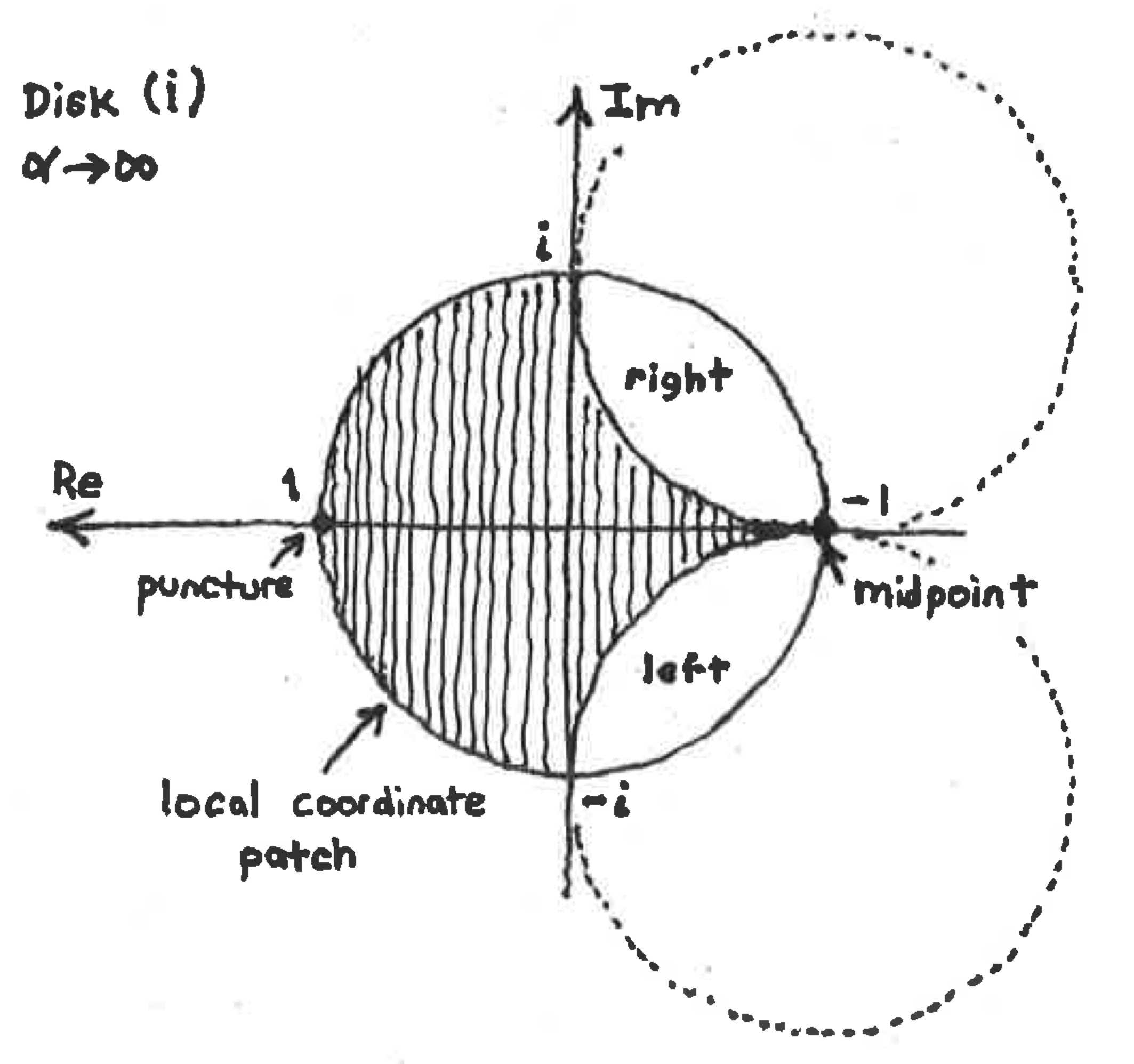}}
\vspace{0cm}
\end{wrapfigure} we remove the local coordinate patch from the unit disk, what remains are two disconnected regions. The worldsheet path integral on each region defines a Schr{\" o}dinger functional which depends respectively on the left or right half of the open string. Therefore the Schr{\" o}dinger functional of the sliver state can be factorized as
\begin{equation}\Omega^\infty[l(\sigma),r(\sigma)] = \Omega[l(\sigma_{\frac{1}{2}}(\sigma))]\times\Omega[r(\sigma_\frac{1}{2}(\sigma))].\end{equation}
The $SL(2,\mathbb{R})$ vacuum functional on the right hand side is evaluated on curves $l(\sigma_\frac{1}{2}(\sigma))$ and $r(\sigma_\frac{1}{2}(\sigma))$, where $\sigma_\frac{1}{2}(\sigma)$ maps the full string $\sigma\in[0,\pi]$ into the half string $\sigma_\frac{1}{2}\in[0,\frac{\pi}{2}]$.
\begin{exercise}
Show that
\begin{equation}
\sigma_\frac{1}{2}(\sigma) = \mathrm{gd}\left(\tan\frac{\sigma}{2}\right).
\end{equation}
modulo a reparameterization which leave the $SL(2,\mathbb{R})$ vacuum invariant.
\end{exercise}
\noindent The factorization of the Schr{\"o}dinger functional suggests an analogy to the projector $|0\rangle\langle 0|$ onto the ground state of the harmonic oscillator. In the position space representation, the ground state projector takes the form
\begin{equation}\langle x|0\rangle\langle 0|y\rangle = \frac{1}{\sqrt{\pi}}e^{-x^2/2}e^{-y^2/2},\end{equation}
which, like to the sliver state, is factorized between $x$ and $y$. This factorization is characteristic of a {\it rank 1 projector}, where ``rank" refers to the dimension of the image of the projector. The sliver state can be viewed as a rank 1 projector in the sense that it formally projects onto a 1-dimensional subspace of the vector space of half-string functionals. The identity string field is also a projector since $I*I = I$. It is a projector of {\it infinite} rank, since it maps onto an infinite dimensional subspace of half-string functionals (in fact, it leaves the entire space invariant). Unlike the sliver state, the Schr{\"o}dinger functional of the identity string field is not factorized between left and right halves of the string. The sliver state is not the only string field that can be viewed as a rank 1 projector. Any surface state which has a local coordinate map on the upper half plane which satisfies 
\begin{equation}f(i) = \infty\end{equation}

\begin{wrapfigure}{l}{.27\linewidth}
\vspace{-1cm}
\centering
\resizebox{2in}{1.5in}{\includegraphics{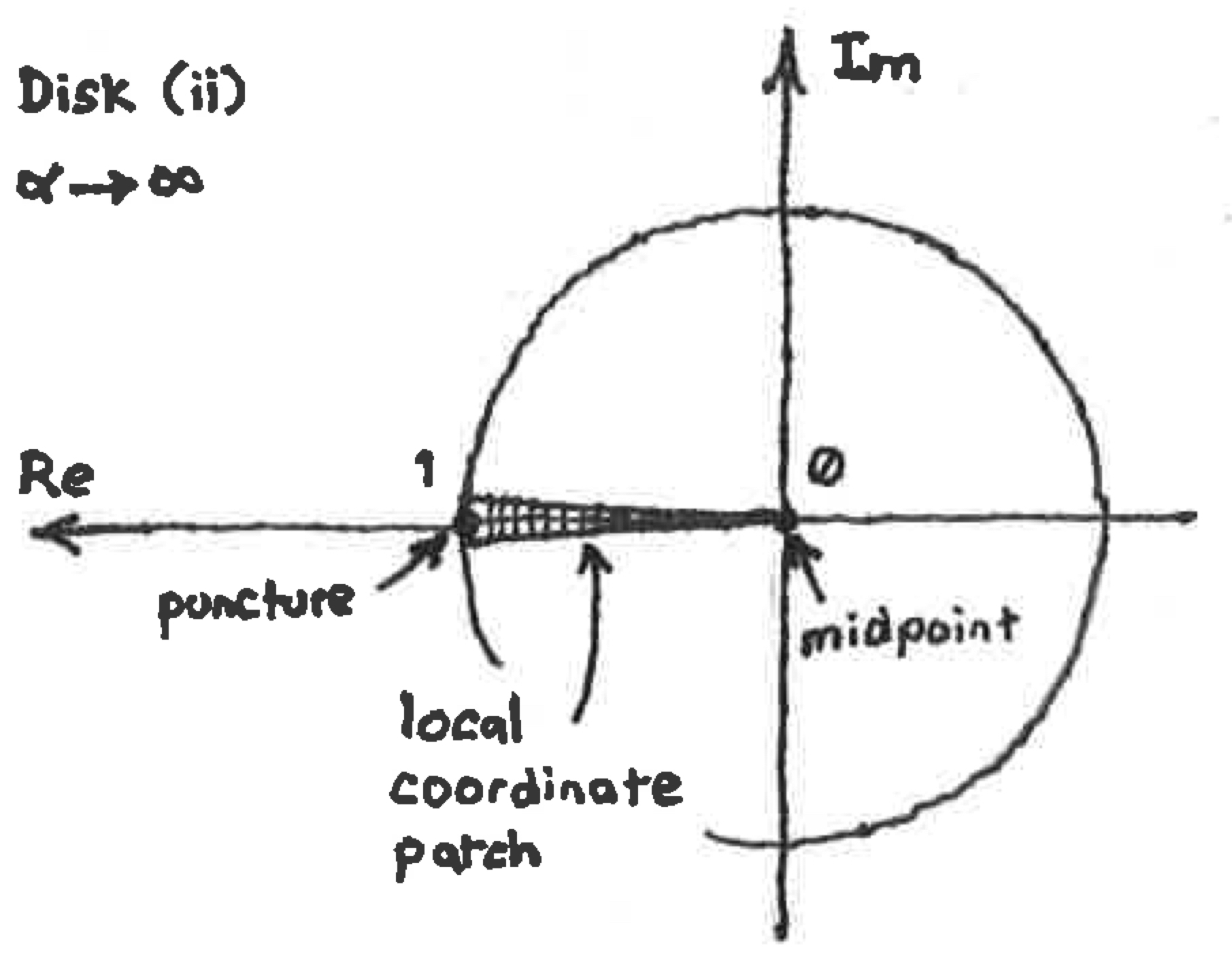}}
\end{wrapfigure}
\noindent will have a left/right factorized Schr{\"o}dinger functional \cite{GRSZ_projectors}, and is formally a rank 1 projector. It is instructive to look at the sliver state in the second representation on the unit disk \eq{(ii)}. In this picture, the angle of the wedge defining the local 
coordinate patch shrinks to zero. This does not suggest a factorized functional, but this is an illusion caused by the fact that the local coordinate patch becomes degenerate. The local coordinate patch looks like an infinitesimally thin ``sliver." This is the origin of the term ``sliver state" \cite{RSZ_classical}.

The sliver state is characterized by a Gaussian functional and looks normalizable. Nevertheless, it is a singular state. In fact, it is much worse than the identity string field. The sliver state can be thought of as a distribution, in that it defines a linear functional on well-behaved states (for example Fock space states)
\begin{equation}\Omega^\infty: \H  \to \mathbb{C};\ \ \Phi \mapsto  \langle \Omega^\infty,\Phi\rangle,\end{equation}
but it is not itself an element of the open string star algebra. To see why this is so, suppose we attempt to multiply two sliver states on either side of the zero momentum tachyon:
\begin{equation}\Omega^\infty*c_1|0\rangle*\Omega^\infty.\end{equation}
The zero momentum tachyon can be viewed as a strip of width $1$ containing a $c$-ghost at the origin. If we glue a strip of infinite width on either side, it looks like we have a 1-point function of the $c$-ghost on the upper half plane. But we need three $c$-ghosts to form a nonvanishing correlator. The other two $c$-ghosts are supposed to appear when contracting this expression with a test state, but in fact they are separated from the $c$ ghost at the origin by an infinite distance. Since the correlator of $c$ ghosts grows linearly with separation, the above state is actually divergent. In fact, even the projector relation $\Omega^\infty*\Omega^\infty = \Omega^\infty$ does not hold without ambiguity (see exercise \ref{ex:Bsliver}).  For this reason, the sliver state cannot be included as part of the open string star algebra. 

\begin{wrapfigure}{l}{.35\linewidth}
\centering
\resizebox{2.5in}{1.1in}{\includegraphics{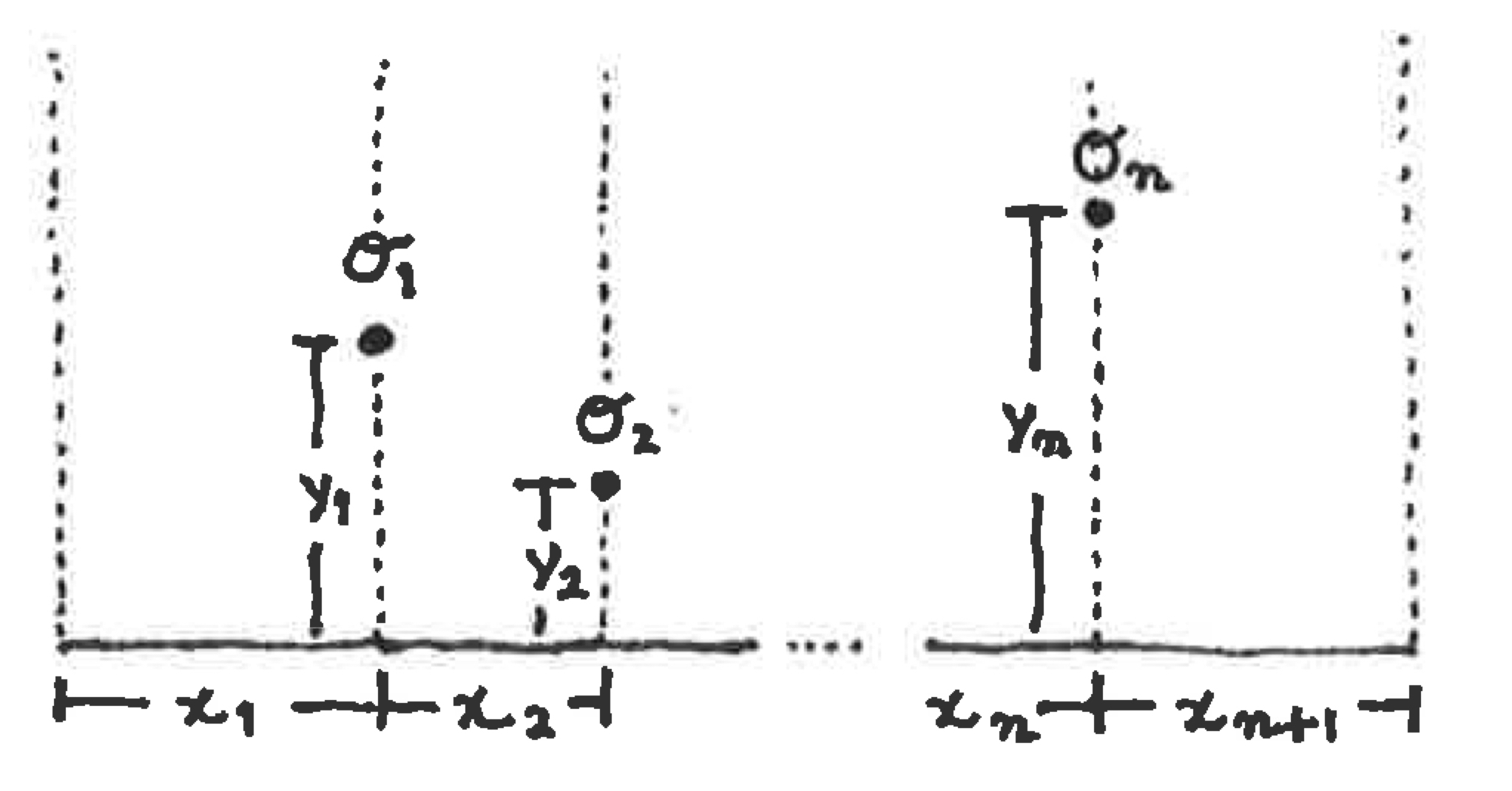}}
\vspace{-1cm}
\end{wrapfigure}

Wedge states are not enough by themselves to create solutions to the equations of motion. This is clear since wedge
states carry ghost number $0$, but a solution has ghost number $1$. To get a richer class of states, we consider strips of worldsheet of varying width containing insertions of local operators. These are often called {\it wedge states with insertions} \cite{Sch_wedge}. It is useful to describe such states as factorized into products of wedge states and fields representing insertions of local operators. Consider for example a state represented as a semi-infinite strip containing an operator $\mathcal{O}_1$ a distance $x_1$ from the leftmost vertical edge and a~distance $y_1$ above the real axis; the operator $\mathcal{O}_2$ is a distance $x_1+x_2$ from the left edge and a~distance  $y_2$ above the real axis, and so on. The idea is to introduce a string field $\mathcal{O}_i$ for every operator insertion. Following  standard practice, we denote the operator and the string field with the same symbol. The string field $\mathcal{O}_i$ is defined by~an  infinitesimally thin strip containing an 

\begin{wrapfigure}{l}{.15\linewidth}
\centering
\resizebox{1.0in}{1.1in}{\includegraphics{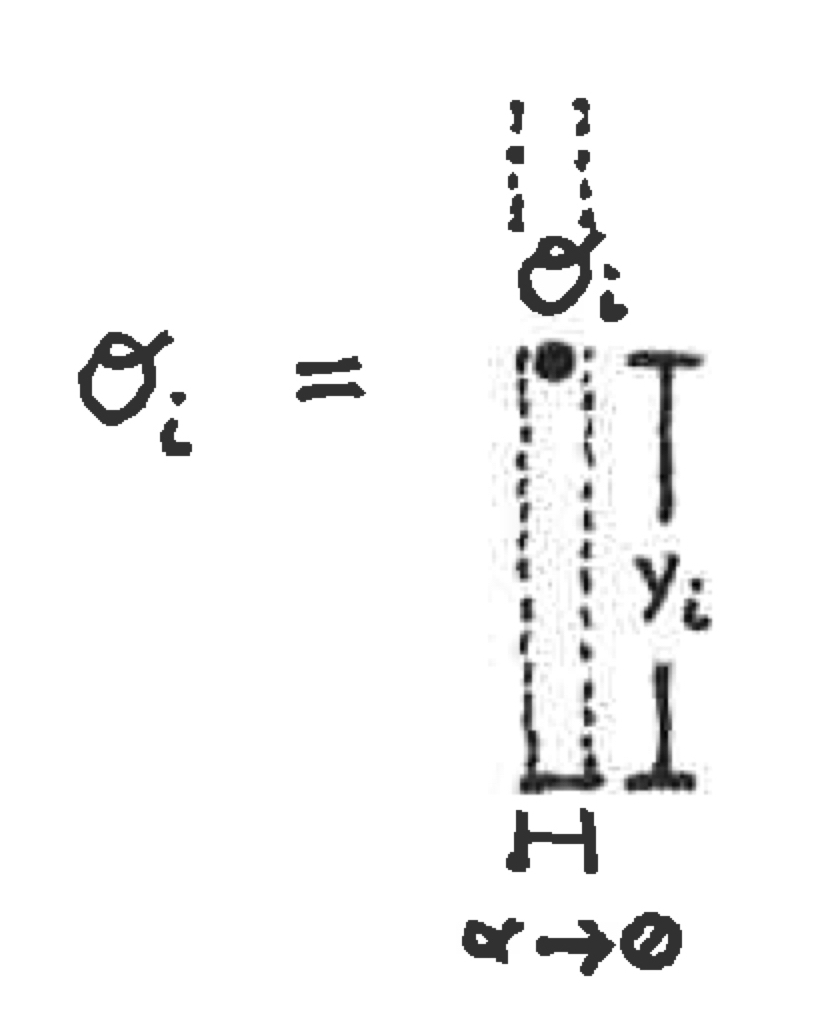}}
\vspace{1cm}
\end{wrapfigure}
\noindent   insertion  of 
the operator $\mathcal{O}_i$ a distance $y_i$ above the real axis. The region of the surface between insertions $\mathcal{O}_i$ and $\mathcal{O}_{i+1}$ can be described as an empty strip of width $x_{i+1}$---in other words, a wedge state. We can then express wedge states with insertions as a product of wedge states and string fields representing operator insertions, as follows: 
\begin{equation}\Omega^{x_1}\mathcal{O}_1\Omega^{x_2} ... \Omega^{x_n}\mathcal{O}_n\Omega^{x_{n+1}}.\end{equation}
\begin{exercise}
Show that the zero momentum tachyon state can be represented as 
\begin{equation}c_1|0\rangle = \frac{\pi}{2}\sqrt{\Omega}c\sqrt{\Omega},\end{equation}
where the string field $c$ is defined by an infinitely thin strip with a boundary insertion of the $c$-ghost.
\label{ex:tach}
\end{exercise}
\begin{exercise}
Let $X(z,\overline{z})$ be a spacelike free boson subject to Neumann boundary conditions. Show that the Schr\"odinger functional of a wedge state depends on this factor of the $\BCFT$ as 
\begin{equation}
\Omega^\alpha[l(y),r(y)] = \mathcal{N}_\alpha \exp\!\Bigg[-\frac{1}{16\pi}\int_{-\infty}^\infty\frac{d\kappa}{2\pi}\kappa \bigg(\!\left(\tanh\frac{\alpha\kappa}{2}\right)(l(\kappa)+r(\kappa))^2 +\left(\coth\frac{\alpha\kappa}{2}\right)(l(\kappa)-r(\kappa))^2\bigg)\Bigg],\label{eq:Schwedge}
\end{equation}
where $l(y),r(y)$ set the boundary conditions for $X(z,\overline{z})$ in the path integral at a height $y$ on the left/right vertical edges of a strip of width $\alpha$, and $l(\kappa),r(\kappa)$ are defined though Fourier transform\footnote{The Fourier modes on the edge of the strip define what is called the {\it kappa basis} \cite{spectroscopy,BelovLovlace}. The normalization of $\kappa$ used here differs from earlier literature by $\kappa_\mathrm{here} =\frac{\pi}{2}\kappa_\mathrm{elsewhere}$.}
\begin{equation}l(y) = \int_{-\infty}^\infty \frac{d\kappa}{2\pi}l(\kappa)e^{i\kappa y},\ \ \ r(y) = \int_{-\infty}^\infty \frac{d\kappa}{2\pi}r(\kappa)e^{i\kappa y},\end{equation}
which are assumed to be even functions of $\kappa$ to ensure the Neumann boundary condition. In addition, confirm the following:
\begin{itemize}
\item The Schr{\"o}dinger functional of the identity string field vanishes unless the left and right halves of the string coincide.
\item The Schr{\"o}dinger functional of the sliver state factorizes between the left and right halves of the string. 
\item The Schr{\"o}dinger functional of a wedge state satisfies 
\begin{equation}\Omega^\alpha[l(y),r(y)] = \Omega[l(\alpha y),r(\alpha y)].\end{equation}
\item The wedge state $\Omega^\alpha$ is characterized by a normalizable Gaussian functional if and only if $\mathrm{Re}(\alpha)>0$.
\item When $\alpha=1$, \eq{Schwedge} agrees with the $SL(2,\mathbb{R})$ vacuum functional derived at the end of subsection \ref{subsec:Psi}.
\end{itemize}
\end{exercise}

\subsection{$K$ and the wedge algebra}
\label{subsec:K}

A wedge state is an exponential whose base is the $SL(2,\mathbb{R})$ vacuum. Since the natural base of the exponential is Euler's number $e$, we can expect that the string field 
\begin{equation}\ln \Omega\end{equation}
plays an important role in understanding wedge states. We can deduce the nature of this state by computing the derivative of a wedge state with respect to the wedge parameter. We will do this following the computation of Okawa \cite{Okawa}. Consider the overlap of $\Omega^\alpha$ with a test state $|\phi\rangle$ given by an insertion $\phi(0)$ at the origin of a semi-infinite strip of unit width. We assume that $\phi(0)$ has definite scaling dimension $h$ in the sliver coordinate frame. The overlap is given by a 1-point function on a cylinder of circumference $\alpha+1$:
\begin{equation}\langle \phi,\Omega^\alpha\rangle = \langle \phi(0)\rangle_{C_{\alpha+1}}.\end{equation}
Through a scale transformation we can adjust the circumference of the cylinder to unity, producing a factor from the conformal transformation of $\phi(0)$:
\begin{equation}\langle \phi,\Omega^\alpha\rangle = \left(\frac{1}{\alpha+1}\right)^h\langle \phi(0)\rangle_{C_1}.\end{equation}
Now take the derivative with respect to $\alpha$ and scale the cylinder back to circumference $\alpha+1$:
\begin{eqnarray}
\left\langle\phi,\frac{d}{d\alpha}\Omega^\alpha\right\rangle \lineup = -h\left(\frac{1}{\alpha+1}\right)^{h+1}\langle\phi(0)\rangle_{C_1}\nonumber\\
\lineup = -h\frac{1}{\alpha+1}\langle\phi(0)\rangle_{C_{\alpha+1}}.
\end{eqnarray}
Since $\phi(0)$ has scaling dimension $h$, its OPE with the energy-momentum tensor takes the form
\begin{equation}T(z)\phi(0) = ... +\frac{h}{z^2}\phi(0)+\frac{1}{z}\d\phi(0)+...\ ,\end{equation}
which implies
\begin{equation}h\phi(0) = \oint_0\frac{dz}{2\pi i}zT(z)\phi(0).\end{equation}
Therefore we can write
\begin{equation}\left\langle\phi,\frac{d}{d\alpha}\Omega^\alpha\right\rangle= -\frac{1}{\alpha+1}\left\langle \oint_0\frac{dz}{2\pi i}zT(z)\phi(0)\right\rangle_{C_{\alpha+1}}.\end{equation}
\begin{wrapfigure}{l}{.56\linewidth}
\centering
\vspace{-.5cm}
\resizebox{4.2in}{1.2in}{\includegraphics{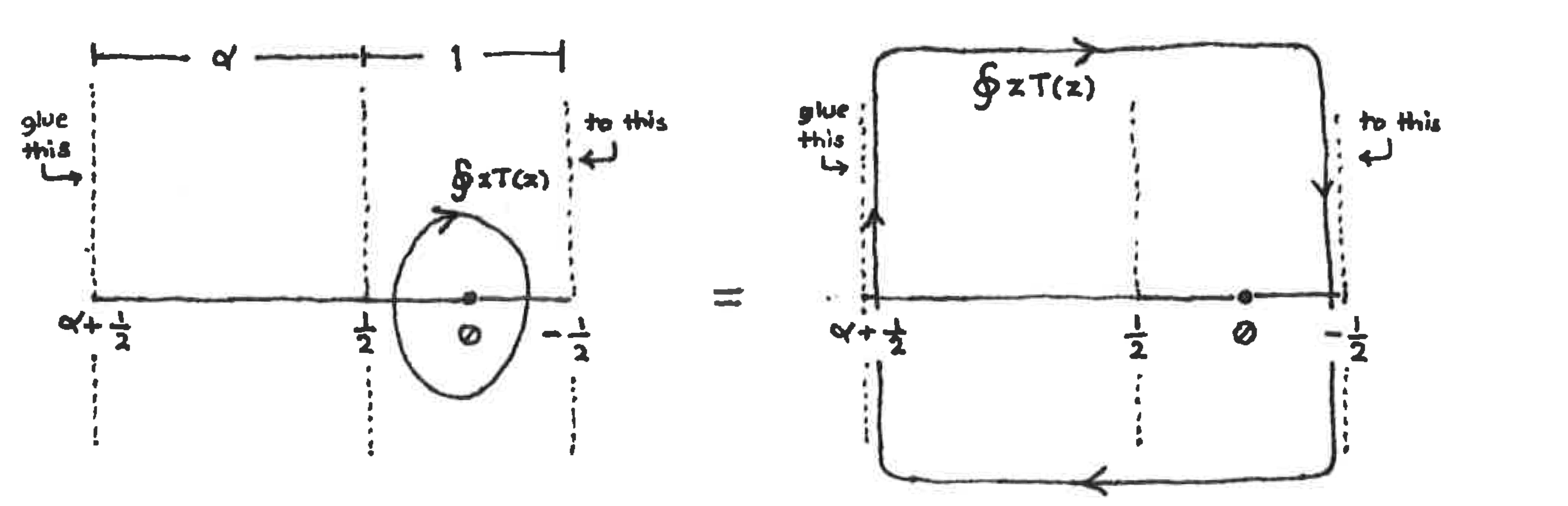}}
\vspace{-1.5cm}
\end{wrapfigure} Next we unravel the energy-momentum contour inside the cylinder. Suppose the cylinder is 
represented as a strip $\frac{1}{2}+\alpha\geq \mathrm{Re}(z)\geq -\frac{1}{2}$, with opposite sides identified. We use the doubling trick, so the semi-infinite cylinder is represented by a holomorphic copy of the full infinite cylinder. Expanding the contour gives a contribution from the left vertical edge of the strip and the right vertical edge:
\begin{equation}
\oint_0\frac{dz}{2\pi i}z T(z) = \int_{-i\infty+\alpha+\frac{1}{2}}^{i\infty +\alpha+\frac{1}{2}}\frac{dz}{2\pi i}z T(z) - \int_{-i\infty-\frac{1}{2}}^{i\infty-\frac{1}{2}}\frac{dz}{2\pi i}z T(z).
\end{equation}
In the second integral we make a substitution $z\to z-(\alpha+1)$ so that both terms share a common integration variable:
\begin{equation}
\oint_0\frac{dz}{2\pi i}z T(z) = \int_{-i\infty+\alpha+\frac{1}{2}}^{i\infty +\alpha+\frac{1}{2}}\frac{dz}{2\pi i}\Big(z T(z)-(z-(\alpha+1))T(z-(\alpha+1))\Big).
\end{equation}
The identification on the vertical edges implies
\begin{equation}T(z)=T(z-(\alpha+1)).\end{equation}
Therefore 
\begin{eqnarray}
\oint_0\frac{dz}{2\pi i}z T(z) \lineup = \int_{-i\infty+\alpha+\frac{1}{2}}^{i\infty +\alpha+\frac{1}{2}}\frac{dz}{2\pi i}\Big(z-(z-(\alpha+1))\Big) T(z)\nonumber\\
\lineup = (\alpha+1)\int_{-i\infty+\alpha+\frac{1}{2}}^{i\infty+\alpha+\frac{1}{2}}\frac{dz}{2\pi i}T(z).
\end{eqnarray}

\noindent Through contour deformation the precise horizontal placement of the vertical energy-momentum 
contour is not very important. So we can write
\begin{equation}\left\langle\phi,\frac{d}{d\alpha}\Omega^\alpha\right\rangle  = -\left\langle\int_{-i\infty}^{i\infty}\frac{dz}{2\pi i} T(z)\phi(0)\right\rangle_{C_{\alpha+1}}.\end{equation}
We introduce the string field $K$, defined as an infinitely thin strip of worldsheet containing an 
\begin{wrapfigure}{l}{.22\linewidth}
\centering
\resizebox{1.6in}{1.3in}{\includegraphics{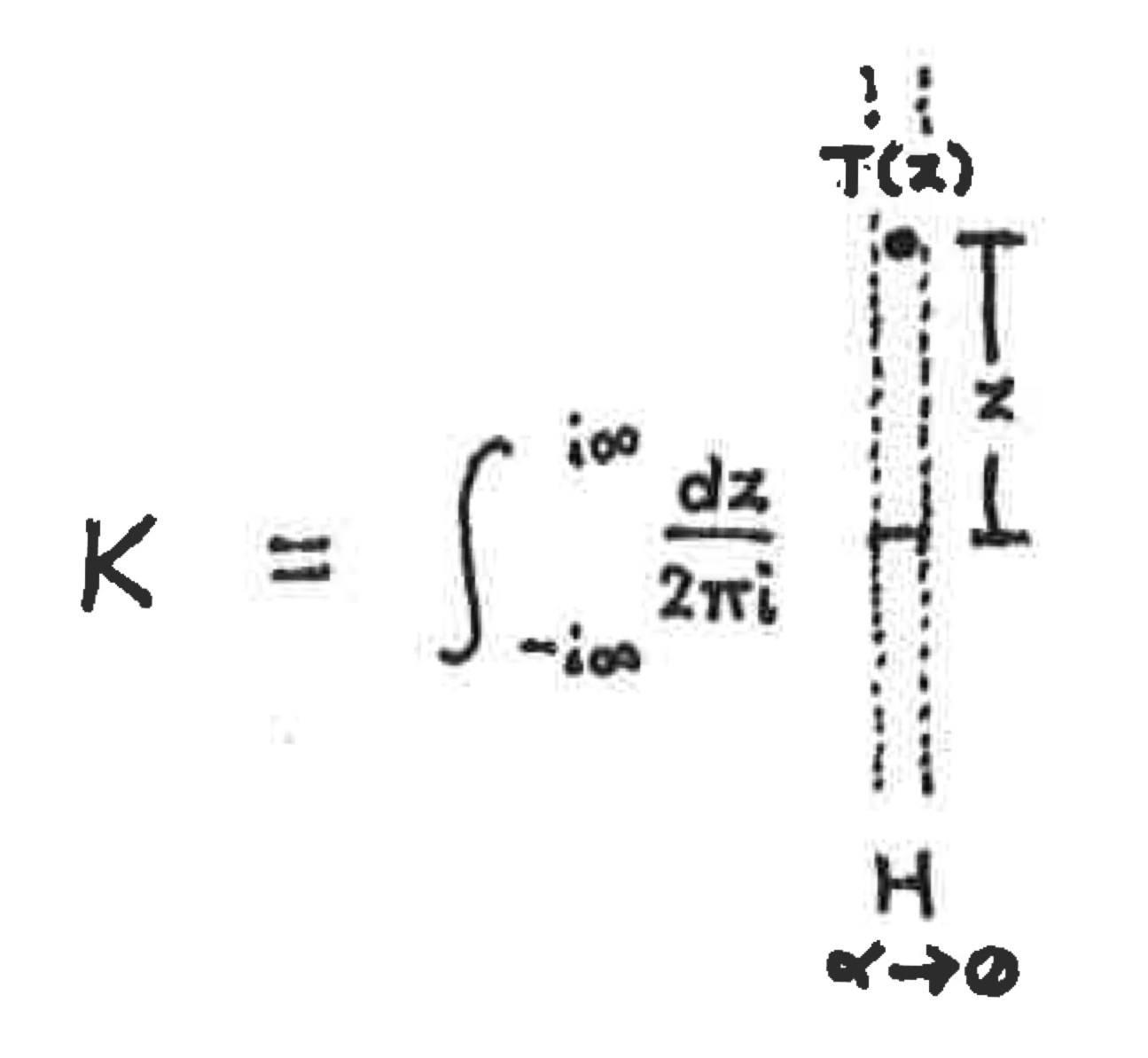}}
\end{wrapfigure} 
insertion of the energy-momentum tensor, integrated vertically on the imaginary axis. The last equation can then be rewritten
\begin{equation}\frac{d}{d\alpha}\Omega^\alpha = -K\Omega^\alpha.\end{equation}
Since $\Omega^0=1$ is the identity string field, the solution of this differential equation implies that wedge states can be written
\begin{equation}\Omega^\alpha = e^{-\alpha K},\end{equation}
and in particular 
\begin{equation}\ln \Omega = -K.\end{equation}
$K$ can be viewed as a Hamiltonian which creates wedge states through Euclidean time evolution. We can derive a vertex operator representing a wedge state by writing
\begin{equation}\Omega^\alpha = \sqrt{\Omega}e^{-(\alpha-1)K}\sqrt{\Omega}.\end{equation}
The state on the right hand side can be viewed as a strip of width 1 containing an infinite number of vertical contour insertions of the energy-momentum tensor. To derive the vertex operator we must map the strip back to the canonical half-disk. Noting that 
\begin{equation}f_\mathcal{S}^{-1}\circ \int_{-i\infty}^{i\infty}\frac{dz}{2\pi i}T(z) = \frac{\pi}{2}\int_{-i}^i \frac{d\xi}{2\pi i}(1+\xi^2)T(\xi),\end{equation}
the vertex operator is given by 
\begin{equation}V_{\Omega^\alpha}(0) = \sum_{n=0}^\infty\frac{1}{n!}\left(-\frac{\pi(\alpha-1)}{2}\int_{-i}^i \frac{d\xi}{2\pi i}(1+\xi^2)T(\xi)\right)^n.\end{equation}
As expected, the vertex operator is nonlocal.

\begin{exercise}
Show that  $K$ is a real string field, that is $K^\ddag= K$.
\end{exercise}

The {\it wedge algebra} is the subalgebra of the open string star algebra defined by taking products and linear combinations of wedge states. Since there are a continuum of wedge states, in general we can form continuous linear combinations:
\begin{equation}F(K) = \int_0^\infty dt f(t)\Omega^t.\label{eq:Laplace}\end{equation}
The right hand side can be viewed as a function of $K$, obtained through Laplace transform of the function $f(t)$ in the ``time domain." Therefore, the algebra of wedge states can be viewed as an algebra of functions of $K$. Since an algebra of functions is a fairly simple thing---especially in comparison to the full open string star algebra---we can imagine defining the wedge algebra precisely. We start by describing a proposal due to Rastelli \cite{Rastelli}. We consider the string field $F(K)$ as isomorphic to a function $F(k)$ of numbers $k$ in the spectrum of $K$. The spectrum is defined by the property that the string field 
\begin{equation}K-k\end{equation} is not invertible. The inverse may be computed using the Schwinger parameterization
\begin{equation}\frac{1}{K-k}=\int_0^\infty  dt\, e^{kt}\Omega^t.\label{eq:spec_int}\end{equation}
Since $\Omega^t$ approaches a constant (the sliver state) for large $t$, the integral is divergent for all non-negative $k$. Therefore we are looking for an algebra of functions of nonnegative real numbers $K$ (we henceforth suppress the distinction between the string field $K$ and an element of its spectrum). We can define convergence in this algebra with the uniform norm 
\begin{equation}||F(K)||_{C^*} = \sup_{K\geq 0}|F(K)|.\label{eq:Cstar}\end{equation}
This has the usual properties of a vector space norm, and in addition satisfies 
\begin{equation}||F(K)G(K)||_{C^*}\leq ||F(K)||_{C^*}\cdot ||G(K)||_{C^*}.\label{eq:norm2}\end{equation}
which implies that multiplication is continuous. Finally, we have the {\it $C^*$ identity}
\begin{equation}||F(K) F(K)^\ddag ||_{C^*} = ||F(K)||_{C^*}^2.\end{equation}
where reality conjugation in the wedge algebra is equivalent to complex conjugation of the associated function of $K$. Requiring completeness with respect to the norm leads to a $C^*$-algebra of bounded, continuous functions of non-negative~$K$. We denote this as $C_0(\mathbb{R}_{\geq 0})$. We make a few observations about this.  First, wedge states with non-negative wedge parameter---including the
\begin{wrapfigure}{l}{.3\linewidth}
\centering
\resizebox{2.1in}{1.3in}{\includegraphics{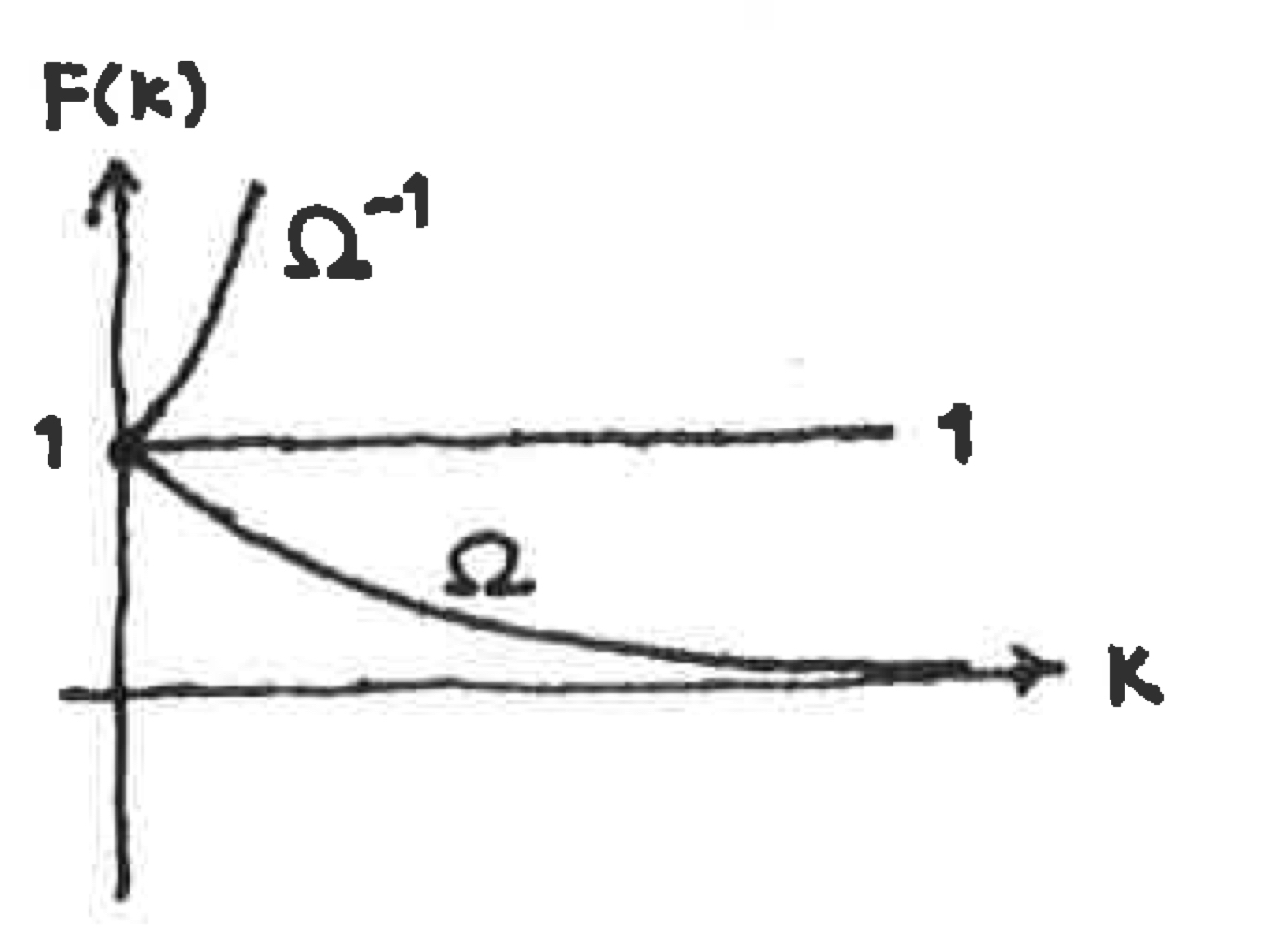}}
\vspace{-1cm}
\end{wrapfigure}
 identity string field---are part of the algebra. Even complex wedge parameter is allowed as long as it has positive real part. Wedge states with negative wedge parameter are excluded since they are not bounded functions of $K\geq 0$. Second, the sliver state is excluded from the wedge algebra since it is not continuous:
\begin{equation}\Omega^\infty = \left\{\begin{matrix} \ 1\ \ \mathrm{at}\ \ K=0\ \ \\ \ 0\ \ \mathrm{for}\ \ K> 0\end{matrix}\right. .\end{equation}
In particular, while the sliver limit converges in the Fock space expansion, it does not converge as a Cauchy sequence with respect to the norm. One can check that 
\begin{equation}||\Omega^{Nn}-\Omega^n||_{C^*} = N^{-\frac{1}{N-1}} - N^{-\frac{N}{N-1}}.\end{equation}
which holds independent of $n$, and in particular is nonvanishing in the $n\to\infty$ limit. The divergence of the sliver limit has important consequences for the physical interpretation of analytic solutions.

A difficulty with the $C^*$-algebra proposal is that bounded continuous functions of $K\geq 0$ do not always have a representation as a Laplace transform. Therefore we cannot fully construct the wedge algebra by taking superpositions of wedge states.  But it is not clear otherwise how the wedge algebra can be constructed.\footnote{A proposal of \cite{exotic} is to construct elements  of the $C^*$ algebra through the replacement 
\begin{equation}
\left(\frac{1}{1+\alpha}\right)^n\rightarrow \frac{1}{(n-1)!}\int_0^\infty dK K^{n-1}e^{-K}F(K)
\end{equation}
inside the polynomials which define the coefficients of wedge states in the Virasoro basis. The integral converges for any bounded and continuous $F(K)$. However, the resulting states are not in general characterized by surfaces, and we cannot rely on the generalized gluing and resmoothing theorem to ensure that the states multiply correctly. This remains an open question.} We therefore mention two other possible definitions of the wedge algebra. When considering the integral \eq{spec_int} we implicitly assumed that the spectrum of $K$ was real. This seems natural since $K$ is a real string field. But on second thought \eq{spec_int} is divergent for any complex $k$ with positive real part. From the point of view of the previous proposal, this divergence indicates the failure of the Schwinger parameterization to define the inverse. But presently we will take the divergence seriously. This suggests that the wedge algebra should be understood as an algebra of functions of a complex variable with non-negative real part. Since we want these functions to be represented by a Laplace transform, we further require that they should be {\it holomorphic} on the positive half of the complex $K$-plane. We can relate $K$ to a coordinate $\zeta$ on the unit disk with the transformation
\begin{equation}K(\zeta) = \frac{1+\zeta}{1-\zeta},\ \ \ \ \zeta\in D_2,\end{equation}
and introduce the norm
\begin{equation}||F(K)||_{D_2} = \sup_{\zeta\in D_2}\left|F\Big(K(\zeta)\Big)\right|.\end{equation}
This satisfies the usual properties of a vector space norm and multiplication is continuous. However, we do not have the $C^*$ identity; there is no $C^*$-algebra of holomorphic functions. However, we can define two Banach $*$-algebras:
\begin{description}
\item{(1)} The {\it Hardy space} $H^\infty$ of bounded, holomorphic functions on the interior of the unit disk.\footnote{This is closely related to the space of {\it $L_0$ safe} states discussed in \cite{lightning}. However, $L_0$ safe states are only assumed to be polynomial bounded for nonnegative $\mathrm{Re}(K)$. }
\item{(2)} The {\it disk algebra} $A(D_2)$ consisting of bounded, holomorphic functions on the interior of the unit disk which extend to continuous functions on the boundary of the unit disk. 
\end{description}
We have the inclusions
\begin{equation}C_0(\mathbb{R}_{\geq 0})\supset H^\infty \supset A(D_2).\end{equation}
Let us mention a few consequences. The Hardy space includes the identity string field and wedge states with positive wedge parameter. But it does not allow for wedge states with complex wedge parameter, even with positive real part, since their absolute value diverges towards $\pm i\infty$ in the complex $K$-plane. The most restrictive definition is the disk algebra. This includes the identity string field but not wedge states with positive wedge parameter, since they do not extend to continuous functions on the boundary of the disk. In particular, the $SL(2,\mathbb{R})$ vacuum is excluded, which seems rather drastic but could have a rationale---for example, if singularities in correlation functions require that $f(t)$ in the time domain is a smooth function. The $D_2$ norm gives useful information about how the wedge algebra is realized in terms of wedge states. But the existence of Schnabl's solution suggests that the $C^*$ norm gives a more physically meaningful topology on the wedge algebra (see for example subsection \ref{subsec:Sch} and exercise \ref{ex:SchSimp}). This could change in light of new understanding.

\subsection{Schnabl's $\mathcal{L}_0$ } 

An important role in the theory is played by the dilatation generator in the sliver coordinate frame, introduced by Schnabl \cite{Schnabl}:
\begin{equation}\mathcal{L}_0 = \oint_0 \frac{dz}{2\pi i} zT(z),\ \ \ \ (\text{sliver frame}). \label{eq:curlL0sliver}\end{equation}
This is different from the usual $L_0$ since the contour is integrated around the vertex operator on the strip of width 1, rather than the unit half-disk. To relate $\mathcal{L}_0$ to ordinary Virasoros, we must map back to the half-disk:
\begin{eqnarray}
\mathcal{L}_0\lineup = f_\mathcal{S}^{-1}\circ\oint_0\frac{dz}{2\pi i} z T(z) = \oint_0 \frac{d\xi}{2\pi i}(1+\xi^2)\tan^{-1}\!\xi \,T(\xi)\ \ \ \ (\text{half-disk})\nonumber\\
\lineup = L_0+\frac{2}{3}L_2-\frac{2}{15}L_4+...\ .
\label{eq:curlL0}\end{eqnarray}
Since $\mathcal{L}_0$ is made from positively moded Virasoros, we have $\mathcal{L}_0|0\rangle = 0$. This indicates that, in the sliver frame, we can shrink the contour without encountering poles since there is no vertex operator at the origin.

The above formulas give a definition of $\mathcal{L}_0$ from the point of view of radial quantization. But this will not be convenient, since we work with wedge states, whose representation in radial quantization requires complicated nonlocal vertex operators. Our first task is to represent the action $\mathcal{L}_0$ on an arbitrary wedge state as an operator insertion in correlation functions on the cylinder. It will be helpful to be clear about the coordinate system where the operator is inserted.  Let $A$ be a wedge state of width $\alpha$ carrying operator insertions. Using the doubling trick, we introduce a {\it natural coordinate} $z_A$ associated with this state as a region in the complex plane
\begin{equation}\frac{L}{2}> \mathrm{Re}(z_A)> -\frac{L}{2},\end{equation}
where $L\geq \alpha$ is the circumference of the cylinder where the strip of $A$ appears. The origin of $z_A$ is on the open string boundary of the strip of $A$ half way between its left and right vertical edges. The operator insertions of $A$ will be written in the coordinate $z_A$ as $\mathcal{O}_A$, and we indicate this~as
\begin{equation}A \ \ \longrightarrow\ \ \mathcal{O}_A.\end{equation}
Let $A$ and $B$ be wedge states with  insertions of  respective widths $\alpha$ and $\beta$. The inner product
\begin{equation}\langle A,B\rangle\end{equation}
can be computed as a correlation function on a cylinder of width $\alpha+\beta$. The states $A$ and $B$ come with natural coordinates  $z_A$ and $z_B$ bounded according to
\begin{equation}-\frac{\alpha+\beta}{2}<\mathrm{Re}(z_A)<\frac{\alpha+\beta}{2},\ \  \ \ \  \ \ -\frac{\alpha+\beta}{2}<\mathrm{Re}(z_B)<\frac{\alpha+\beta}{2}.\end{equation}
\begin{wrapfigure}{l}{1\linewidth}
\centering
\resizebox{6in}{1.8in}{\includegraphics{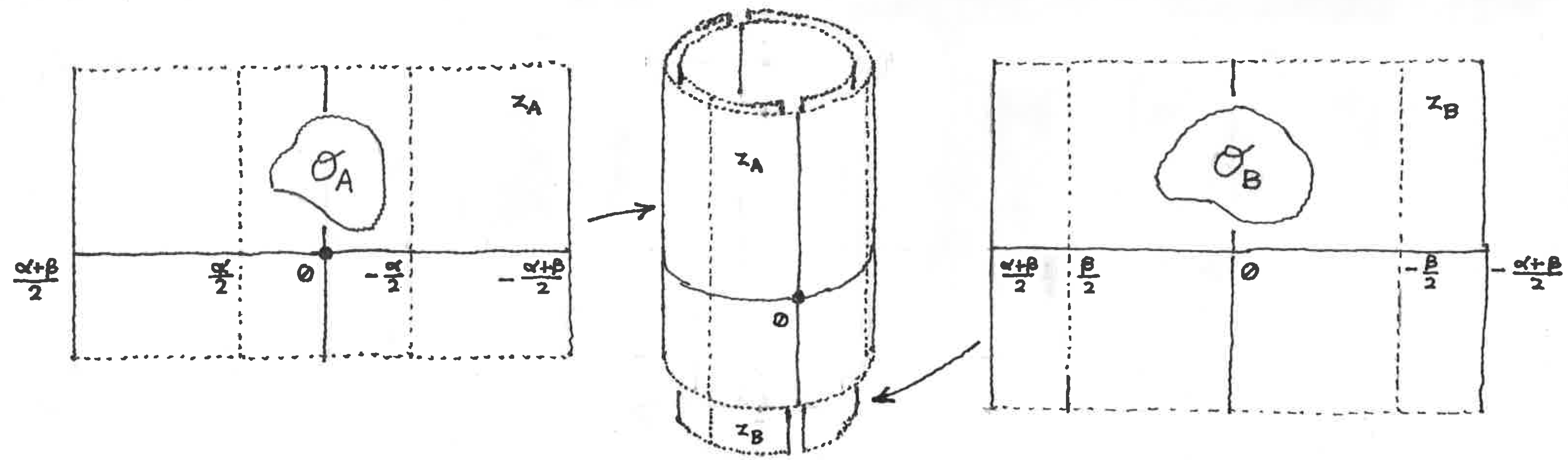}}
\end{wrapfigure}\\ \\ \\ \\ \\ \\ \\ \\  \\ \\ \\ \\ 
The coordinates differ in the location of the origin. The origin of $z_A$ is on the open string boundary of the strip of $A$ half way between its vertical edges, and the origin of $z_B$ is on the open string boundary of the strip of $B$ halfway between its vertical edges. This implies that $z_A$ and $z_B$ are related by a nontrivial transition function
\begin{eqnarray}z_B \lineup = I_{\alpha+\beta}(z_A)  = z_A - \frac{\alpha+\beta}{2}\mathrm{sgn}(z_A), \\
z_A \lineup = I_{\alpha+\beta}(z_B)  = z_B - \frac{\alpha+\beta}{2}\mathrm{sgn}(z_B) ,
\end{eqnarray}
where $\mathrm{sgn}(z)$ denotes the sign of the real part of $z$. The transition function is its own inverse
\begin{equation} I_{\alpha+\beta}\circ I_{\alpha+\beta}(z)=z,\end{equation}
and is closely related to the BPZ conformal map. If $\mathcal{O}_A$ and $\mathcal{O}_B$ are the operator insertions of the states $A$ and $B$ in the respective natural coordinates, the BPZ inner product can be represented as a correlation function on the cylinder
\begin{equation}\langle A,B\rangle = \big\langle (I_{\alpha+\beta}\circ \mathcal{O}_A) \, \mathcal{O}_B\big\rangle_{C_{\alpha+\beta}}= \big\langle \mathcal{O}_A \,( I_{\alpha+\beta}\circ \mathcal{O}_B)\big\rangle_{C_{\alpha+\beta}}.\end{equation}
The first equality represents the correlation function in the  coordinate $z_B$ and the second in the coordinate $z_A$.

Now consider the inner product
\begin{equation}\langle A,\mathcal{L}_0B\rangle.\end{equation}
The state $A$ will be represented as in the previous paragraph as a strip of width $\alpha$ containing the operators $\mathcal{O}_A$ in the natural coordinate $z_A$. But, to make use of the definition of $\mathcal{L}_0$ given in \eq{curlL0sliver}, we assume that $B$ is represented from the point of view of radial quantization in the sliver frame. That is, $B$ will be a strip of width 1 containing a (possibly nonlocal) vertex operator $f_\mathcal{S}\circ B(0)$ inserted at the origin of the natural coordinate $z_B$. The goal will be to derive the operator insertion corresponding to the state $\mathcal{L}_0^\bigstar A$, where $\mathcal{L}_0^\bigstar$ is the BPZ conjugate of $\mathcal{L}_0$:
\begin{equation}\mathcal{L}_0^\bigstar =   L_0+\frac{2}{3}L_{-2}-\frac{2}{15}L_{-4}+...\ .\end{equation}
The inner product can be expressed as a correlation function on the cylinder of circumference $\alpha+1$,
\begin{equation}\langle A,\mathcal{L}_0B\rangle = \left\langle (I_{\alpha+1}\circ \mathcal{O}_A) \, \oint_{\d B}\frac{dz_B}{2\pi i}z_B T(z_B)(f_\mathcal{S}\circ B(0))\right\rangle_{C_{\alpha+1}},\end{equation}
where we used the definition of  $\mathcal{L}_0$ given in \eq{curlL0sliver}. The insertions in the correlator are expressed in the coordinate $z_B$, and $\d B$ is the closed contour which follows the left and right  vertical boundaries of the strip of $B$ in the standard orientation. We now transform the  correlator into the coordinate $z_A$:
\begin{eqnarray}
\langle A,\mathcal{L}_0 B\rangle \lineup = 
\left\langle \mathcal{O}_A \,\oint_{\d B}\frac{dz_B}{2\pi i}z_B \big(I_{\alpha+1}\circ T(z_B)\big)\big(I_{\alpha+1}\circ f_\mathcal{S}\circ B(0)\big)\right\rangle_{C_{\alpha+1}}\nonumber\\
\lineup = \left\langle \mathcal{O}_A \,\oint_{\d B}\frac{dz_B}{2\pi i}z_B T(I_{\alpha+1}(z_B))\big(I_{\alpha+1}\circ f_\mathcal{S}\circ B(0)\big)\right\rangle_{C_{\alpha+1}}\nonumber\\
\lineup = -(-1)^{|A||B|}\left\langle (I_{\alpha+1}\circ f_\mathcal{S}\circ B(0))\, \oint_{\d A}\frac{dz_A}{2\pi i}I_{\alpha+1}(z_A) T(z_A)\mathcal{O}_A \right\rangle_{C_{\alpha+1}}.
\end{eqnarray}
In the first step we noted that $I_{\alpha+1}\circ T(z) = T(I_{\alpha+1}(z))$ if we avoid the branch cut of the $\mathrm{sgn}$ function. The  location of the  contour is on the boundary of the strip of $B$ and does not encounter the branch cut, so this is justified. In the  second step we reordered the insertions and relabeled the  integration variable to the natural coordinate  $z_A$. This produces a sign from commuting $\mathcal{O}_A$ and $\mathcal{O}_B$ in addition to a sign from orientation reversal of the contour. The final equation should be interpreted as $(-1)^{|A||B|}\langle B,\mathcal{L}_0^\bigstar A\rangle$. In this way we find that the  insertions of the state $\mathcal{L}_0^\bigstar A$ in the coordinate $z_A$ are
\begin{equation}\mathcal{L}_0^\bigstar A \ \ \longrightarrow\ \ -\oint_{\d A}\frac{dz_A}{2\pi i} \left(z_A - \frac{\alpha+1}{2}\mathrm{sgn}(z_A)\right) T(z_A)\,\mathcal{O}_A,\label{eq:curlyL0starins}
\end{equation}
where we  substituted the explicit form of $I_{\alpha+1}(z_A)$ in  the  integrand.

Next we can determine the operator insertions corresponding to $\mathcal{L}_0A$. This can be derived by considering the inner product
\begin{equation}
\langle A,\mathcal{L}_0^\bigstar B\rangle = \left\langle (I_{\alpha+1}\circ\mathcal{O}_A) \left(-\oint_{\d B} \frac{dz_B}{2\pi i}I_2(z_B) T(z_B)(f_\mathcal{S}\circ B(0))\right)\right\rangle_{C_{\alpha+1}},
\end{equation}
where again $B$ is represented from the point of view of radial quantization in the sliver frame, and the insertion of $\mathcal{L}_0^\bigstar$ follows from \eq{curlyL0starins}. Following the steps of the previous paragraph we find that 
\begin{equation}\mathcal{L}_0 A \ \ \longrightarrow\ \ \oint_{\d A}\frac{dz_A}{2\pi i} \left(z_A - \frac{\alpha-1}{2}\mathrm{sgn}(z_A)\right) T(z_A)\,\mathcal{O}_A.\label{eq:curlyL0ins}
\end{equation}
Note that if $A$ is a strip of width 1, the $\mathrm{sgn}$ function cancels out and we recover \eq{curlL0sliver}.

We have completed the preliminary task of representing $\mathcal{L}_0$ and $\mathcal{L}_0^\bigstar$ as operator insertions on the cylinder. The next task is to understand what these operators are doing. For this it is useful to consider BPZ even and odd combinations 
\begin{eqnarray}
\mathcal{L}^+ \lineup = \mathcal{L}_0+\mathcal{L}_0^\bigstar,\label{eq:Lpeven}\\
\mathcal{L}^- \lineup = \mathcal{L}_0-\mathcal{L}_0^\bigstar.\label{eq:Lminusodd}
\end{eqnarray}
Using \eq{curlyL0starins} and \eq{curlyL0ins} it follows that the insertions corresponding to these operators are
\begin{eqnarray}
\mathcal{L}^+ A\lineup  \ \ \longrightarrow\ \ \oint_{\d A}\frac{dz_A}{2\pi i} \mathrm{sgn}(z_A) T(z_A)\,\mathcal{O}_A,\label{eq:Lplusop}\\
\frac{1}{2}\mathcal{L}^- A\lineup  \ \ \longrightarrow\ \ \oint_{\d A}\frac{dz_A}{2\pi i}\left(z_A-\frac{\alpha}{2}\mathrm{sgn}(z_A) \right)T(z_A)\,\mathcal{O}_A.\label{eq:Lminusop}
\end{eqnarray}
We will see that $\mathcal{L}^-$ is naturally normalized by a factor of $1/2$. It is instructive to make a change of integration variable in \eq{Lplusop} and \eq{Lminusop} to coordinates $z_l$ and $z_r$ defined by 
\begin{eqnarray}
z_A \lineup = \frac{\alpha}{2}+z_l,\label{eq:zl}\\
z_A \lineup = -\frac{\alpha}{2}+z_r.\label{eq:zr}
\end{eqnarray} 
On the left facing vertical edge of the strip of $A$, $z_\ell$ is purely imaginary, while on the right facing vertical edge, $z_r$ is purely imaginary. In this way the operator insertions of $\mathcal{L}_0^+ A$ and $\mathcal{L}_0^- A$ can be represented as
\begin{eqnarray}
\mathcal{L}^+ A\lineup  \ \ \longrightarrow\ \ \int_{-i\infty}^{i\infty}\frac{dz_l}{2\pi i} T\Big(z_l+\frac{\alpha}{2}\Big)\,\mathcal{O}_A\,+\,\mathcal{O}_A\int_{-i\infty}^{i\infty}\frac{dz_r}{2\pi i} T\Big(z_r-\frac{\alpha}{2}\Big),\\
\frac{1}{2}\mathcal{L}^- A\lineup  \ \ \longrightarrow\ \ \int_{-i\infty}^{i\infty}\frac{dz_l}{2\pi i} z_l T\Big(z_l+\frac{\alpha}{2}\Big)\,\mathcal{O}_A\,-\,\mathcal{O}_A\int_{-i\infty}^{i\infty}\frac{dz_r}{2\pi i} z_r T\Big(z_r-\frac{\alpha}{2}\Big).\label{eq:Lmins}
\end{eqnarray}
Here we split the contour $\d A$ into contributions on the left and right vertical edges, and changed the integration variable in each component according to \eq{zl} and \eq{zr}. In the first equation we have a pair of vertical contour insertions of the energy momentum tensor on the left and right edges of the strip of $A$. These insertions define the string field $K$. In this way we find
\begin{equation}\mathcal{L}^+ A = K A+ A K.\label{eq:LpK}\end{equation}
Apparently, the BPZ even part of $\mathcal{L}_0$ is not a new object. What is new is the BPZ odd part, $\mathcal{L}^-$. 

First we note that $\mathcal{L}^-$ is a derivation of the open string star product. To see this, consider the product $AB$. We consider $A$ and $B$ to be strips of respective widths $\alpha$ and $\beta$ containing operator insertions $\mathcal{O}_A$ and $\mathcal{O}_B$ in the respective natural coordinates. The product $AB$ is a strip of width $\alpha+\beta$ containing the operator insertions
\begin{equation}AB\ \ \ \longrightarrow\ \ \  (T_{\beta/2}\circ\mathcal{O}_A)(\,T_{-\alpha/2}\circ \mathcal{O}_B)\end{equation}
in the natural coordinate of $AB$, where $T_a$ is the translation map \eq{transTa}. Using \eq{Lmins} we find
\begin{eqnarray}
 \frac{1}{2}\mathcal{L}^-(AB)\ \ \ \longrightarrow\ \ \ \lineup \int_{-i\infty}^{i\infty}\frac{dz_l}{2\pi i} z_l T\Big(z_l+\frac{\alpha+\beta}{2}\Big) (T_{\beta/2}\circ\mathcal{O}_A)(T_{-\alpha/2}\circ \mathcal{O}_B)\nonumber\\
\lineup \ \ \ -(T_{\beta/2}\circ\mathcal{O}_A)(T_{-\alpha/2}\circ \mathcal{O}_B)\int_{-i\infty}^{i\infty}\frac{dz_r}{2\pi i} z_r T\Big(z_r-\frac{\alpha+\beta}{2}\Big).
\end{eqnarray}
Now on the right hand side we add and subtract an energy momentum contour insertion which passes on the vertical line separating the strips of $A$ and $B$:
\begin{eqnarray}
 \frac{1}{2}\mathcal{L}^-(AB)\ \ \ \longrightarrow\ \ \ \lineup \int_{-i\infty}^{i\infty}\frac{dz_l}{2\pi i} z_l T\Big(z_l+\frac{\alpha+\beta}{2}\Big) (T_{\beta/2}\circ\mathcal{O}_A)(T_{-\alpha/2}\circ \mathcal{O}_B) \nonumber\\
 \lineup -  (T_{\beta/2}\circ\mathcal{O}_A)\int_{-i\infty}^{i\infty}\frac{dz_r}{2\pi i} z_r T\Big(z_r-\frac{\alpha-\beta}{2}\Big) (T_{-\alpha/2}\circ \mathcal{O}_B)\nonumber\\
 \lineup + (T_{\beta/2}\circ\mathcal{O}_A)\int_{-i\infty}^{i\infty}\frac{dz_l}{2\pi i} z_l T\Big(z_l+\frac{\beta-\alpha}{2}\Big) (T_{-\alpha/2}\circ \mathcal{O}_B)\nonumber\\
\lineup \ \ \ -(T_{\beta/2}\circ\mathcal{O}_A)(T_{-\alpha/2}\circ \mathcal{O}_B)\int_{-i\infty}^{i\infty}\frac{dz_r}{2\pi i} z_r T\Big(z_r-\frac{\alpha+\beta}{2}\Big).
\end{eqnarray}
We pull the contours of the first two terms inside the translation map acting on $\mathcal{O}_A$, and the contours in the second two terms inside the translation map acting on $\mathcal{O}_B$:
\begin{eqnarray}
\lineup  \frac{1}{2}\mathcal{L}^-(AB)\ \ \ \longrightarrow\nonumber\\
\lineup \ \ \ \ \ \ \ \ \ \ T_{\beta/2}\circ\left(\int_{-i\infty}^{i\infty}\frac{dz_l}{2\pi i} z_l T\Big(z_l+\frac{\alpha}{2}\Big) \mathcal{O}_A- \mathcal{O}_A \int_{-i\infty}^{i\infty}\frac{dz_r}{2\pi i} z_r T\Big(z_r-\frac{\alpha}{2}\Big)\right)(T_{-\alpha/2}\circ \mathcal{O}_B) \nonumber\\
 \lineup\ \ \ \ \ \ \ \ \ \  + (T_{\beta/2}\circ\mathcal{O}_A)T_{-\alpha/2}\circ \left(\int_{-i\infty}^{i\infty}\frac{dz_l}{2\pi i} z_l T\Big(z_l+\frac{\beta}{2}\Big) \mathcal{O}_B- \mathcal{O}_B\int_{-i\infty}^{i\infty}\frac{dz_r}{2\pi i} z_r T\Big(z_r-\frac{\beta}{2}\Big)\right).
\ \ \ \ \ \ \ \ \ \end{eqnarray}
Here we recognize the operator insertions corresponding to $\frac{1}{2}\mathcal{L}^- A$ and $\frac{1}{2}\mathcal{L}^- B$. Therefore
\begin{equation}
\mathcal{L}^-(AB) = (\mathcal{L}^- A) B+A(\mathcal{L}^- B),
\end{equation}
and $\mathcal{L}^-$ is a derivation.

Next consider how $\mathcal{L}^-$ acts on a string field $\mathcal{O}$ defined by a strip of vanishing width containing a boundary insertion $\mathcal{O}(0)$ in the natural coordinate. Using \eq{Lminusop} with $\alpha = 0$, we learn that
\begin{equation}\frac{1}{2}\mathcal{L}^- \mathcal{O}\ \ \longrightarrow\ \ \oint_0\frac{dz}{2\pi i} zT(z)\mathcal{O}(0) = h\mathcal{O}(0),\end{equation}
where $h$ is the scaling dimension of $\mathcal{O}(0)$. Therefore
\begin{equation}\frac{1}{2}\mathcal{L}^- \mathcal{O} = h\mathcal{O}.\label{eq:LmO}\end{equation}
We also find that
\begin{equation}\frac{1}{2}\mathcal{L}^- K = K.\label{eq:LmK}\end{equation}
\begin{exercise} Demonstrate \eq{LmK} by evaluating the operator insertion corresponding to $\frac{1}{2}\mathcal{L}^-K$ using the self-OPE of the energy momentum tensor. Alternatively, show that
\begin{equation}\frac{1}{2}\mathcal{L}^- \Omega^\alpha = -\alpha K\Omega^\alpha\end{equation}
by evaluating \eq{Lminusop} in the case that $\mathcal{O}_A$ is the identity operator. Taking the derivative with respect to $\alpha$ and setting $\alpha=0$ then implies \eq{LmK}.\end{exercise}
\noindent If $s_\lambda(z) = \lambda z$ is a scale transformation by a factor $\lambda$, we have
\begin{equation}s_\lambda\circ \int_{-i\infty}^{i\infty} \frac{dz}{2\pi i} T(z) =  \lambda \int_{-i\infty}^{i\infty} \frac{dz}{2\pi i} T(z),
\end{equation}
which means that the operator insertion corresponding to $K$ has scaling dimension 1. From this it is clear that  \eq{LmO} and \eq{LmK} have a consistent interpretation: $\frac{1}{2}\mathcal{L}^-$ computes the scaling dimension of operators in correlation functions on the cylinder. Since $\mathcal{L}^-$ is a derivation, this allows us to determine the action of $\mathcal{L}^- $ on the whole subalgebra of wedge states with insertions. Consider for example the state 
\begin{equation}A = \Omega^{\alpha_1}\mathcal{O}_1\Omega^{\alpha_2} ... \Omega^{\alpha_n}\mathcal{O}_n\Omega^{\alpha_{n+1}},\end{equation}
where $\mathcal{O}_i$ represent boundary operator insertions of scaling dimension~$h_i$. A finite transformation generated by $\mathcal{L}^-$ satisfies
\begin{equation}\lambda^{\frac{1}{2}\mathcal{L}^-}K = \lambda K,\ \ \ \ \lambda^{\frac{1}{2}\mathcal{L}^-}\mathcal{O}_i = \lambda^{h_i}\mathcal{O}_i,\end{equation}
and the derivation property of $\mathcal{L}^-$ then implies 
\begin{equation}
\lambda^{\frac{1}{2}\mathcal{L}^-}A = \lambda^{h_1+...+h_n}\,\Omega^{\lambda\alpha_1}\mathcal{O}_1\Omega^{\lambda\alpha_2} ... \Omega^{\lambda\alpha_n}\mathcal{O}_n\Omega^{\lambda\alpha_{n+1}}.
\end{equation}
The result is simply the original state after a scale transformation by $\lambda$ on the cylinder. As discussed at the beginning of subsection \ref{subsec:wedge}, a scale transformation might not appear to change the state. But in comparing the strip with insertions before and after scale transformation, we must be mindful of the fact that the boundary conditions for the worldsheet path integral are modified by the scale transformation. If $A[l(y),r(y)]$ is the split string Schr{\"o}dinger functional of $A$, the split string Schr{\"o}dinger functional of $\lambda^{\frac{1}{2}\mathcal{L}^-}A$ is
\begin{equation}\lambda^{\frac{1}{2}\mathcal{L}^-}A[l(y),r(y)] = A[l(\lambda y),r(\lambda y)],\end{equation}
where $y$ is the vertical coordinate on the left or right edge of the strip of $A$.

Having understood $\mathcal{L}^-$, let us return to $\mathcal{L}_0$. From \eq{Lpeven}, \eq{Lminusodd} and \eq{LpK} we have the relation
\begin{equation}\mathcal{L}_0 A = \frac{1}{2}\mathcal{L}^- A + \frac{1}{2}(KA+AK).\label{eq:L0Lm}\end{equation}
Now consider
\begin{eqnarray}
\frac{1}{2}\mathcal{L}^- \Big(\sqrt{\Omega}A\sqrt{\Omega}\Big) \lineup = \left(\frac{1}{2}\mathcal{L}^-\sqrt{\Omega}\right)A\sqrt{\Omega}+\sqrt{\Omega}\left(\frac{1}{2}\mathcal{L}^-A\right)\sqrt{\Omega}+\sqrt{\Omega}A\left(\frac{1}{2}\mathcal{L}^-\sqrt{\Omega}\right)\nonumber\\
\lineup = -\frac{1}{2}K\sqrt{\Omega}A\sqrt{\Omega} +\sqrt{\Omega}\left(\frac{1}{2}\mathcal{L}^-A\right)\sqrt{\Omega}-\frac{1}{2}\sqrt{\Omega}AK\sqrt{\Omega}.
\end{eqnarray}
Bringing the first and last terms to the other side of the equation and using \eq{L0Lm} implies
\begin{equation}\mathcal{L}_0\Big(\sqrt{\Omega}A\sqrt{\Omega}\Big)=\sqrt{\Omega}\left(\frac{1}{2}\mathcal{L}^-A\right)\sqrt{\Omega}.\end{equation}
Therefore, $\mathcal{L}_0$ generates scale transformations of whatever appears inside a pair of strips of width $1/2$ on either side of the state. If the state does not come with empty strips of width $1/2$ on both sides, we can still apply the above relation by formally multiplying and dividing by $\sqrt{\Omega}$. But in most situations this is not necessary. The $\sqrt{\Omega}$ factors are called {\it security strips}, and appear in most analytic solutions (sometimes with $\Omega$ replaced by some other function of $K$). ``Security" refers to the fact that these strips preclude OPE divergence from operator collisions when computing star products between wedge states with insertions. 

As a sample calculation, consider the string field $c$ introduced in exercise \ref{ex:tach}. Since the $c$-ghost has weight $-1$, it follows that
\begin{equation}\frac{1}{2}\mathcal{L}^- c= -c,\end{equation}
from which we learn that 
\begin{equation}\mathcal{L}_0\big(\sqrt{\Omega}c\sqrt{\Omega}\big) = -\sqrt{\Omega}c\sqrt{\Omega}.\end{equation}
Then exercise \ref{ex:tach} then implies that the zero momentum tachyon state $c_1|0\rangle$ has $\mathcal{L}_0$ eigenvalue $-1$. This can be readily checked from the expansion of $\mathcal{L}_0$ in terms of Virasoros, \eq{curlL0}.

The operator $\mathcal{L}^-$ generates what is known as a {\it midpoint preserving reparameterization} of the string field \cite{WittenSuper}. A midpoint preserving reparameterization generator $\mathcal{K}$ is a linear combination of Virasoros that is BPZ odd and anti-Hermitian:
\begin{equation}\mathcal{K}^\star = -\mathcal{K},\ \ \ \ \mathcal{K}^\dag = -\mathcal{K}.\label{eq:KKoddanti}\end{equation}
In the coordinate $\xi$ of radial quantization, such an operator can be expressed in the form
\begin{equation}\mathcal{K} = \oint_{|\xi|=1} \frac{d\xi}{2\pi i}v(\xi)T(\xi),\end{equation}
where $v(\xi)$ is a chosen vector field (function of $\xi$) subject to the conditions \eq{KKoddanti}. The operator $\mathcal{K}$ generates an infinitesimal reparameterization of the left and right halves of the string in the Schr{\"o}dinger functional:
\begin{equation}(1+\eps \mathcal{K})A[l(\sigma),r(\sigma)] = A\Big[l\big(\sigma+\eps V(\sigma)\big),r\big(\sigma+\eps V(\sigma)\big)\Big],\end{equation}
where the vector field $V(\sigma)$ is related to $v(\xi)$ through
\begin{equation}V(\sigma) = -i e^{-i\sigma}v(e^{i\sigma}).\end{equation}
\begin{exercise}
Use \eq{KKoddanti} to show that $V(\sigma)$ is real and preserves the midpoint, i.e. $V(\pi/2)=0$
\end{exercise}
\noindent Midpoint preserving reparameterization generators commute with the BRST operator, are derivations of the open string star product, and annihilate the trace
\begin{equation}[Q,\mathcal{K}]=0,\ \ \ \mathcal{K}(AB) = (\mathcal{K}A)B+A(\mathcal{K}B),\ \ \ \Tr[\mathcal{K}A]= 0.\end{equation}
This implies that they generate symmetries of Witten's open bosonic SFT---the transformation $\delta \Psi = \eps\mathcal{K}\Psi$ leaves the action invariant. The operator $\mathcal{L}^-$ generates one of only two midpoint preserving reparameterizations which preserve the subalgebra of wedge states with insertions. The other is generated by an operator we write as $\d$:
\begin{equation}\d A = [K,A].\end{equation}
This operator can be expressed in terms of Virasoros as
\begin{equation}\d A = \frac{\pi}{2}K_1 A,\ \ \ \ K_1= L_1+L_{-1}.\end{equation}
Technically, this is a midpoint preserving reparameterization generator only after multiplying by $i$ to make it anti-Hermitian. It generates a shift in the coordinate $y$ on the left and right edges of the strip, $y\to y+\eps$. The notation $\d$ is natural for the following reason. Consider the action of $\d$ on a string field $\mathcal{O}$ defined by a strip of vanishing width containing a boundary insertion $\mathcal{O}(0)$ in the natural coordinate. We find
\begin{eqnarray}
\d \mathcal{O}=[K,\mathcal{O}]\ \ \ \longrightarrow\ \ \ \lineup \int_{-i \infty}^{i\infty} \frac{dz_l}{2\pi i}T(z_l+\eps)\mathcal{O}(0)-\mathcal{O}(0)\int_{-i \infty}^{i\infty} \frac{dz_r}{2\pi i}T(z_l-\eps)\nonumber\\
\lineup = \oint_0\frac{dz}{2\pi i}T(z)\mathcal{O}(0) = \d\mathcal{O}(0).
\end{eqnarray}
Therefore the $\d$ computes the derivative of an operator with respect to its position in the sliver coordinate frame. 

The generators $\mathcal{L}^-$ and $\d$ satisfy a two dimensional Lie algebra
\begin{equation}[\mathcal{L}^-,\d] = 2\d. \end{equation}
This is the Lie algebra of translations and dilatations of the real line. In the present context, the real line is the coordinate $y$ on the left and right vertical edges of the strip of a wedge state. The Lie algebra follows directly from the fact that $K$ carries $\frac{1}{2}\mathcal{L}^-$ eigenvalue 1. For this reason, one can see that the same Lie algebra is satisfied after replacing $\d$ with $\mathcal{L}^+$.  This Lie algebra is sometimes equivalently written as
\begin{equation}[\mathcal{L}_0,\mathcal{L}_0^\bigstar ] = s(\mathcal{L}_0+\mathcal{L}_0^\bigstar),\end{equation}
where presently $s=1$ is a parameter. This is called the {\it special projector algebra} \cite{RZspecial}. Its significance is that it can be used to determine abelian subalgebras of surface states, other than wedge states, which can be used in the construction of analytic solutions. The abelian subalgebra consists of a 1-parameter family of surface states which interpolate between the identity string field and a rank 1 surface state projector, called a {\it special projector}. The parameter $s$ of the special projector algebra becomes significant if we require that $\mathcal{L}_0$ is normalized 
\begin{equation}\mathcal{L}_0 = L_0 + \text{positive mode Virasoros}.\end{equation}
A nontrivial example is given by
\begin{equation}\mathcal{L}_0 = L_0+L_2.\end{equation}
for $s=2$. This defines a family of surface states which interpolate between the identity string field and a surface state projector known as the {\it butterfly} \cite{Gaiotto}. There is a lot of interesting theory behind special projectors that we will not describe here. As far as it has been pursued, the upshot is that every analytic solution constructed using wedge states has an analogue based on states which interpolate to another chosen special projector. The solutions are related through midpoint preserving reparameterizations \cite{RZO}.

Derivations of an associative algebra come in two kinds: {\it inner derivations}, which take the form of a commutator with some element of the algebra, and {\it outer derivations}, which do not. The operator $\d$ can be called an inner derivation, since it can be represented as a star algebra commutator with the string field $K$.\footnote{According to the discussion of subsection \ref{subsec:K}, $K$ would not be considered as part of the wedge algebra since it does not have finite $C^*$ norm. For the present discussion we nevertheless consider it to generate an inner derivation.} It might appear from \eq{Lmins}  that $\mathcal{L}^-$ can also be realized as an inner derivation,
\begin{equation}\mathcal{L}^-A = [\mathcal{L},A],\end{equation}
where $\frac{1}{2}\mathcal{L}$ is an infinitesimally thin strip containing $z T(z)$ integrated vertically on the imaginary axis. An equivalent definition is 
\begin{equation}\mathcal{L} = (\mathcal{L}^-)_L|I\rangle,\end{equation}
where the subscript $L$ indicates that we take the {\it left half} of the operator. To explain, let $\mathcal{O}$ be an operator defined by a contour integral of a holomorphic operator $\phi(\xi)$ around the unit circle:
\begin{equation}\mathcal{O} = \oint_{|\xi|=1}\frac{d\xi}{2\pi i}w(\xi)\phi(\xi),\end{equation}
where $w(\xi)$ is a weight function which is holomorphic in the vicinity of the unit circle. The {\it left half} of this operator is defined by the portion of the contour on the positive half of the unit circle:
\begin{equation}\mathcal{O}_L=\int_{|\xi|=1,\mathrm{Re}(\xi)>0}\frac{d\xi}{2\pi i}w(\xi)\phi(\xi).\end{equation}
We may define the right half similarly. 
\begin{exercise}
Show that $K=\frac{\pi}{2}(K_1)_L|I\rangle$. 
\end{exercise}
\noindent There is an important complication with extracting the left half of an operator. The contour on the positive half of the unit circle is pinned to the midpoint at $\pm i$. This means that repeated application of $\mathcal{O}_L$ is not guaranteed to be well-defined, since contours cannot be deformed away from each other at the midpoint in case of singularities in the OPE of $\phi(\xi)$ with itself. A notable example of this problem occurs with the BRST operator. We can try to represent it as an inner derivation using a string field $Q_L|I\rangle$:
\begin{equation}Q A = \Big[Q_L|I\rangle,A\Big].\end{equation}
In the sliver frame, $Q_L|I\rangle$ is an infinitesimally thin strip containing a contour insertion of the BRST current along the imaginary axis. However, the OPE of two BRST currents contains a third order pole, and this renders repeated application of $Q_L$ undefined. The state $Q_L|I\rangle$ was proposed long ago as an analytic solution of open bosonic SFT \cite{purelycubic}, but for this reason the solution is not well-behaved and does not appear to be physically meaningful. The BRST operator should really be understood as an outer derivation.  However, in more general cases splitting operators into halves may be unproblematic if the weight function $w(\xi)$ vanishes fast enough at the midpoint to compensate for singular OPEs. How quickly it needs to vanish is a delicate question in general, but it appears that splitting $K_1$---as needed for the string field $K$---does not encounter problems. The weight function of $K_1$ vanishes quadratically at the midpoint. The weight function for $\mathcal{L}^-$ does not vanish quite as fast, only as $x^2\ln x$. Finite powers of $(\mathcal{L}^-)_L$ appear to be well-defined. However, ambiguities appear if we consider nonpolynomial combinations. Consider the object
\begin{equation} \lambda^{\frac{1}{2}\mathcal{L}}\Omega^\alpha,\ \ \ \ \lambda>0.\end{equation}
This is called a {\it slanted wedge} \cite{KZloop}. In the sliver frame it can be visualized as a strip of width~$\alpha$, like a wedge state, but the parameterization differs between the left and right vertical edges. On the left edge the parameterization has been scaled relative to the right by a factor of $\lambda$. The split string Sch{\"o}dinger functional would be expressed as
\begin{equation}\lambda^{\frac{1}{2}\mathcal{L}}\Omega^\alpha[l(y),r(y)]=\Omega[l(\lambda\alpha y),r(\alpha y)].\end{equation}
However, the split string representation is problematic at the midpoint, and presently is missing an important subtlety. If a slanted wedge is a string field, it must at least be possible to extract its coefficients in the Fock space expansion. This requires evaluating the trace, which glues 
\begin{wrapfigure}{l}{.5\linewidth}
\centering
\resizebox{3.5in}{1.7in}{\includegraphics{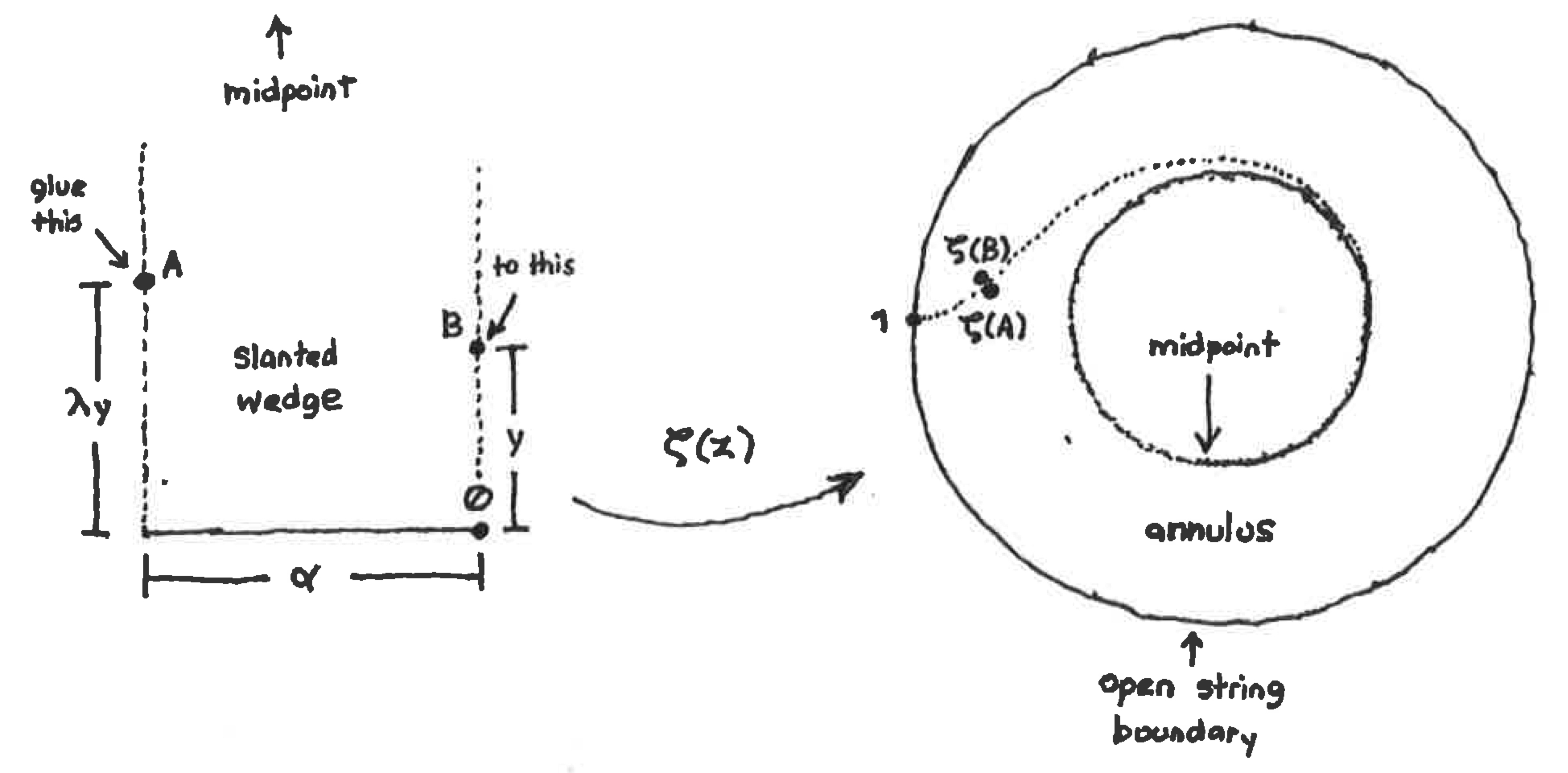}}
\end{wrapfigure} 
the left and right edges of the semi-infinite strip with a ``slanted" identification. The resulting surface, however, is not conformally equivalent to the upper half plane, but rather to an {\it annulus}. If we fix the origin of the sliver coordinate $z$ on the slanted wedge to coincide with the intersection between the right vertical edge and the open string boundary, the conformal transformation to the annulus is given by \cite{KZloop}
\begin{equation}\zeta(z) = \exp\left[\frac{2\pi i}{\ln \lambda}\ln\left(\frac{(\lambda-1)z}{\alpha}+1\right)\right].\end{equation}
The annulus is bounded by two concentric circles. If $\lambda>1$, the outer circle has radius $1$ and represents the open string boundary, whereas the inner circle has radius $e^{-\frac{\pi^2}{\ln\lambda}}$ and is the image of the midpoint at $+i\infty$ in the sliver coordinate frame. In particular, the ``midpoint" of a slanted wedge is actually a nontrivial closed curve. The Fock space expansion of a slanted wedge is not defined unless we fix boundary conditions for the worldsheet fields on the inner circle, which effectively requires specifying a closed string state. Therefore a slanted wedge state is not a string field, but actually lives in a tensor product of open string and closed string state spaces. Or, after BPZ conjugation, it effectively defines an open/closed string vertex. The implication is that the object $\mathcal{L}$ is also not really a string field, and $\mathcal{L}^-$ should be viewed as an {\it outer} derivation, like the BRST operator.

\subsection{$KBc$ subalgebra}
\label{subsec:KBc}

Now we introduce a subalgebra of wedge states with insertions which is sufficient to find analytic solutions for the tachyon vacuum. It is natural to guess that the subalgebra should include the zero momentum tachyon state
\begin{equation}c_1|0\rangle = \frac{\pi}{2}\sqrt{\Omega}c\sqrt{\Omega},\end{equation}
since this is the most important fluctuation field of the D-brane which acquires expectation value after tachyon condensation. Therefore we can consider a subalgebra given by products of string fields $K$ and $c$. However, this subalgebra is not rich enough to describe interesting tachyon vacuum solutions. The crucial additional ingredient was introduced through considerations of gauge fixing. In level truncation, the tachyon vacuum is found after fixing Siegel gauge
\begin{equation}b_0\Psi = 0.\end{equation}
The problem is that once we have $b_0$ we must also consider $L_0$. $L_0$ does not operate inside the subalgebra of wedge states, but generates much larger algebra which is poorly understood. However, we have seen that the analogue of $L_0$ in the sliver frame does operate within the wedge algebra. This suggests that we consider the gauge
\begin{equation}\mathcal{B}_0\Psi=0, \end{equation}
where $\mathcal{B}_0=f_\mathcal{S}\circ b_0$ is the $b$-ghost analogue of $\mathcal{L}_0$. This is called {\it Schnabl gauge}. The BPZ odd combination
\begin{equation}\mathcal{B}^- = \mathcal{B}_0-\mathcal{B}_0^\bigstar\end{equation}
is a derivation of the open string star product and annihilates the trace, for essentially the same reasons as $\mathcal{L}^-$ does. One can check that
\begin{equation}\frac{1}{2}\mathcal{B}^- K = B,\end{equation}
where $B$ is a new string field analogous to $K$ but defined by a vertical contour insertion of the $b$-ghost. We have the relations
\begin{eqnarray}
\lineup \mathcal{B}_0 X= \frac{1}{2}\mathcal{B}^- X + \frac{1}{2}(BX+(-1)^{|X|}XB),\label{eq:B0Bm}\\
\lineup \ \mathcal{B}_0\Big(\sqrt{\Omega}X\sqrt{\Omega}\Big) = \sqrt{\Omega}\left(\frac{1}{2}\mathcal{B}^- X\right)\sqrt{\Omega}.
\end{eqnarray}
The fields $K$, $B$ and $c$ are together generate the $KBc$ {\it subalgebra} \cite{Okawa}. States in the $KBc$ subalgebra live in the universal sector, so we can hope to find solutions for the tachyon vacuum.

The fields have Grassmann and ghost number assignments 
\begin{eqnarray}
K \lineup =\mathrm{Grassmann\ even},\ \mathrm{gh}\#\ 0,\nonumber\\
B \lineup =\mathrm{Grassmann\ odd},\ \mathrm{gh}\#\ -1,\nonumber\\  
c \lineup =\mathrm{Grassmann\ odd},\ \mathrm{gh}\#\ 1,
\end{eqnarray}
and satisfy the identities
\begin{eqnarray}
\lineup [K,B] = 0,\ \ \ \ B^2=c^2=0, \ \ \ \ [B,c] = 1,\label{eq:KBcId1}\\
\lineup \ \ QK=0,\ \ \ \ \ \ QB=K,\ \ \ \ \ \ Qc=cKc.\label{eq:KBcId2}
\end{eqnarray}
Further relations involving $\mathcal{L}^-$ and $\mathcal{B}^-$ are
\begin{eqnarray}
\frac{1}{2}\mathcal{L}^-K = K,\lineup \ \ \ \ \ \ \frac{1}{2}\mathcal{B}^- K=B,\nonumber\\
\frac{1}{2}\mathcal{L}^-B= B,\lineup\ \ \ \ \ \ \, \frac{1}{2}\mathcal{B}^- B=0,\nonumber\\
\frac{1}{2}\mathcal{L}^-c = -c,\lineup \ \ \ \ \ \ \ \, \frac{1}{2}\mathcal{B}^- c=0.
\end{eqnarray}
The fields are real:
\begin{equation}K^\ddag= K,\ \ \ \ B^\ddag = B,\ \ \ \ c^\ddag = c .\end{equation}
These identities are easily verified by the appropriate contour deformations inside correlation functions on the cylinder. The fields $K$ and $c$ additionally satisfy a hierarchy of relations called {\it auxiliary identities}
\begin{equation}(\d^m c)^2 =  0, \ \ \ \ m\geq 1.\label{eq:aux}\end{equation}
These are ``auxiliary" since they are not necessary for important known analytic solutions the $KBc$ subalgebra. In addition, the $KBc$ subalgebra has a useful automorphism structure (see subsection \ref{subsec:dualL}) in which auxiliary identities are not preserved. 

The general state in the $KBc$ subalgebra at each ghost number takes the form
\begin{eqnarray}
\text{gh\# } -1: \lineup\ \  BF(K),\\
\text{gh\# }0: \lineup\ \  F(K) + \int_0^\infty d\alpha_1d\alpha_2\, f(\alpha_1,\alpha_2) \Omega^{\alpha_1} cB\Omega^{\alpha_2},\\
\text{gh\# }1: \lineup \ \ \int_0^\infty d\alpha_1d\alpha_2d\alpha_3\, f(\alpha_1,\alpha_2,\alpha_3)\Omega^{\alpha_1}cB\Omega^{\alpha_2}c\Omega^{\alpha_3},\\
\text{gh\# }2: \lineup \ \ \int_0^\infty d\alpha_1d\alpha_2d\alpha_3 d\alpha_4\,f(\alpha_1,\alpha_2,\alpha_3,\alpha_4)\Omega^{\alpha_1}cB\Omega^{\alpha_2}c\Omega^{\alpha_3}c\Omega^{\alpha_4},\\
\text{gh\# }3: \lineup \ \ \int_0^\infty d\alpha_1d\alpha_2d\alpha_3 d\alpha_4 d\alpha_5\, f(\alpha_1,\alpha_2,\alpha_3,\alpha_4,\alpha_5)\Omega^{\alpha_1}cB\Omega^{\alpha_2}c\Omega^{\alpha_3}c\Omega^{\alpha_4}c\Omega^{\alpha_5},\label{eq:KBc3}\\
\lineup \vdots\ \ .\nonumber
\end{eqnarray}
There are no states at ghost number less than $-1$. At higher ghost numbers, the $c$ ghost can become entangled inside integrals over wedge states, so the general state is characterized by a function of several variables in the time domain. After Laplace transform, these can be thought of as functions of several $K$ variables, so the state at ghost number zero is specified by an $F(K)$  and an $F(K_1,K_2)$, the state at ghost number one by $F(K_1,K_2,K_3)$, and so on. Therefore the $KBc$ subalgebra is a type of algebra of functions of several variables. Presently it is unclear how it should be understood precisely as an infinite dimensional algebra. However, our understanding of the wedge algebra will be sufficient for most purposes.  

Let us take a first look at classical solutions. One might notice that the field $c$ itself almost looks like a solution, except for the factor of $K$ which appears between the two $c$s when computing the BRST variation. This can be remedied by multiplying by $K$, so  that 
\begin{equation}\Psi = -cK\end{equation} 
is a solution to the equations of motion. This is a so-called {\it residual solution} \cite{IdSing}, and is not physically very interesting (see lecture 4).  However, if we add $c$ we get something more significant:
\begin{equation}\Psi = c(1-K).\label{eq:Idtv}\end{equation}
This is a solution for the tachyon vacuum. Unfortunately the solution is not normalizable, in a similar way as the identity string field. To verify Sen's  conjecture we should substitute the solution into the action, but doing this leads to a correlation function on a  ``needle-like" cylinder whose circumference is strictly zero. A cylinder with vanishing circumference cannot be mapped to the upper half plane, so there is no way to unambiguously compute the correlation function. However there is another way to check that the solution represents the tachyon vacuum. We can show that linearized fluctuations do not contain physical open string states. This requires investigating the cohomology of the shifted kinetic operator
\begin{equation}Q_\Psi = Q + [c(1-K),\,\cdot\,].\end{equation}
It is interesting to consider how this operator acts on the string field $B$:
\begin{eqnarray}
Q_\Psi B \lineup = K + [c,B](1-K)\nonumber\\
\lineup = K + 1-K\nonumber\\
\lineup = 1.\label{eq:QtvB}
\end{eqnarray}
Given an on-shell fluctuation of the solution
\begin{equation}Q_\Psi \varphi = 0,\end{equation}
we can therefore write
\begin{equation}\varphi = 1*\varphi = \big(Q_\Psi B\big)\varphi = Q_\Psi\big(B\varphi\big).\end{equation}
This implies that all $Q_\Psi$-closed states are $Q_\Psi$ exact, and all on-shell fluctuations of the solution are gauge equivalent to no fluctuation at all. Note that the cohomology is empty at {\it all} ghost numbers, not just ghost number $1$ which would be enough to confirm the absence of open string states.  For tachyon vacuum solutions in the $KBc$ subalgebra, the absence of cohomology is generally demonstrated by finding a string field $A$ satisfying
\begin{equation}Q_\Psi A = 1.\end{equation}
The string field $A$ is often called a {\it homotopy operator}. For the tachyon vacuum \eq{Idtv},  the homotopy operator is simply $B$.

It will be useful to have formulas for the trace of string fields in the $KBc$ subalgebra. The trace is nonvanishing only at ghost number 3, and from \eq{KBc3} it is clear that the result is determined by the expression
\begin{equation}\Tr(\Omega^{\alpha_1}c\Omega^{\alpha_2}c\Omega^{\alpha_3}c\Omega^{\alpha_4}cB)\label{eq:genca}\end{equation}
For simplicity let us start with the special case
\begin{equation}
\Tr(\Omega^{\alpha_1}c\Omega^{\alpha_2}c\Omega^{\alpha_3}c)=\langle c(z_1) c(z_2) c(z_3)\rangle_{C_L}.
\end{equation}
which follows from \eq{genca} after setting $\alpha_1=0$ and relabeling the $\alpha$s. On the right hand side we expressed the trace as a correlation function on the cylinder. The circumference of the cylinder and the position of the $c$-ghost insertions is related to the wedge parameters through
\begin{eqnarray}
L\lineup  = \alpha_1+\alpha_2+\alpha_3,\\
z_2\lineup = \alpha_3+z_3,\\
z_1\lineup = \alpha_2+\alpha_3+z_3.\label{eq:wedge_corr}
\end{eqnarray}
The position $z_3$ is determined by a choice of origin on the cylinder, which is not specified by the wedge parameters. However, rotational symmetry of the cylinder ensures that the correlator is independent of the choice of origin. Using \eq{CL} we map the cylinder to the upper half plane, which leads to
\begin{eqnarray}
\lineup \!\!\!\!\!\!\!\!\langle c(z_1) c(z_2) c(z_3)\rangle_{C_L} = \Big\langle C_L^{-1}\circ\big(c(z_1) c(z_2) c(z_3)\big)\Big\rangle_{\mathrm{UHP}}\nonumber\\
\lineup = \left(\cos^2\frac{\pi z_1}{L}\right)\left(\cos^2\frac{\pi z_2}{L}\right)\left(\cos^2\frac{\pi z_3}{L}\right)\left\langle c\left(\frac{L}{\pi}\tan\frac{\pi z_1}{L}\right) c\left(\frac{L}{\pi}\tan\frac{\pi z_2}{L}\right) c\left(\frac{L}{\pi}\tan\frac{\pi z_3}{L}\right)\right\rangle_\mathrm{UHP}.\nonumber\\
\end{eqnarray}
The correlator of three $c$s on the upper half plane can be evaluated with \eq{3cs}, and using various trigonometric identities we arrive at the result
\begin{equation}\langle c(z_1) c(z_2) c(z_3)\rangle_{C_L} = \left(\frac{L}{\pi}\right)^3\sin\frac{\pi z_{12}}{L}\sin\frac{\pi z_{13}}{L}\sin\frac{\pi z_{23}}{L}, \ \ \ \ (z_{ij}=z_i-z_j),\label{eq:ccc}\end{equation}
assuming the matter correlator is normalized to unity. Turning to the more general case \eq{genca} we may write
\begin{equation}\Tr(\Omega^{\alpha_1}c\Omega^{\alpha_2}c\Omega^{\alpha_3}c\Omega^{\alpha_4}cB)=\langle c(z_1)c(z_2)c(z_3)c(z_4)B\rangle_{C_L},\end{equation}
where the circumference and $c$-ghost positions are related to the wedge parameters in a similar way as \eq{wedge_corr}. The insertion $B$ on the right hand side represents a vertical line integral of the $b$-ghost passing directly on the negative side of $z_4$. One way to compute the correlator is to use $\mathcal{B}^-$ invariance of the trace
\begin{equation}\Tr\left(\frac{1}{2}\mathcal{B^-}(\Omega^{\alpha_1}c\Omega^{\alpha_2}c\Omega^{\alpha_3}c\Omega^{\alpha_4}c)\right) = 0.\end{equation}
Acting with $\frac{1}{2}\mathcal{B}^-$ produces four terms with a $B$ accompanying each wedge state. Commuting all of the $B$s to the right then gives a formula for the correlator with four $c$s and one $B$ in terms of correlators with only three $c$s. In this way we obtain \cite{Schnabl}
\begin{eqnarray}
\langle c(z_1)c(z_2)c(z_3)c(z_4)B\rangle_{C_L}\lineup = \frac{L^2}{\pi^3}\left(z_1 \sin\frac{\pi z_{23}}{L}\sin\frac{\pi z_{24}}{L}\sin\frac{\pi z_{34}}{L} - z_2 \sin\frac{\pi z_{13}}{L}\sin\frac{\pi z_{14}}{L}\sin\frac{\pi z_{34}}{L}\right.\nonumber\\
\lineup\ \ \ \ \left. + z_3 \sin\frac{\pi z_{12}}{L}\sin\frac{\pi z_{14}}{L}\sin\frac{\pi z_{24}}{L} -z_4\sin\frac{\pi z_{12}}{L}\sin\frac{\pi z_{13}}{L}\sin\frac{\pi z_{23}}{L}\right).\ \ \ \ \ \ \ \ \ \label{eq:ccccB}
\end{eqnarray}
\begin{exercise}
Do this calculation.
\end{exercise}
\noindent  An equivalent useful formula (corrected from \cite{SSFII}) is 
\begin{eqnarray}
\langle c(z_1)c(z_2)c(z_3)c(z_4)B\rangle_{C_L}\lineup = \frac{L^2}{4\pi^3}\left(z_{14} \sin\frac{2\pi z_{23}}{L}+z_{23}\sin\frac{2\pi z_{14}}{L}-z_{13}\sin\frac{2\pi z_{24}}{L}\right.\nonumber\\
\lineup\ \ \ \ \left. -z_{24}\sin\frac{2\pi z_{13}}{L}+z_{12}\sin\frac{2\pi z_{34}}{L} +z_{34}\sin\frac{2\pi z_{12}}{L}\right).\ \ \ \ \ \ \ \ \ 
\end{eqnarray}
This is manifestly independent of the choice of origin on the cylinder and is only linear in sines. A similar formula without the $B$ insertion can be found by setting $z_{14}=L$:
\begin{equation}\langle c(z_1) c(z_2) c(z_3)\rangle_{C_L} = \frac{1}{4}\left(\frac{L}{\pi}\right)^3\left(\sin\frac{2\pi z_{12}}{L}-\sin\frac{2\pi z_{13}}{L}+\sin\frac{2\pi z_{23}}{L}\right).\end{equation}

\section{Lecture 3: Analytic solutions}

In this lecture we describe some of the significant analytic solutions of Witten's open bosonic SFT. This includes Schnabl's solution for the tachyon vacuum \cite{Schnabl}; Schnabl gauge solutions for marginal deformations \cite{KORZ,Schnabl_marg}; the simple tachyon vacuum \cite{simple}; the solution of Kiermaier, Okawa, and Soler \cite{KOS}, further generalized to arbitrary time-independent backgrounds in \cite{KOSsing}; and the solution for the Wilson line deformation introduced by Fuchs, Kroyter, and Potting \cite{FKP} and further generalized to arbitrary marginal deformations by Kiermaier and Okawa \cite{KO}. 

We will not discuss the marginal and tachyon vacuum solutions of Takahashi and Tanimoto \cite{TT1,TT2}. These are somewhat different from the solutions we discuss, since they are not in a significant way built from wedge states with insertions. Instead, they are built from acting certain operators on the identity string field. An advantage of this approach is that it is relatively easy to study the string field theory expanded around the new background, even in the level truncation scheme, since the shifted kinetic operator is given by a contour integral of a local operator around the origin, much like the BRST operator. A drawback is that the solutions are not normalizable. However, using the Zeze map (see subsection \ref{subsec:dualL}) it is possible to find variants  which are normalizable \cite{MaccaferriTT,IshibashiTT}.

\subsection{Schnabl's solution}
\label{subsec:Sch}

We will give a derivation of Schnabl's solution which is different from the original approach, but is more direct from the perspective of our development. We look for solutions among states in the $KBc$ subalgebra satisfying the Schnabl gauge condition 
\begin{equation}\mathcal{B}_0\Psi = 0.\end{equation}
A class of such states takes the form
\begin{equation}\Psi = \sqrt{\Omega}cBG(K)c\sqrt{\Omega}.\end{equation}
The fully general state in Schnabl gauge is slightly more complicated than this. To make discussion simpler we focus on this ansatz, which is sufficient to find the Schnabl gauge solutions.  
\begin{exercise}
Find the most general state in Schnabl gauge in the  $KBc$ subalgebra, and show that it does not support  more solutions  than  those described below.
\end{exercise}
\noindent  To verify the Schnabl gauge condition, note
\begin{equation}\mathcal{B}_0\big(\sqrt{\Omega}cBG(K)c\sqrt{\Omega}\big) =\sqrt{\Omega}\frac{1}{2}\mathcal{B}^-(cBG(K)c)\sqrt{\Omega}.\end{equation}
Recall that $\mathcal{B}^-$ acts as a derivation and annihilates $B$ and $c$. The only possible contribution appears when $\mathcal{B}^-$ acts on $G(K)$, giving
\begin{equation}\frac{1}{2}\mathcal{B}^- G(K) = BG'(K).\end{equation}
However, this does not contribute due to $B^2=0$. Next we plug the ansatz into the equations of motion to fix the form of $G(K)$:
\begin{eqnarray}
Q\Psi \lineup = -\sqrt{\Omega}cKBc G(K)c\sqrt{\Omega} + \sqrt{\Omega}cB G(K) cKc \sqrt{\Omega},\\
\Psi^2\lineup = \sqrt{\Omega}cB G(K) c\Omega G(K) c\sqrt{\Omega} - \sqrt{\Omega}cB \Omega G(K)cG(K)c\sqrt{\Omega}.
\end{eqnarray}
A moment's thought reveals that the equations of motion are equivalent to the following functional equation for $G(K)$:
\begin{equation} - K_1 G(K_2) +G(K_1)K_2+G(K_1)e^{-K_2}G(K_2)- e^{-K_1}G(K_1)G(K_2) = 0.\end{equation}
Since there are two variables $K_1,K_2$ and only one undetermined function $G(K)$, this equation looks over determined. Still there is a solution. After some algebra we can rewrite this as
\begin{equation}\frac{G(K_1)}{K_1+e^{-K_1}G(K_1)}=\frac{G(K_2)}{K_2+e^{-K_2}G(K_2)}.\end{equation}
Since the left hand side is a function only of $K_1$, and the right hand side a function of $K_2$, the only way this can be consistent is if both sides are equal to a constant, which we call $\lambda$:
\begin{equation}\frac{G(K)}{K+\Omega G(K)} = \lambda.\label{eq:Grel}\end{equation}
This implies 
\begin{equation}G(K) = \frac{\lambda K}{1-\lambda \Omega},\end{equation}
and
\begin{equation}\Psi_\lambda = \lambda \sqrt{\Omega}c \frac{KB}{1-\lambda \Omega}c\sqrt{\Omega}.\label{eq:Psil}\end{equation}
We have a 1-parameter family of $KBc$ solutions in Schnabl gauge. Incidentally, the solution satisfies the reality condition
\begin{equation}\Psi_\lambda^\ddag = \Psi_\lambda\end{equation}
if $\lambda$ is real. Since $K,B$ and $c$ are real string fields, the reality condition amounts to the statement that the solution reads the same way from the left as from the right.

If $\lambda = 0$ we obtain the trivial solution
\begin{equation}\Psi_{\lambda=0} = 0.\end{equation}
This is the perturbative vacuum. If $\lambda$ is small we can expand the solution perturbatively:
\begin{equation}\Psi_\lambda = \lambda \sqrt{\Omega}cKBc\sqrt{\Omega} +\mathcal{O}(\lambda^2).\end{equation}
The leading order contribution is BRST exact:
\begin{equation}\sqrt{\Omega}cKBc\sqrt{\Omega} = Q\big(\sqrt{\Omega}Bc\sqrt{\Omega}\big).\end{equation}
This means that, for sufficiently small $\lambda$, the solution represents a deformation of the perturbative vacuum by a trivial element of the BRST cohomology. Physically, this represents no deformation at all, and for small enough $\lambda$ the solution is gauge equivalent to the perturbative vacuum. This is somewhat unexpected, since we have supposedly fixed the gauge. Apparently, the Schnabl gauge condition does not fully fix the gauge. 

What we wanted to find is a solution for the tachyon vacuum. We can hope that the tachyon vacuum will appear for large enough $\lambda$. We can test this by searching for a homotopy operator,
\begin{equation}Q_{\Psi_\lambda} A_\lambda =1,\end{equation}
which trivializes the cohomology at some $\lambda$. Assuming the homotopy operator can be found in the $KBc$ subalgebra, it must take the form
\begin{equation}A_\lambda = B H(K)\end{equation}
for some $H(K)$. Plugging into the previous equation we find 
\begin{equation}H(K)= \frac{1-\lambda\Omega}{K}.\end{equation}
\begin{wrapfigure}{l}{.3\linewidth}
\centering
\resizebox{2in}{1.4in}{\includegraphics{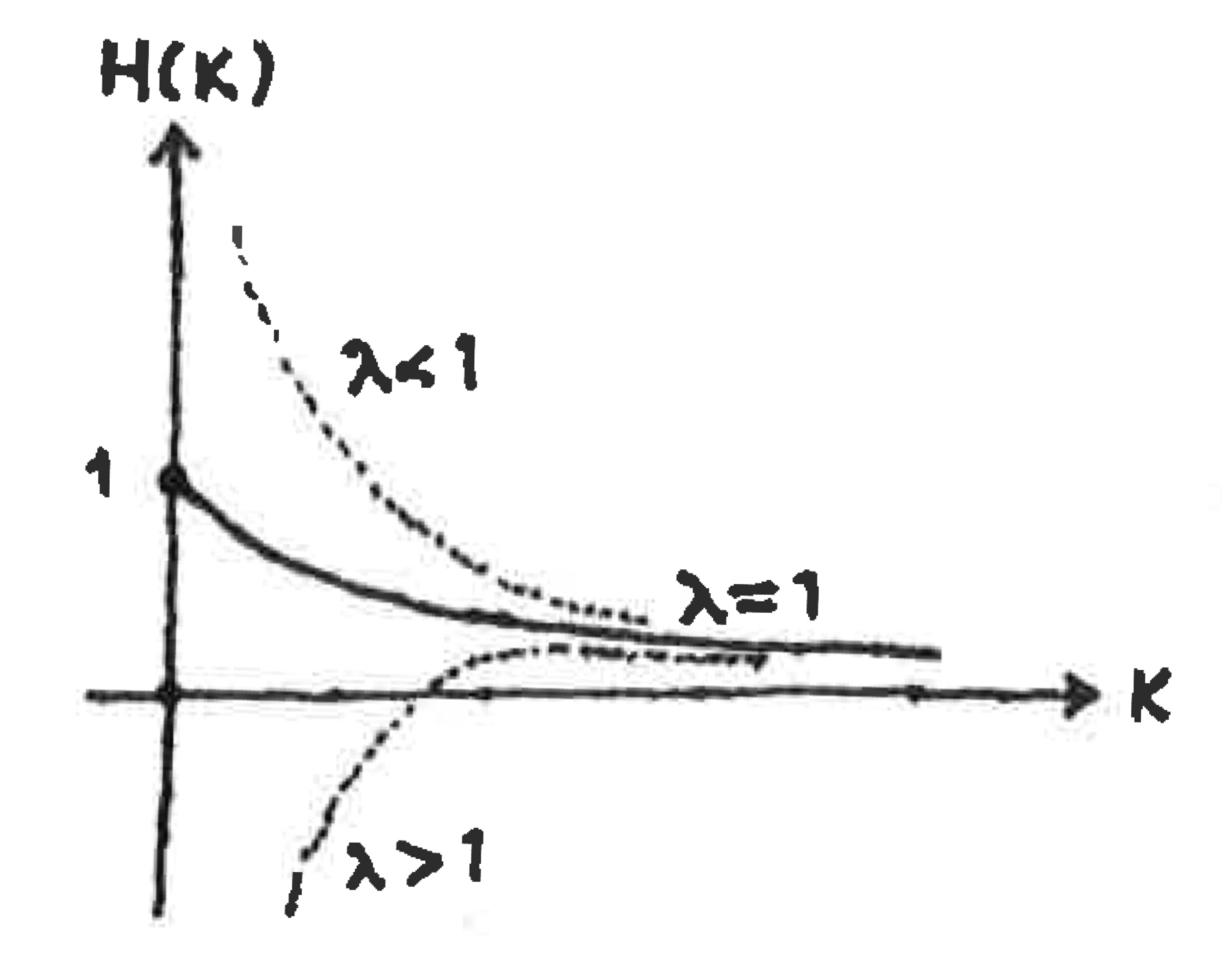}}
\end{wrapfigure}
This is a meaningful function of a real variable $K$, but it does not necessarily exist as a string field. In fact, $H(K)$ is divergent in the $C^*$ norm due to a pole at $K=0$. The exception is at $\lambda=1$, where we find
\begin{equation}||H(K)||_{C^*} = 1,\ \ \ \ (\lambda=1).\end{equation}
To see that the pole should be taken seriously, consider the homotopy operator at $\lambda=0$:
\begin{equation}A_{\lambda=0}=\frac{B}{K}.\end{equation}
We already know this string field can't exist since the BRST operator supports cohomology. But it is instructive to try to construct it anyway. If we think of $K$ as a positive real variable, we have the equality
\begin{equation}\frac{1}{K} = \int_0^\infty d\alpha\, e^{-\alpha K}\end{equation}
This indicates that the homotopy operator at $\lambda=0$ should be expressed as a superposition of wedge states in the form 
\begin{equation}\frac{B}{K} \stackrel{?}{=} B\int_0^\infty d\alpha\, \Omega^\alpha.\label{eq:BoK}\end{equation}
The integral by itself is divergent, since wedge states approach a constant---the sliver state---for large wedge parameter. But the divergence is canceled by $B$, as follows from a useful result: 
\begin{claim} 
\label{claim:Bsliver}
The state $B\Omega^\alpha$ tends to zero in the Fock space expansion as $1/\alpha^3$ for large $\alpha$. In particular, $B$ annihilates the sliver state.
\end{claim}
\noindent Therefore \eq{BoK} is a well defined string field, at least in the Fock space expansion. The difficulty however is that it does not define a homotopy operator for $Q$:
\begin{equation}
Q\left(\frac{B}{K}\right) = \int_0^\infty d\alpha K\Omega^\alpha = -\int_0^\infty d\alpha \frac{d}{d\alpha}\Omega^\alpha = 1-\Omega^\infty.\label{eq:QBovK}
\end{equation}
We find an unwanted boundary contribution from the sliver state. The reason it appears is precisely the expected obstruction from cohomology.  The Fock space expansion of a wedge state takes the form 
\begin{equation}\Omega^\alpha = |0\rangle + (\text{total Virasoro descendants of the vacuum}).\end{equation}
The descendant terms are BRST exact since total Virasoros can be derived by BRST variation of $b$ ghosts. The only piece which is nontrivial in the cohomology is the vacuum $|0\rangle$. Subtracting the sliver state therefore removes the only part of the identity string field which cannot be expressed in BRST exact form. Therefore \eq{BoK} and presumably all homotopy operators for $\lambda\neq 1$ do not imply the absence of cohomology. For $\lambda=1$ however the $C^*$ norm is finite and we have a well-defined homotopy operator,
\begin{equation}A_\text{Sch} = B\frac{1-\Omega}{K}=B\int_0^1 d\alpha\,\Omega^\alpha,\end{equation}
and the solution
\begin{equation}\Psi_\text{Sch} = \sqrt{\Omega}c\frac{KB}{1-\Omega}c\sqrt{\Omega},\end{equation}
does not support open string states. This is Schnabl's solution for the tachyon vacuum~\cite{Schnabl}. 

\begin{exercise} Prove claim \ref{claim:Bsliver}.\end{exercise}
\begin{exercise}
The result of claim \ref{claim:Bsliver} implies that the sliver state can be defined as a limit
\begin{equation}\Omega^\infty = \lim_{\alpha\to\infty}\big(\Omega^\alpha cB\big).\end{equation}
Show that the relation
\begin{equation} \lim_{\alpha\to\infty}\big(\Omega^\alpha cB\big)*\big(\Omega^\alpha cB\big) = 0\end{equation}
holds in the Fock space expansion. From this point of view it appears that the sliver state is not a projector, but rather squares to zero. 
\label{ex:Bsliver}
\end{exercise}

To define Schnabl's solution we need to realize the function
\begin{equation}\frac{K}{1-\Omega}\end{equation}
concretely as a state in the wedge algebra. This turns out to be tricky. For general $\lambda$ the Schnabl gauge solution involves the state 
\begin{equation}G(K) = \lambda K + \frac{\lambda^2 K\Omega}{1-\lambda \Omega}.\end{equation}
This state has infinite $C^*$ norm for any nonzero $\lambda$ due to the linear growth of the first term towards $K=\infty$. Indeed, the string field $K$ is singular in a similar way as the identity string field. In the present context this is not a concern, since the state (together with ghost insertions) appears multiplied by the $SL(2,\mathbb{R})$ vacuum in the solution, which effectively tames the singularity (see subsection \ref{subsec:dualL} for explanation). Therefore we focus on the second term in $G(K)$. It is clear that the $C^*$ norm will be finite if the denominator does not have a zero for positive $K$. Since $e^{-K}$ is less than one, a zero can appear only if $\lambda$ is a real number greater than one. Therefore 
\begin{equation}\left|\left|\frac{\lambda^2 K\Omega}{1-\lambda \Omega}\right|\right|_{C^*} = \mathrm{finite},\ \ \ \ \ \text{iff }\lambda\ngtr 1.\end{equation}
The tachyon vacuum sits just on the edge of singularity at $\lambda=1$, but the norm is still finite:
\begin{equation}
\left|\left|\frac{K\Omega}{1-\Omega}\right|\right|_{C^*} = 1.
\end{equation}
A curious fact is that the Fock space coefficients of the Schnabl gauge solutions vary continuously with $\lambda$. Since there is a discontinuous change of background at $\lambda=1$, this implies that the Fock space expansion does not capture the physically relevant notion of ``proximity" of string fields. If we strip off the ghost insertions we can measure the proximity of pure gauge and tachyon vacuum solutions using the $C^*$ norm. We find that 
\begin{equation}
\left|\left|\frac{K\Omega}{1-\Omega}-\frac{\lambda^2 K\Omega}{1-\lambda \Omega}\right|\right|_{C^*} \geq 1.
\end{equation}
Even though the functions converge to each other pointwise for $K>0$, at $K=0$ the first is always $1$ and the second always zero. From this point of view the pure gauge and tachyon vacuum solutions are never close. In fact, the limit $\lambda\to 1$ does not converge in the $C^*$ norm, in the same way as the sliver limit does not converge. 

From the point of view of bounded, continuous functions of $K\geq 0$ the Schnabl gauge solution exists for any $\lambda$ which is not real and greater than $1$. But from the point of view of bounded, {\it analytic} functions of $\mathrm{Re}(K)> 0$ there are further restrictions and a complication. To have finite $D_2$ norm we must require that $1-\lambda\Omega$ has no zeros for non-negative $\mathrm{Re}(K)$. There are an infinite number of zeros  located at
\begin{equation}K_\mathrm{zero} = \ln|\lambda|+i\arg\lambda + 2\pi i n,\ \ \ \ n\in \mathbb{Z}.\end{equation}
If $|\lambda|$ is strictly less than $1$, the zeros do not enter the non-negative half of the complex $K$ plane. Therefore
\begin{equation}\left|\left|\frac{\lambda^2 K\Omega}{1-\lambda \Omega}\right|\right|_{D_2} = \mathrm{finite},\ \ \ \ \ \text{iff }|\lambda|<1. \end{equation}
Unfortunately the tachyon vacuum is excluded. The function $\frac{K\Omega}{1-\Omega}$ has an  infinite number of poles on the imaginary axis, and is not bounded for $\mathrm{Re}(K)\geq 0$. It is not clear whether this indicates that Schnabl's solution is singular. But the infinite $D_2$ norm does imply that there will be complications defining Schnabl's solution as a superposition of wedge states.

One possible workaround is to define $K/(1-\Omega)$ through its Taylor series around the origin. In fact, this is the generating function for Bernoulli numbers, so the solution can be written 
\begin{eqnarray}
\Psi_\text{Sch}\lineup = \sum_{n=0}^\infty \frac{(-1)^n B_n}{n!}\sqrt{\Omega}cBK^nc\sqrt{\Omega}\\
\lineup = \sqrt{\Omega}c\sqrt{\Omega} - \frac{1}{2}\sqrt{\Omega}cKBc\sqrt{\Omega} +\frac{1}{12}\sqrt{\Omega}cK^2 Bc\sqrt{\Omega} +...\ .
\end{eqnarray}
Each term in this expansion is an eigenstate of $\mathcal{L}_0$:
\begin{equation}\mathcal{L}_0\Big(\sqrt{\Omega}cK^nBc\sqrt{\Omega}\Big) = (n-1)\sqrt{\Omega}cBK^n c\sqrt{\Omega}.\end{equation}
This defines the so-called {\it $\mathcal{L}_0$ level expansion} of the solution. This can be seen as analogous to an expansion into a basis $L_0$ eigenstates, but formulated in the sliver frame. We will define the {\it level} of a state in the $\mathcal{L}_0$ level expansion to be its $\mathcal{L}_0$ eigenvalue. Therefore the $\mathcal{L}_0$ level expansion of Schnabl's solution starts at level $-1$ with the zero momentum tachyon state $c_1|0\rangle$, multiplied by a coefficient $2/\pi \approx 0.64$. For comparison, the coefficient of the zero momentum tachyon in the ordinary level expansion of the Siegel gauge tachyon vacuum is $\approx 0.54$. It is interesting to investigate the $\mathcal{L}_0$ level expansion of the Schnabl gauge solutions when $\lambda\neq 1$:
\begin{equation}\Psi_\lambda = \frac{\lambda}{1-\lambda}\sqrt{\Omega}cKBc\sqrt{\Omega} - \frac{\lambda^2}{(1-\lambda)^2}\sqrt{\Omega}c K^2 Bc\sqrt{\Omega}+...\ .\end{equation}
The zero momentum tachyon is now absent from the expansion, and the leading state at level $0$ is trivial in the BRST cohomology. More interestingly, the $\mathcal{L}_0$ level expansion is divergent in the limit $\lambda\to 1$. This is another way to see that the tachyon vacuum and pure gauge solutions are not close. 

Unfortunately the $\mathcal{L}_0$ level expansion does not give a fully adequate definition of the solution. To see why,  consider the $\mathcal{L}_0$ level expansion of an inverse wedge state:
\begin{equation}\Omega^{-1} = \sum_{n=0}^\infty \frac{2^n}{n!}\sqrt{\Omega}K^n\sqrt{\Omega}.\end{equation}
Aside from signs, this is the same as the $\mathcal{L}_0$ expansion of $\Omega^3$. Obviously, $\Omega^3$ is an ordinary wedge state while $\Omega^{-1}$ is not normalizable. So the existence of the $\mathcal{L}_0$ level expansion is not enough to tell us that a string field is well-behaved. Therefore we look for a different way to define Schnabl's solution. One possibility is to define it through the geometric series
\begin{equation}\frac{K}{1-\Omega}  \stackrel{?}{=} \sum_{n=0}^\infty K\Omega^n .\end{equation}
The sum converges in the Fock space expansion, since $K\Omega^\alpha$ vanishes as $1/\alpha^3$ for large $\alpha$ (in a similar way as  $B\Omega^\alpha$). But the $\mathcal{L}_0$ level expansion on the right hand side does not work correctly. The leading level contribution to each term in the geometric series is $\sqrt{\Omega}K\sqrt{\Omega}$, and performing the sum gives a divergent result. This suggests that the right hand side really represents the pure gauge solution in the limit $\lambda\to 1$. To see how to fix this, we note the identity
\begin{equation}\frac{K}{1-\Omega} = \sum_{n=0}^N K\Omega^n +\frac{K}{1-\Omega}\Omega^{N+1}.\end{equation}
We truncated the sum, leaving a finite remainder. For a function of a positive real number $K$, the remainder would be ignorable for large $N$. But considered as a string field, the remainder is nonzero and approaches the sliver state. We can expand the coefficient of the remainder in terms of Bernoulli numbers
\begin{equation}
\frac{K}{1-\Omega} = \sum_{n=0}^N K\Omega^n + \sum_{n=0}^\infty \frac{(-1)^n B_n}{n!}K^n\Omega^{N+1}.
\end{equation}
For large $N$ the higher powers of $K$ acting on $\Omega^{N+1}$ are suppressed. In practice it appears to be enough to keep the leading term  and make the identification
\begin{equation}
\frac{K}{1-\Omega} = \lim_{N\to\infty}\left[\sum_{n=0}^N K\Omega^n + \Omega^{N}\right].
\end{equation}
This limit should be understood in a special sense. Given an expression which depends on Schnabl's solution,  each appearance of $K/(1-\Omega)$ should be replaced by the expression in brackets above {\it for a single common $N$}. One then performs the calculation at finite $N$, and the limit $N\to\infty$ is taken only as a final step. This prescription is good enough to get the correct value for the on-shell action as predicted by Sen's conjecture,  but in more general contexts it is not clear if it is sufficient. In as far as this prescription is applicable, this leads to an expression for  Schnabl's solution
\begin{equation}
\Psi_\text{Sch} = \lim_{N\to\infty}\left[\sum_{n=0}^N \sqrt{\Omega}cKB\Omega^n c\sqrt{\Omega}+ \sqrt{\Omega}cB\Omega^{N}c\sqrt{\Omega}\right].
\end{equation}
This can be represented pictorially as a combination of semi-infinite strips with operator insertions:
\begin{wrapfigure}{l}{1\linewidth}
\centering
\resizebox{5in}{1.1in}{\includegraphics{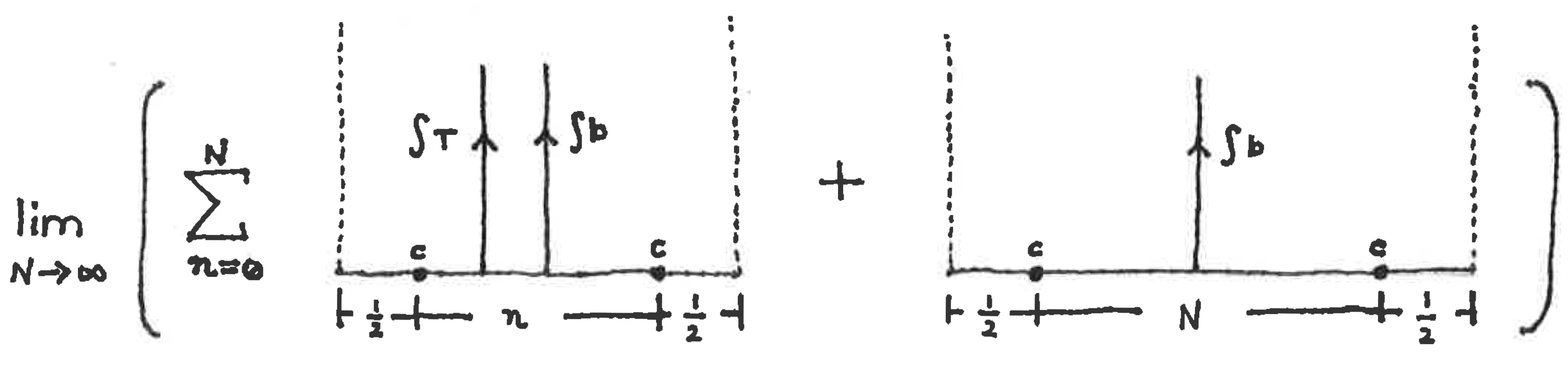}}
\end{wrapfigure} \\ \\ \\ \\ \\ \\ \\ \\ 
\noindent To make contact with the original notation of \cite{Schnabl}, define the state
\begin{equation}\psi_n = \sqrt{\Omega}cB\Omega^nc\sqrt{\Omega}.\end{equation}
Schnabl's solution is then written as 
\begin{equation}
\Psi_\text{Sch} = \lim_{N\to\infty}\left[\psi_N - \sum_{n=0}^N \frac{d}{dn}\psi_n\right].\label{eq:Schpsin}
\end{equation}
The first piece $\psi_N$ is the famous {\it phantom term} of Schnabl's solution. The result of claim \ref{claim:Bsliver} implies that the phantom term vanishes in the Fock space expansion when $N$ is taken to infinity. Nevertheless, it is the most physically important part of Schnabl's solution. It gives the sole contribution to the zero momentum tachyon in the $\mathcal{L}_0$ level expansion, and the remaining terms can be seen as remnants of a pure gauge solution. 

It was shown by Okawa \cite{Okawa} that the Schnabl gauge solutions for $\lambda\neq 1$ can be explicitly represented as a finite gauge transformation of the perturbative vacuum:
\begin{equation}
\Psi_\lambda = U^{-1}QU,\ \ \ \ U=1-\lambda\sqrt{\Omega}cB\sqrt{\Omega}.\label{eq:Okform}
\end{equation}
\begin{exercise}
Confirm this formula.
\end{exercise}
\noindent This expression must somehow break down $\lambda = 1$. Computing $U^{-1}$ gives 
\begin{equation}U^{-1} = 1+\lambda\sqrt{\Omega}cB\frac{1}{1-\lambda \Omega}\sqrt{\Omega}.\end{equation}
The factor $1/(1-\lambda\Omega)$ develops a pole at $K=0$ in the limit $\lambda\to 1$. Therefore the gauge transformation becomes singular. This singularity is closely related to the presence of a phantom term in the solution, as we will discuss in subsection \ref{subsec:singularGT}. 

With proper care for the regularization and phantom term, we may evaluate the action for Schnabl's solution analytically, and it gives the correct D-brane tension as predicted by Sen's conjecture. The calculation is technical and there are better ways of getting the result (see subsections \ref{subsec:simple} and \ref{subsec:singularGT}), so instead we describe the computation of the Ellwood invariant. Since there are no D-branes at the tachyon vacuum, the closed string 1-point function on a disk should vanish. This means that the Ellwood invariant should evaluate to
\begin{equation}\Tr_\mathcal{V}(\Psi_\text{Sch}) = -\mathcal{A}_0(\mathcal{V}).\end{equation}
To demonstrate this we need to evaluate the trace of the $\psi_n$ terms in Schnabl's solution:
\begin{equation}\Tr_\mathcal{V}(\psi_n) = \Tr_\mathcal{V}(\Omega cB\Omega^n c).\end{equation}
To simplify the ghosts we use $\mathcal{B}^-$ invariance of the trace:
\begin{eqnarray}
0\lineup = \Tr_\mathcal{V}\left(\frac{1}{2}\mathcal{B}^-(\Omega c\Omega^n c)\right)\nonumber\\
\lineup = -\Tr_\mathcal{V}(\Omega Bc\Omega^n c)+ n \Tr_\mathcal{V}(\Omega cB\Omega^n c)\nonumber\\
\lineup = -\Tr_\mathcal{V}(\Omega^{n+1} c)+(n+1)\Tr_\mathcal{V}(\Omega cB\Omega^n c).
\end{eqnarray}
Next we scale the cylinder in the first term down to unit circumference:
\begin{equation}\Tr_\mathcal{V}(\Omega^{n+1} c)=\Tr_\mathcal{V}\left(\left(\frac{1}{n+1}\right)^{\frac{1}{2}\mathcal{L}^-}(\Omega^{n+1} c)\right) = (n+1) \Tr_\mathcal{V}(\Omega c).\label{eq:Ellpsin}
\end{equation}
Together with the previous equation, this implies 
\begin{equation}\Tr_\mathcal{V}(\psi_n) = \Tr_\mathcal{V}(\Omega c).\end{equation}
The Ellwood invariant for Schnabl's solution is 
\begin{equation}
\Tr_\mathcal{V}(\Psi_\text{Sch}) = \lim_{N\to\infty}\left[\Tr_\mathcal{V}(\psi_N) - \sum_{n=0}^n\frac{d}{dn} \Tr_\mathcal{V}(\psi_n)\right].
\end{equation}
From \eq{Ellpsin}, the trace of $\psi_n$ is independent of $n$. Therefore the sum vanishes identically, and the sole contribution comes from the phantom term:
\begin{equation}
\Tr_\mathcal{V}(\Psi_\text{Sch}) = \Tr_\mathcal{V}(\Omega c). \label{eq:SchEll1}
\end{equation}
Note that the pure gauge solutions only have the sum (with terms multiplied by $\lambda^{n+1}$), and no phantom contribution. In this case the Ellwood invariant evaluates to zero, as expected for the perturbative vacuum. We also see that the phantom term gives the physically important contribution to the Ellwood invariant, and while it vanishes in the Fock space expansion, it can be nonzero in the context of some calculations. To compare \eq{SchEll1} to the disk 1-point function, we map the cylinder to the unit disk with the transformation
\begin{equation}f(z) = e^{2\pi i z}.\end{equation}
This gives
\begin{eqnarray}
\Tr_\mathcal{V}(\Psi_\text{Sch}) \lineup = \langle c\overline{c}V^\text{m}(i\infty,-i\infty)c(0)\rangle_{C_1}\nonumber\\
\lineup = \frac{1}{2\pi i} \langle c\overline{c}V^\text{m}(0,0)c(1)\rangle_\mathrm{disk}\nonumber\\
\lineup = - \frac{1}{2\pi i} \langle V^\text{m}(0,0)\rangle_\mathrm{disk}^\mathrm{m}\nonumber\\
\lineup = -\mathcal{A}_0(\mathcal{V}),
\end{eqnarray}
where in the last step we noted that the $c$ ghost correlator evaluates to $-1$.

Given a Fock space basis $|\phi_i\rangle$ and a dual basis $|\phi^i\rangle$, the Fock space expansion of Schnabl's solution is given by 
\begin{equation}
\Psi_\text{Sch} = \sum_{i}|\phi_i\rangle \Tr\Big(\sqrt{\Omega}(f_\mathcal{S}\circ\phi^i)\sqrt{\Omega}\Psi_\text{Sch}\Big).
\end{equation}
The computation of the coefficients of highly descendant states is technically complicated. The systematics can be dealt with in the  operator formalism used in Schnabl's original paper. The motivation for such computations is to compare to level truncation results in Siegel gauge and to test convergence of the action level by level given the exact coefficients of the tachyon condensate. We do not enter into the details of these calculations here. The result is that Schnabl's solution appears to behave well. The infinite $D_2$ norm and the associated subtleties with the phantom term do not translate to noticeable problems in the Fock space expansion. In fact, the pure gauge solutions for $|\lambda|<1$ appear to be less well-behaved. The action evaluated on such solutions should tend to zero level by level, but convergence is at best extremely slow \cite{Takahashi_trunc}, especially as $\lambda$ approaches $1$. In a different direction, it is also possible to derive a numerical solution for the tachyon vacuum in Schnabl gauge using the level truncation scheme~\cite{AldoMatjej}. The action converges to the expected value very well, and the numerical coefficients are in plausible agreement with the analytic coefficients, though not with the accuracy that could have been hoped for given that the computation was taken out to level 24. Interestingly, the pure gauge solutions in Schnabl gauge have not been seen in level truncation.

While computation of descendant states is involved, for primaries it is straightforward. For the tachyon vacuum, the only primary operator which acquires expectation value is the zero momentum tachyon $c(0)$. The dual vertex operator is $-c\d c(0)$. 
Therefore the coefficient of the tachyon state
\begin{equation}\Psi_\text{Sch} = T c_1|0\rangle + \text{higher levels}\end{equation}
is given by
\begin{equation}T = -\frac{\pi}{2}\Tr\big(\sqrt{\Omega}c\d c\sqrt{\Omega}\Psi_\text{Sch}\big).\end{equation}
To compute this we evaluate the overlap with $\psi_n$ using \eq{ccccB}:
\begin{eqnarray}
-\frac{\pi}{2}\Tr(\sqrt{\Omega}c\d c\sqrt{\Omega}\psi_n) \lineup = -\frac{\pi}{2}\Tr\big(\Omega^n c\Omega c\d c\Omega cB\big)\nonumber\\
\lineup = -\frac{n+2}{\pi}\sin^2\frac{\pi}{n+2}\left(\frac{n+2}{2\pi}\sin \frac{2\pi}{n+2}-1\right).
\end{eqnarray}
From the Taylor series of the sine, the first factor goes as $1/n$ for large $n$, while the second factor vanishes as $1/n^2$. In total the tachyon coefficient for $\psi_n$ vanishes as $1/n^3$ for large $n$, consistent with claim \ref{claim:Bsliver}. Summing the derivatives of $\psi_n$ then gives the tachyon coefficient for Schnabl's solution:
\begin{equation}
T = \sum_{n=0}^\infty \frac{d}{dn}\left[\frac{n+2}{\pi}\sin^2\frac{\pi}{n+2}\left(\frac{n+2}{2\pi}\sin \frac{2\pi}{n+2}-1\right)\right]\approx 0.55.
\end{equation}
This is very similar to $T\approx 0.54$ derived for the Siegel gauge condensate in level truncation. We can similarly derive the tachyon coefficient for pure gauge solutions:
\begin{equation}
T(\lambda) = \sum_{n=0}^\infty \lambda^{n+1}\frac{d}{dn}\left[\frac{n+2}{\pi}\sin^2\frac{\pi}{n+2}\left(\frac{n+2}{2\pi}\sin \frac{2\pi}{n+2}-1\right)\right].
\end{equation}
The sum converges for $|\lambda|\leq 1$, but the rate of convergence for $|\lambda|$ strictly less than $1$ is exponential while at $|\lambda|=1$ it converges only as a sum of $1/n^4$. This has interesting consequence for the nature of the limit $\lambda\to 1$. While the limit is continuous, it is not differentiable; in particular
\begin{equation}\frac{d^3}{d\lambda^3}T(\lambda)\end{equation}
diverges as a harmonic series as $\lambda$ approaches 1. In this sense, even in the Fock space expansion it can be seen that the tachyon vacuum is a special configuration among Schnabl gauge solutions in the $KBc$ subalgebra.  

\subsection{Schnabl gauge marginal deformations}
\label{subsec:SchMarg}

We now describe analytic solutions for marginal deformations in Schnabl gauge \cite{KORZ,Schnabl_marg}. These correspond to deformations of the reference D-brane given by moving along flat directions in the string field potential. At linearized order, such solutions are represented by a nontrivial element of the BRST cohomology, which we assume takes the form
\begin{equation}\Psi_\text{marg} = cV(0)|0\rangle +\text{nonlinear corrections},\end{equation}
where $V(x)$ is a boundary matter primary of weight $1$. If we introduce a string field $V$ defined by an infinitesimally thin strip containing $V(0)$ at the origin, we can write
\begin{equation}\Psi_\text{marg} = \sqrt{\Omega}cV\sqrt{\Omega} + \text{nonlinear corrections}.\end{equation}
Some important properties of $V$ are 
\begin{equation}Q(cV) = 0,\ \ \ \ \frac{1}{2}\mathcal{L}^- V = V,\ \ \ \ \frac{1}{2}\mathcal{B}^- V = 0.\end{equation}
The second property says that $V$ has scaling dimension $1$ in the sliver frame, and the third property holds because $V(x)$ is a matter operator. 
For the same reason 
\begin{equation}
[B,V] = [c,V] = 0.
\end{equation}
Often we multiply $V$ by a constant $\lambda$ corresponding to the expectation value of the field generated by the vertex operator. To avoid proliferation of $\lambda$s in formulas, we will absorb this constant into the normalization of $V$ by writing
\begin{equation}V = \lambda \widehat{V},\end{equation}
where $\widehat{V}$ is defined with a fixed normalization. The solution can be expanded perturbatively
\begin{equation}
\Psi_\text{marg} = \Psi_1+\Psi_2+\Psi_3+... ,
\end{equation}
where $\Psi_n$ contains $n$ insertions of $V$. These represent nonlinear corrections that account for the fact that the field generated by $cV$ has finite expectation value. Matching terms that contain the same number of $V$s, the equations of motion imply
\begin{eqnarray}
\lineup Q\Psi_1= 0,\\
\lineup Q\Psi_2+\Psi_1^2= 0,\\
\lineup Q\Psi_3 +\Psi_1\Psi_2+\Psi_2\Psi_1 = 0,\\
\lineup\ \ \ \ \ \ \ \ \vdots\nonumber\ \ .
\end{eqnarray}
There may be obstruction to solution of these equations if the quadratic terms containing lower order corrections are not BRST exact (one can check that they are automatically BRST closed). The physical interpretation of this obstruction is that the potential for the field $cV$ is not exactly flat, so a generic constant expectation value is not a solution. If the construction fails to give a solution for $\Psi_n$, this means that the potential vanishes as $\lambda^{n+1}$ for small $\lambda$. If the obstruction is absent for all $n$, then the deformation generated by $cV$ is called {\it exactly marginal}.

We look for a solution for marginal deformations in Schnabl gauge:
\begin{equation}\mathcal{B}_0\Psi_n=0.\end{equation}
Acting $\mathcal{B}_0$ on the equations of motion and using $[Q,\mathcal{B}_0]=\mathcal{L}_0$, we obtain a recursive set of equations for the corrections of the form
\begin{equation}\mathcal{L}_0\Psi_n+\mathcal{B}_0(\text{lower order corrections}) = 0.\end{equation}
If the second term does not produce states in the kernel of $\mathcal{L}_0$, we can invert $\mathcal{L}_0$ to obtain an explicit formula for $\Psi_n$. Let us work this out for the second order correction:
\begin{equation}\Psi_2 = -\frac{\mathcal{B}_0}{\mathcal{L}_0}\Psi_1^2.\end{equation}
A convenient representation of the inverse of $\mathcal{L}_0$ is through the Schwinger parameterization
\begin{equation}\frac{1}{\mathcal{L}_0} = \int_0^\infty dt\, e^{-t\mathcal{L}_0}.\end{equation} 
The integration variable $t$ can be interpreted as a coordinate on part of the moduli space of a Riemann surface defining an open string amplitude where $1/\mathcal{L}_0$ appears in the propagator.\footnote{In Schnabl gauge the full propagator is $\frac{\mathcal{B}_0}{\mathcal{L}_0}Q\frac{\mathcal{B}_0^*}{\mathcal{L}_0^*}$ \cite{Schnabl,RZVeneziano,KSZ}. Since $\mathcal{L}_0$ is not BPZ even, the propagator contains two moduli integrals, which makes the connection between the Schnabl gauge amplitude and integration over the moduli space somewhat less direct than in Siegel gauge.}
Substituting $\Psi_1$ we may compute
\begin{eqnarray}
\Psi_2\lineup = -\frac{\mathcal{B}_0}{\mathcal{L}_0}\sqrt{\Omega}cV\Omega cV\sqrt{\Omega}\nonumber\\
\lineup = - \frac{1}{\mathcal{L}_0}\sqrt{\Omega}cV B\Omega cV\sqrt{\Omega}\nonumber\\
\lineup = -\int_0^\infty dt\, e^{-t\mathcal{L}_0}\Big(\sqrt{\Omega}cV B\Omega cV\sqrt{\Omega}\Big)\nonumber\\
\lineup = -\sqrt{\Omega}\left[\int_0^\infty dt\,e^{-\frac{t}{2}\mathcal{L}^-}\Big(cV B\Omega cV\Big)\right]\sqrt{\Omega}\nonumber\\
\lineup = -\sqrt{\Omega}cV B\left[\int_0^\infty dt\,e^{-t} \Omega^{e^{-t}}\right] cV\sqrt{\Omega}.
\end{eqnarray}
Substituting $\alpha = e^{-t}$ gives
\begin{equation}\Psi_2 = -\sqrt{\Omega}cVB \left[\int_0^1 d\alpha\,\Omega^{\alpha}\right]cV\sqrt{\Omega}.\label{eq:Psi2int}\end{equation}
We recognize the integral from the homotopy operator for Schnabl's solution. Therefore we can write
\begin{equation}\Psi_2 =  -\sqrt{\Omega}cVB \frac{1-\Omega}{K}cV\sqrt{\Omega}.\end{equation}
This result raises a puzzle. Usually marginal operators, being dimension 1 primaries, have a double pole in their OPE: 
\begin{equation}V(x)V(0) = \frac{\mathcal{N}}{x^2} +\text{regular},\end{equation}
where $\mathcal{N}$ is a constant. It is known that a subleading single pole cannot appear, otherwise the deformation is not exactly marginal. The trouble is that the wedge state separating two $V$s under the integral in \eq{Psi2int} can be arbitrarily thin, and the double pole in the  OPE will create a divergence towards the lower limit of integration. Part of the problem is caused by the Schwinger representation of $1/\mathcal{L}_0$. The Schwinger representation is only valid operating on states with positive $\mathcal{L}_0$ eigenvalue. The zero momentum tachyon, however, has negative eigenvalue. This is actually a fairly common problem in string perturbation theory. For example, the representation of the Veneziano amplitude as an integral over moduli space likewise suffers from divergences from collisions of tachyon vertex operators. From the SFT point of view, such divergences originate from the failure of the Schwinger representation to correctly define the propagator. There are various possible remedies. In the Veneziano amplitude, the most common is to analytically continue to unphysical momenta where OPEs of tachyon vertex operators are not divergent. Presently, we can proceed by simply dividing by the eigenvalue of $\mathcal{L}_0$ in the $\mathcal{L}_0$ level expansion. To do this we define a ``normal ordered" string field through the relation
\begin{equation}V\Omega^\alpha V =\, :\!\!V\Omega^\alpha V\!\!: +\frac{\mathcal{N}}{\alpha^2}\Omega^\alpha.\end{equation}
The OPE divergence is absorbed in the second term, so that the first term is finite as $\alpha\to 0$. In this way we can write
\begin{equation}
\mathcal{B}_0\Psi_1^2 = \sqrt{\Omega}cB\Big(:\!\!V\Omega V\!\!: +\mathcal{N}\Omega\Big)c\sqrt{\Omega}.\label{eq:normord}
\end{equation}
We now expand this into eigenstates of $\mathcal{L}_0$:
\begin{eqnarray}
\mathcal{B}_0\Psi_1^2  \lineup =\underbrace{\phantom{\Big)}\!\!\mathcal{N}\sqrt{\Omega}c\sqrt{\Omega}}_{\text{level } -1} \, - \underbrace{\phantom{\Big)}\!\!\mathcal{N}\sqrt{\Omega}cKBc\sqrt{\Omega}}_{\text{level } 0} \,+ \underbrace{\sqrt{\Omega}\left(\frac{\mathcal{N}}{2!}cK^2 Bc +c:\!\! V^2\!\!:\right)\sqrt{\Omega}}_{\text{level }1}\,+\,\text{higher levels}.\ \ \ \ \ \ \ \ 
\end{eqnarray}
The  second order correction is given by dividing each term on  the right hand side  by  its level. But we encounter a problem: there is a state at level zero. In Siegel gauge, a state in the kernel of $L_0$ normally indicates that the deformation is not exactly marginal. Typically it would arise from acting $b_0$ on 
\begin{equation}\Psi_1^2 = ...+\underbrace{V'c\d c|0\rangle}_{L_0\text{ level }0} + ...,\end{equation}
where $V'$ is a marginal operator appearing as the coefficient of the single pole in the $V\dash  V$ OPE. This state is nontrivial in the BRST cohomology, and so cannot appear from computing $Q\Psi_2$. Therefore there is no solution at second order, and the deformation is not exactly marginal. But presently we are assuming that the single pole in the $V\dash  V$ OPE is absent, so this cannot be the problem with the Schnabl gauge solution. The problem comes from acting $\mathcal{B}_0$ on the state
\begin{equation}\Psi_1^2=...+\underbrace{\mathcal{N}\sqrt{\Omega}cK^2c\sqrt{\Omega}}_{\mathcal{L}_0\text{ level }0}+...\ \ .\end{equation}
This state, however, is BRST exact,
\begin{equation}\sqrt{\Omega}cK^2c\sqrt{\Omega}=Q\Big(\sqrt{\Omega}Bc\d c\sqrt{\Omega}+\sqrt{\Omega}c\d cB\sqrt{\Omega}\Big),\end{equation}
so does not present an obstruction to solution. But the solution cannot  be found in Schnabl gauge.

If we want to proceed in Schnabl gauge, our only choice is to assume that the OPE of marginal operators is not singular:
\begin{equation}\lim_{x\to 0}V(x) V(0) =\text{finite}.\end{equation}
To simplify the analysis  at higher orders, we will in fact assume that all powers of the string field $V$ are finite. This will be called a {\it regular marginal deformation}. If higher order products of $V$ are divergent, new obstructions can appear at third order or higher. Depending on whether the obstruction is BRST exact, this would either indicate that the deformation is not exactly marginal or that the solution cannot be found in Schnabl gauge. The restriction to regular marginal deformations is not as limiting as one might think. One example of $V$ with regular OPE is the exponential rolling tachyon deformation
\begin{equation}V(x) = e^{X^0(x)},\ \ \ \ e^{X^0(x)}e^{X^0(0)} = x^2 e^{2X^0(0)}+...\ .\end{equation}
This leads to a time-dependent solution where the reference D-brane decays starting from an infinitesimal, homogeneous tachyon fluctuation in the infinite past. At late times the solution oscillates violently with diverging amplitude, a phenomenon which has been the subject of much discussion \cite{MoellerZwiebach_rolling,Hata_rolling}. Another interesting example is the lightlike rolling tachyon deformation \cite{Hellerman}
\begin{equation}V(x) = e^{X^+(x)},\ \ \ \ e^{X^+(x)}e^{X^+(0)} = e^{2X^+(0)}+...,\end{equation}
which in a linear dilaton background can be made marginal. This represents a solution where the reference D-brane decays starting from an infinitesimal homogeneous tachyon fluctuation in the infinite lightcone past. Unlike the timelike decay process, the solution does not oscillate at late times and smoothly approaches the tachyon vacuum \cite{Hellerman,rollingvac}. This is notable as an example of an exact solution representing a time-dependent transition between open string vacua. Regular marginal deformations also include a number of exactly marginal deformations which are independent of the $X^0$ component of the $\BCFT$ \cite{KOSsing}.  Given $\widehat{V}_\mathrm{bare}(x)$ with a double pole (and {\it only} a double pole) in the OPE with itself with unit coefficient,  we may consider a modified marginal operator
\begin{equation}
V(x) =\lambda\left( \widehat{V}_\mathrm{bare}(x)+\frac{i}{\sqrt{2}}\d_\parallel X^0(x)\right).
\end{equation}
This operator has regular self-OPE since the double pole of the $\d X^0$-$\d X^0$ OPE cancels that of $\widehat{V}_\mathrm{bare}\text{-}\widehat{V}_\mathrm{bare}$. The cross terms do not create singularity since $\widehat{V}_\mathrm{bare}$ is independent of the $X^0$ BCFT. If we are lucky, higher order products of $V$ will also be finite. This occurs for the Wilson line deformation, the deformation representing transverse displacement of the D-brane, and the cosine tachyon deformation \cite{Ludwig}. The modified marginal operator gives an expectation value to the field of $\widehat{V}_\mathrm{bare}$ in addition to turning on a timelike gauge potential $A_0$. However, in physical situations the timelike direction on the D-brane worldvolume is noncompact, and the timelike gauge potential does not produce a physical deformation of the BCFT. This idea will play a very important role in the solution of subsection \ref{subsec:KOS}. Still there are deformations which cannot be easily implemented using marginal operators with regular OPE. An important example is the hyperbolic cosine rolling tachyon deformation \cite{SenRolling}.

Assuming that collisions of $V$ cause no problems, we may construct higher order corrections to the solution. We simply quote the result:
\begin{equation}
\Psi_{n+1} = (-1)^{n+1}\sqrt{\Omega}cV B\left(\frac{1-\Omega}{K}V\right)^n c\sqrt{\Omega}.
\end{equation}
We can sum over $n$ to find the full solution
\begin{equation}
\Psi_\text{marg} = \sqrt{\Omega} cV \frac{B}{1+\frac{1-\Omega}{K}V}c\sqrt{\Omega}.\label{eq:SchMarg}
\end{equation}
\begin{wrapfigure}{l}{.5\linewidth}
\centering
\resizebox{3.6in}{1.1in}{\includegraphics{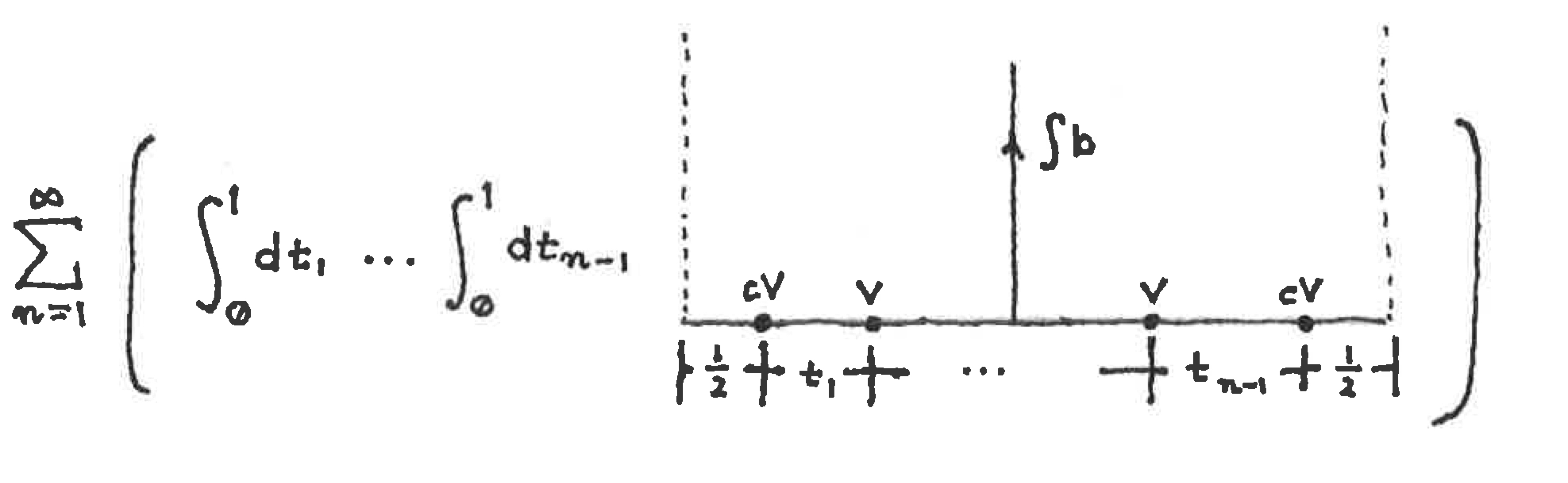}}
\end{wrapfigure}
The solution can be represented as a combination of semi-infinite strips with operator insertions, as shown left.
\begin{exercise}
Prove that the solution satisfies the equations of motion.
\end{exercise}
\begin{exercise} The solution can be written in terms of $J=cV$ as 
\begin{equation}\Psi_\text{marg} = \sqrt{\Omega} J \frac{1}{1+B\frac{1-\Omega}{K}J}\sqrt{\Omega}.\end{equation}
Prove that the equation of motion hold in this form. Show that if $J = \lambda cKBc$ this reduces to the pure gauge solution discussed in the previous section. Find the relation between the marginal parameter $\lambda$ in front of $cKBc$ and the gauge parameter $\lambda$ in \eq{Psil}.
\end{exercise}

Now we can try to compute some observables. The action is not very interesting, since a marginal deformation moves the string field along a flat direction in the potential, and the action is unchanged. More interesting is the Ellwood invariant. Let us compute the leading order contribution
\begin{equation}\Tr_\mathcal{V}(\Psi_1) = \Tr_\mathcal{V}(\Omega cV).\end{equation}
First we introduce a trivial integration $\int_0^1 dt\, = 1$, and use cyclicity of the trace to write
\begin{equation}\Tr_\mathcal{V}(\Psi_1) = \int_0^1 dt\Tr_\mathcal{V}(\Omega^t cV\Omega^{1-t}).\end{equation}
Second we insert a trivial factor $[B,c]=1$:
\begin{eqnarray}\Tr_\mathcal{V}(\Psi_1)\lineup = \int_0^1 dt\Tr_\mathcal{V}(\Omega^{1-t} cV\Omega^{t}[B,c])\nonumber\\
\lineup= \int_0^1 dt\Tr_\mathcal{V}(\Omega^{1-t}[B, cV]\Omega^{t}c)\nonumber\\
\lineup = \int_0^1 dt\Tr_\mathcal{V}(\Omega^{1-t} V\Omega^{t}c)\nonumber\\
\lineup =\left\langle \left( \int_0^1dt\,V(t)\right) c(0)\, c\overline{c}V^\text{m}(i\infty,-i\infty)\right\rangle_{C_1}.\end{eqnarray}
Mapping the cylinder to the unit disk and evaluating the ghost correlator gives
\begin{equation}\Tr_\mathcal{V}(\Psi_1) = -\frac{1}{2\pi i}\left\langle\left(\int_0^{2\pi} d\theta \, V(\theta)\right) V^\text{m}(0,0)\right\rangle_\mathrm{disk}^\mathrm{m},\end{equation}
where $V(\theta)$ is inserted on the boundary of the unit circle at an angle $\theta$. This represents the first order change of the boundary condition in the closed string 1-point function implemented by the marginal deformation.
\begin{exercise}
By a similar manipulation show that 
\begin{equation}\Tr_\mathcal{V}(\Psi_2) = \frac{1}{2\pi i}\left\langle \frac{1}{2!}\left(\int_0^{2\pi} d\theta\, V(\theta)\right)^{\!\!2} V^\mathrm{m}(0,0)\right\rangle_\mathrm{disk}^\mathrm{m}.\end{equation}
\end{exercise}
\noindent The general result was derived by Kishimoto \cite{Kishimoto_tad}:
\begin{equation}\Tr_\mathcal{V}(\Psi_\text{marg}) = \frac{1}{2\pi i}\left\langle \left(e^{-\int_0^{2\pi} d\theta\, V(\theta)}-1\right) V^\text{m}(0,0)\right\rangle_\mathrm{disk}^\mathrm{m}.\end{equation}
By exponentiating the integration of $V$ around the boundary we are effectively deforming the open string boundary condition. The correlation function can be represented through a worldsheet path integral, and in this context the exponential operator adds a boundary term to the worldsheet action whose effect is to modify the open string boundary condition of the original D-brane. Therefore the Ellwood invariant computes the shift in the closed string 1-point function on the disk. Note that since we are assuming that all products of $V$ are regular, the exponential operator is defined without renormalization. 

Now we can ask about coefficients of the solution in the Fock space expansion. For this purpose it is helpful to restore the explicit dependence on $\lambda$:
\begin{equation}V = \lambda \widehat{V}.\end{equation}
Through the Ellwood invariant, $\lambda$ can be identified as the coupling constant of the boundary deformation of the worldsheet action. The most interesting coefficient of the solution is the expectation value of the D-brane fluctuation field generating the marginal deformation:
\begin{equation}\lambda_\mathrm{SFT} c\widehat{V}(0)|0\rangle .\end{equation}
To linearized order the coefficient $\lambda_\mathrm{SFT}$ is equal to the coupling constant $\lambda$, but at the nonlinear level these parameters may be different. The coefficient $\lambda_\mathrm{SFT}$ can be extracted by contracting the solution with a dual state defined by vertex operator $-c\d c \widehat{V}_\text{dual}$, where $\widehat{V}_\text{dual}$ is a weight 1 primary field satisfying
\begin{equation}\langle  I\circ \widehat{V}_\text{dual}(0) \widehat{V}(0)\rangle^\text{m}_\text{UHP} = 1.\end{equation}
The order $\lambda^n$ contribution to $\lambda_\mathrm{SFT}$ is given by evaluating the trace
\begin{equation}-\Tr\big(\sqrt{\Omega}c\d c \widehat{V}_\text{dual}\sqrt{\Omega}\Psi_n\big).\end{equation}
This can be reduced to the computation of a matter $n+1$-point function on the upper half plane:
\begin{equation}\Big\langle \big(I\circ \widehat{V}_\text{dual}(0)\big) \widehat{V}(1)\Big[\widehat{V}(x_2)...\widehat{V}(x_{n-1})\Big] \widehat{V}(0)\Big\rangle_\mathrm{UHP}^\mathrm{m},\end{equation}
where after $SL(2,\mathbb{R})$ transformation we can bound the $x_i$s between $1$ and $0$. Now suppose that we scale the correlator with a factor of $\eps$ through the conformal transformation $s_\eps(\xi)=\eps \xi$. The BPZ conformal map inverts the scaling
\begin{equation}s_\eps\circ I(\xi) = I\circ s_{1/\eps}(\xi).\end{equation}
Since all of the insertions carry weight $1$, we find
\begin{eqnarray}
\lineup \Big\langle \big(I\circ \widehat{V}_\text{dual}(0)\big) \widehat{V}(1)\Big[\widehat{V}(x_2)...\widehat{V}(x_{n-1})\Big] \widehat{V}(0)\Big\rangle_\mathrm{UHP}^\mathrm{m}\nonumber\\
\lineup\ \ \ \ \ \ \ \ \ \ \ \ \ \ \ \ \ \ \ \ \ \ \ \ \ \ \ \ \ \ \ \ \ \ =\eps^{n-1}\Big\langle \big(I\circ \widehat{V}_\text{dual}(0)\big) \widehat{V}(\eps)\Big[\widehat{V}(\eps x_2)...\widehat{V}(\eps x_{n-1})\Big] \widehat{V}(0)\Big\rangle_\mathrm{UHP}^\mathrm{m}.\ \ \ \ \ 
\end{eqnarray}
This relation is valid for any $\eps$, and in particular we can take $\eps$ to zero. The right hand side has a vanishing factor for $n>1$, and ordinarily this would be compensated by divergent OPEs as the marginal operators are squeezed to the origin. But presently we are assuming that divergences in products of $\widehat{V}$ are absent. Therefore the $n+1$ point function must vanish identically for $n>1$, which implies that the $\lambda^n$ contribution to $\lambda_\mathrm{SFT}$ must vanish for $n>1$. Therefore, for Schnabl gauge marginal deformations the equality
\begin{equation}\lambda_\mathrm{SFT}=\lambda\end{equation}
holds at the fully nonlinear level. In Siegel gauge, the relation between $\lambda_\mathrm{SFT}$ and $\lambda$ is nontrivial and has been an open problem for many years. For the Wilson line deformation, early work in 
level truncation indicated that $\lambda_\text{SFT}$ could only reach a finite maximum value even though 
\begin{wrapfigure}{l}{.15\linewidth}
\centering
\resizebox{1in}{1.6in}{\includegraphics{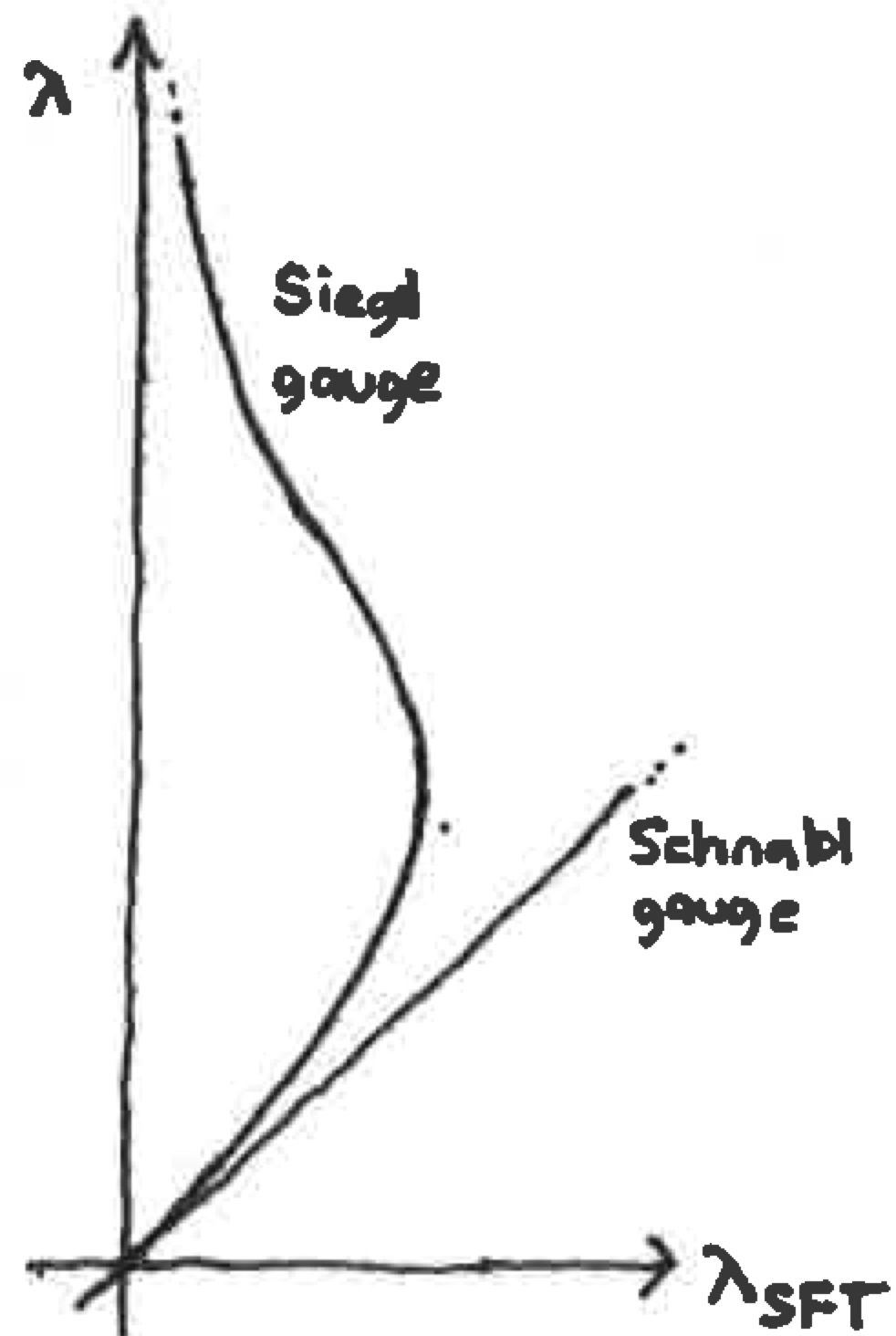}}
\end{wrapfigure} 
the boundary coupling is unbounded. More recent work \cite{MaccaferriMatjej} has shown that the boundary 
coupling constant is in fact not a single valued function of $\lambda_\text{SFT}$; it consists of (at least) two 
branches which join at the maximum value of $\lambda_\text{SFT}$. Though this has not been fully confirmed in Siegel gauge level truncation, analysis of analytic solutions capable of describing singular OPEs \cite{MaccaferriSchnabl,KOSsingII} indicates that the limit of large boundary coupling is represented by solutions where $\lambda_\text{SFT}$ tends to zero. A sketch of the conjectured relation between $\lambda_\text{SFT}$ and $\lambda$ is shown left. Therefore, in Siegel gauge a given expectation value of the marginal field can represent more than one marginally deformed background. To distinguish the backgrounds it is apparently necessary to look at other coefficients of the solution. The story in Schnabl gauge is apparently simpler. While the tachyon vacuum in Schnabl gauge in some ways seems comparable to the Siegel gauge solution, the solutions for marginal deformations are quite different. 

\subsection{Simple tachyon vacuum}
\label{subsec:simple}

We now describe another solution for the tachyon vacuum in the $KBc$ subalgebra which is simpler than Schnabl's solution \cite{simple}. While technical simplifications are always welcome, the structure of the solution appears to connect with something deeper which has allowed progress in several directions.

It was noticed by Okawa \cite{Okawa} that Schnabl gauge solutions in the $KBc$ subalgebra can be readily generalized to depend on an arbitrary state in the wedge algebra $F=F(K)$:\footnote{The solution satisfies the gauge condition $\mathcal{B}_{\sqrt{F},\sqrt{F}}\Big(\Psi_\text{Ok}\Big) = 0$, where
\begin{equation}\mathcal{B}_{\sqrt{F},\sqrt{F}} = \sqrt{F}\frac{1}{2}\mathcal{B}^{-}\left(\frac{1}{\sqrt{F}}(\ \cdot\ )\frac{1}{\sqrt{F}}\right)\sqrt{F}\label{eq:BsqrtFsqrtF}\end{equation}
This is referred to as a {\it dressed Schnabl gauge} \cite{simple}. If $F=\Omega$, this reduces to the ordinary Schnabl gauge condition.}
\begin{equation}\Psi_\text{Ok} = \sqrt{F}c\frac{KB}{1-F}c\sqrt{F}.\label{eq:Oktype}\end{equation}
In this solution, $\sqrt{F}$ plays the role of the ``security strip."
The solution is real if $\sqrt{F}$ is real. It was shown in \cite{SSFII} that solutions of this kind may represent two gauge orbits:
\begin{eqnarray}
\text{Perturbative vacuum}:\lineup \ \ \ \ \ F(0)=\lambda\neq 1.\nonumber\\
\text{Tachyon vacuum}:\lineup\ \ \ \ \ F(0)=1,\ \ \ F'(0)\neq 0.
\end{eqnarray}
In the second case, the derivative of $F(K)$ cannot vanish at $K=0$ since otherwise the state between the $c$s contains a pole at $K=0$. Several other necessary conditions on $F(K)$ are known, but the complete set of necessary and sufficient conditions for the viability of the solution are not fully clear. We will leave it as it is for now. If we have a tachyon vacuum solution in this form, the homotopy operator is 
\begin{equation}
A_\text{Ok} = B\frac{1-F}{K}.
\end{equation}
One might observe that $F(K)$ which defines the security strip and $\frac{1-F(K)}{K}$ which defines the homotopy operator both play an important role in the solution. This suggests that  some simplification may occur if the two states are equal:
\begin{equation}F=\frac{1-F}{K}.\end{equation}
Solving for $F$ leads to 
\begin{equation}F = \frac{1}{1+K} = \int_0^\infty d\alpha \,e^{-\alpha}\Omega^\alpha.\end{equation}
Using the Schwinger representation this is expressed as a continuous superposition of wedge states from the identity out to the sliver. To write the solution we also need the square root
\begin{equation}\frac{1}{\sqrt{1+K}} = \frac{1}{\sqrt{\pi}}\int_0^\infty d\alpha\,\frac{e^{-\alpha}}{\sqrt{\alpha}}\Omega^\alpha .\end{equation}
This leads to the {\it simple tachyon vacuum}:
\begin{equation}\Psi_\text{simp} = \frac{1}{\sqrt{1+K}}c(1+K)Bc\frac{1}{\sqrt{1+K}},\end{equation}
\begin{wrapfigure}{l}{.5\linewidth}
\vspace{-.5cm}
\centering
\resizebox{3.3in}{1.1in}{\includegraphics{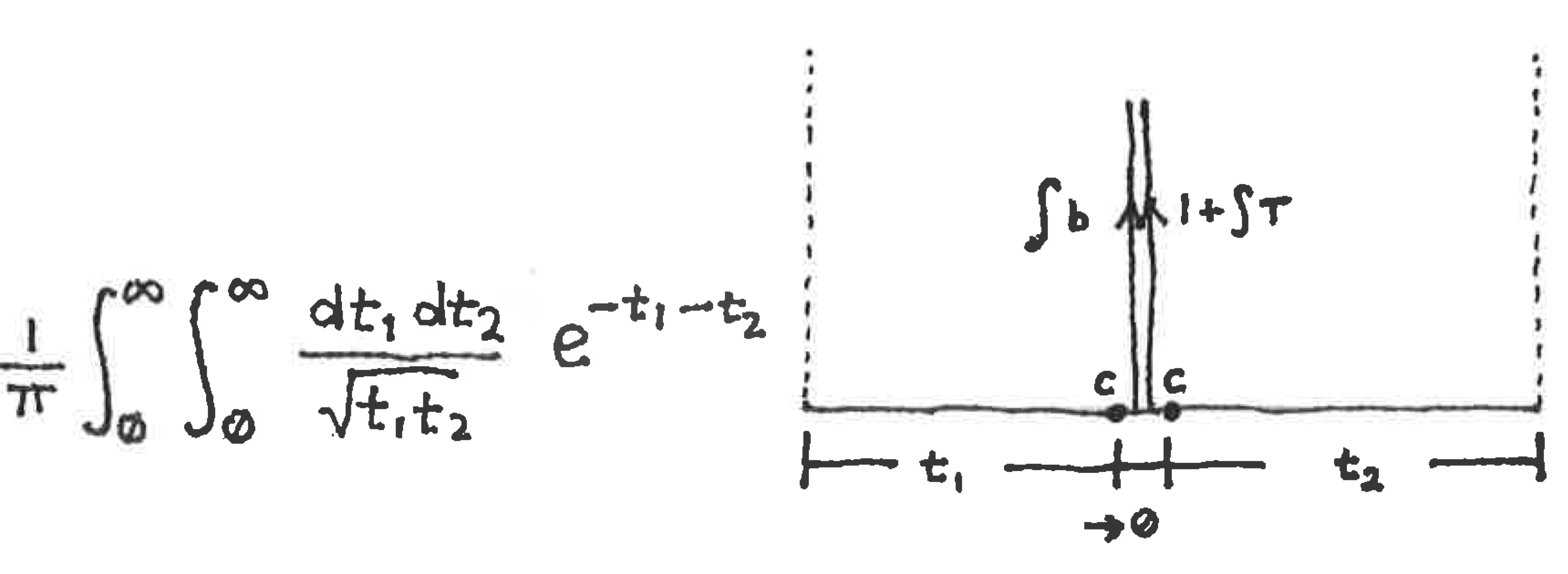}}
\vspace{-.5cm}
\end{wrapfigure}
The homotopy operator is
\begin{equation}A_\text{simp} = \frac{B}{1+K}.\end{equation}
The worldsheet picture is a very different from Schnabl's solution. Instead of a discrete sum of wedge states with positive integer width, we  have a continuous superposition of wedge states from the identity string field out to the sliver. 

Another difference from Schnabl's solution is that there is  no need for  a regularization or phantom term. This is because  both states
\begin{equation}
F,\ \ \ \ \frac{KF}{1-F},\label{eq:D2finite}
\end{equation}
have finite $D_2$ norm, and therefore have a direct representation as superpositions of wedge states. For Schnabl's solution, the first state $F=\Omega$ has finite $D_2$ norm, but $K\Omega/(1-\Omega)$ does not. Contrary to what is sometimes suggested, the simple tachyon vacuum is not the only tachyon vacuum solution which can be defined without regularization and  phantom  term. One example is given by 
\begin{equation}F(K) = \frac{1}{1+K}\Omega.\end{equation}
Unlike the simple tachyon vacuum, this solution does not receive contribution from states close to the identity string field. 

This however raises a puzzle. In Schnabl's solution, the phantom term could be justified as necessary to distinguish the tachyon vacuum from a  pure gauge configuration. One might ask why is this not also the case for the simple tachyon vacuum. There is a ``simple" analogue of the pure gauge solutions defined by
\begin{equation}F = \frac{\lambda}{1+K}.\end{equation}
The pure gauge solutions can be expressed as 
\begin{eqnarray}\Psi_\lambda\lineup =\frac{1}{\sqrt{1+K}}c(1+K)Bc\frac{1}{\sqrt{1+K}}  \,-\,  \frac{1}{\sqrt{1+K}}c(1+K)\frac{\eps}{\eps + K}(1+K)Bc\frac{1}{\sqrt{1+K}},\ \ \ \ \  \label{eq:simp_gauge}\end{eqnarray}
where $\eps = 1-\lambda$. The first term by itself is the simple tachyon vacuum. Therefore the second term is a solution to the equations of motion expanded around the simple tachyon vacuum. Physically, it represents recreation of the reference D-brane after it has been annihilated by tachyon condensation. This foreshadows a structure which we will develop in the next subsection. If $\lambda\neq 0$ the second term does not cancel the first, so the perturbative vacuum is represented by a nontrivial field configuration. If we consider the limit $\lambda\to  1^-$ or  $\eps\to 0^+$, the second term looks like it vanishes since it is proportional to $\eps$. But note that 
\begin{equation}\frac{\eps}{\eps+K} = \eps\int_0^\infty d\alpha e^{-\eps \alpha}\Omega^\alpha = \int_0^\infty dt\, e^{-t}\Omega^{t/\eps}.
\end{equation}
In the limit $\eps\to 0$ the wedge state in the integrand approaches the sliver, and the integral over $t$ evaluates to $1$. Therefore the second term in the pure gauge solution becomes sliver-like, and since the sliver is multiplied by $B$ it will vanish in the Fock space expansion. This looks precisely like a phantom term, but it appears in a limit of the pure gauge solution, rather than in the tachyon vacuum. 

The computation of the action for the simple tachyon vacuum is  straightforward. Using the equations of motion, the on-shell action can be expressed
\begin{equation}S = -\frac{1}{6}\Tr\big(\Psi_\text{simp} Q\Psi_\text{simp}\big).\end{equation}
We write the solution in the form
\begin{equation}\Psi_\text{simp} = \frac{1}{\sqrt{1+K}}c\frac{1}{\sqrt{1+K}} + Q\left(\frac{1}{\sqrt{1+K}}Bc\frac{1}{\sqrt{1+K}}\right).\end{equation}
The second term is BRST exact and drops out when we plug into the action. Then we find
\begin{eqnarray}
S \lineup = -\frac{1}{6}\Tr\left(\frac{1}{1+K}c\frac{1}{1+K}c\d c\right)\nonumber\\
\lineup =-\frac{1}{6}\int_0^\infty d\alpha\int_0^\infty d\beta \,e^{-\alpha-\beta}\Tr\big(\Omega^\alpha c\Omega^\beta c\d c\big).
\end{eqnarray}
The trace can be evaluated by taking the derivative of the correlator of three $c$s on the cylinder, \eq{ccc}. Accounting for vacuum normalization of the matter correlator, this gives
\begin{equation}\Tr\big(\Omega^\alpha c\Omega^\beta c\d c\big) = -g_0 \left(\frac{\alpha+\beta}{\pi}\right)^2\sin^2\frac{\pi \alpha}{\alpha+\beta}.\end{equation}
With our chosen normalization of the action, $g_0$ should be identified with the spacetime volume of the D-brane. The value of the action is then given by the double integral
\begin{equation}
S = \frac{g_0}{6}\int_0^\infty d\alpha\int_0^\infty d\beta \,e^{-\alpha-\beta}  \left(\frac{\alpha+\beta}{\pi}\right)^2\sin^2\frac{\pi \alpha}{\alpha+\beta}.
\end{equation}
To evaluate the integral we make the substitution
\begin{eqnarray}
L=\alpha+\beta,\ \ \ \ \theta=\frac{\pi\alpha}{\alpha+\beta},\ \ \ \ d\alpha d\beta = \frac{1}{\pi}LdL d\theta.
\end{eqnarray}
The double integral then factorizes into a product of two single integrals:
\begin{eqnarray}
S = \frac{g_0}{6}\frac{1}{\pi}\int_0^\infty dL \frac{L^3}{\pi^2}e^{-L}\int_0^\pi d\theta \sin^2\theta.
\end{eqnarray}
The integral of $\sin^2$ gives half the period. The integral over $L$ produces $3!$. Thus in total
\begin{equation}S = \frac{g_0}{6}\frac{1}{\pi}\frac{3!}{\pi^2}\frac{\pi}{2}=\frac{g_0}{2\pi^2}.\end{equation}
Multiplying by $-1$ and dividing by the volume of the time coordinate gives the energy of the D-brane in agreement with Sen's conjecture.

\begin{exercise}
Consider the tachyon vacuum solution
\begin{equation}\Psi = c(i+K)Bc\frac{i}{i+K}.\end{equation}
Without deforming the solution, find a way to define it as a superposition of wedge states and compute the energy to confirm Sen's conjecture.
\end{exercise}

\begin{exercise} 
\label{ex:SchSimp}
Consider a tachyon vacuum solution defined by \cite{genericF(K)}
\begin{equation}F(K) = \left(1-\frac{1}{\nu}K\right)^\nu \end{equation}
where $\nu<0$ is a parameter. Determine appropriate functions $h_\nu(\alpha)$ and $g_\nu(\alpha)$ so that 
\begin{eqnarray}
\frac{1-F(K)}{K} \lineup = \int_0^\infty d\alpha\, h_\nu(\alpha)\Omega^\alpha\\
\frac{KF(K)}{1-F(K)} \lineup = \int_0^\infty d\alpha\, g_\nu(\alpha)\Omega^\alpha
\end{eqnarray}
and show that as $\nu\in(-1,-\infty)$ the functions $h_\nu(\alpha)$ and $g_\nu(\alpha)$ interpolate between those of the simple tachyon vacuum and Schnabl's solution. Further show that the sequence of states $\frac{KF(K)}{1-F(K)}$ converges in the $C^*$ topology as $\nu\to-\infty$ but diverges in the $D_2$ topology. 
\end{exercise}

The tachyon coefficient of the simple solution is given by
\begin{eqnarray}
T \lineup =\frac{1}{2\pi^2}\int_0^\infty du\int_{-1}^1 dw\, e^{-u}\frac{(u+1)^2}{\sqrt{1-w^2}}\cos^2\left(\frac{\pi}{2}\frac{uw}{u+1}\right)\nonumber\\
\lineup \approx 0.51.\label{eq:Tsimp}
\end{eqnarray}
This is again similar to the tachyon expectation value in Siegel gauge and Schnabl gauge. One surprise, however, is that if we compute the energy of the simple tachyon vacuum level by level given the exact coefficients, we obtain a divergent series. The reasons for this are discussed in detail in \cite{simple}, but ultimately it comes down to the fact that the simple solution receives contributions from wedge states all the way down to the identity string field, which is marginally non-normalizable. For this reason the simple tachyon vacuum is less  well-behaved in the level expansion than Schnabl's solution, which only contains wedge states of strictly positive width. However, the series for the energy of the simple tachyon vacuum can be resummed to give good agreement with Sen's conjecture.

\subsection{Simple intertwining solution}
\label{subsec:KOS}

The analogue of the simple tachyon vacuum for marginal deformations was investigated by Kiermaier, Okawa, and Soler \cite{KOS}, who showed that the solution can be written in a surprisingly simple way in terms of boundary condition changing operators. Later it was realized that the solution could be generalized to describe any time independent D-brane system \cite{KOSsing}, and comes very close to providing a proof of background independence in Witten's open bosonic string field theory. We call it the {\it simple intertwining solution}. Another version of the ``intertwining solution" was recently described in \cite{KOSsingII}, and is proposed to address some of the limitations of the solution described here.  

We consider string field theory on a reference D-brane whose boundary conformal field theory we write as $\BCFT_0$. We want a solution of this string field theory which represents a target D-brane, whose boundary conformal field theory we write as $\BCFT_*$. Initially, we will assume that $\BCFT_0$ and $\BCFT_*$ are related by regular marginal deformation, as was done by Kiermaier, Okawa, and Soler. We start with the Schnabl gauge solution of subsection \ref{subsec:SchMarg}
and make the replacement
\begin{equation}\Omega\to \frac{1}{1+K},\end{equation}
so that we obtain a solution for regular marginal deformations 
\begin{equation}\Psi_* = \frac{1}{\sqrt{1+K}}cV\frac{B}{1+\frac{1}{1+K}V}c\frac{1}{\sqrt{1+K}}.\end{equation}
The subscript indicates that the solution describes the D-brane of $\BCFT_*$. We now explain how this can be rewritten so as to reveal a more general structure. We start with 
\begin{eqnarray}
\Psi_* \lineup = \frac{1}{\sqrt{1+K}}cV\frac{B}{\frac{1}{1+K}(1+K+V)}c\frac{1}{\sqrt{1+K}}\nonumber\\
\lineup = \frac{1}{\sqrt{1+K}}cV\frac{B}{1+K+V}(1+K)c\frac{1}{\sqrt{1+K}}\nonumber\\
\lineup = \frac{1}{\sqrt{1+K}}c(1+K+V-(1+K))\frac{B}{1+K+V}(1+K)c\frac{1}{\sqrt{1+K}}\nonumber\\
\lineup = \underbrace{\frac{1}{\sqrt{1+K}}c(1+K)Bc\frac{1}{\sqrt{1+K}}}_{\Psi_\tv}\ \underbrace{-\frac{1}{\sqrt{1+K}}c(1+K)\frac{B}{1+K+V}(1+K)c\frac{1}{\sqrt{1+K}}}_{\Psi^\tv_*}.\nonumber\\
\lineup = \Psi_\tv+\Psi^\tv_*
\end{eqnarray}
The first term is the simple tachyon vacuum, which we denote as $\Psi_\tv$. The second term $\Psi^\tv_*$ must be a solution to the equations of motion expanded around the simple tachyon vacuum:
\begin{equation}Q_{\Psi_\tv}\Psi^\tv_ *+ (\Psi^\tv_*)^2 = 0.\end{equation}
The first term destroys the reference D-brane, and the second term creates a marginally deformed D-brane out of the tachyon vacuum. This generalizes the structure of the pure gauge solutions described in the previous subsection.

The solution $\Psi^\tv_*$ involves the state 
\begin{equation}
\frac{1}{1+K+V} = \int_0^\infty d\alpha \, e^{-\alpha}\Omega_*^\alpha,
\end{equation}
where
\begin{equation}\Omega_*^\alpha = e^{-\alpha(K+V)}.\label{eq:Omstar1}\end{equation}
This looks like a wedge state, but $V$ appears in addition to $K$ in the exponential. It is important to understand what this means at the level of correlation functions. We claim is that $\Omega_*^\alpha$ is a 
\begin{wrapfigure}{l}{.4\linewidth}
\vspace{-.5cm}
\centering
\resizebox{2.7in}{1.5in}{\includegraphics{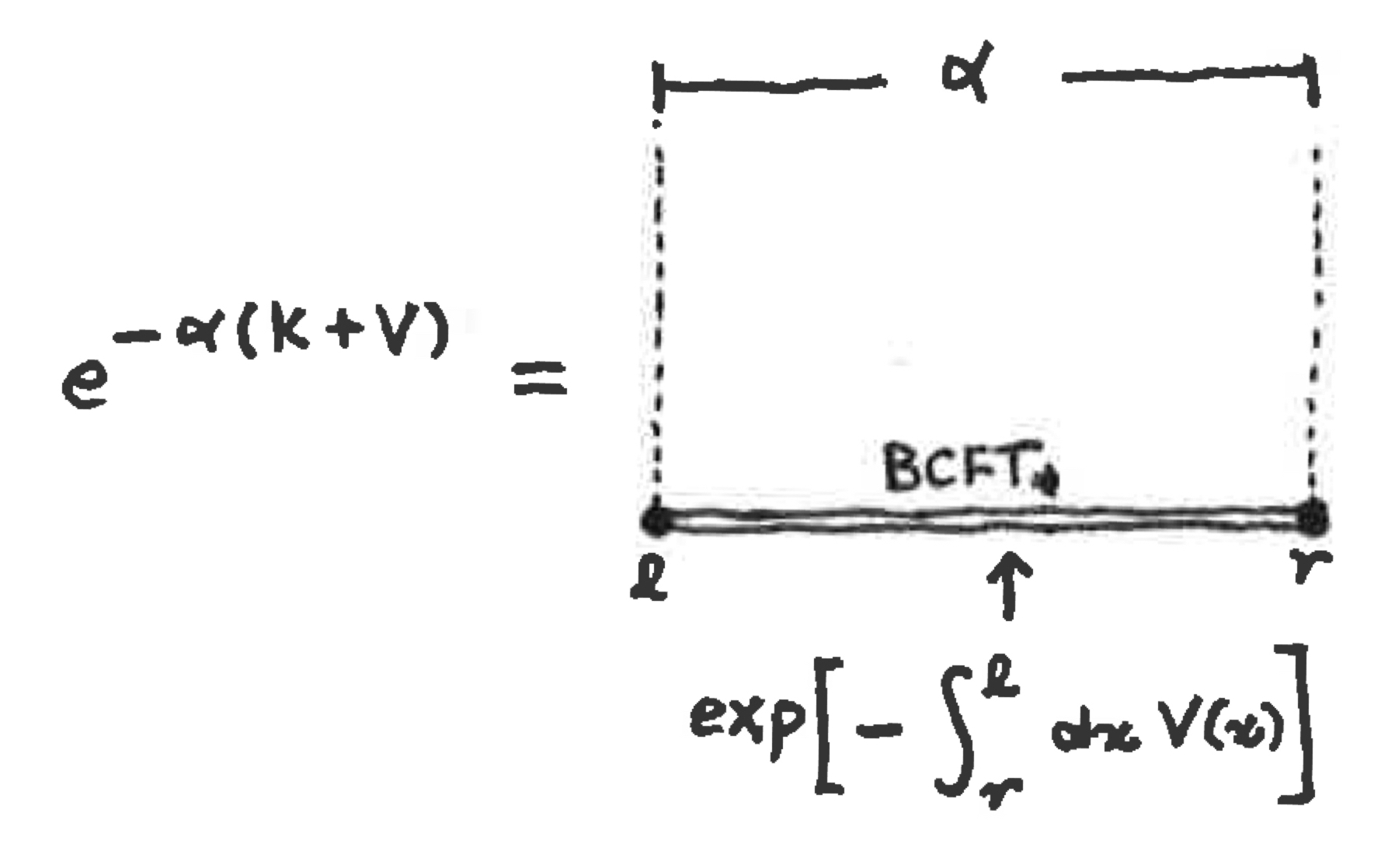}}
\vspace{-.5cm}
\end{wrapfigure} 
strip of width $\alpha$ containing an exponential of line integral insertions of $V$  on the open string boundary:
\begin{equation}\langle \phi,\Omega_*^\alpha\rangle = \left\langle e^{-\int_{1/2}^{\alpha+1/2} dx\, V(x)}f_\mathcal{S}\circ\phi(0)\right\rangle_{C_{\alpha+1}}.\label{eq:Omstar}
\end{equation}
The line integrals of $V$ have the effect of deforming the open string boundary condition from $\BCFT_0$ to $\BCFT_*$. Therefore $\Omega_*^\alpha$ is essentially the wedge state of the marginally deformed background. 

\begin{wrapfigure}{l}{.21\linewidth}
\centering
\vspace{-.7cm}
\resizebox{1.3in}{2.9in}{\includegraphics{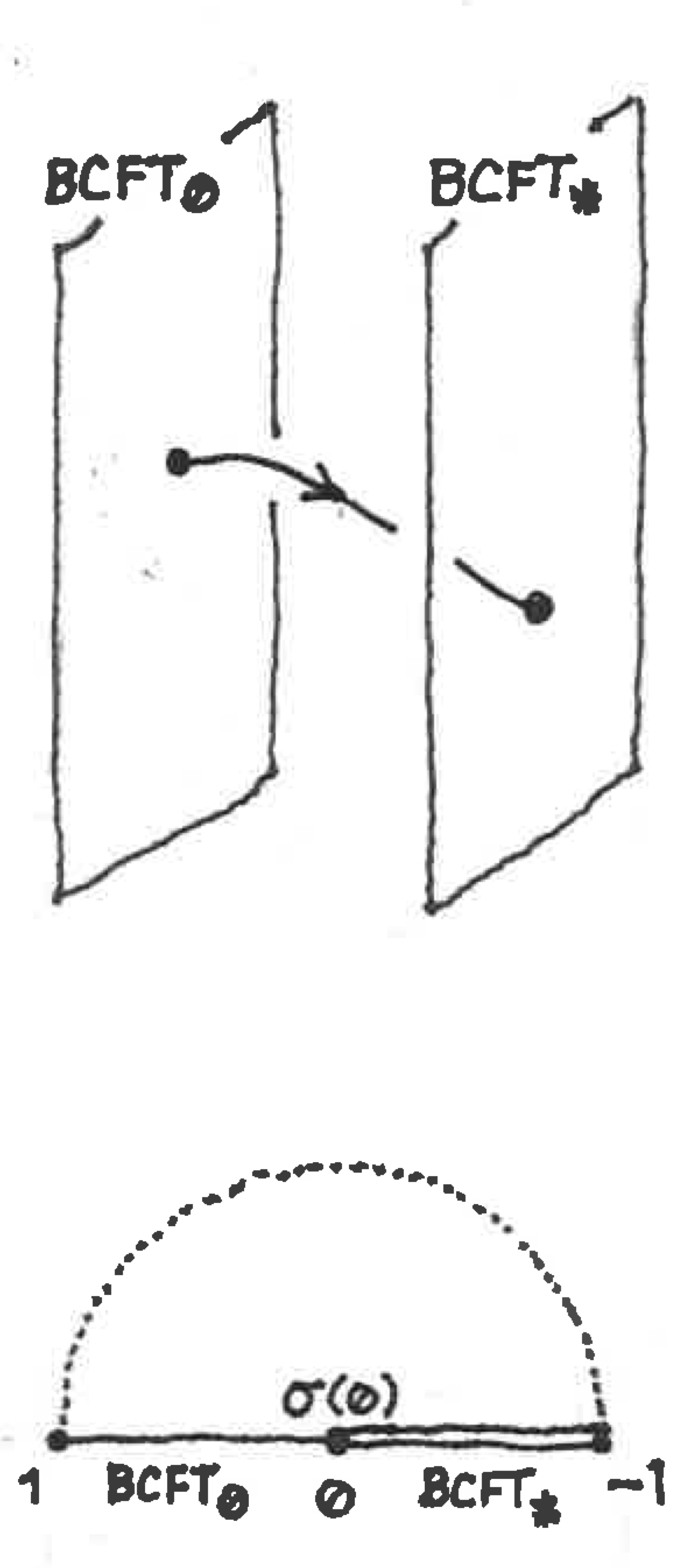}}
\end{wrapfigure}

\begin{exercise}
\label{ex:Omstar}
Show that \eq{Omstar} implies that $\Omega_*^\alpha$ satisfies the differential equation
\begin{equation}\frac{d}{d\alpha}\Omega_*^\alpha = -(K+V)\Omega_*^\alpha\end{equation}
Together with the boundary condition $\Omega_*^0 = 1$, this demonstrates that $\Omega_*^\alpha$ can be expressed as \eq{Omstar1}
\end{exercise}

It will be helpful to adopt a different language for describing the change of boundary condition between $\BCFT_0$ and $\BCFT_*$. Consider an open string connecting a D-brane $\BCFT_0$ and a D-brane $\BCFT_*$. From the 
point of view of radial quantization, such an open string can be associated to a unit half-disk with $\BCFT_0$ boundary conditions on the positive real axis and $\BCFT_*$ boundary conditions on 
the negative real axis. It is natural to think of this state as being created by a vertex operator which somehow changes the boundary condition from $\BCFT_0$ to $\BCFT_*$. This is called a {\it boundary condition changing operator}. We denote this as $\sigma(0)$. There is also a boundary condition changing operator which shifts the boundary condition from $\BCFT_*$ back to $\BCFT_0$, which we denote as $\sigmabar(0)$. Boundary condition changing operators are not really local operators from the point 
of view of $\BCFT_0$ or $\BCFT_*$, since they must always appear in conjugate pairs on the boundary. But since the boundary conditions on either side are conformal, they behave in many ways like local operators. They have OPEs and comparable conformal transformation properties. 

Consider a disk with two boundary components $C_0$ and $C_*$, carrying $\BCFT_0$ and $\BCFT_*$ boundary conditions. Tracing a clockwise path around the boundary, let $a$ and $b$ be the points at 
the junction of $C_0$ and $C_*$, and $C_*$ and $C_0$, respectively. The boundary condition changing operators for 
regular marginal deformations are related to the exponential insertion of line integrals of $V$ through
\begin{wrapfigure}{l}{.23\linewidth}
\centering
\resizebox{1.5in}{1.3in}{\includegraphics{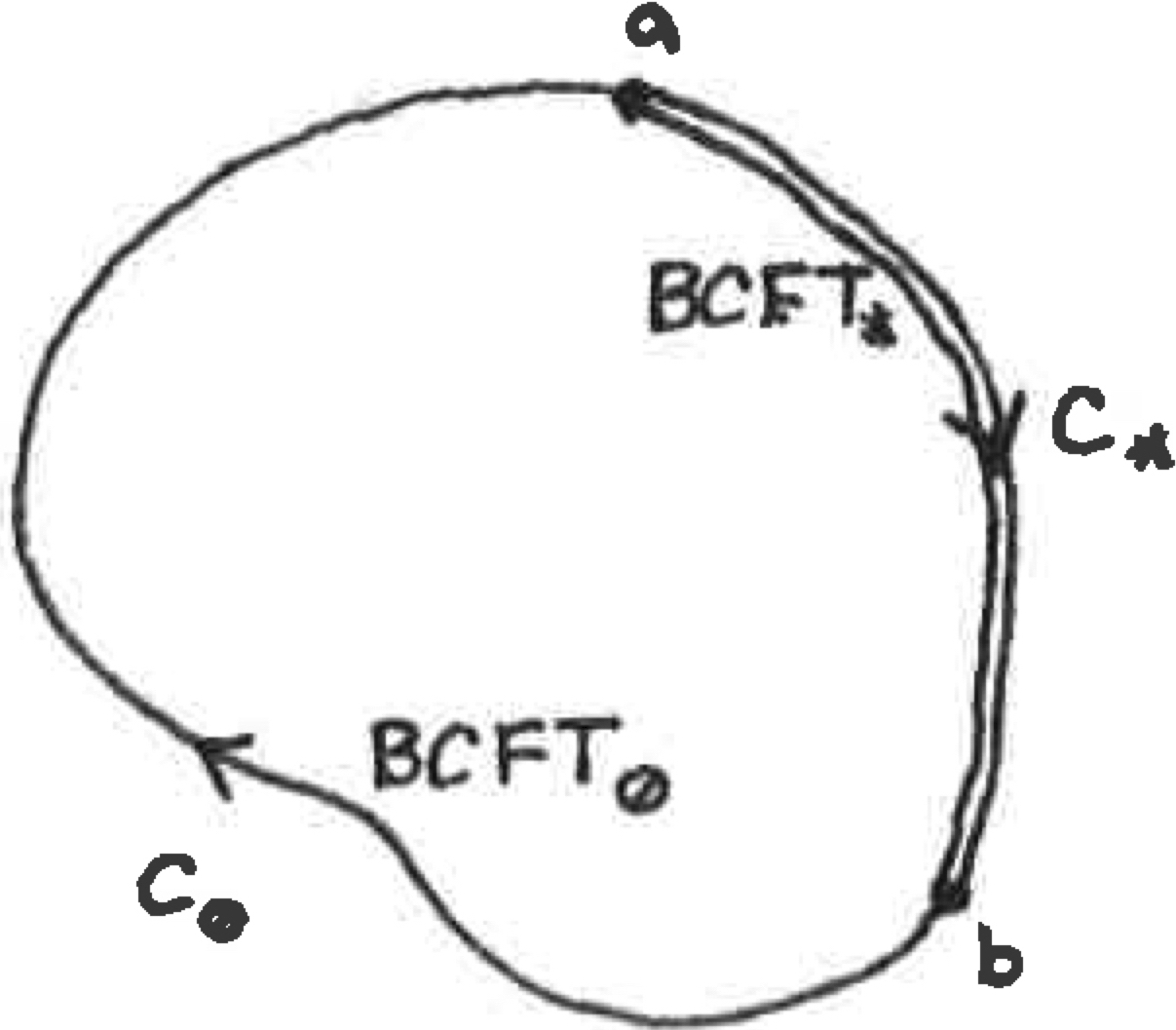}}
\vspace{1cm}
\end{wrapfigure} 
\begin{equation}\sigma(a)\sigmabar(b) = e^{-\int_{C_*}dz V(z)}.\end{equation}
We can learn a few things from this identification. If $b$ approaches $a$ from a clockwise direction, the whole boundary of the disk carries an exponential insertion of line integrals of $V$. From the point of view of an open string in $\BCFT_*$, this is simply a trivial insertion of the identity operator:
\begin{equation}\lim_{{b\to a \atop \text{clockwise}}}\sigmabar(b)\sigma(a) =  e^{-\int_{\d\text{disk}}dz V(z)}=1_\mathrm{BCFT_*}.\end{equation}
On the other hand, if $a$ approaches $b$ from a clockwise direction, the exponential insertion of $V$s disappears. Therefore
\begin{equation}\lim_{{a\to b \atop \text{clockwise}}}\sigma(a)\sigmabar(b) =  1_\mathrm{BCFT_0}.\end{equation}
These properties hold because the OPEs of $V(x)$ are finite. If they were not finite, the exponential insertion of line integrals of $V$ would need to be defined with some renormalization, and any prescription consistent with conformal invariance will lead to divergence in the limits where $a$ and $b$ collide. In a sense, $\sigma$ and $\sigmabar$ develop singularities in their OPE. It is also clear that the exponential insertion of $V$s map in a trivial way under conformal transformation:
\begin{equation}f\circ e^{-\int_{C_*}dz V(z)} = e^{-\int_{f\circ C_*}dz V(z)}.\end{equation}
This implies that 
\begin{equation}f\circ \Big(\sigma(a)\sigmabar(b)\Big) = \sigma(f(a))\sigmabar(f(b)).\end{equation}
Apparently $\sigma$ and $\sigmabar$ are primary operators of weight $0$. Again,  if the exponential insertion of $V$s required  renormalization it would not map trivially under conformal transformation, and the boundary condition changing operators would acquire some conformal weight. An important advantage of the boundary condition changing operator point of view is that it is universal. For any two D-brane systems, regardless of whether they are related by marginal deformation, there are always open strings connecting them. The vertex operators of these open strings are boundary condition changing operators. 

\begin{wrapfigure}{l}{.35\linewidth}
\centering
\resizebox{2.5in}{1in}{\includegraphics{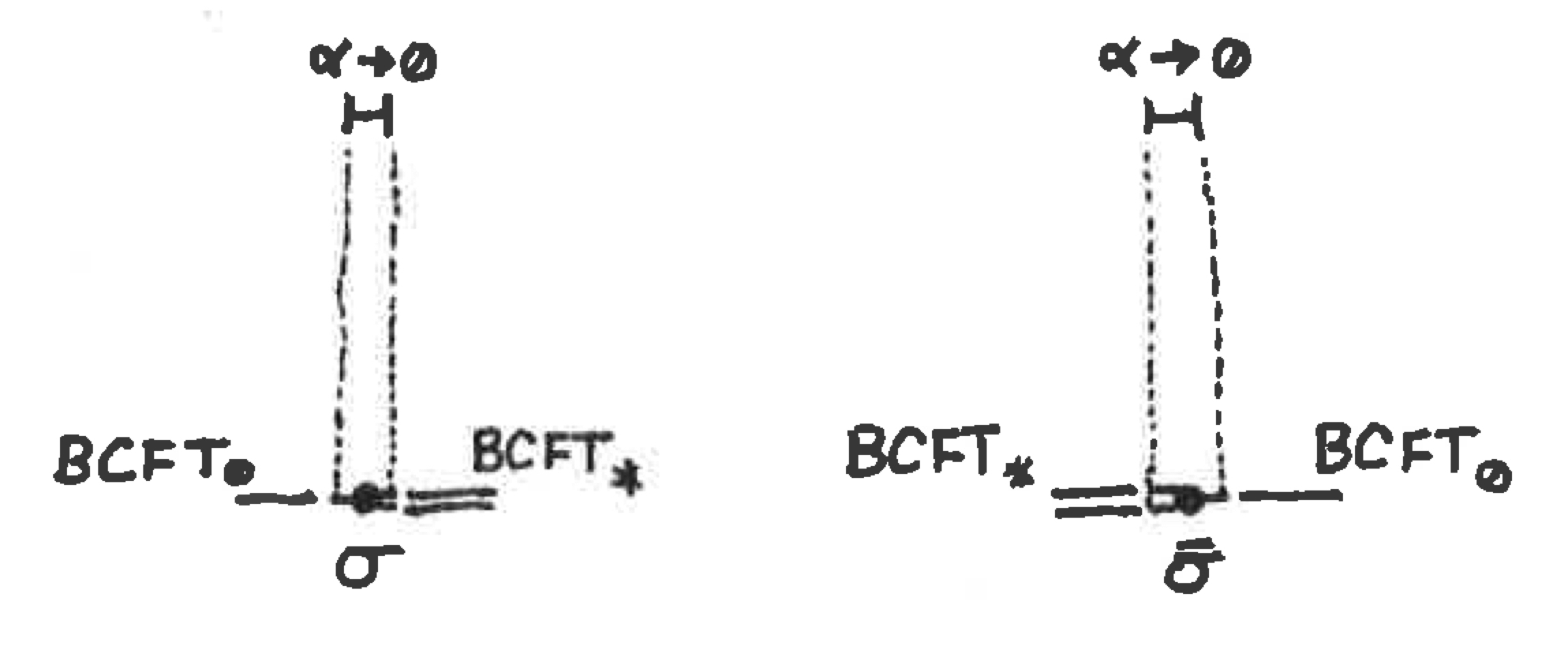}}
\end{wrapfigure}
With this motivation it is natural to describe the solution in terms of $\sigma$ and $\sigmabar$ rather than~$V$. We introduce string fields $\sigma$ and $\sigmabar$ as infinitesimally thin strips containing $\sigma(0)$ and $\sigmabar(0)$ on the 
boundary. Assuming that $\sigma$ and $\sigmabar$ change the boundary condition by a regular marginal deformation, we have the relations
\begin{equation}\sigmabar\sigma = 1,\ \ \ \ \ \sigma\sigmabar = 1.\end{equation}
Note that in the first equation, the right hand side is the identity string field of $\BCFT_*$, while in the second equation, it is the identity string field of $\BCFT_0$. To avoid cumbersome notation, we use the placement of $\sigma$ and $\sigmabar$ in expressions to indicate which state space a string field occupies. In addition we have the properties
\begin{eqnarray}
\lineup [B,\sigma]= [c,\sigma]= 0,\ \ \ \ [B,\sigmabar]= [c,\sigmabar]= 0; \\
\lineup \ \ \ \ \ \ \ \ \ \ \ \frac{1}{2}\mathcal{B}^-\sigma = \frac{1}{2}\mathcal{B}^-\sigmabar = 0;
\end{eqnarray}
and
\begin{eqnarray}
\lineup Q\sigma = c\d \sigma,\ \ \ \ Q\sigmabar = c\d\sigmabar;\\
\lineup \ \ \ \ \frac{1}{2}\mathcal{L}^-\sigma = \frac{1}{2}\mathcal{L}^-\sigmabar = 0.
\end{eqnarray}
The last properties follow because the boundary condition changing operators are primaries of weight $0$. Wedge states in $\BCFT_*$ can therefore be represented in two equivalent ways:
\begin{equation}\Omega_*^\alpha = e^{-\alpha(K+V)} = \sigma\Omega^\alpha\sigmabar.\end{equation}
\begin{exercise}\label{ex:Vsigma}
Show that $V = \sigma\d \sigmabar$. Use this to prove the above relation.
\end{exercise}
\noindent The simple intertwining solution can then be written
\begin{equation}\Psi_* = \underbrace{\frac{1}{\sqrt{1+K}}c(1+K)c\frac{1}{\sqrt{1+K}}}_{\Psi_\tv}\ \underbrace{-\frac{1}{\sqrt{1+K}}c(1+K)\sigma\frac{B}{1+K}\sigmabar (1+K)c\frac{1}{\sqrt{1+K}}}_{\Psi^\tv_*}.\end{equation}
We are still not done. It is helpful to extract a factor of the simple tachyon vacuum between $\sigma$ and $\sigmabar$ using the relation
\begin{equation}\frac{B}{1+K} = \frac{B}{\sqrt{1+K}}\Psi_\tv\frac{B}{\sqrt{1+K}}.\end{equation}
Then
\begin{equation}\Psi_* = \Psi_\tv + \underbrace{\left(\frac{1}{\sqrt{1+K}}cB(1+K)\sigma\frac{1}{\sqrt{1+K}}\right)\bigg(-\Psi_\tv\bigg)\left(\frac{1}{\sqrt{1+K}}\sigmabar (1+K)Bc\frac{1}{\sqrt{1+K}}\right)}_{\Psi^\tv_*}.\label{eq:KOS1}\end{equation}
Now we make some observations. The state $\Psi^\tv_*$ represents the creation of the D-brane $\BCFT_*$ out of the tachyon vacuum. On the other hand, the factor $-\Psi_\tv$ in between the boundary condition changing operators also represents the creation of $\BCFT_*$ out of the tachyon vacuum. The difference is that $\Psi^\tv_*$ lives in $\BCFT_0$, while $-\Psi_\tv$ between $\sigma$ and $\sigmabar$ lives in $\BCFT_*$. This suggests that the additional factors provide a kind of dictionary between the degrees of freedom of $\BCFT_0$ and $\BCFT_*$. We call these {\it intertwining fields}, denoted $\Sigma$ and $\Sigmabar$. The simple intertwining solution is finally written
\begin{equation}\Psi_* = \Psi_\tv -\Sigma\Psi_\tv\Sigmabar.\end{equation}
The solution satisfies the equations of motion provided that 
\begin{eqnarray}\lineup  Q_{\Psi_\tv}\Sigma  = Q_{\Psi_\tv}\Sigmabar = 0,\label{eq:Qtvinv}\\
\lineup\  \ \  \  \  \ \phantom{\Big)} \Sigmabar\Sigma  = 1.\end{eqnarray}
Since the cohomology around the tachyon vacuum is trivial, the first relations imply that the intertwining fields are $Q_{\Psi_\tv}$-exact. A little guesswork leads to the expressions
\begin{eqnarray}
\Sigma \lineup = Q_{\Psi_\tv}\left(\frac{B}{\sqrt{1+K}}\sigma\frac{1}{\sqrt{1+K}}\right) \nonumber\\
\lineup = \frac{1}{\sqrt{1+K}}cB(1+K)\sigma\frac{1}{\sqrt{1+K}}\ +\ \frac{B}{\sqrt{1+K}}\sigma c\sqrt{1+K},\phantom{\Bigg)} \ \ \ \ \ \ \ \ \\
\Sigmabar\lineup = Q_{\Psi_\tv}\left(\frac{1}{\sqrt{1+K}}\sigmabar\frac{B}{\sqrt{1+K}}\right) \nonumber\\
\lineup =\frac{1}{\sqrt{1+K}}\sigmabar(1+K)Bc\frac{1}{\sqrt{1+K}}\  + \ \sqrt{1+K}\sigmabar c\frac{B}{\sqrt{1+K}}.\phantom{\Bigg)}
\end{eqnarray}
Only the first terms appear in \eq{KOS1}. This is because the second terms vanish when multiplied with the tachyon vacuum due to $c^2=0$. The second crucial property of the intertwining fields is $\Sigmabar\Sigma = 1$. This can be demonstrated as follows:
\begin{eqnarray}
\Sigmabar\Sigma \lineup = Q_{\Psi_\tv}\left(\frac{1}{\sqrt{1+K}}\sigmabar\frac{B}{\sqrt{1+K}}\right)Q_{\Psi_\tv}\left(\frac{B}{\sqrt{1+K}}\sigma\frac{1}{\sqrt{1+K}}\right)\nonumber\\
\lineup = Q_{\Psi_\tv}\left(\frac{1}{\sqrt{1+K}}\sigmabar\frac{B}{\sqrt{1+K}}Q_{\Psi_\tv}\left(\frac{B}{\sqrt{1+K}}\sigma\frac{1}{\sqrt{1+K}}\right)\right)\nonumber\\
\lineup = Q_{\Psi_\tv}\bigg(\frac{1}{\sqrt{1+K}}\sigmabar\frac{B}{\sqrt{1+K}}\underbrace{Q_{\Psi_\tv}\left(\frac{B}{1+K}\right)}_{1}\sqrt{1+K}\sigma\frac{1}{\sqrt{1+K}}\nonumber\\
\lineup\ \ \ \ \ \ \ \ - \underbrace{\frac{1}{\sqrt{1+K}}\sigmabar\frac{B}{\sqrt{1+K}}\frac{B}{1+K}Q_{\Psi_\tv}\left(\sqrt{1+K}\sigma\frac{1}{\sqrt{1+K}}\right)}_0\bigg)\nonumber\\
\lineup = Q_{\Psi_\tv}\left(\frac{1}{\sqrt{1+K}}\sigmabar B\sigma \frac{1}{\sqrt{1+K}}\right)\nonumber\\
\lineup = Q_{\Psi_\tv}\left(\frac{B}{1+K}\right)\nonumber\\
\lineup = 1.
\end{eqnarray}
Note that we have a similar relation in reverse order 
\begin{equation}\Sigma\Sigmabar=1,\label{eq:wrongord}\end{equation}
but this is not necessary for the equations of motion.

Up to now we have been discussing regular marginal deformations. But we can generalize as follows. Suppose $\BCFT_0$ and $\BCFT_*$ represent arbitrary time-independent D-brane systems. Then it should be possible to relate them by a pair of boundary condition changing operators $\sigma_\text{bare}(x),\sigmabar_\text{bare}(x)$ which are independent of the timelike free boson factor of the $\BCFT$ and moreover are primaries of weight $h$ with OPE
\begin{equation}\sigmabar_\text{bare}(x)\sigma_\text{bare}(0) = \frac{1}{x^{2h}}+... \ .\end{equation}
An intertwining solution can be constructed using a modified pair of boundary condition changing operators 
\begin{eqnarray}
\sigma(x) \lineup = \sigma_\text{bare}e^{i\sqrt{h}X^0(x)},\label{eq:modsigma}\\
\sigmabar(x) \lineup = \sigmabar_\text{bare}e^{-i\sqrt{h}X^0(x)}\label{eq:modsigmabar}.
\end{eqnarray}
which satisfy
\begin{equation}\sigmabar(x)\sigma(0) = 1+\text{vanishing}.\ \label{eq:regOPE}\end{equation}
The singular OPE of $\sigmabar_\text{bare}$ and $\sigma_\text{bare}$ is canceled by the vanishing  OPE of the timelike plane wave vertex operators.  The timelike plane-wave vertex operators have the effect of turning on a gauge potential $A_0$ on the target D-brane, but since the time direction is noncompact, this is not physically observable.  Therefore, the modified boundary condition changing operators still connect the boundary conditions  of $\BCFT_0$ and $\BCFT_*$. In this way we are able to construct an analytic solution representing any time independent D-brane configuration. 

When we generalize in this way, however, the reverse relation \eq{wrongord} no longer holds:
\begin{equation}\Sigma\Sigmabar\neq 1.\end{equation}
To see why, consider the matter 2-point function of boundary condition changing operators on the unit disk:\ \ \ \ \ \ \ \ \ \ \ \ \ \ \ \ \ \ \ \ \ \ \ \ \ \ \ \ \ \ \ \ \ \ \ \ \ \ \ \ \ \ \ \ \ \ \ \ \ \ \ \ \ \ \ \ \ \ \ \ \ \ \ \ \ \ \ \ \ \ \ \ \ \ \ \ \ \ \ 
\vspace{-.4cm}
\begin{wrapfigure}{l}{.2\linewidth}
\centering
\resizebox{1.5in}{1.2in}{\includegraphics{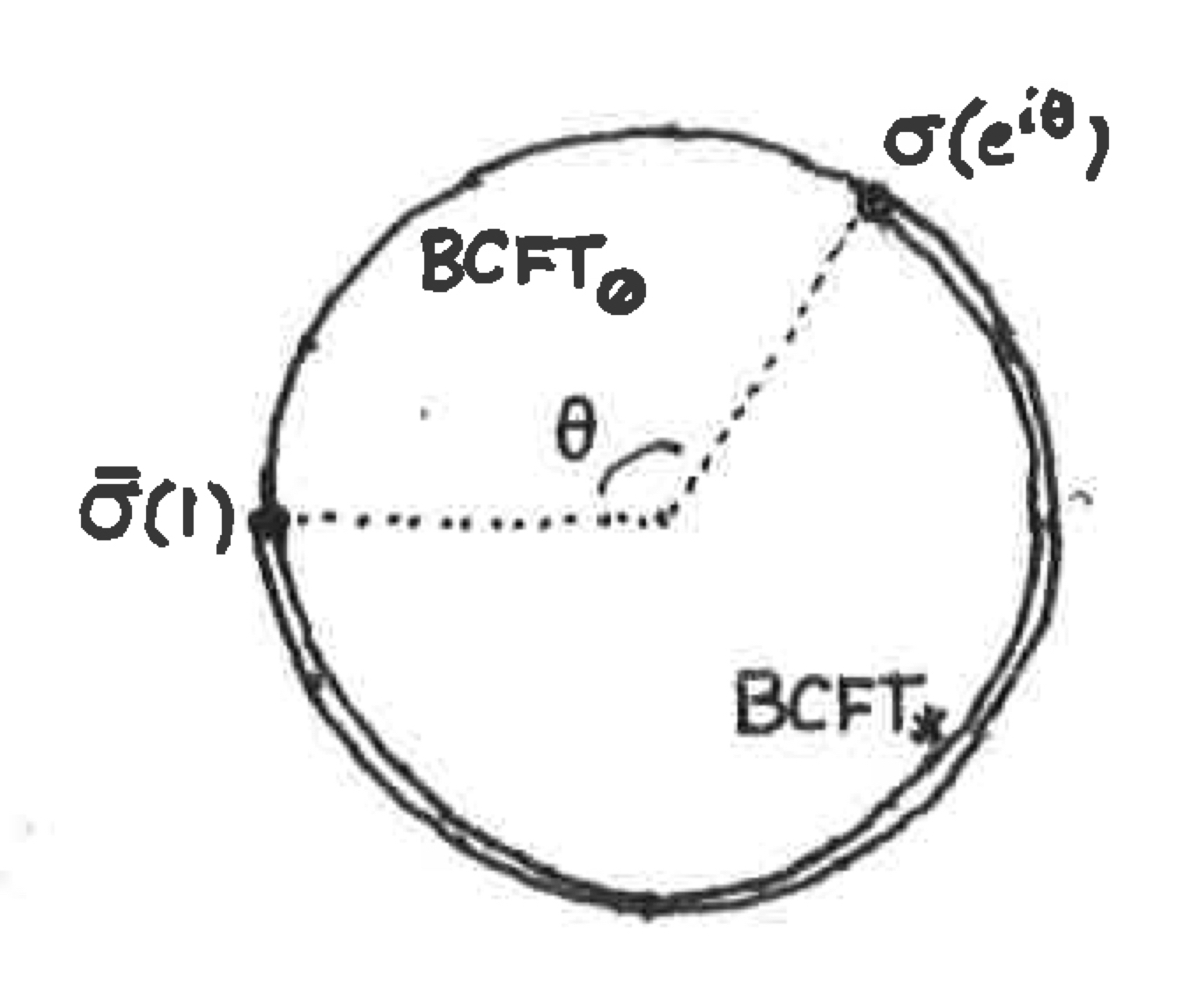}}
\end{wrapfigure}
\begin{equation}\langle\sigmabar(1)\sigma(e^{i\theta})\rangle_\text{disk}^\text{m}.\end{equation}
For angles between $0$ and $\theta$ the boundary of the disk carries $\BCFT_0$ boundary conditions, and outside that range it carries $\BCFT_*$ boundary conditions. Since $\sigma$ and $\sigmabar$ are weight zero primaries, the correlator is independent of $\theta$. We evaluate the correlator by taking the limit $\theta\to 0^+$, and using the OPE \eq{regOPE} we obtain
\begin{equation}\langle\sigmabar(1)\sigma(e^{i\theta})\rangle_{\text{disk}}^{\text{m}} = \langle 1\rangle_{\text{disk}}^{\text{m},\BCFT_*}=g_*,\end{equation}
where $g_*$ is the norm of the matter $SL(2,\mathbb{R})$ vacuum in $\BCFT_*$. We can also take the limit $\theta\to 2\pi^-$, where the boundary condition on the disk is $\BCFT_0$. This produces $g_0$, the norm of the matter $SL(2,\mathbb{R})$ vacuum in $\BCFT_0$. This leads to a puzzle: generally the reference and target D-brane systems will not have the same energy, so $g_0\neq g_*$. But conformal invariance requires that the correlator is independent of $\theta$. The resolution to this puzzle is that the OPE between $\sigma$ and $\sigmabar$ depends on which boundary condition is squeezed between the operators. By choice of normalization we already have
\begin{equation}\lim_{x\to 0^+}\sigmabar(x)\sigma(0) = 1,\end{equation}
so in the opposite order we must have
\begin{equation}\lim_{x\to 0^+}\sigma(x)\sigmabar(0) = \frac{g_*}{g_0}.\end{equation}
This implies that the corresponding intertwining fields satisfy, 
\begin{equation}\Sigmabar\Sigma=1,\ \ \ \ \ \Sigma\Sigmabar = \frac{g_*}{g_0}.\end{equation}
For regular marginal deformations $g_*=g_0$, and the second equality reduces to \eq{wrongord}. This result, however, creates a problem. Apparently, products of intertwining fields are not associative: 
\begin{equation}(\Sigmabar\Sigma)\Sigmabar\neq \Sigmabar(\Sigma\Sigmabar).\end{equation}
This reflects the fact that correlators do not have a well-defined limit when three boundary condition changing operators collide. Such ambiguous products do not appear when evaluating the equations of motion, the action, or the Ellwood invariant. However, they do lead to some complications in understanding background independence.

To prove background independence, at least for time-independent vacua, we need to find a field redefinition which relates the string field theories of the reference  and target D-branes. We can approach this as follows. The string field $\Psi^{(0)}$ of  $\BCFT_0$ can be written in terms of a fluctuation field around the simple intertwining solution:
\begin{equation}\Psi^{(0)}=\Psi_* +\varphi.\label{eq:varphi}\end{equation}
The action for the fluctuation field is 
\begin{equation}
S_0[\Psi^{(0)}]- S_0[\Psi_*]  = -\frac{1}{2}\Tr\big(\varphi Q_{\Psi_*} \varphi\big)-\frac{1}{3}\Tr\big(\varphi^3\big),
\end{equation}
where $S_0$ is the action for the reference D-brane and $Q_{\Psi_*}$ is the shifted kinetic operator around the simple intertwining solution. It is natural to guess that the fluctuation field can be related to the string field $\Psi^{(*)}$ of the target D-brane through
\begin{equation}\varphi = f(\Psi^{(*)})=\Sigma\Psi^{(*)}\Sigmabar.\label{eq:field_red}\end{equation}
To confirm this identification, we need to show that it transforms the action of the fluctuation field into the action of the target D-brane. The main aspect of this is how $Q_{\Psi_*}$ is replaced by the BRST operator of the target D-brane. To explain it is helpful to consider the operator
\begin{equation}Q_{\Psi_1\Psi_2}X = QX +\Psi_1X-(-1)^{|X|}X\Psi_2.\end{equation}
This is nilpotent 
\begin{equation}Q_{\Psi_1\Psi_2}^2=0.\end{equation}
provided that $\Psi_1$ and $\Psi_2$ satisfy the equation of motion. We also have a version of the Leibniz rule:
\begin{equation}Q_{\Psi_1\Psi_3}(XY) = (Q_{\Psi_1\Psi_2} X)Y + (-1)^{|X|}X(Q_{\Psi_2\Psi_3}Y),\label{eq:ShiftLeibniz}\end{equation}
\begin{wrapfigure}{l}{.26\linewidth}
\centering
\resizebox{2in}{1.4in}{\includegraphics{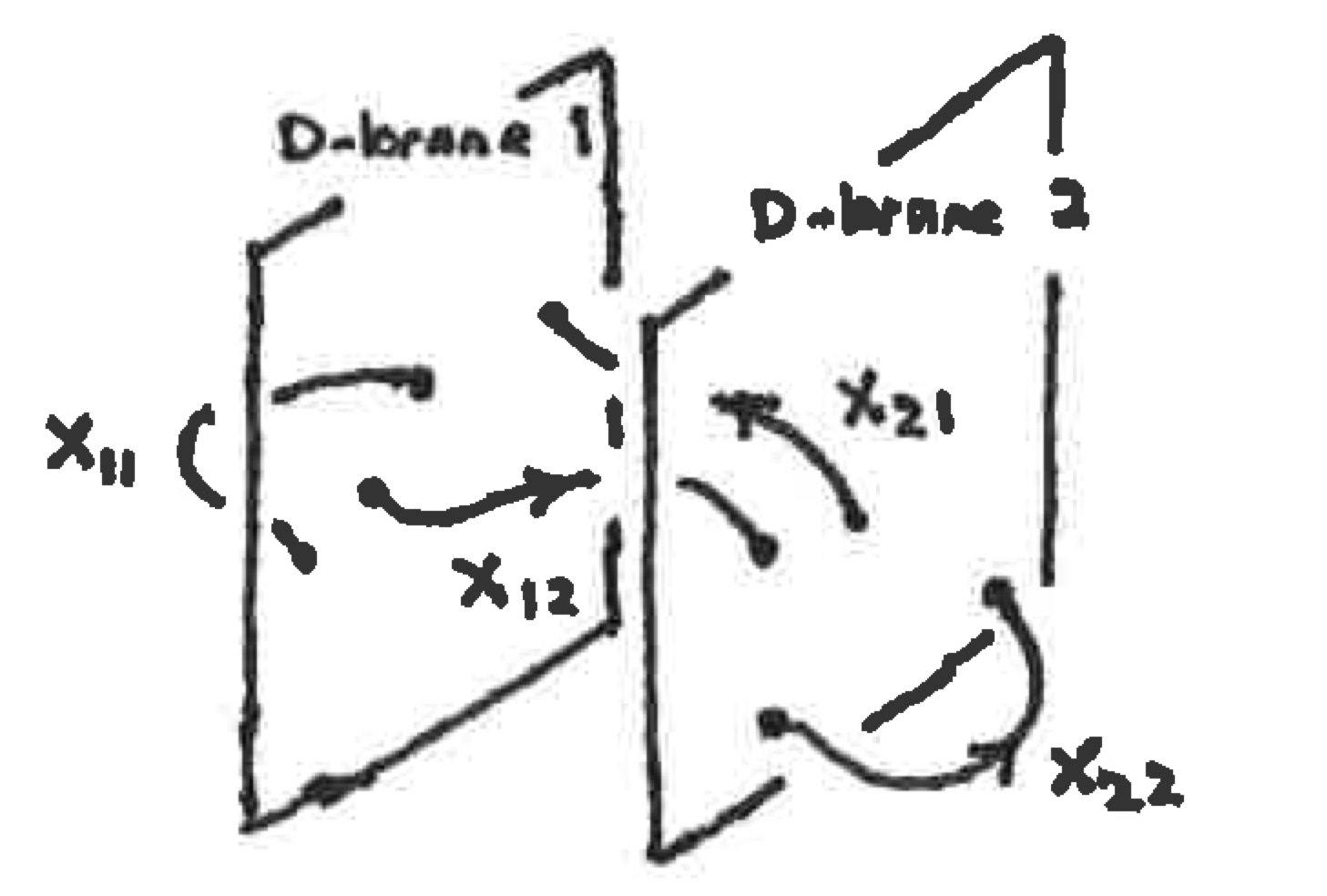}}
\end{wrapfigure}
where $\Psi_2$ on the right hand side is any solution. This operator naturally appears in open SFT formulated on a pair of D-branes. In this setup the string field contains $2\times 2$ Chan-Paton factors and can be arranged into a $2\times 2$ matrix
\begin{equation}X=\left(\begin{matrix} X_{11}\ \ X_{12} \\ X_{21} \ \ X_{22}\end{matrix}\right).\end{equation}
If we condense the first D-brane to a solution $\Psi_1$ and the second D-brane to a solution $\Psi_2$, the solution of the combined system is
\begin{equation}\Psi = \left(\begin{matrix}\Psi_1 & 0 \\  0  & \Psi_2\end{matrix}\right),\end{equation}
and the kinetic operator expanded around $\Psi$ is
\begin{equation}Q_\Psi X = \left(\begin{matrix} Q_{\Psi_1} X_{11} & Q_{\Psi_1\Psi_2}X_{12} \\ Q_{\Psi_2\Psi_1}X_{21} & Q_{\Psi_2}X_{22}\end{matrix}\right).\end{equation}
Therefore $Q_{\Psi_1\Psi_2}$ is the shifted kinetic operator for a stretched string connecting a D-brane condensed to a solution $\Psi_1$ and a D-brane condensed to a solution $\Psi_2$. Having explained this, we note that the intertwining  fields satisfy 
\begin{equation}Q_{\Psi_* 0}\Sigma = 0,\ \ \ \ Q_{0\Psi_*}\Sigmabar = 0 .\end{equation}
\begin{exercise}
Show that this follows from \eq{Qtvinv}. 
\end{exercise}  
\noindent The interpretation is that $\Sigma$ is annihilated by the kinetic operator for a stretched string connecting a $\BCFT_0$ D-brane condensed to the simple intertwining solution and a $\BCFT_*$ D-brane at the perturbative vacuum. However, these two configurations physically represent the same thing. In fact, the intertwining fields are representatives of the cohomology class of the identity operator in $\BCFT_*$. Now we can check that
\begin{eqnarray}
Q_{\Psi_*}\varphi \lineup = Q_{\Psi_*}\big(\Sigma\Psi^{(*)}\Sigmabar\big)\nonumber\\
\lineup = \big(Q_{\Psi_*0}\Sigma\big)\Psi^{(*)}\Sigmabar + \Sigma\big(Q\Psi^{(*)}\big)\Sigmabar + \Sigma\Psi^{(*)}\big(Q_{0\Psi_*}\Sigmabar\big)\nonumber\\
\lineup = \Sigma\big(Q\Psi^{(*)}\big)\Sigmabar.
\end{eqnarray}
where we used the Leibniz rule \eq{ShiftLeibniz}. Therefore the kinetic operator around the simple intertwining solution is directly related to the BRST operator of $\BCFT_*$. We can now substitute \eq{field_red} into the action and find 
\begin{equation}S_0[\Psi^{(0)}] - S_0[\Psi_\tv] = S_*[\Psi^{(*)}]-S_*[\Psi_\tv].\end{equation}
where $\Psi^{(0)}$ and $\Psi^{(*)}$ are related by \eq{varphi} and \eq{field_red}, and $S_*$ is the action for the target D-brane. This seems to establish that the string field theories are related by field redefinition. 

An important subtlety, however, is that \eq{field_red} is not necessarily invertible. Therefore it is not properly a field redefinition. One can try to find an inverse transformation of the form
\begin{equation}\Psi^{(*)} = f^{-1}(\varphi) = \Sigmabar\varphi\Sigma .\label{eq:inv_field_red}\end{equation}
Consistently, we have
\begin{equation}f^{-1}(f(\Psi^{(*)}))=\Psi^{(*)}. \end{equation}
Unfortunately, composing in the opposite order does not give the identity transformation:
\begin{equation}f(f^{-1}(\varphi))=\left(\frac{g_*}{g_0}\right)^2\varphi.\end{equation}
Even more confusing, $f$ and $f^{-1}$ do not compose associatively:
\begin{eqnarray}
f\circ (f^{-1}\circ f) \lineup = f,\\ (f\circ f^{-1})\circ f \lineup= \left(\frac{g_*}{g_0}\right)^2 f.\label{eq:fanom}
\end{eqnarray} 
Well defined mappings will always compose associatively. The problem can be understood as an inconsistency in the domains where $f$ and $f^{-1}$ are defined. Either the image of $f^{-1}$ is not contained in the domain of $f$, or vice versa. In either case, $f^{-1}$ is not an inverse map to $f$, and there is no isomorphism of field variables.  A more satisfactory account of background independence is given by the solution of \cite{KOSsingII}. But even here we do not find an isomorphism of the state spaces of different D-brane systems. We only find an isomorphism modulo gauge transformation, which is arguably enough. 

Let us discuss multiple D-brane backgrounds. A surprising feature of D-branes is that they can be superimposed to create composite D-brane systems; it is almost as though D-branes do not interact.  In ordinary field theories, adding soliton solutions does not give a multi-soliton solution since the theory is nonlinear. Suppose we superimpose simple intertwining solutions which create D-branes $\BCFT_1$ and $\BCFT_2$ out of the tachyon vacuum:
\begin{equation}\Psi = \Psi_\tv - \Sigma_1\Psi_\tv\Sigmabar_1 - \Sigma_2\Psi_\tv\Sigmabar_2.\label{eq:dd}\end{equation}
Since the theory is nonlinear, there is no  reason to  expect this to satisfy the  equations of  motion. However, it {\it does} satisfy the equations of motion if the intertwining fields are chosen to satisfy
\begin{equation}\Sigmabar_1\Sigma_2 = 0,\ \ \ \ \Sigmabar_2\Sigma_1 = 0. \label{eq:ddid}\end{equation}
It is interesting to assemble the intertwining fields into row and column vectors
\begin{equation}\Sigma  = \big(\Sigma_1\ \ \Sigma_2\big),\ \ \ \ \Sigmabar = \left(\begin{matrix}\Sigmabar_1 \\ \Sigmabar_2\end{matrix}\right).\end{equation}
Then the double D-brane solution \eq{dd} can be written 
\begin{equation}\Psi = \Psi_\tv - \Sigma\Psi_\tv\Sigmabar,\end{equation}
where the $\BCFT_1$ and $\BCFT_2$ terms arise upon contracting the row and column vectors. Now we observe that 
\begin{equation}\Sigmabar\Sigma = \left(\begin{matrix} 1 \ \ 0\\ 0\ \ 1\end{matrix}\right)\end{equation}
as a consequence of \eq{ddid}. This is the  identity string field of the double D-brane system with $2\times 2$ Chan-Paton factors. The result is that the solution for two D-branes is a special case of the simple intertwining solution. The possibility of superimposing D-brane solutions is directly connected with the Chan-Paton structure of the composite system. 

Next we discuss the Fock space expansion. The simple intertwining solution contains only two matter operator insertions, namely $\sigma$ and $\sigmabar$, and as a result the Fock space coefficients are determined simply by matter 3-point functions of a pair of boundary condition changing operators with a probe vertex operator. By contrast, marginal solutions in Schnabl gauge contain an infinite number of matter operator insertions, and determining coefficients requires computing all $n+1$-point functions of $n$  marginal operators with a probe vertex operator. The easiest coefficients to compute are for matter primary operators acting on the zero momentum tachyon state. These can be extracted  by  contracting with a test state $-\phi c\d c(0)|0\rangle$, where $\phi(0)$ is a matter  primary of weight $h$. If $h\neq 0$ the first term of the solution does not contribute and we find explicitly
\begin{equation}
-\langle \phi c\d c,\Psi_*\rangle = \big\langle I\circ\phi(0)\sigma(1)\sigmabar(0)\big\rangle_\mathrm{UHP}^\mathrm{m}\, \times \, g(h), \ \ \ \ \ (h\neq 0),
\end{equation}
where $g(h)$ is a ``universal" function of $h$ which is the same for all reference and target backgrounds, and can be computed once and for all. All of the information about the reference and target backgrounds has been factored out into the three point correlator. The universal function is given by
\begin{eqnarray}
g(h) \lineup = \frac{2}{\pi^3}\frac{\d}{\d\eps_1}\frac{\d}{\d\eps_2}\int_{1/2}^\infty dx\int_0^\infty ds\int_{1/2}^\infty dy\frac{e^{-L}}{\sqrt{(x-\frac{1}{2})(y-\frac{1}{2})}}\left(\frac{2}{L}\right)^{h-2}\Big(\cot \theta_{x+\eps_1}+\cot\theta_{y+\eps_2}\Big)^{h}\nonumber\\
\lineup\ \ \ \ \ \ \ \ \ \ \ \ \ \times\Big(\theta_x\sin^2\theta_y+\theta_y\sin^2\theta_x-\sin\theta_x\sin\theta_{s+\eps_1+\eps_2}\sin\theta_y\Big),
\end{eqnarray}
where in the integrand we introduce the variables
\begin{equation}
L = x+s+y+\eps_1+\eps_2,\ \ \ \ \theta_a = \frac{\pi a}{L}.
\end{equation}
At $h=0$ we have the zero momentum tachyon coefficient, which also receives contribution from the first term of the solution. We find
\begin{equation}T = T_\text{simp}\left(1-\frac{g_*}{g_0}\right),\end{equation}
where $T_\text{simp}$ is the tachyon coefficient of the simple tachyon vacuum, given in \eq{Tsimp}. With these formulas one can compute, for example, the position space profile of the tachyon lump, representing the formation of a D$(p-1)$-brane in the worldvolume of a D$p$-brane \cite{KOSsing}. The results show qualitative agreement with the Siegel gauge lump \cite{MoellerSenZwiebach}. Presently there have been no systematic efforts to compute the coefficients of descendant fields, which could be interesting for the purpose of testing convergence of the energy and other observables in the level expansion. The expectation, however, is that convergence will be less than ideal, similarly to the simple tachyon vacuum in the level expansion. 

\subsection{The solutions of Fuchs, Kroyter and Potting and of Kiermaier and Okawa}
\label{subsec:singular}

The solutions described so far share a certain genetic relationship. They satisfy the Schnabl gauge condition or a closely related gauge condition \eq{BsqrtFsqrtF}. A common feature of such solutions is that they have difficulty dealing with singular OPEs between matter operators. Here we describe a different solution which can handle marginal deformations with singular operator products. While it is built from wedge states with insertions, the structure otherwise has little in common with solutions described so far. The solution was discovered in its basic form by Fuchs, Kroyter and Potting \cite{FKP} in the context of the Wilson line deformation. The structure was clarified by Kiermaier and Okawa, who further generalized it to describe arbitrary marginal deformations \cite{KO}. 

We start by describing the solution for the Wilson line. This corresponds to giving a constant expectation value to a gauge field along some direction, say $x^1$. In Maxwell theory, such a background appears to be trivial since the gauge field can be removed by gauge transformation
\begin{equation}A_\mu = \d_\mu\big(A_1 x^1),\end{equation}
where $A_1$ is the value of the gauge field along the $x^1$-direction. However, this gauge transformation fails to be well-defined if the $x^1$ direction is compactified on a circle of radius $R$, since $x^1$ does not respect the periodicity of the circle. Moreover, while the field strength vanishes, the constant gauge field generates a nontrivial Wilson line around the circle
\begin{equation}e^{i \int_0^{2\pi R}dx^1 A_1} = e^{2\pi i R A_1}.\end{equation}
This is why the solution is called a Wilson line deformation.

For open bosonic strings the Wilson line deformation is generated by the exactly marginal boundary operator $i\d_\parallel X^1(y)$, where $\d_\parallel$ is the derivative along the open string boundary. If the $x^1$ direction is noncompact, the corresponding solution of the linearized equations of motion can be written in BRST exact form 
\begin{equation}\lambda\, i c \d_\parallel X^1(0)|0\rangle = Q\Big(\lambda\, i X^1(0)|0\rangle\Big).\end{equation}
If $x^1$ is compactified on a circle, the linearized solution is not BRST trivial since the operator $X^1(0)$ does not respect the periodicity of the circle. For convenience we parameterize the deformation in terms of $\lambda$; at linearized order this is related to the expectation value of the gauge field through
\begin{equation}A_1 = \sqrt{2}\lambda +\mathcal{O}(\lambda^2).\end{equation}
The strategy of Fuchs, Kroyter and Potting is based on the observation that pure gauge solutions are easier to find than nontrivial solutions. You simply choose a ghost number $0$ string field $\Lambda$ and compute
\begin{equation}\Psi = \Lambda Q \Lambda^{-1}.\end{equation}
The idea is to find a finite nonlinear gauge transformation which generalizes the BRST exact form of the vertex operator for the gauge field. If the finite gauge parameter $\Lambda$ can be chosen so that the resulting solution carries zero momentum along the $x^1$ direction,
\begin{equation}p_1\Big(\Lambda Q\Lambda^{-1}\Big) = 0,\label{eq:0p}\end{equation}
then the solution is well-defined even after compactifying $x^1$ on a circle. The gauge parameter $\Lambda$, however, is not well defined, and in this way we generate a nontrivial solution through gauge transformation. Interestingly, the zero momentum condition \eq{0p} is structurally very similar to the equations of motion of open superstring field theory in the Wess-Zumino-Witten-like formulation~\cite{Berkovits}.

We expand $\Lambda$ in powers of $\lambda$:
\begin{equation}\Lambda = 1-\Big(\lambda\Lambda_1 +\lambda^2\Lambda_2 + \lambda^3\Lambda_3+...\Big),\end{equation}
where
\begin{equation}\Lambda_1 = \sqrt{\Omega}\big(iX^1\big)\sqrt{\Omega},\end{equation}
and $X^1$ is defined by an infinitesimally thin strip containing $X^1(0)$ at the origin. We write the solution as
\begin{equation}\Psi =-(Q\Lambda)\Lambda^{-1}.\end{equation}
Expanding $\Lambda^{-1}$ as a geometric series gives the order $\lambda^n$ contribution to the solution
\begin{equation}\Psi_N = \sum_{n=1}^N\sum_{{k_1+k_2+...+k_n=N \atop k_i\geq 1}}\big(Q\Lambda_{k_1}\big)\Lambda_{k_2}...\Lambda_{k_i}.\end{equation}
For example
\begin{equation}\Psi_1 = Q\Big(\sqrt{\Omega}\big(iX^1\big)\sqrt{\Omega}\Big) = \sqrt{\Omega}\big(i c\d_\parallel X^1\big)\sqrt{\Omega}.\end{equation}
This satisfies 
\begin{equation}p_1 \Psi_1 = 0,\end{equation}
since the zero momentum photon vertex operator is independent of the position zero mode. At second order we have
\begin{eqnarray}
\Psi_2\lineup = Q\Lambda_2 + \big(Q\Lambda_1\big)\Lambda_1\nonumber\\
\lineup = Q\Lambda_2 + \sqrt{\Omega} Q(iX^1)\Omega(iX^1)\sqrt{\Omega}.
\end{eqnarray}
We determine $\Lambda_2$ so that  
\begin{equation}p_1\Psi_2 = 0.\end{equation}
The momentum operator is the zero mode of a weight 1 primary field, 
\begin{equation}p_1 = \oint \frac{dz}{2\pi i} i\d X^1(z),\end{equation}
and is a derivation of the open string star product, for the same reason as the BRST operator. We have
\begin{equation}p_1(iX^1) = 1.\label{eq:pX1} \end{equation}
It is useful to introduce the string field $(iX^1)^n$ as an infinitesimally thin strip containing the operator $(iX^1(0))^n$ at the origin, where the power of $X^1(0)$ is defined with boundary normal ordering. Note that $(iX^1)^n$ is not the same as the $n$ star products of $iX^1$, which would be divergent due to the logarithm in the $X^1$-$X^1$ OPE. Generalizing \eq{pX1}, 
\begin{equation}p_1 (iX^1)^n = n (iX^1)^{n-1}.\label{eq:pX1n}\end{equation}
\begin{exercise}
Prove this.
\end{exercise}
\noindent Now we can compute the action of the momentum operator on $\Psi_2$. Setting this to zero leads to a condition on $\Lambda_2$:
\begin{equation}p_1\big(Q\Lambda_2) + \sqrt{\Omega}Q( iX^1)\Omega\sqrt{\Omega}=0.\end{equation}
Noting \eq{pX1n}, the obvious solution is
\begin{equation}\Lambda_2 = -\frac{1}{2!}\sqrt{\Omega}(iX^1)^2 \Omega\sqrt{\Omega}.\end{equation}
\begin{exercise}
The third order contribution to the solution is
\begin{equation}\Psi_3 = Q\Lambda_3 + \big(Q\Lambda_2\big)\Lambda_1 + \big(Q\Lambda_1\big)\Lambda_2 + \big(Q\Lambda_1)\Lambda_1\Lambda_1.\end{equation}
Requiring that $\Psi_3$ has zero momentum, show that $\Lambda_3$ can be chosen as
\begin{equation}\Lambda_3 = \frac{1}{3!}\sqrt{\Omega} (iX^1)^3\Omega^2\sqrt{\Omega}.\end{equation}
\end{exercise}
\noindent The general result is now easy to guess:
\begin{equation}\Lambda_n = \frac{(-1)^{n+1}}{n!}\sqrt{\Omega}(iX^1)^n \Omega^{n-1}\sqrt{\Omega}.\end{equation}
This is the solution as characterized by Fuchs, Kroyter, and Potting \cite{FKP}. The operator insertions are separated by wedge states with positive integer width, and there is no question of OPE divergence. One distinctive feature of the solution is the absence of $b$-ghosts, as one finds in Schnabl gauge or Siegel gauge. This is because the solution is not characterized by a gauge condition, and is not constructed from a propagator. This is surprising since there is a close connection between the perturbative construction of marginal deformations and the computation of the tree-level $S$-matrix. In this context one typically expects $b$-ghosts to provide the correct measure for integration over the moduli space of disks with boundary punctures. Another unusual feature of the solution is that it does not satisfy the reality condition, as can be readily seen by inspecting the second order contribution
\begin{equation}\Psi_2 = -\frac{1}{2!}\sqrt{\Omega}Q(iX^1)^2\Omega\sqrt{\Omega}+\sqrt{\Omega}Q(iX^1)\Omega(iX^1)\sqrt{\Omega}\neq\Psi_2^\ddag.\end{equation}
This is not a deep concern since the solution can be made real by gauge transformation. A recipe for achieving this is described in \cite{KO}.  For physical questions, the solution is equivalent to a real solution. 

The construction so far does not immediately apply to other marginal deformations, since it is not clear what should be the analogue of the compactification and zero momentum constraint. For this it is helpful to adopt a different point of view on the operator $X^1(y)$. The boundary condition changing operators which turn on the Wilson line are  
\begin{equation}\sigma_\mu=e^{i\mu X^1(y)},\ \ \ \ \sigmabar_\mu = e^{-i\mu X^1(y)}.\end{equation}
For general $\mu$ these operators do not respect the periodicity of compactification on the circle, and appear to be undefined. But this is actually expected, since boundary condition changing operators are not well defined local operators in the reference $\BCFT$. From this point of view the issue with $X^1(y)$ is not necessarily that it does not respect the periodicity of compactification, but that it implements an infinitesimal change in the open string boundary condition. We can write
\begin{equation}(iX^1)^n=\left.\frac{d^n}{d\mu^n}\sigma_\mu\right|_{\mu=0}.\end{equation}
It is helpful to streamline notation somewhat by introducing the operator 
\begin{equation}d^n =\left.\frac{d^n}{d\mu^n}\right|_{\mu=0},\end{equation}
where we take $\mu=0$ after all derivatives are evaluated. We will also leave the dependence on $\mu$ in the argument of the derivative operator implicit, so for example 
\begin{equation}d\sigma = \left.\frac{d}{d\mu}\sigma_\mu\right|_{\mu=0}.\end{equation}
The $n$th contribution to the gauge parameter is then 
\begin{equation}
\Lambda_n = \frac{(-1)^{n+1}}{n!}\sqrt{\Omega}d^n\sigma \Omega^{n-1}\sqrt{\Omega}.
\end{equation}
Summing over $n$ gives a formula for the complete finite gauge transformation
\begin{equation}\Lambda = \sqrt{\Omega}\Big(\sigma e^{-\lambda\overleftarrow{d}\Omega}\Big)\frac{1}{\sqrt{\Omega}}.\end{equation}
The arrow over $d$ indicates that it is acting on $\sigma$ to the left. The inverse wedge state here is formal; it cancels out when we evaluate the expression.

The goal is to solve the zero momentum constraint by showing that a change of boundary condition implemented by $\sigma$ inside the solution is always undone by a $\sigmabar$. The second order deformation takes the form 
\begin{equation}
\Psi_2 = -\frac{1}{2!}\sqrt{\Omega}Q(d^2\sigma)\Omega\sqrt{\Omega}+\sqrt{\Omega}Q(d\sigma)\Omega(d\sigma)\sqrt{\Omega}.
\end{equation}
This expression assumes that $x^1$ is noncompact, since there is no $\sigmabar$ to undo the change of boundary condition. In the noncompact case we have the equality 
\begin{equation}\sigmabar_\mu = \sigma_{-\mu},\ \ \ \ \sigma_{\mu=0}=1,\end{equation}
which implies
\begin{equation}d\sigma = -d\sigmabar,\ \ \ \ d^0\sigmabar = 1.\end{equation}
This allows us to rewrite the second order deformation as 
\begin{eqnarray}
\Psi_2 \lineup = -\frac{1}{2!}\sqrt{\Omega}Q(d^2\sigma)\Omega (d^0\sigmabar)\sqrt{\Omega}-\sqrt{\Omega}Q(d\sigma)\Omega(d\sigmabar)\sqrt{\Omega}\nonumber\\
\lineup = -\frac{1}{2!}\Big(\sqrt{\Omega}Q(d^2\sigma)\Omega (d^0\sigmabar)\sqrt{\Omega}+2\sqrt{\Omega}Q(d\sigma)\Omega(d\sigmabar)\sqrt{\Omega}+\sqrt{\Omega}Q(d^0\sigma)\Omega (d^2\sigmabar)\sqrt{\Omega}\Big)\nonumber\\
\lineup = -\frac{1}{2!}d^2\Big(\sqrt{\Omega}Q\sigma\Omega \sigmabar\sqrt{\Omega}\Big).
\end{eqnarray}
In this final form the change of boundary condition is canceled, and the second order deformation is meaningful even if $x^1$ is compact. In fact, the second order deformation is defined for any marginal deformation with an associated one parameter family of boundary condition changing operators~$\sigma_\mu,\sigmabar_\mu$. This is how Kiermaier and Okawa manage to generalize the Wilson line solution. 

The systematics of how the change of boundary condition is undone at higher orders could be complicated. To see how to deal with it, consider the state
\begin{equation}\Lambda\Lambda^\ddag = \sqrt{\Omega}\Big(\sigma e^{-\lambda\overleftarrow{d}\Omega}\Big)\frac{1}{\sqrt{\Omega}}\frac{1}{\sqrt{\Omega}}\Big(e^{-\lambda\overrightarrow{d}\Omega}\sigmabar\Big)\sqrt{\Omega}.\end{equation}
Expanding the exponentials in Taylor series gives
\begin{eqnarray}
\Lambda\Lambda^\ddag \lineup = \sum_{m,n=0}^\infty\frac{ (-\lambda)^{m+n}}{m!n!}\sqrt{\Omega}(d^m\sigma)\Omega^{m+n-1}(d^n\sigmabar)\sqrt{\Omega}\nonumber\\
\lineup = \sum_{N=0}^\infty \frac{(-\lambda)^N}{N!}\sum_{k=0}^N \left({ N\atop k}\right)\sqrt{\Omega}(d^{N-k}\sigma)\Omega^{N-1}(d^k\sigmabar) \sqrt{\Omega}\nonumber\\
\lineup = \sum_{N=0}^\infty \frac{(-\lambda)^N}{N!}d^N\Big(\sqrt{\Omega}\sigma\Omega^{N-1}\sigmabar\sqrt{\Omega}\Big)\nonumber\\
\lineup = \sqrt{\Omega}\left(\sigma e^{-\lambda \overleftrightarrow{d}\Omega}\frac{1}{\Omega}\sigmabar \right)\sqrt{\Omega}.
\end{eqnarray}
The double arrow over $d$ indicates that it acts as a total derivative on the boundary condition changing operators to the left and right. The final expression is a well-defined state for any marginal deformation. This leads us to modify the pure gauge ansatz by writing 
\begin{equation}\Psi = - (Q\Lambda)\Lambda^{-1} = -(Q\Lambda)\Lambda^\ddag\frac{1}{\Lambda\Lambda^\ddag}.\end{equation}
This leads to the expression
\begin{eqnarray}
\Psi \lineup = -A U^{-1},
\end{eqnarray}
where
\begin{eqnarray}
A \lineup = \sqrt{\Omega}\left(Q\sigma e^{-\lambda \overleftrightarrow{d}\Omega}\frac{1}{\Omega}\sigmabar \right)\sqrt{\Omega},\\
U \lineup = \sqrt{\Omega}\left(\sigma e^{-\lambda \overleftrightarrow{d}\Omega}\frac{1}{\Omega}\sigmabar \right)\sqrt{\Omega}.\label{eq:singularU}
\end{eqnarray}
This is the form of the solution found by Kiermaier and Okawa \cite{KO}, and is defined for arbitrary marginal deformations. It is amusing to note the appearance of a translation operator which formally shifts the marginal coupling constant in the direction of the $SL(2,\mathbb{R})$ vacuum. 

\begin{exercise}
Demonstrate that the Kiermaier-Okawa solution satisfies the equations of motion.
\end{exercise}

Kiermaier and Okawa mostly do not use the language of boundary condition changing operators, but describe the solution in terms of an appropriately renormalized exponential insertion of line integrals of a marginal operator. The connection between these descriptions is given in terms of correlation functions on the cylinder:
\begin{eqnarray}\langle \phi,\sigma_\mu\Omega^n\sigmabar_\mu\rangle \lineup = \big\langle \sigma_\mu(n+1/2)\sigmabar_\mu(1/2)f_\mathcal{S}\circ\phi(0)\big\rangle_{C_{n+1}}\nonumber\\
\lineup = \Big\langle\Big[e^{\mu \int_{1/2}^{n+1/2} dy V(y)}\Big]_r f_\mathcal{S}\circ \phi(0)\Big\rangle_{C_{n+1}},
\end{eqnarray}
where $V(y)$ is the marginal operator. The bracket $[\cdot]_r$ is a reminder that powers of the line integral are defined with the appropriate renormalization. The nature of the renormalization scheme is a major aspect of the discussion of \cite{KO}. This is an important point because the boundary condition changing operators for a generic marginal deformation are usually not known, and to characterize the solution it is necessary to construct the boundary condition changing operators ``from scratch" by figuring out how to renormalize the exponential insertion. Further analysis of this aspect appears in \cite{Longton1,Longton2}. It is interesting to mention that the line integrals of $V$ can be interpreted in terms of integration over the moduli spaces of disks with boundary punctures appearing in tree level amplitudes. The $b$-ghost which provides the measure is effectively hidden in the fact that $V$ is an {\it integrated} vertex operator, related to the usual on-shell vertex operator $cV$ through a contour integral of the $b$-ghost around the puncture. The $b$-ghost deletes the $c$ and effectively disappears from the amplitude, and likewise the solution.

Many important questions about this solution remain unanswered. Not much is known about the nonperturbative behavior of the solution for finite $\lambda$. It should also be possible to understand perturbative background independence by expanding the action around the solution, but this has not been investigated. 

\section{Lecture 4: Toolbox}
\label{sec:tool}

In the previous lecture we discussed a number of specific analytic solutions. Now we describe some techniques which give a broader view of how analytic solutions work more generally.   

\subsection{$\mathcal{L}^-$ level expansion}
\label{subsec:L}

Often it is useful to extract information about a string field by probing it with a test state. Interesting things can happen if the test state is sliver-like. Given a Fock space state $|\phi\rangle$, we can consider a sliver-like test state
\begin{equation}\left(\frac{1}{\eps}\right)^{\frac{1}{2}\mathcal{L}^-}\!\!|\phi\rangle,\ \ \ \ \eps\text{ small}.\end{equation}
This is a strip of worldsheet of width $1/\eps$ containing a vertex operator in the middle creating the state $|\phi\rangle$. If we probe a string field $X$ with a test state of this form, it may happen that the result can be expanded as a power series in $\eps$:
\begin{equation}\left\langle \left(\frac{1}{\eps}\right)^{\frac{1}{2}\mathcal{L}^-} \!\!|\phi\rangle ,X\right\rangle = \eps^{h_1}\langle\phi, X_{h_1}\rangle + \eps^{h_2}\langle \phi, X_{h_2}\rangle + \eps^{h_3}\langle \phi,X_{h_3}\rangle + ...\ \ \ \ h_1<h_2<h_3< ... \ .\end{equation}
The coefficients of the series define the overlap of $\phi$ with a sequence of string fields $X_{h_1},X_{h_2},X_{h_3},...$. The states $X_h$ must be eigenstates of $\frac{1}{2}\mathcal{L}^-$ with eigenvalue $h$:
\begin{equation}\frac{1}{2}\mathcal{L}^-X_h = hX_h,\end{equation}
and we can formally write 
\begin{equation}X = X_{h_1}+X_{h_2} + X_{h_3} + ... ,\ \ \ \ h_1<h_2<h_3< ...\ . \label{eq:LmX}\end{equation}
This is called the $\mathcal{L}^-$ {\it level expansion} \cite{exotic}. ``Level" refers to the $\frac{1}{2}\mathcal{L}^-$ eigenvalue of a state in the expansion.  The leading level in the expansion is the lowest level, since it makes the most important contribution to the overlap with a sliver-like test state in the limit $\eps\to 0$. Higher levels are subleading. The $\mathcal{L}^-$ level expansion is a variant of the $\mathcal{L}_0$ level expansion briefly described in subsection \ref{subsec:Sch}. Given the $\mathcal{L}^-$ level expansion of $X$ in \eq{LmX}, the $\mathcal{L}_0$ level expansion of $\sqrt{\Omega}X\sqrt{\Omega}$ is determined by
\begin{equation}\sqrt{\Omega}X\sqrt{\Omega} = \sqrt{\Omega}X_{h_1}\sqrt{\Omega}\, +\, \sqrt{\Omega}X_{h_2}\sqrt{\Omega} \, +\,  \sqrt{\Omega}X_{h_3}\sqrt{\Omega} \, +\,  ... ,\ \ \ \ h_1<h_2<h_3< ... \ .\end{equation}
In this sense the expansions are equivalent. The $\mathcal{L}_0$ level expansion selects a preferred wedge state of positive width  (the $SL(2,\mathbb{R})$ vacuum), and this choice is to some degree  arbitrary. For this reason, the  $\mathcal{L}^-$   level expansion  is  slightly simpler to work with. However, an advantage of the $\mathcal{L}_0$ level expansion is that its eigenstates are normalizable, so we can  compute the energy by substituting the expansion into the action. Typically, this will express the energy as an asymptotic series \cite{Schnabl,simple}. In any case it is straightforward to translate between $\mathcal{L}^-$ and $\mathcal{L}_0$ level expansions. 

The $\mathcal{L}^-$ level expansion is most useful in the context of a {\it singularity free subalgebra}. This is a subalgebra of wedge states with insertions where products of all fields are finite. The $KBc$ subalgebra is a singularity free subalgebra, as is its extension to regular marginal deformations. Other singularity free subalgebras appear in superstring field theory. We make two important claims:
\begin{claim}
The $\mathcal{L}^-$ level expansion of a state in a singularity free subalgebra can be computed by expanding the state in powers of $K$ around $K=0$ and ordering terms in sequence of increasing $\frac{1}{2}\mathcal{L}^-$ eigenvalue.
\end{claim}
\begin{claim}
In a singularity free subalgebra, the $\mathcal{L}^-$ level expansion of a product of states is given by multiplying the $\mathcal{L}^-$ level expansions of the states individually. Level is additive under star multiplication.
\end{claim}
\noindent These properties hold only if we are working with a singularity free subalgebra. To appreciate claim 1, consider the state 
\begin{equation}\sqrt{\Omega}V\Omega V\sqrt{\Omega},\label{eq:VOmV}\end{equation}
where $V$ is a marginal operator. If we expand around $K=0$ we obtain
\begin{equation}
\sqrt{\Omega}V\Omega V\sqrt{\Omega}= \underbrace{\phantom{\Big)}\!\!V^2}_{\text{level }2} -\underbrace{VKV -\frac{1}{2}KV^2 -\frac{1}{2}V^2 K}_{\text{level }3}\, + \, \text{higher levels},\ \ \ \ \ \ \text{(regular OPE)}.\label{eq:LmVOmV}
\end{equation}
This expression is meaningful only if $V$ has regular OPE with itself. Otherwise the terms are divergent. To appreciate claim 2, note that \eq{VOmV} is the square of another state
\begin{equation}\sqrt{\Omega}V\Omega V\sqrt{\Omega} = (\sqrt{\Omega}V\sqrt{\Omega})^2,\end{equation}
which has $\mathcal{L}^-$ expansion
\begin{equation}\sqrt{\Omega}V\sqrt{\Omega} = \underbrace{\phantom{\Big)}\!\!\!V}_{\text{level }1}- \underbrace{\frac{1}{2}KV -\frac{1}{2}V K}_{\text{level }2}\, +\, \text{higher levels}.\label{eq:LmV}\end{equation}
However, we cannot directly compute square of the right hand side unless $V$ has regular OPE. In the presence of singular OPEs the $\mathcal{L}^-$ level expansion can still be computed but we must take care to subtract divergences before expanding in $K$.  In the present example, it can be found by normal ordering  following \eq{normord}
\begin{equation}\sqrt{\Omega}V\Omega V\sqrt{\Omega} = \mathcal{N}\Omega^2+\sqrt{\Omega}\!:\!V\Omega V\!:\!\sqrt{\Omega}.\end{equation}
The $\mathcal{L}^-$ level expansion is then 
\begin{equation}
\sqrt{\Omega}V\Omega V\sqrt{\Omega} = \underbrace{\phantom{\Big)}\!\!\mathcal{N}}_{\text{level }0} -\underbrace{\phantom{\Big)}\!\!2\mathcal{N}K}_{\text{level }1} + \underbrace{\phantom{\Big)}\!\!2\mathcal{N}K^2+\!:\!V^2\!:\!}_{\text{level }2}\, +\, \text{higher levels},\ \ \ \ \ \ \text{(singular OPE)},
\end{equation}
which is quite different from \eq{LmVOmV}. 

If we are lucky enough to have a singularity free subalgebra, the $\mathcal{L}^-$ level expansion allows us to solve the equations of motion perturbatively in level. This is essentially what lead to the discovery of Schnabl's solution. Let us see how it works. A general state at ghost number 1 in the $KBc$ subalgebra can be expanded in level
\begin{equation}\Psi = \Psi_{-1}+\Psi_0 + \Psi_1 + \text{higher levels},\end{equation}
where the index on $\Psi_n$ indicates the level, and the general ghost number 1 state at each level can be written
\begin{eqnarray}
\Psi_{-1} \lineup= \varphi_1 c,\\
\Psi_0 \lineup = \varphi_2 cK +\varphi_3 Kc + \varphi_4 cKBc ,\\
\Psi_1 \lineup = \varphi_5 K^2 c + \varphi_6 KcK + \varphi_7 cK^2 + \varphi_8 KcKBc + \varphi_9 cKBcK + \varphi_{10} cK^2 Bc, \\
\lineup \vdots\ \ \ .\nonumber 
\end{eqnarray}
The coefficients $\varphi_i$ can be interpreted as expectation values of fields in this basis, and should be determined by solving the equations of motion. The equations of motion imply
\begin{eqnarray}
0 \lineup= \Psi_{-1}^2,  \\
0\lineup = Q\Psi_{-1} + [\Psi_{-1},\Psi_0] ,\\
0 \lineup = Q\Psi_0 + [\Psi_{-1},\Psi_1]+\Psi_0^2, \\
\lineup \vdots\ \ \ . \nonumber
\end{eqnarray}
We solve the first two equations:
\begin{eqnarray}
0 \lineup = (\varphi_1 c)^2,\\
0 \lineup = Q(\varphi_1 c)+ [ \varphi_1 c,\varphi_2 cK + \varphi_3 Kc + \varphi_4 cKBc]\\
\lineup = \varphi_1(1+\varphi_2+\varphi_3) cKc.
\end{eqnarray}
The first equation holds trivially for any $\varphi_1$. The second equation only implies nontrivial conditions if $\varphi_1$ is nonzero, in which case we find a three parameter family of solutions
\begin{equation}\varphi_1 \neq 0,\ \ \ \ \varphi_2+\varphi_3 = -1,\ \ \ \ \varphi_4\ \text{arbitrary}.\end{equation}
One can check that the higher level equations of motion do not impose further constraints on these coefficients. This family of solutions, constructed here up to level zero, are gauge equivalent representatives of the tachyon vacuum. To see this, we may use the $\mathcal{L}^-$ level expansion to construct the homotopy operator. The general state at ghost number $-1$ in the $KBc$ subalgebra can be expanded
\begin{equation} A = A_1+A_2+\text{higher levels},\end{equation}
where
\begin{eqnarray}
A_1\lineup  = h_1 B,\\
A_2\lineup = h_2 KB,\\
\lineup \vdots \ \ .\nonumber
\end{eqnarray}
The homotopy operator should satisfy $Q_\Psi A=1$, which expanded in level requires
\begin{eqnarray}
1\lineup =[\Psi_{-1},A_1],\\
0\lineup = Q A_1 +[\Psi_{-1},A_2]+[\Psi_0,A_1],\\
\lineup \ \ \ \vdots.
\end{eqnarray}
We find that 
\begin{equation}1=\varphi_1 h_1,\ \ \ 0= h_1(1+\varphi_2+\varphi_3),\ \ \ 0= \varphi_1h_2+\varphi_4 h_1.\end{equation}
The second relation holds identically due to the equations of motion. Thus the coefficients of the homotopy operator at the first two levels are given by
\begin{equation}h_1 = \frac{1}{\varphi_1},\ \ \ \ h_2 =- \frac{\varphi_4}{\varphi_1^2}.\end{equation}
Finally, we can understand why the tachyon vacuum is not gauge equivalent to the perturbative vacuum. Proving gauge equivalence requires finding a finite gauge parameter $U$ which satisfies 
\begin{equation}\Psi = U^{-1}Q U.\end{equation}
Multiplying by $U^{-1}$ gives a linear equation
\begin{equation}
QU - U\Psi= 0.\label{eq:LmPsitvU}
\end{equation}
The general state at ghost number $0$ in the $KBc$ subalgebra can be expanded
\begin{equation}U = U_0 + U_1+\text{higher levels},\end{equation}
where
\begin{eqnarray}
U_0 \lineup = \lambda_1  + \lambda_2 Bc,\\
U_1\lineup = \lambda_3 K + \lambda_4 K Bc + \lambda_5 BcK,\\
\lineup \vdots\ \ \ . \nonumber 
\end{eqnarray}
Expanding \eq{LmPsitvU} gives 
\begin{eqnarray}
0\lineup = -U_0\Psi_{-1},\\
0\lineup = QU_0 - U_1\Psi_{-1} - U_0\Psi_0,\\
\lineup\vdots\ \ \ .\nonumber
\end{eqnarray}
Focus on the first equation:
\begin{equation}U_0\Psi_{-1} =\varphi_1\lambda_1 c = 0 .\end{equation}
Since $\varphi_1\neq 0$, this requires $\lambda_1=0$ and at leading level the gauge parameter must take the form 
\begin{equation}U_0 = \lambda_2 Bc.\end{equation}
Unfortunately this does not have a star algebra inverse, as can be seen by multiplying by $B$ from the left. Therefore the perturbative vacuum and the tachyon vacuum are not gauge equivalent. 

One can systematize this kind of analysis to give a classification of gauge orbits in the $KBc$ subalgebra. The classification from \cite{IdSing} rests on the following assumptions:
\begin{description}
\item{(1)} Only nonnegative integer powers of $K$ appear in the $\mathcal{L}^-$ level expansion. We in fact implicitly assumed this in the previous paragraph. Noninteger powers of $K$ can be defined through
\begin{equation} K^{\nu} = \frac{1}{\Gamma(-\nu)}\int_0^\infty d\alpha\alpha^{-\nu-1}\Omega^\alpha.\label{eq:Knu}\end{equation}
where for $\nu>0$ the integrand should be interpreted as a distribution
\begin{equation}\alpha^{-\nu-1}\Omega^\alpha = (-1)^{m}\frac{\Gamma(-\nu)}{\Gamma(m-\nu)} \alpha^{m-\nu-1} \frac{d^m}{d\alpha^m}\Omega^\alpha,\ \ \ \ 0<m-\nu<1,\end{equation}
so the seemingly divergent integration towards $\alpha=0$ is subtracted. What is notable about non-integer  $\nu$ is that the integrand gives at best inverse power suppression to the sliver state. It is therefore expected that non-analytic powers of $K$ lead to somewhat singular solutions. 
\item{(2)} The defining relations of the $KBc$ subalgebra \eq{KBcId1}-\eq{KBcId2} are assumed  to hold,  but not the auxiliary identities \eq{aux}.  Auxiliary identities give more ways for the equations of motion to be satisfied, and the resulting gauge orbits have not been classified. 
\item{(3)} For regular solutions in the $KBc$ subalgebra, gauge equivalence level-by-level in the $\mathcal{L}^-$ level expansion is both necessary and sufficient to establish true gauge equivalence.
\item{(4)} Solutions which are inequivalent through gauge transformation in the $KBc$ subalgebra are inequivalent in the whole open string star algebra.
\end{description}
It is not known whether relaxing these assumptions can lead to additional gauge orbits representing physically interesting backgrounds. If they exist, they have not been found. A longstanding question is whether the $KBc$ subalgebra contains multiple D-brane solutions, representing two or more copies of the perturbative vacuum. There are interesting candidates for such solutions if we relax assumption (1) \cite{MurataSchnabl}, but the resulting sliver-like singularities lead to inconsistencies in the equations of motion. Dropping assumption~(2) leads to an elaborate spectrum of new gauge orbits, but known solutions of this kind are also singular.  Presently there is no clear reason to doubt assumption (3), and assumption (4) is taken for the sake of discussion.\footnote{If (4) is not valid, some of the gauge orbits in the $KBc$ subalgebra will be identified in the full theory. Clearly the perturbative vacuum and tachyon vacuum will not be identified in the full theory. What might happen is that some of the residual solutions may prove to be gauge equivalent to each other, or to the perturbative vacuum or tachyon vacuum.} The resulting classification is as follows: 
\begin{claim}
\label{claim:gauge_orbits}
Assuming (1)-(4), there are six gauge orbits of solutions in the $KBc$ subalgebra, and they are uniquely characterized by their leading contribution to the $\mathcal{L}^-$ level expansion:
\begin{itemize}
\item Perturbative vacuum: $\Psi = \lambda cKBc + \text{higher levels},\ \ \ \ (\lambda\neq -1)$.
\item Tachyon vacuum: $\Psi = \alpha c +\text{higher levels},\ \ \ \ (\alpha\neq 0)$.
\item Residual perturbative vacuum: $\Psi = -cKBc + \text{higher levels}$.
\item Residual tachyon vacuum: $\Psi = -cK + \text{higher levels}$.
\item Residual conjugate tachyon vacuum: $\Psi = -Kc +\text{higher levels}$,
\item Masuda-Nuomi-Takahashi (MNT) ghost brane \cite{MNT}: $\Psi = -cK-Kc +cKBc + \text{higher levels}$.
\end{itemize}
\end{claim}
\noindent The last four gauge orbits are called {\it residual solutions}. The residual tachyon vacuum was the first solution we found in subsection \ref{subsec:KBc}. Some motivation for the terminology of the gauge orbits comes from the following \cite{IdSing}: 
\begin{claim}
The sum $\varphi_2+\varphi_3$ of coefficients of $cK$ and $Kc$ is a gauge invariant quantity in the $KBc$ subalgebra (even off shell). It can informally be identified with the tension relative to the perturbative vacuum in units of $\frac{1}{2\pi^2}$.
\end{claim}
\noindent For the perturbative vacuum and residual perturbative vacuum, $\varphi_2+\varphi_3$ vanishes and the background can be interpreted as a state of zero energy; for the tachyon vacuum and residual tachyon vacuum solutions,  $\varphi_2+\varphi_3 = -1$ so the energy of the reference D-brane is canceled. For the MNT ghost brane, $\varphi_2+\varphi_3 = -2$, and so the solution appears to represent a D-brane with negative energy. 
\begin{exercise}\label{ex:go} Prove that the following states
\begin{eqnarray}
\Psi \lineup = c\frac{KB}{1-\Omega}c(\Omega-1).\\
\Psi \lineup = -c\frac{(1+K)^2+1}{2+K}Bc+\frac{1}{(1+K)^2}c\frac{(1+K)^2}{2+K} Bc+c\frac{(1+K)^2}{2+K}Bc\frac{1}{(1+K)^2}.\ \ \ \ \ \ \ \ \\
\Psi \lineup = -\frac{1}{\sqrt{1+K}}cK(1+K)Bc\frac{1}{\sqrt{1+K}}.\\
\Psi \lineup = c-\frac{1}{2}\left(c\d c\frac{B}{1+K}+\frac{B}{1+K}c\d c\right)-\frac{1}{4}\frac{B}{1+K}c\d c\frac{B}{1+K}c\d c\frac{B}{1+K}.
\end{eqnarray}
satisfy the equations of motion, and identify their gauge orbits.
\end{exercise}
\noindent We have found several gauge orbits of solutions in $KBc$ subalgebra besides the perturbative vacuum and the tachyon vacuum. However, their physical significance is obscure. This is an important issue since the consistency of string field theory demands that every solution represents a physical string background. We are not free to ignore solutions we do not like. As explained in the next subsection,  residual solutions and other exotic gauge orbits have so far been excluded since they do not contain representatives which are sufficiently regular to allow for unambiguous computation of physical observables. This cannot be readily seen in the $\mathcal{L}^-$ level expansion, but the $\mathcal{L}^-$ level expansion does not reveal all features relevant to the regularity of a string field at finite level.

What it does clearly display, however, is problematic behavior related to the sliver state. This will be manifested through singularities in the  expansion around $K=0$. For example, consider the homotopy operator for $KBc$ solutions in Schnabl gauge:
\begin{equation}A_\lambda = B\frac{1-\lambda\Omega}{K}.\end{equation}
The homotopy operator is well-defined only if $\lambda=1$. This can be seen through the $\mathcal{L}^-$ level expansion:
\begin{equation}A_\lambda = \underbrace{(1-\lambda)\frac{B}{K} }_{\text{level }0}+ \underbrace{\phantom{\Big)}\!\!\!\lambda B}_{\text{level }1} +\text{higher levels}.\label{eq:LmAl}\end{equation}
We immediately see that the leading level is undefined due to the inverse of $K$. The leading level is absent when $\lambda=1$. Problems may also appear with other non-analytic powers of $K$. For example, we can consider a class of solutions whose $\mathcal{L}^-$ level expansion starts as
\begin{equation}\Psi = \underbrace{\phantom{\big)}\!\! cK^{1-\nu} Bc}_{\text{level }-\nu} + \text{higher levels},\ \ \ \ 0\leq \nu\leq 1.\end{equation}
Here at least negative powers of $K$ do not appear. Nevertheless, such solutions are singular. Formally, they imply the existence of a 1-parameter family of stationary points of the action which connect the perturbative vacuum to the tachyon vacuum. But this is in contradiction with the fact that the perturbative vacuum and tachyon vacuum have different energy.

\subsection{Dual $\mathcal{L}^-$ level expansion}
\label{subsec:dualL}

It is interesting to think about what happens when we probe a string field with an identity-like test state
\begin{equation}\eps^{\frac{1}{2}\mathcal{L}^-}|\phi\rangle,\ \ \ \ \eps\text{ small}.\end{equation}
This can be represented as a strip of width $\eps$ containing a vertex operator in the middle creating the state $|\phi\rangle$. Relative to such a test state, a typical string field will look sliver-like. This is not true, however, for the identity string field, since regardless of how small we take the width of the test state, the width of the identity is vanishing, and so is always negligible by comparison. In fact, presently we are not interested in wedge states whose relative width exceeds a certain finite cutoff $\Lambda>0$ as the test state shrinks. For this reason we introduce a projection operator 
\begin{equation}P_\Lambda, \end{equation}
which acts as the identity on wedge states with insertions whose total width is less than $\Lambda$, and acts a zero otherwise. With this projection operator inserted, it may happen that the overlap of an identity-like test state with a string field $X$ can be expanded as a power series in $1/\eps$:
\begin{eqnarray}
\left\langle \phi, P_\Lambda\left(\frac{1}{\eps}\right)^{\frac{1}{2}\mathcal{L}^-} X\right\rangle \lineup = \left(\frac{1}{\eps}\right)^{h_1}\langle \phi,P_\Lambda X_{h_1}\rangle + \left(\frac{1}{\eps}\right)^{h_2}\langle \phi,P_\Lambda X_{h_2}\rangle+\left(\frac{1}{\eps}\right)^{h_3}\langle \phi,P_\Lambda X_{h_3}\rangle+...\nonumber\\
\lineup\ \ \ \ \ \ \ \ \ \ \ \ \ \ \ \ \ \ \ \ \ \ \ \ \ \ \ \ \ \ \ \ \ \ \ \ \ \ \ \ \ \ \ \ \ \ \ \ \ \ \ \ \ \ \ \ \ \ \ \ \ \ \ \   h_1>h_2>h_3> ...\ .\ \ \ \ \ \ \ \ \ \label{eq:dualL}
\end{eqnarray}
Leaving the projection implicit, formally we can write
\begin{equation}X = X_{h_1}+X_{h_2}+X_{h_3} + ...\ \ \ \  h_1>h_2>h_3> ...\ ,\end{equation}
where $X_{h}$ are eigenstates of $\mathcal{L}^-$:
\begin{equation}\frac{1}{2}\mathcal{L}^- X_h =h X_h.\end{equation}
This defines the {\it dual $\mathcal{L}^-$ level expansion} \cite{IdSing}. As before, level is identified with the $\frac{1}{2}\mathcal{L}^-$ eigenvalue. In this case the leading level is the {\it highest} level, and subleading levels are increasingly negative.

The nature of the eigenstates in this expansion is a little surprising, so it is worth seeing how they arise from the field $\frac{1}{1+K}$:
\begin{eqnarray}
\left\langle \phi, P_\Lambda\left(\frac{1}{\eps}\right)^{\frac{1}{2}\mathcal{L}^-} \frac{1}{1+K}\right\rangle \lineup = \left\langle \phi, P_\Lambda\frac{\eps}{\eps+K}\right\rangle \nonumber\\
\lineup = \eps\int_0^\Lambda d\alpha\, e^{-\eps \alpha}\langle\phi,\Omega^\alpha\rangle.
\end{eqnarray}
For fixed $\Lambda$ and $\eps$ sufficiently small, we can expand the integrand in powers of $\eps$:
\begin{equation}
\left\langle \phi, P_\Lambda\left(\frac{1}{\eps}\right)^{\frac{1}{2}\mathcal{L}^-} \frac{1}{1+K}\right\rangle = \underbrace{\eps \left\langle \phi,\int_0^\Lambda d\alpha\, \Omega^\alpha\right\rangle}_{\text{level }-1}- \underbrace{\eps^2\left\langle \phi,\int_0^\Lambda d\alpha\, \alpha\Omega^\alpha\right\rangle}_{\text{level }-2}+\underbrace{\frac{\eps^3}{2!}\left\langle \phi,\int_0^\Lambda d\alpha\, \alpha^2\Omega^\alpha\right\rangle}_{\text{level }-3}+...\ .
\end{equation}
Comparing to \eq{dualL} we see that the leading level in the expansion is $-1$, followed by subleading levels for all negative integers. To see the $\mathcal{L}^-$ eigenstates we make the identification
\begin{equation}\frac{1}{\Gamma(-\nu)}\int_0^\Lambda d\alpha\,\alpha^{-\nu-1}\Omega^\alpha = P_\Lambda  K^{\nu},\end{equation}
where the power of $K$ is defined as a continuous superposition of wedge states through \eq{Knu}. If $\nu$ is negative the integral defining $K^\nu$ produces a divergence proportional to the sliver state. In the present situation this does not matter, since the sliver divergence is discarded by the projection. Therefore in the present context we can assume that all powers of $K$ are allowed, and all are eigenstates of $\mathcal{L}^-$:
\begin{equation}\frac{1}{2}\mathcal{L}^- K^\nu = \nu K^\nu.\end{equation}
This leads us to identify the dual $\mathcal{L}^-$ expansion of $\frac{1}{1+K}$ as 
\begin{equation}\frac{1}{1+K} = \underbrace{\frac{1}{K}}_{\text{level }-1}-\underbrace{\frac{1}{K^2}}_{\text{level }-2} +\underbrace{\frac{1}{K^3}}_{\text{level }-3} - \,\text{lower levels}.\end{equation}
This is simply an expansion around $K=\infty$. This anticipates the general result:
\begin{claim}
The dual $\mathcal{L}^-$ level expansion in the subalgebra of wedge states with insertions can be computed by expanding the state in powers of $K$ around $K=\infty$ and ordering terms in sequence of decreasing $\frac{1}{2}\mathcal{L}^-$ eigenvalue.
\end{claim}
\begin{claim}
In the subalgebra of wedge states with insertions, the dual $\mathcal{L}^-$ level expansion of a product of states is given by multiplying the dual $\mathcal{L}^-$ level expansions of the states individually. Level is additive under star multiplication.
\end{claim}
\noindent We make two comments:
\begin{description}
\item{(1)} It is not necessarily true that a string field in the subalgebra of wedge states with insertions will have an expansion in powers of $K$ around $K=\infty$. For example, this is not possible for the $SL(2,\mathbb{R})$ vacuum $\Omega=e^{-K}$. By definition, a state which falls off faster than any inverse power of $K$ is considered to have level $-\infty$. It can also happen that expansion around $K=\infty$ produces contributions which are bounded by finite powers of $K$ but are not themselves powers of $K$. For example we could find $1/\ln(K)$. In this case the state does not have a dual $\mathcal{L}^-$ expansion. This is uncommon in practice, and many aspects of the present discussion generalize in such situations. 
\item{(2)} We did not qualify these claims as holding within a singularity free subalgebra. This is because an expansion around $K=\infty$ never produces OPE divergence. Consider for example the state
\begin{equation}V\frac{1}{(1+K)^3}V = \frac{1}{2!}\int_0^\infty d\alpha\, \alpha^2e^{-\alpha} V\Omega^\alpha V,\end{equation}
where $V$ is a weight 1 primary with double pole OPE. The double pole does not lead to divergence in the state since the $\alpha^2$ factor in the integrand cancels the $1/\alpha^2$ from the contraction of $V$s in the limit $\alpha\to 0$. The dual $\mathcal{L}^-$ expansion of this state is
\begin{eqnarray}
V\frac{1}{(1+K)^3}V \lineup =\, \underbrace{V\frac{1}{K^3}V}_{\text{level }-1} \, -\,  \underbrace{3 V\frac{1}{K^4}V}_{\text{level }-2}\, + \,\underbrace{6 V\frac{1}{K^5}V}_{\text{level }-3}\, +\, \text{lower levels}\nonumber\\
\lineup = \frac{1}{2!}\int_0^\Lambda d\alpha \, \alpha^2 V\Omega^\alpha V - \frac{3}{3!}\int_0^\Lambda d\alpha \alpha^3 V\Omega^\alpha V + \frac{6}{4!}\int_0^\Lambda d\alpha\, \alpha^4 V\Omega^\alpha V +...\ .\ \ \ \ \ \ \ \ \ 
\end{eqnarray}
In all eigenstates the double pole in the $V$-$V$ OPE is canceled in the integrand towards $\alpha=0$, and more than canceled as the level becomes progressively negative.
\item{(3)} There is a close connection between the level in the dual $\mathcal{L}^-$ level expansion and the magnitude of the contribution of the identity string field. Increasingly positive levels receive stronger contribution from states close to the identity, while increasingly negative levels receive weaker contribution.
\end{description}
The dual $\mathcal{L}^-$ is in a sense complementary to the (ordinary) $\mathcal{L}^-$ level expansion, since it reveals behavior related to the identity string field but obscures behavior related to the sliver state. In the $\mathcal{L}^-$ level expansion the situation is opposite. For example, the inverse of the $SL(2,\mathbb{R})$ vacuum is well-defined in the $\mathcal{L}^-$ level expansion, but in the dual $\mathcal{L}^-$ level expansion it is singular since it diverges at $K=\infty$ faster than any power of $K$. By contrast, the homotopy operator for $KBc$ solutions in Schnabl gauge has an acceptable dual $\mathcal{L}^-$ level expansion
\begin{equation}
A_\lambda = \underbrace{\frac{B}{K}}_{\text{level }0}-\underbrace{\lambda \frac{B\Omega}{K}}_{\text{level }-\infty}.
\end{equation}
The difficulty for $\lambda\neq 1$ is not related to the identity string field but to the sliver state, as is manifest when written in the $\mathcal{L}^-$ level expansion as \eq{LmAl}. Together the $\mathcal{L}^-$ and dual $\mathcal{L}^-$ level expansions give a lot of information about the regularity of a string field. But this is not necessarily sufficient. Consider for example
\begin{eqnarray}
\frac{1}{1-K} \lineup =- \underbrace{\frac{1}{K}}_{\text{\ level -1}} -\underbrace{ \frac{1}{K^2}}_{\text{level }-2} +\, \text{lower levels},\ \ \ \ (\text{dual }\mathcal{L}^-\text{ level expansion}),\nonumber\\
\lineup = \underbrace{1}_{\text{level }0} + \underbrace{K}_{\text{level } 1}\, +\, \text{higher levels},\ \ \ \ (\mathcal{L}^-\ \text{level expansion}).
\end{eqnarray}
The $\mathcal{L}^-$ and dual $\mathcal{L}^-$ level expansions are acceptable, but nevertheless this state is singular due to the pole at $K=1$. The problem in this state is not related to either the identity string field or the sliver state. 

We now state a central result of the formalism:
\begin{claim}
Let $X$ be a string field which admits a dual $\mathcal{L}^-$ level expansion. Then $\Tr[X]$ is well-defined only if the leading level of $X$ in the dual $\mathcal{L}^-$ level expansion is strictly negative.
\end{claim}
\noindent The argument is that integration over the circumference of the cylinder in correlators must be absolutely convergent as the circumference shrinks to zero.  For further explanation see \cite{IdSing}. This statement has an important corollary: 
\begin{claim}
Let $\Psi$ be a regular solution that admits a dual $\mathcal{L}^-$ level expansion. Then the leading level of $\Psi$ in the dual $\mathcal{L}^-$ level expansion must be strictly negative. 
\end{claim}
\noindent This follows from the assumption that a regular solution should have well-defined action and Ellwood invariant. Schnabl's solution is regular by this criterion since the only contribution to the dual $\mathcal{L}^-$ level expansion occurs at level~$-\infty$. The simple tachyon vacuum and intertwining solution are also regular by this criterion, but by a narrow margin: 
\begin{eqnarray}
\Psi_\text{simp}\lineup =\underbrace{ Q\left(\frac{1}{\sqrt{K}}Bc\frac{1}{\sqrt{K}}\right)}_{\text{level }-1}+\,\text{lower levels},\nonumber\\
\Psi_*\lineup = \underbrace{Q\left(\frac{1}{\sqrt{K}}Bc\frac{1}{\sqrt{K}} - \frac{1}{\sqrt{K}}\sigma\frac{B}{K}\sigmabar Kc\frac{1}{\sqrt{K}}\right)}_{\text{level }-1}+\,\text{lower levels}.\label{eq:dualLmsimpKOS}
\end{eqnarray}
Both solutions have leading level $-1$, which is the highest integer level consistent with a regular solution. Going a step further, we may consider the identity-like tachyon vacuum \eq{Idtv} and the solution from exercise \ref{ex:go}:
\begin{eqnarray}
\Psi \lineup = c(1-K) =  \underbrace{\phantom{)}\!\!\!-cK}_{\text{level }0} + \underbrace{\phantom{)}\!\!\!c}_{\text{level }-1},\\
\Psi \lineup = c\frac{KB}{1-\Omega}c(\Omega-1) = \underbrace{-cKBc}_{\text{level }0} \, + \ \underbrace{c\frac{KB}{1-\Omega}[c,\Omega]}_{\text{level }-\infty}.
\end{eqnarray}
The leading level is $0$ in both cases, so neither solution is regular.   In \cite{IdSing} it was shown that all residual solutions satisfy
\begin{equation}B\Psi B = -BK.\end{equation}
Since $BK$ has $\frac{1}{2}\mathcal{L}^-$ eigenvalue $+2$, it follows that all residual solutions must contain a level $0$ state in the dual $\mathcal{L}^-$ level expansion. Therefore residual solutions are singular. 

We mention another useful fact:
\begin{claim}
Let $\Psi$ be a regular solution that admits a dual $\mathcal{L}^-$ expansion. Then all finite levels in the dual $\mathcal{L}^-$ expansion of $\Psi$ can be removed by gauge transformation.
\end{claim}
\noindent Since the leading level of a regular solution is negative, the equations of motion imply that it must be BRST closed. Moreover, BRST closed states are always BRST exact in the dual $\mathcal{L}^-$ level expansion, since we have a homotopy operator 
\begin{equation}Q\left(\frac{B}{K}\right) = 1.\end{equation}
This implies that the highest level state can be removed by gauge transformation, and by iteration all finite levels can be removed. Therefore, the dual $\mathcal{L}^-$ level expansion does not reveal gauge invariant information. In this it contrasts with the ordinary $\mathcal{L}^-$ level expansion. All physical information about a regular solution can be transferred to states at level $-\infty$. 

There are many ways to remove levels in the dual $\mathcal{L}^-$ level expansion, but two approaches are notable. The first is using the so-called {\it Zeze map} \cite{Zeze}. Given a solution $\Psi$ and an element of the wedge algebra $F(K)$ satisfying $F(0)=1$, we can form a new solution $\Psi'$ with the transformation
\begin{equation}
\Psi' =\sqrt{F}\Psi \frac{1}{1+B\frac{1-F}{K}\Psi}\sqrt{F}.
\end{equation}
If the leading levels of $\Psi$ and $F$ are both negative, the leading level of $\Psi'$ is the sum of the leading levels of $\Psi$ and $F$. One might notice a close resemblance to the Schnabl gauge solution for marginal deformations. In fact, the Schnabl gauge marginal solution is obtained by applying the Zeze map with $F = \Omega$ to
\begin{equation}\Psi = cV.\label{eq:IdMarg}\end{equation}
Surprisingly, this is a solution by itself. It is separately BRST invariant and multiplies with itself to give zero (assuming regular OPE). However, it is singular in the dual $\mathcal{L}^-$ level expansion. Still, it is of interest due to its close relation to the sigma model understanding of marginal deformations as boundary deformations of the worldsheet action. This can be seen from the nature of perturbation theory around the solution. For example, in Siegel gauge the propagator is given by 
\begin{equation}
\frac{b_0}{[Q_{cV},b_0]} = b_0\int_0^\infty dt\, e^{-t( L_0 + V(1) + V(-1))},
\end{equation}
where $V(1)$ is inserted at $\xi=1$ on the unit disk and $V(-1)$ is inserted at $\xi=-1$. Following exercise \ref{ex:Omstar}, it is clear that the integrand represents a Siegel gauge propagator strip containing exponentials of line integrals of $V$ on the open string boundaries. These can be understood to arise from a boundary deformation of the worldsheet action. It is unclear if some generalization of this solution for singular OPEs (perhaps defined in a singular limit) would give renormalized exponential insertions of line integrals of $V$. The identity-like marginal solution satisfies the gauge condition
\begin{equation}\mathcal{B}^-\Psi = 0.\end{equation}
There is no finite $KBc$ solution for the tachyon vacuum in this gauge, though approaching it as a limit leads formally to $\infty\times c$. This also has a nice interpretation from the sigma model point of view. It corresponds to turning on a relevant boundary coupling to the identity operator and running the RG flow to the infrared fixed point, where the coupling is divergent and all BCFT correlators are multiplied by a vanishing normalization. 

\begin{exercise} Show that Schnabl's solution arises from applying the Zeze map to the state $\varphi_1 c$ with $F=\Omega$ and taking the limit $\varphi_1\to\infty$.\
\end{exercise}

\begin{exercise}
In claim \ref{claim:gauge_orbits} there are five gauge orbits that start at level 0 in the $\mathcal{L}^-$ level expansion. The leading level in this case is by itself a solution to the equations of motion. Apply the Zeze map to the leading level and show that the result is regular in the dual $\mathcal{L}^-$ level expansion only for the pure gauge solutions. 
\end{exercise}

Another method for removing levels in the dual $\mathcal{L}^-$ expansion is through reparameterization symmetries of the spectrum of $K$. These symmetries can be realized through certain automorphisms of the $KBc$ subalgebra \cite{Erler_simple,genericF(K)}. We discuss this under the assumption that the spectrum consists of nonnegative real numbers; if it does not, than some automorphisms should more properly be called endomorphisms. A diffeomorphism of the spectrum is defined by a real, continuous, and monotonically increasing function $f(K)$ satisfying $f(0)=0$ and $f(+\infty)=+\infty$. Under such a diffeomorphism, we postulate that $K,B,c$ transform as 
\begin{eqnarray} 
K' \lineup = f\circ K = f(K),\phantom{\bigg)}\\
B'\lineup  = f\circ B = B\frac{f(K)}{K},\\
c'\lineup= \, f\circ c \, = c\frac{KB}{f(K)}c.
 \end{eqnarray}
The transformed fields satisfy the same relations as $K,B$ and $c$ (excepting auxiliary identities). If we wish to transform from a $KBc$ solution whose leading level in the dual $\mathcal{L}^-$ level expansion is $\nu<0$ to another $KBc$ solution whose leading level is $\nu'<0$, the leading level of the diffeomorphism $f(K)$ must be
\begin{equation}\frac{\nu'}{\nu}>0.\end{equation}
For example, we can write any Okawa-type tachyon vacuum solution
\begin{equation}\Psi = \sqrt{F}c\frac{KB}{1-F}c\sqrt{F}\end{equation}
with monotonically decreasing $F(K)$ as the simple tachyon vacuum
\begin{equation}
\Psi = f\circ\left(\frac{1}{\sqrt{1+K}}c(1+K)Bc\frac{1}{\sqrt{1+K}}\right),
\end{equation}
with the diffeomorphism
\begin{equation}f(K) = 1-\frac{1}{F(K)}.\end{equation}
The simple tachyon vacuum can itself be written as a transformation of the identity-like tachyon vacuum:
\begin{equation}\Psi_\text{simp} = f\circ\Big(c(1-K)\Big),\end{equation}
where
\begin{equation}f(K) = \frac{K}{1+K}.\end{equation}
However this is not a diffeomorphism of the spectrum. It maps the spectrum into the interval $[0,1]$. This is related to the fact that the identity-like tachyon vacuum is singular in the dual $\mathcal{L}^-$ level expansion. This automorphism structure appears to be specific to the $KBc$ subalgebra, though it can be extended in an interesting way to include boundary condition changing operators \cite{genericF(K)}.

With these tools we can gain insight into certain topological properties of string field theory solutions. Such considerations are likely to be especially important in the context of open superstring field theory, where they are relevant for understanding the origin of D-brane charges. However, purely in the context of open bosonic SFT there is an interesting fact which involves related considerations: that the sign of the tachyon coefficient at the tachyon vacuum is positive. Here we concentrate on a  limited version of this statement:
\begin{claim}
Given a real and nonsingular tachyon vacuum solution in the $KBc$ subalgebra, the coefficient of the tachyon state in the $\mathcal{L}^-$ level expansion is positive. 
\end{claim}
\noindent To demonstrate this, we observe that Okawa-type tachyon vacuum solutions \eq{Oktype} satisfy 
\begin{equation}B\Psi B = B\frac{KF}{1-F}.\label{eq:BPsiB}\end{equation}
This relation in fact holds for arbitrary tachyon vacuum solutions in the $KBc$ subalgebra \cite{Jokel}. The $\mathcal{L}^-$ level expansion of $\Psi$ takes the form
\begin{equation}\Psi = \varphi_1 c+ \text{higher levels},\end{equation}
where $\varphi_1$ is the coefficient of the tachyon field in this basis. Plugging this into \eq{BPsiB} and expanding the right hand side around $K=0$ we learn that 
\begin{equation}\varphi_1 = \frac{1}{F'(0)}.\end{equation}
Meanwhile, we know that the leading level of $\Psi$ in the dual $\mathcal{L}^-$ expansion must be negative. This implies that the leading level of the right hand side of \eq{BPsiB} must be less than $2$. This is only possible if $F(K)$ vanishes at infinity:
\begin{equation}F(\infty) = 0.\end{equation}
Finally, consider the state 
\begin{equation}H(K) = \frac{1-F}{K},\end{equation}
which appears in the homotopy operator for the tachyon vacuum. From the behavior of $F$ we know that $H$ satisfies the boundary conditions \ \ \ \ \ \ \ \ \ \ \ \ \ \ \ \ \ \ \  \ \ \ \ \ \ \ \ \ \ \ \ \ \ \ \ \ \ \ \ \ \ \ \ \ \ \ \ \ \ \ \ 
\begin{equation}H(0) = \frac{1}{\varphi_1},\ \ \ \ \lim_{K\to\infty}KH(K) = 1.\end{equation}
Now suppose that $\varphi_1$ is negative. Since the homotopy operator should be well-defined, $H(K)$ must be continuous for non-negative $K$. Then the above boundary conditions imply that $H(K)$ must 
\begin{wrapfigure}{l}{.34\linewidth}
\centering
\resizebox{2.4in}{1.5in}{\includegraphics{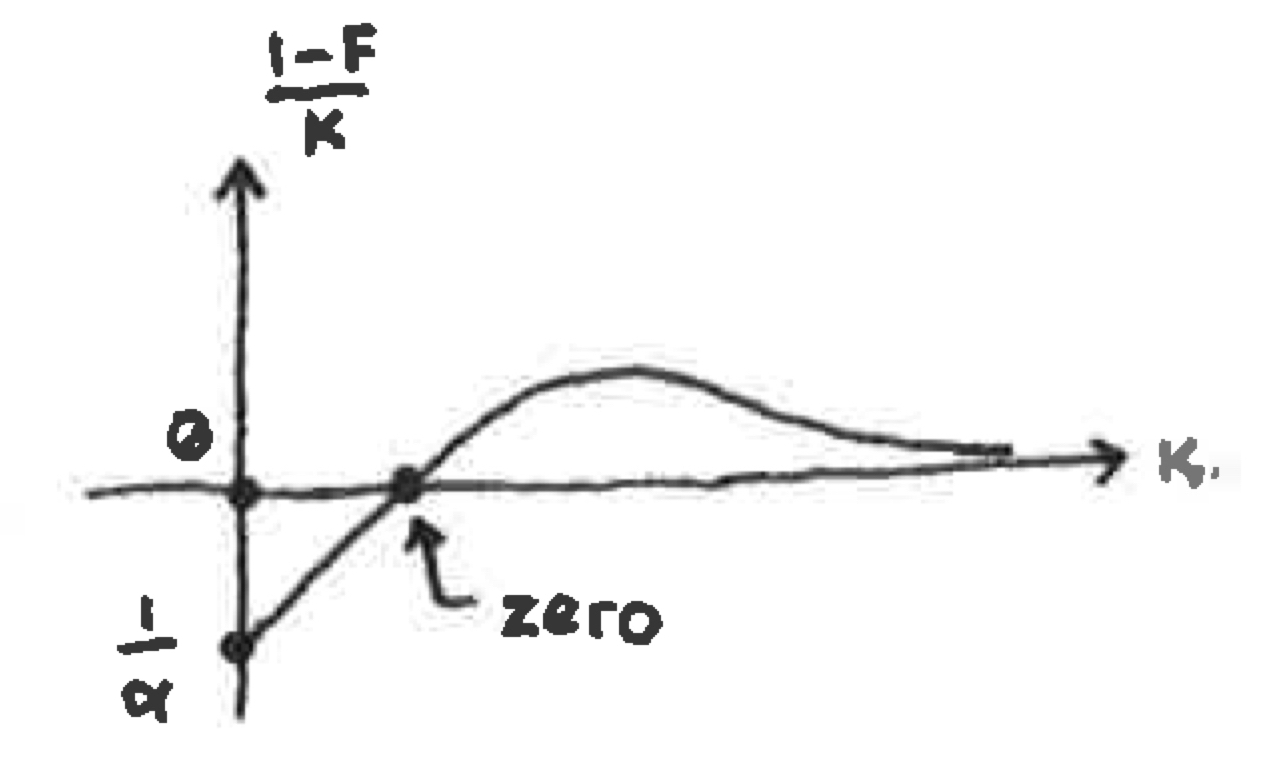}}
\end{wrapfigure}
have at least one zero for positive $K$. This is the ``topological" part of the argument. On the other hand, the right hand side of \eq{BPsiB} contains the state
\begin{equation}\frac{F(K)}{H(K)}.\end{equation}
This must be discontinuous at a pole corresponding to the zero of $H(K)$, and is therefore singular. Since multiplication by $B$ cannot turn a regular solution into a singular state, we conclude that $\varphi_1$ must be positive.

\subsection{Singular gauge transformations}
\label{subsec:singularGT}

Schnabl's phantom term is one of the most striking mysteries in the study of analytic solutions. It strictly vanishes as a state in the Fock space expansion, but at the same time provides the main distinction between the perturbative vacuum and the tachyon vacuum. This is part of a more general phenomenon. Formally, every classical solution in open SFT can be expressed as a finite gauge transformation of the tachyon vacuum
\begin{equation}\Psi = \frac{1}{1+A(\Psi-\Psi_\text{tv})}\Big(Q+\Psi_\text{tv}\Big)\Big(1+A(\Psi-\Psi_\text{tv})\Big),\end{equation}
and therefore all solutions appear to be gauge transformations of each other. Apparently this is not correct, and something like a phantom term must distinguish between physically inequivalent solutions. Here we describe a general picture of how this works \cite{Ellsingular,EMsingular,EMphantom}. This topic has a reputation for being difficult to make precise. We attempt a slightly original presentation which aims to clarify some points. 

Consider the equation 
\begin{equation}Q_{\Psi_1\Psi_2}U = 0,\ \ \ \ \mathrm{gh}(U)=0,\end{equation}
where $\Psi_1$ and $\Psi_2$ are classical solutions and $Q_{\Psi_1\Psi_2}$ is the shifted kinetic operator for a stretched string connecting $\Psi_1$ and $\Psi_2$, as described in subsection \ref{subsec:KOS}. A solution to this equation (at ghost number zero) will be called a {\it morphism} from $\Psi_1$ to $\Psi_2$. $\Psi_1$ and $\Psi_2$ may be called the {\it source} and {\it target} solutions of the morphism. As the terminology suggests, morphisms define a category. The objects of the category are classical solutions, and composition of morphisms is defined with the open string star product. To make the category structure clear, we note the following:
\begin{itemize}
\item  Let $U_{12}$ be a morphism from $\Psi_1$ to $\Psi_2$ and $U_{23}$ be a morphism from $\Psi_2$ to $\Psi_3$. Then the Leibniz rule \eq{ShiftLeibniz} implies that the product $U_{12}U_{23}$ is a morphism from $\Psi_1$ to~$\Psi_3$:
\begin{eqnarray}
Q_{\Psi_1\Psi_3}\big(U_{12}U_{23}\big)\lineup = \big(Q_{\Psi_1\Psi_2}U_{12}\big)U_{23}+U_{12}\big(Q_{\Psi_2\Psi_3}U_{23}\big) = 0 .
\end{eqnarray}
Composition of morphisms is associative since the star product is associative.
\item Between any source and target solution there is a zero morphism $U=0$.
\item Between any solution and itself there is an identity morphism $U=1$.
\item A special kind of morphism is an {\it isomorphism}; this is a morphism $U$ from $\Psi_1$ to $\Psi_2$ which has an inverse morphism $U^{-1}$ from $\Psi_2$ to $\Psi_1$. In particular $UU^{-1} = U^{-1}U = 1$. In fact, an isomorphism is the same thing as a {\it gauge transformation} between $\Psi_1$ and $\Psi_2$. In particular, if $U$ has an inverse the following relations are equivalent: 
\begin{equation}Q_{\Psi_1\Psi_2}U =0 \ \ \ \longleftrightarrow \ \ \ \Psi_2 = U^{-1}(Q +\Psi_1)U.
\end{equation}
\end{itemize}
Therefore, the gauge group of open SFT naturally extends to a category which connects all classical solutions. This is a particular instance of the general concept of ``D-brane categories" which appears in various guises, especially in discussions of open topological strings. In the most basic form, a D-brane category consists of D-branes as objects, and open strings connecting D-branes as morphisms.

What makes the category structure interesting is that morphisms connect solutions which are not physically equivalent. Given any source and target solutions we can find a nonzero morphism of the form 
\begin{equation}U = Q_{\Psi_1\Psi_2}b,\ \ \ \ \mathrm{gh}(b) = -1.\end{equation}
This is called an {\it exact morphism}. If the source and target solutions represent fully distinct open string boundary conditions, any morphism connecting them is expected to be exact. This is because the cohomology of $Q_{\Psi_1\Psi_2}$  is expected to be isomorphic to the cohomology of the BRST operator of a string connecting the boundary conditions represented by $\Psi_1$ and $\Psi_2$. If the boundary conditions are distinct, the BRST operator has no cohomology at ghost number zero. Exact morphisms can also connect solutions representing the same background, but typically they are not gauge transformations. The exception is the following circumstance: 
\begin{claim}\label{claim:10}
A gauge transformation is an exact morphism if and only if it connects two solutions for the tachyon vacuum. 
\end{claim}
\noindent For the ``if" part of this statement, a gauge transformation between tachyon vacuum solutions $\Psi_1$ and $\Psi_2$ will satisfy $Q_{\Psi_1\Psi_2} U=0$. Moreover, $\Psi_1$ will have a homotopy operator satisfying $Q_{\Psi_1}A = 1$. Then we can write
\begin{equation}U = Q_{\Psi_1\Psi_2}(AU),\end{equation}
and the morphism is exact. For the ``only if" part, suppose $U$ is exact and possesses an inverse. Using the Leibniz rule \eq{ShiftLeibniz} this implies
\begin{eqnarray}
1\lineup =U^{-1}U = U^{-1}Q_{\Psi_1\Psi_2}b = Q_{\Psi_2}\big(U^{-1}b\big), \\
1\lineup = UU^{-1} = \big(Q_{\Psi_1\Psi_2}b\big)U^{-1}= Q_{\Psi_1}\big(bU^{-1}\big).
\end{eqnarray}
Therefore the identity is trivial in the cohomology around $\Psi_1$ and $\Psi_2$, so both solutions must represent the tachyon vacuum. 

We introduce another class of morphisms called {\it resolvable}. Resolvable morphisms can be thought of as ``infinitesimally close" to gauge transformations. This can be articulated as follows. Every morphism $U$ has a spectrum, consisting of all (complex) numbers $u$ with the property that 
\begin{equation}
U-u
\end{equation}
has no star algebra inverse. 
\begin{exercise}
Find the spectrum of Okawa's gauge transformation \eq{Okform},
\begin{equation}U = 1-\lambda\sqrt{\Omega}cB\sqrt{\Omega},\end{equation}
as a function of $\lambda$ under the assumption that wedge algebra factors should have finite $C^*$ norm. Also determine the spectrum assuming finite $D_2$ norm.
\end{exercise}
\noindent A {\it resolvable morphism} $U$ has the property that every neighborhood of the origin $u=0$ has a point which is outside of the spectrum of $U$. This means that the string field 
\begin{equation}\frac{1}{U+\eps}\end{equation}
exists for some $\eps$ in any (arbitrarily small) vicinity of the origin. For convenience we introduce $\eps$ so that $-\eps$ is outside the spectrum of $U$. Gauge transformations are resolvable morphisms since $\eps=0$ is outside of the spectrum. The more interesting case is when $\eps$ can be arbitrarily small but not zero. In this case the resolvable morphism is not a gauge transformation, but is ``infinitesimally close" to one. Exact morphisms are sometimes resolvable morphisms. An example is the string field $K$:
\begin{equation}U = K = QB,\ \ \ \ QU= 0.\end{equation}
This can be viewed as an exact morphism connecting the perturbative vacuum to itself. From the $C^*$ point of view the spectrum consists of nonnegative real numbers, and from the $D_2$ point of view the spectrum consists of numbers with nonnegative real part. In either case every neighborhood of $K=0$ leaves the spectrum, so $K$ is an exact and resolvable morphism. The string field 
\begin{wrapfigure}{l}{.37\linewidth}
\centering
\resizebox{2.5in}{2in}{\includegraphics{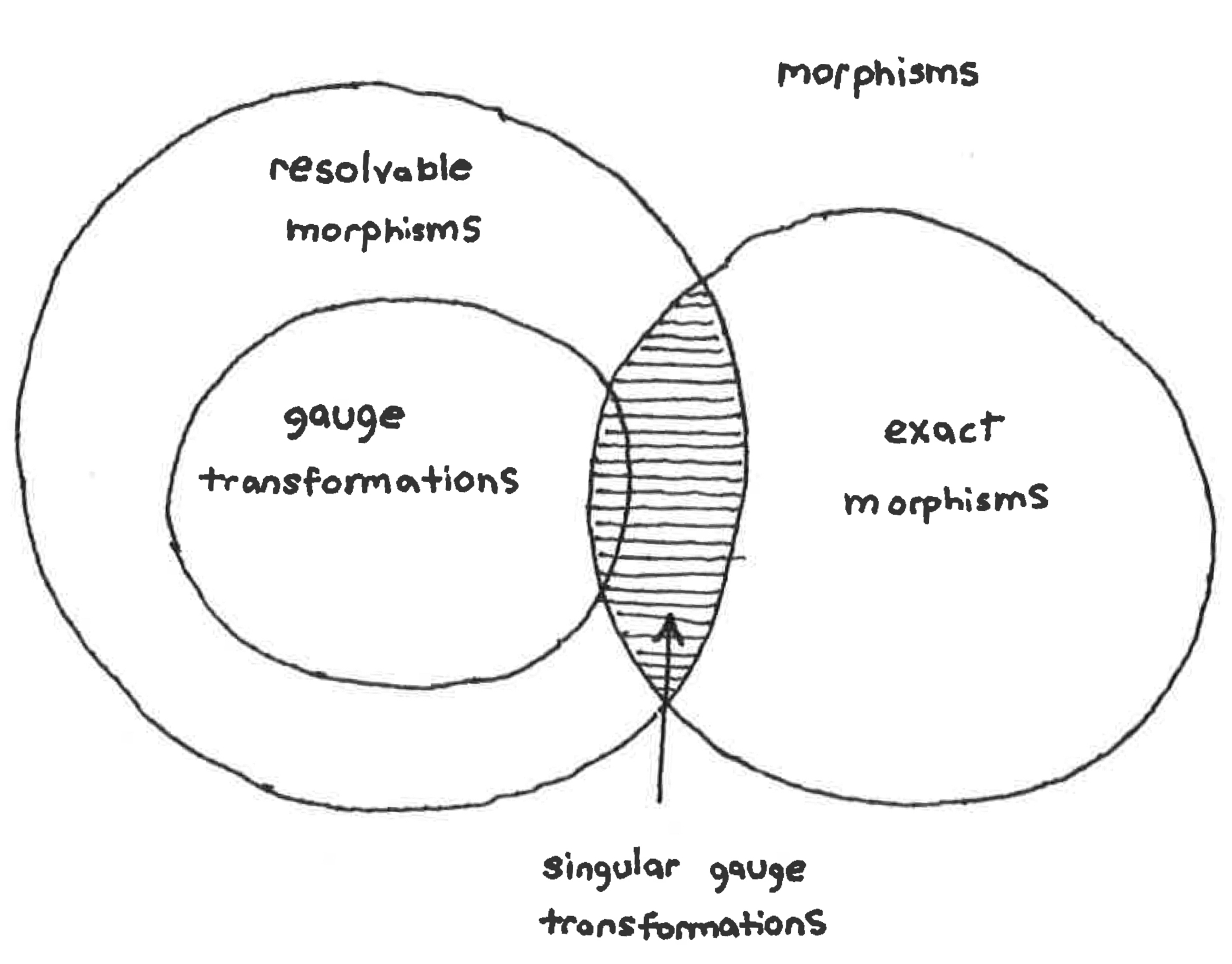}}
\vspace{.5cm}
\end{wrapfigure}
\begin{equation}\frac{1}{\eps+ K} = \int_0^\infty d\alpha e^{-\eps \alpha} \Omega^\alpha.\end{equation}
exists for any $\eps$ with positive real part, no matter how small, though it is divergent in  
the $\eps\to 0$ limit. A morphism which is exact and resolvable will be called a {\it singular gauge transformation}. By claim \ref{claim:10}, an ordinary gauge transformation is a singular gauge transformation if and only if it connects two solutions for the tachyon vacuum. Depending on whether a morphism is a gauge transformation, is exact, or is resolvable it can belong to one of six classes, as illustrated to the left. 

\begin{exercise}
Find a representative example of a morphism for all six classes.
\end{exercise}

Singular gauge transformations are interesting since they can connect physically distinct solutions, but are infinitesimally close to gauge transformations. If $U$ is a singular gauge transformation from $\Psi_1$ to $\Psi_2$, we can {\it almost} relate the solutions by gauge transformation through the identity 
\begin{equation}\Psi_2 = \frac{1}{\eps+U}(Q+\Psi_1)(\eps+U)\ + \ \frac{\eps}{\eps+U}(\Psi_2- \Psi_1).\label{eq:phantom}\end{equation}
The first term is a gauge transformation of $\Psi_1$ for any $\eps$, no matter how small, assuming that $-\eps$ is outside the spectrum of $U$. The second term looks like it will vanish in the $\eps\to 0$ limit. But if $\Psi_1$ and $\Psi_2$ are not gauge equivalent, the second term must be nontrivial. This is a generalization of the {\it phantom term} of Schnabl's solution. 

A key aspect of understanding the phantom term is determining the limit\footnote{The limit is understood to be directed along a path the complex plane which avoids the spectrum of $U$.}
\begin{equation}
\delta(U) = \lim_{\eps\to 0} \frac{\eps}{\eps+U}.\label{eq:deltaU}
\end{equation}
This limit is understood in the sense of distributions---that is, as defining a linear functional on a class of well-behaved test states (for example Fock space states). If the limit exists, it defines the {\it boundary condition changing projector} associated to $U$. Typically, the boundary condition changing projector will be a rank 1 projector, like the sliver state. Therefore the limit $\eps\to 0$ will not exist in the open string star algebra. In fact, the boundary condition changing projector for $U=K$ is precisely the sliver state:
\begin{equation}
\delta(K) = \lim_{\eps\to 0} \frac{\eps}{\eps+K} = \Omega^\infty.
\end{equation}
Formally we should have the relation 
\begin{equation}\delta(U)*\delta(U) = \delta(U),\label{eq:deltaU2}\end{equation}
but the star product is not really well defined. Nevertheless, the boundary condition changing projector can be understood as a projector in the following sense. The ratio $\frac{\eps}{\eps+U}$ defines a function on the spectrum of $U$. The pointwise limit of this function as $\eps\to 0$ vanishes everywhere on the spectrum except at $U=0$, where its value is $1$. Therefore the boundary condition projector can be viewed as a projector onto the kernel of $U$. The notation $\delta(U)$ is meant to refer to a Kronecker delta, rather than a Dirac delta. Note that the limit \eq{deltaU} is not the only way to derive a projector onto the kernel of $U$. The limit
\begin{equation}\delta(U) = \lim_{\alpha\to\infty}(1-U)^\alpha\label{eq:deltaUproj}\end{equation}
should compute the same thing if the spectrum of $1-U$ is bounded in absolute value by $1$. In this presentation the projector property \eq{deltaU2} is formally manifest. The limits \eq{deltaU} and \eq{deltaUproj} agree in known cases where both are applicable.\footnote{The boundary condition changing projector may be well-defined for exact morphisms which are not resolvable. However, it cannot be computed through the limits \eq{deltaU} and \eq{deltaUproj} which assume that every neighborhood of the origin leaves the spectrum.}

The boundary condition changing projector has been computed in several of examples, and an interesting pattern arises. To describe it, consider a singular gauge transformation $U$ connecting solutions $\Psi_1$ and $\Psi_2$ representing backgrounds $\BCFT_1$ and $\BCFT_2$. We assume the string field theory is formulated in $\BCFT_0$. Typically, the boundary condition changing projector takes the form
\begin{equation}\delta(U) = \lim_{\alpha\to\infty}\Big(\Sigma_{02}\,S^{\alpha}\,\Sigma_{21}\, S^\alpha\,\Sigma_{10}\Big).\label{eq:PU}\end{equation}
The string fields on the right hand side are given as follows:
\begin{itemize}
\item $S^\alpha$ is a surface state $S$ multiplied with itself $\alpha$ times, where $\alpha$ is a regulating parameter for the purpose of making the star products on the right hand side unambiguous. The boundary condition changing projector is therefore mainly characterized by the surface state projector $S^\infty$. The solutions discussed in this review are formulated in the sliver frame, where $S^\infty$ will be the sliver state. We may also obtain other rank 1 projectors though midpoint-preserving reparameterizations. 
\item $\Sigma_{02}$ lives in the state space of a stretched string connecting the perturbative vacuum and the background $\BCFT_2$. 
\item$\Sigma_{21}$ lives in the state space of a stretched string connecting the backgrounds $\BCFT_2$ and $\BCFT_1$. 
\item Finally, $\Sigma_{10}$ lives in the state space of a stretched string connecting $\BCFT_1$ back to the perturbative vacuum. 
\end{itemize}

\noindent Probing with a test state, the right hand side of \eq{PU} can be computed as a correlation function on the unit disk. In the limit $\alpha\to\infty$ we will generally obtain a rank 1 projector, where the local coordinate patch touches the boundary of the unit disk at the midpoint. Meanwhile, the state
\begin{wrapfigure}{l}{.35\linewidth}
\vspace{-.5cm}
\centering
\resizebox{2.4in}{2in}{\includegraphics{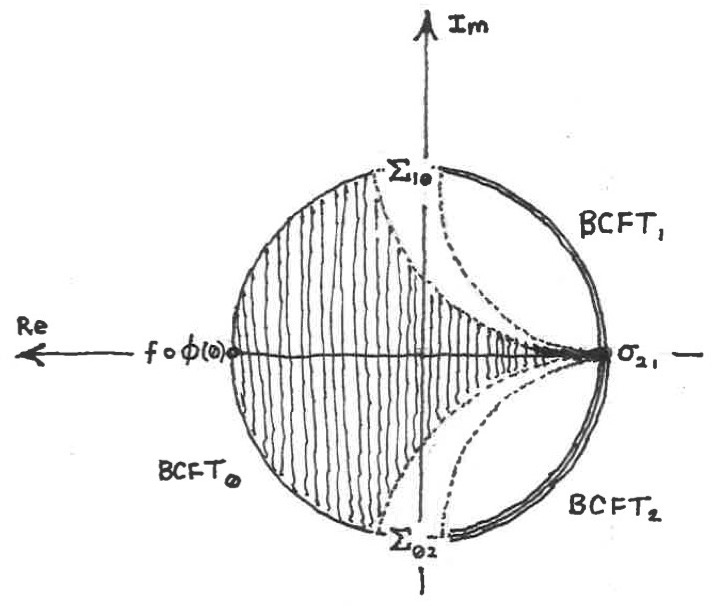}}
\vspace{-1cm}
\end{wrapfigure} 
$\Sigma_{21}$ is squeezed between a pair of $S^\alpha$s until it degenerates to a boundary condition changing operator between $\BCFT_2$ and $\BCFT_1$ inserted at the open string midpoint. This is the origin of the term ``boundary condition changing projector."\footnote{For a singular gauge transformation from a solution to itself, there will be no change of boundary condition at the midpoint.  In this case, the boundary condition changing projector is sometimes called the {\it characteristic projector} \cite{Ellsingular}.} In the end we obtain a Schr{\"o}dinger functional which is factorized between the left and right halves of the string. The left half functional carries $\BCFT_2$ boundary conditions near the midpoint, while the right half string functional carries $\BCFT_1$ boundary conditions near the midpoint. In this way, the boundary condition changing projector is able to reconstruct the boundary conformal field theories ``hidden" inside the classical solutions $\Psi_1$ and $\Psi_2$. There is no proof that this must always happen under reasonable conditions. But the expectation can be motivated through the {\it BRST identity}
\begin{equation}
Q\left(\frac{\eps}{\eps+U}\right) \ =\  -\Psi_2\frac{\eps}{\eps+U}\ + \ \frac{\eps}{\eps+U}(\Psi_2-\Psi_1)\frac{\eps}{\eps+U} \ + \ \frac{\eps}{\eps+U}\Psi_1.\label{eq:BRSTid}
\end{equation}
This relation is exact for all $\eps>0$, and we can consider the limit $\eps\to 0$. Right hand side has three terms. The first term can be seen to arise from the BRST variation of $\Sigma_{02}$; the second term can be seen to arise from BRST variation of the boundary condition changing operator at the midpoint; and finally the last term arises from BRST variation of $\Sigma_{10}$.

Let us confirm this picture in a few ``trivial" examples:
\begin{itemize}
\item The zero morphism $U=0$ is a singular gauge transformation. The boundary condition changing projector is
\begin{equation}\delta(0) = 1.\end{equation}
This can be seen as a degenerate example of \eq{PU} where $S^\alpha$ is the identity string field and the shifts of boundary condition all cancel out. This example does not reveal any information about the $\BCFT$s of the source and target solutions. This is expected since the zero morphism takes the same form regardless of what the solutions are. 
\item The morphism $U=K$ is a singular gauge transformation from the perturbative vacuum $\BCFT_0$ to itself, and the boundary condition changing projector,
\begin{equation}\delta(K)=\Omega^\infty,\end{equation}
supports $\BCFT_0$ boundary conditions near the midpoint in both left and right half string functionals. In this case, there is no change of boundary condition at the midpoint since the source and target solutions are the same.
\item Okawa's gauge transformation \eq{Okform} evaluated at $\lambda=1$,
\begin{equation}U_\mathrm{Ok} = 1-\sqrt{\Omega}cB\sqrt{\Omega}=Q_{0\Psi_\mathrm{Sch}}\left(\frac{1-\Omega}{K}B\right),\label{eq:UOk}\end{equation}
is a singular gauge transformation relating the perturbative vacuum to Schnabl's solution. The boundary condition changing projector is
\begin{equation}\delta(U_\mathrm{Ok}) = \sqrt{\Omega}cB\Omega^\infty = 0,\end{equation}
as a consequence of claim \ref{claim:Bsliver}. This should be interpreted as a rank 1 projector where the left half Schr{\"o}dinger functional has tachyon vacuum boundary conditions near the midpoint. But open strings cannot end on the tachyon vacuum, so it is consistent to set all correlation functions with boundaries on the tachyon vacuum equal to zero. This gives a physical explanation for why the phantom term of Schnabl's solution vanishes in the Fock space expansion. Additionally, the implication is that phantom terms do not vanish when relating backgrounds containing D-branes. It is interesting to mention that any gauge transformation between two tachyon vacuum solutions is also a singular gauge transformation. The boundary condition changing projector will vanish for the trivial reason that gauge transformations are invertible. This again can be interpreted as an expression of the fact that correlation functions with boundaries on the tachyon vacuum vanish. It is important to emphasize that the physical interpretation of the boundary condition changing projector assumes it is derived from an exact morphism. Gauge transformations always lead to a vanishing ``boundary condition changing projector," but this is only considered physically significant when the gauge transformation is also an exact morphism. 
\end{itemize}
\noindent Let us compute the boundary condition changing projector in the simplest nontrivial example. Consider two identity-like solutions for marginal deformations:
\begin{equation}\Psi_1 = cV_1,\ \ \ \ \ \Psi_2 = cV_2.\end{equation}
We can find an exact morphism connecting them in the form
\begin{equation}U = Q_{\Psi_1\Psi_2}B = K+ cBV_1 + BcV_2.\end{equation}
This morphism is only resolvable if $V_1$ and $V_2$ have regular OPE with each other. This limitation can be avoided with less identity-like marginal solutions, but presently we want to focus on the simplest example. We first compute  
\begin{eqnarray}
\frac{\eps}{\eps+U} \lineup = \frac{\eps}{\eps + K+ cBV_1 + BcV_2}\nonumber\\
\lineup = \frac{\eps}{\eps + cB(K+V_1)+Bc(K+V_2)}\nonumber\\
\lineup = \left(\frac{\eps}{\eps + Bc(K+V_2)}\right)\left(\frac{\eps}{\eps+cB(K+V_1)}
\right).\label{eq:bccproj1}
\end{eqnarray}
We introduce boundary condition changing operators so that
\begin{equation}\sigma_{01}K\sigma_{10} = K+V_1,\ \ \ \ \sigma_{02}K\sigma_{20} = K+V_2.\end{equation}
Since we consider regular marginal deformations, the boundary condition changing operators satisfy
\begin{eqnarray}\sigma_{01}\sigma_{10} = 1,\lineup \ \ \ \ \sigma_{10}\sigma_{01} = 1,\nonumber\\
\sigma_{02}\sigma_{20} = 1,\lineup \ \ \ \ \sigma_{20}\sigma_{02} = 1.
\end{eqnarray}
In \eq{bccproj1} we also encounter the product
\begin{equation}\sigma_{21} = \sigma_{20}\sigma_{01}.\end{equation}
This product is finite since we assume that $V_1$ and $V_2$ have regular OPE with each other. Then \eq{bccproj1} can be written
\begin{equation}\frac{\eps}{\eps+U} = \sigma_{02}\frac{\eps}{\eps+Bc K}\sigma_{21}\frac{\eps}{\eps + cB K}\sigma_{10}.\end{equation}
We already see the expected structure emerging, but we have to deal with the ghosts. The first factor can be written as $1$ minus a term proportional to $Bc$ and the second factor can be written as $1$ minus a term proportional to $cB$: 
\begin{equation}\frac{\eps}{\eps+U} = \sigma_{02}\left(1-\frac{1}{\eps+K}BcK\right)\sigma_{21}\left(1- cBK \frac{1}{\eps + K}\right)\sigma_{10}.\end{equation}
Multiplying everything out we arrive at 
\begin{eqnarray}
\frac{\eps}{\eps+U}  \, = \,\sigma_{02}\left(\frac{\eps}{\eps+K}\right)Bc\sigma_{20}\, +\,\sigma_{01}cB\left(\frac{\eps}{\eps+K}\right)\sigma_{10}\,+\,\sigma_{02}\left(\frac{\eps}{\eps+K}\sigma_{21} B\d c\frac{1}{\eps+K}\right)\sigma_{10}.\ \ \ \ 
\end{eqnarray}
The first two terms vanish in the $\eps\to 0$ limit because $B$ annihilates the sliver. The third term is nontrivial. Expanding as a superposition of wedge states gives
\begin{equation}
\sigma_{02}\left(\frac{\eps}{\eps+K}\sigma_{21} B\d c\frac{1}{\eps+K}\right)\sigma_{10} = \int_0^\infty dL\, e^{-L}\int_0^{L/\eps} d x \, \sigma_{02}\Big(\Omega^x(\sigma_{21}B\d c) \Omega^{L/\eps-x}\Big)\sigma_{10}.
\end{equation}
In the integrand is a wedge surface of total width $L/\eps$, which for fixed $L$ is approaching the sliver state as $\eps\to 0$. The boundary condition changing operator $\sigma_{21}$ together with $B\d c$ is integrated across the entire width of the wedge surface from $x=0$ to $x=L/\eps$. It is helpful to decompose this integration into three pieces; near the left edge, through the middle, and near the right edge:
\begin{eqnarray}
\int_0^{L/\eps} d x \, \Omega^x (\sigma_{21}B\d c ) \Omega^{L/\eps-x}\lineup = \int_0^{\Lambda} d x \, \Omega^x (\sigma_{21}B\d c ) \Omega^{L/\eps-x}+\int_\Lambda^{L/\eps -\Lambda} d x \, \Omega^x (\sigma_{21}B\d c ) \Omega^{L/\eps-x}\nonumber\\
\lineup\ \ \ \ +\int_0^{\Lambda} d x \, \Omega^{L/\eps-x} (\sigma_{21}B\d c ) \Omega^{x},
\end{eqnarray}
where $\Lambda<L/(2\eps)$ is a parameter (independent of $\eps$) which defines what it means to be ``near" the edges of the wedge surface.  The contributions near the left and right edges vanish in the $\eps\to 0$ limit because $B$ annihilates the sliver, and due to the finite range of integration there can be no compensating divergence. Since $\Lambda$ can be arbitrarily large, we learn that that $\sigma_{21}B\d c$ cannot be a finite distance from the left or right edges of the wedge surface in the $\eps\to 0$ limit. In the conformal frame of the unit disk \eq{(i)}, this implies that the insertion $\sigma_{21}B\d c$ will only give contribution where the midpoint of the local coordinate patch touches the boundary of the unit disk. In this limit the $B\d c$ insertion can be eliminated by evaluating the $bc$ OPE. Doing this carefully provides a normalization factor of unity (see appendix A of \cite{EMsingular} for more detail). In total we find that 
\begin{equation}\delta(U)  = \lim_{\alpha\to\infty}\sigma_{02}\Omega^\alpha\sigma_{21}\Omega^\alpha\sigma_{10}.\end{equation}
This is precisely matches the expected structure of a boundary condition changing projector.

\begin{exercise}
Consider singular gauge transformations from a residual solution to itself within the $KBc$ subalgebra. Show that the boundary condition changing projector either fails to exist or is equal to that of the zero morphism, i.e. the identity string field.
\end{exercise}

One of the initial motivations for the study of singular gauge transformations was as a strategy for constructing new analytic solutions. The idea is  start with a solution $\Psi_1$ (typically the tachyon vacuum) and guess a singular gauge transformation $U$ such that $\delta(U^\ddag U)$ realizes the boundary conditions of the desired background near the midpoint.\footnote{If $U$ is a singular gauge transformation from $\Psi_1$ to $\Psi_2$, and the solutions are real, $U^\ddag U$ will be an exact morphism from $\Psi_2$ to itself. If it is also resolvable, the boundary condition changing projector  $\delta(U^\ddag U)$ is expected to realize the $\BCFT$ of $\Psi_2$ towards the midpoint.} One can construct the solution by writing 
\begin{equation}\Psi_2 =\frac{1}{U+\eps}(Q+\Psi_1)(U+\eps)+\text{phantom term},\label{eq:phantomguess}\end{equation}
and trying to make sense of the phantom term. For a long time this was one of the few ideas for creating nonperturbative analytic solutions besides the tachyon vacuum. But it is unclear how to turn it into a systematic procedure. There was a concrete proposal for $U$ which could create tachyon lumps \cite{Ellsingular}, but attempts to determine a solution through \eq{phantomguess} did not succeed. Instead \eq{phantomguess} leads to a formal solution \cite{BMT} which satisfies the equations of motion up to sliver-like terms~\cite{lumps}. Strangely, the formal solution nevertheless reproduced the correct on-shell action and Ellwood invariant of a tachyon lump. The significance of this development continues to be unclear. 

The formalism has more convincing application as a tool for computation of physical observables. The boundary condition changing projector itself appears to be a kind of observable. However, singular gauge transformations also simplify the computation of the on-shell action and Ellwood invariant. The idea is that a singular gauge transformation allows one to absorb physically irrelevant information into a pure gauge solution, leaving the phantom term to describe the shift in background in the most transparent possible manner. Let us illustrate this by computing the action evaluated on Schnabl's solution. Other applications are described in \cite{EMphantom}. Consider Okawa's singular gauge transformation \eq{UOk} and use \eq{phantom} to write Schnabl's solution as
\begin{eqnarray}\Psi_\text{Sch}\lineup = \frac{1}{\eps+U_\text{Ok}}Q(\eps+U_\text{Ok}) +\frac{\eps}{\eps+U_\text{Ok}}\Psi_\text{Sch}\nonumber\\
\lineup = \lambda \sqrt{\Omega}c\frac{KB}{1-\lambda\Omega}c\sqrt{\Omega}\,+\,\sqrt{\Omega}cB\frac{K}{1-\Omega}\frac{1-\lambda}{1-\lambda\Omega}c\sqrt{\Omega}\nonumber\\
\lineup = \Psi_\lambda + \Delta_\lambda,\phantom{\bigg)}\end{eqnarray}
where $\lambda = \frac{1}{1+\eps}$. The first term is the familiar pure gauge solution in Schnabl gauge, and the second term $\Delta_\lambda$ is what we are calling the phantom term. This is somewhat different from the original phantom term in \eq{Schpsin}, which was derived as a remainder after truncating the geometric series expansion into a discrete sum over wedge states. The difference amounts to a choice of regularization. Presently we are regularizing with the gauge parameter $\lambda$, instead of by truncating the geometric series. If we assume that $\frac{K}{1-\Omega}$ can be represented by its leading term in the $\mathcal{L}^-$ level expansion in the limit $\lambda\to 1^-$, the regularization we are presently using can be represented as 
\begin{equation}
\Psi_\text{Sch} = \lim_{\lambda\to 1^-}\sum_{n=0}^\infty\lambda^n\left((1-\lambda)\psi_n - \lambda\frac{d}{dn}\psi_n\right).
\end{equation}
The advantage of this regularization is that the phantom term satisfies the equations of motion expanded around a pure gauge configuration, and this  is true independent of $\lambda$. The truncated geometric series expansion, however, is not a solution and is therefore not really pure gauge. In particular, it contributes to the energy, which complicates the calculation. The action evaluated on Schnabl's solution  can be written
\begin{eqnarray}
S[\Psi_\text{Sch}] \lineup = S[\Psi_\lambda] + \frac{1}{2}\Tr\Big(\Delta_\lambda (Q_{\Psi_\lambda}\Delta_\lambda)\Big)+\frac{1}{3}\Tr\Big((\Delta_\lambda)^3\Big)\nonumber\\
\lineup = -\frac{1}{6}\Tr\Big((\Delta_\lambda)^3\Big).
\end{eqnarray}
Here we used the equations of motion for the phantom term and dropped the contribution from the pure gauge solution. Substituting the phantom term, the action is 
\begin{equation}
S[\Psi_\text{Sch}]  = -\frac{1}{6}\Tr\left(\left(c\Omega cB\frac{K}{1-\Omega}\frac{1-\lambda}{1-\lambda\Omega}\right)^3\right).
\end{equation}
The action should be independent of $\lambda$, so we can evaluate it in the $\lambda\to 1^-$ limit where some simplifications occur. We perform a scale transformation inside the trace 
\begin{equation}\Tr\Big( ...\Big) = \Tr\Big((1-\lambda)^{\frac{1}{2}\mathcal{L}^-}(...)\Big)\end{equation}
so that the circumference of the cylinder remains finite in the $\lambda\to 1^-$ limit. This gives
\begin{eqnarray}
S[\Psi_\text{Sch}]  \lineup = -\frac{1}{6}\Tr\left(\left(\left[c\Omega^{1-\lambda}cB\frac{K}{1-\Omega^{1-\lambda}}\right]\left[\frac{1-\lambda}{1-\lambda\Omega^{1-\lambda}}\right]\right)^3\right).
\end{eqnarray}
The first factor in square brackets is somewhat tricky to define as a superposition of wedge states, especially in the $\lambda\to 1^-$ limit, since the state $\frac{K}{1-\Omega^{1-\lambda}}$ has infinite $D_2$ norm. The key is to characterize the state in the $\mathcal{L}^-$ level expansion,
\begin{equation}c\Omega^{1-\lambda}cB\frac{K}{1-\Omega^{1-\lambda}} \,= \,-c\d c B\, +\, (1-\lambda)\left(\frac{1}{2}cK^2 c B -\frac{1}{2}c\d c BK\right)\,+\, \mathcal{O}(1-\lambda)^2,\end{equation}
and realize that only the leading term survives in the $\lambda\to 1^-$ limit. The second factor in square brackets can be characterized as a Riemann sum
\begin{eqnarray}
\frac{1-\lambda}{1-\lambda\Omega^{1-\lambda}} \lineup = (1-\lambda)\sum_{n=0}^\infty \lambda^n \Omega^{(1-\lambda)n} \nonumber\\
\lineup = \sum_{n=0}^\infty (\alpha_{n+1}-\alpha_n)\left(1-\frac{\alpha_n}{n}\right)^n \Omega^{\alpha_n},
\end{eqnarray}
where $\alpha_n = (1-\lambda)n$. In the limit the sum converges to an integral:
\begin{eqnarray}
\lim_{\lambda\to 1^-}(1-\lambda)\sum_{n=0}^\infty \lambda^n \Omega^{(1-\lambda)n}\lineup =\int_0^\infty d\alpha e^{-\alpha}\Omega^\alpha 
\nonumber\\
\lineup = \frac{1}{1+K}.
\end{eqnarray}
Substituting these results back into the action gives 
\begin{equation}
S[\Psi_\text{Sch}]  = -\frac{1}{6}\Tr\left(\left(-c\d c \frac{B}{1+K}\right)^3\right).
\end{equation}
This is the cubic term in the action evaluated on the simple tachyon vacuum, as can be seen by expressing it in the form
\begin{equation}
\Psi_\text{simp} = \frac{1}{\sqrt{1+K}}\left(c-c\d c\frac{B}{1+K}\right)\sqrt{1+K}.
\end{equation}
Following the calculation of subsection \ref{subsec:simple}, we reproduce the expected result in accordance with Sen's conjecture. It is interesting how consideration of singular gauge transformations in the context of Schnabl's solution leads to the ``discovery" of the simple tachyon vacuum.

\section{Outlook}

The techniques developed in these lectures have produced analytic solutions representing all D-brane systems in bosonic string theory which have a known BCFT description. They have also given serious evidence that SFT is a fully background independent description of the classical dynamics of D-branes in bosonic string theory. These results are some of the most important achievements of string field theory.

The story for open superstrings is less complete. The most suitable framework to consider is the non-polynomial Wess-Zumino-Witten-like string field theory of Berkovits \cite{Berkovits}. In this setting one has to contend with solutions of greater complexity than in the bosonic string. The problem is the commuting nature of the  $\beta\gamma$  system together with the nonpolynomial structure of the field  equations. This leads to  formal analytic solutions which typically involve an unbounded number of superghost insertions, and we do not really know how to deal with such states. This is not a major problem for marginal deformations \cite{supermarg,Ok_supermarg,Ok_real_supermarg,FK_supermarg,KO_supermarg,NO_supermarg}, since the complexity is controlled by the marginal parameter which can be taken to be parametrically small. It is a significant issue for nonperturbative solutions. Nevertheless, an analytic solution for the tachyon vacuum on non-BPS D-branes and coincident brane/anti-brane pairs has been found \cite{supervac}. Modulo some technical complications \cite{genericF(K)} it should be possible to formulate an analogue of the intertwining solution which connects unstable D-brane systems. However, what is most interesting about the superstring is that D-brane systems are not necessarily unstable. For example, it is believed that string field theory on the D$p$/anti-D$p$-brane system will admit a codimension $2$ solitonic solution representing a BPS D$(p-2)$-brane. This cannot be easily described as an intertwining solution, since it is questionable whether the string field theory on a BPS D-brane admits a tachyon vacuum solution. If it does, it cannot be described as a condensate of the tachyon, and would have to be fundamentally different from the tachyon vacuum solutions discovered so far. For the superstring many basic questions remain unanswered. There is the possibility of major technical and conceptual advances.

Usually, the main reason to study classical solutions is to characterize new vacua.  But in string theory we are in an unusual situation since we already think we know what the vacuum configurations are. They are CFTs. Often the internal consistency conditions of CFTs are a more suitable tool for their elucidation than solving the string field equations. However this is not always the case. Using level truncation in Siegel gauge it is possible to search for boundary states systematically, some of which may represent so-called ``exotic D-branes" which are not defined by linear gluing conditions for the fundamental worldsheet fields, and are not easily described using standard CFT technology. This has been used to construct the so-called ``honeycomb" D-brane on a 2-torus with  hexagonal symmetry \cite{Kudrna}, and, through the folding trick, conformal defects in Virasoro minimal models \cite{Rapcak2}. Analytic methods are far behind level truncation in this direction. It should be possible to use analytic solutions for marginal deformations---the Kiermaier-Okawa solution (subsection  \ref{subsec:singular}), pseudo-Schnabl gauge solutions \cite{Schnabl_marg,pseudoSch}, or possibly even Schnabl gauge solutions (subsection \ref{subsec:SchMarg})---to characterize new backgrounds at least perturbatively. One recent application of this kind, though carried out in Siegel gauge, is turning on the  modulus which dissolves a D-instanton inside a D3-brane in superstring theory~\cite{Vosmera}. See also \cite{MV1,MV2,MV3}. For nonperturbative backgrounds we do not  know how to make a systematic search for analytic solutions. (A minor exception is the $\mathcal{L}^-$ level expansion in a singularity free subalgebra, as discussed in subsection \ref{subsec:L}). If we are given some information about a background to look for, for example a relevant operator which triggers an RG flow to an unknown infrared fixed point, with  sufficient  ingenuity it may be  possible to construct the background as an analytic solution.

If a target background represents a known BCFT we can still ask about the utility of representing it as a SFT solution. A solution establishes that the reference and target backgrounds are connected as part of the same theory. But we should hope to learn something interesting from this connection. One class of questions where the SFT description seems essential is understanding the physics of the tachyon vacuum, since the tachyon vacuum by itself does not have a useful BCFT description. Particularly interesting is understanding D-brane decay and the phenomenon of tachyon matter \cite{Sen_matter}. Some work in this direction appears in \cite{Hellerman,rollingvac}. Another question where it seems necessary to understand the relation between string vacua is explaining the topological origin of D-brane charges.  Some discussion of this appears in~\cite{supervac}.  For the bosonic string, we can ask about magnetic flux on a 2-torus \cite{flux}. What is the string theory analogue of the first Chern class which explains magnetic flux quantization?                    

These lectures have been concerned with the classical dynamics of D-branes. It is natural to ask what happens when we include quantum effects. Quantum effects are difficult to understand due to infrared divergences resulting from the closed string tachyon \cite{Taylor}, and this problem has remained mostly untouched in recent years. See however \cite{Okawa_spec}. One thing we can hope for is an understanding of how closed strings are encoded in the open string field at the quantum level. In this connection it could be worthwhile to consider loop amplitudes in Schnabl gauge \cite{KZloop}, which are simpler than in Siegel gauge. Perturbation theory in Schnabl gauge however is subtle, and its consistency has not been established \cite{RZVeneziano,KSZ}. In the end a satisfactory treatment of quantum effects may require a setting where the closed string tachyon is absent. One may consider, for example, the $c=1$ noncritical string. But ultimately one would like to consider the superstring. So far, however, it seems that quantum effects pose an even greater problem in superstring field theories based on the Witten vertex than they do for the  bosonic string. In the small Hilbert space formulation \cite{WittenSS}, one encounters difficulties from spurious poles; see the fourth lecture of \cite{ClosedSFT_Erler} for more discussion. In the Wess-Zumino-Witten like formulation, this problem seems likely to remain, and we encounter the additional difficulty of giving a tractable definition of the gauge-fixed path integral \cite{WZW_BV}. On the other hand, if we abandon the Witten vertex we are inevitably required to represent closed strings as separate off-shell degrees of freedom. The formulation of closed string field theory connects to some beautiful mathematics, and the formalism  has had application in the treatment of infrared divergences in string perturbation theory \cite{SenErbin}. But the theory is nonpolynomial, and concrete off-shell calculations are usually impractical to carry out beyond cubic order. So far there is not much evidence that closed string field theory contains information about nonperturbative string theory. Understanding closed strings is presently the most important challenge facing string field theory.

\subsubsection*{Acknowledgments}

I would like to thank A. Sen for inviting me to give these lectures and providing ample slots to go through the material. I also thank C. Maccaferri for comments on the typesetted notes. This work is supported by ERDF and M\v{S}MT (Project CoGraDS -CZ.02.1.01/0.0/0.0/15\_ 003/0000437) and the GA{\v C}R project 18-07776S and RVO: 67985840.

\end{document}